\documentclass[3p,times,amssymb,amsmath]{elsarticle}

\usepackage{graphicx,amsfonts,amssymb,amsmath} 
\usepackage{natbib}
\usepackage{textcomp}
\usepackage{esint}
\usepackage{soul}
\usepackage{graphics}
\usepackage{subfigure}
\usepackage{bm}
\usepackage{bbm}
\usepackage[colorlinks,linktocpage,bookmarks=false,citecolor=blue,linkcolor=red,urlcolor=blue]{hyperref}

\usepackage{isomath}

\usepackage{graphicx,amssymb}
\usepackage{color}

\begin{document}

\title{Topological Matter and Fractional Entangled Quantum Geometry through Light}

\author[paris]{Karyn Le Hur}
\address[paris]{CPHT, CNRS, Institut Polytechnique de Paris, Route de Saclay, 91128 Palaiseau, France}
\date{\today}

\begin{abstract}
Here, we reveal our recent progress on a geometrical approach of quantum physics and topological crystals linking with Dirac magnetic monopoles and gauge fields through classical electrodynamics. The Bloch sphere of a quantum spin-$\frac{1}{2}$ particle acquires an integer topological charge in the presence of a radial magnetic field. We show that global topological properties are encoded through the poles of the surface allowing a correspondence between smooth fields, metric, quantum distance and the square of the topological number. The induced information is transported from each pole to the equatorial plane inside a thin Dirac string. We develop the theory, ``quantum topometry" in space and time, and present applications on transport from a Newtonian approach and on a quantized photo-electric effect from circular dichroism of light towards topological band structures in crystals. Edge modes related to topological lattice models are resolved analytically when deforming the sphere or ellipse onto a cylinder. Topological properties of the quantum Hall effect, quantum anomalous Hall effect and quantum spin Hall effect on the honeycomb lattice can be measured locally in the Brillouin zone from light-matter coupling. The formalism allows us to include interaction effects from the momentum space through a simple analytical understanding of the Mott transition and through the elaboration of a variational stochastic approach.  We show that interactions may result in fractional entangled geometry within the curved space measurable in quantum systems.  We elaborate on a relation between entangled wavefunction in quantum mechanics, a way to one-half topological numbers and Majorana fermions which may find applications in quantum information. We show realizations using relevant mathematical methods and quantum field theory associated to quantum transport and light-matter coupling in topological semimetals, proximity effects and bilayer systems, in an assembly of topological planks in a cube related to the Ramanujan infinite series, topological insulators in three dimensions and topological superconductors. We present a link between axion electrodynamics, topological insulators on a surface of a cube and the two-spheres' model via half-Skyrmions.
\end{abstract}

\maketitle
\tableofcontents

\section{Introduction}\label{Introduction}
\label{intro}

\subsection{General Topics}

A charge produces radial electric field lines in accordance with Gauss law. The divergence of the electric field vector is singular at the position of this charge. The magnetic correspondence has attracted a lot of attention with the quest of magnetic monopoles \cite{Curie,Dirac} showing interesting developments \cite{Smooth} related to quantum physics \cite{AharonovBohm,Berry}. Engineering such radial magnetic fields on the surface of a Poincar\' e, Riemann, Bloch sphere for a spin-$\frac{1}{2}$ particle  is now possible in quantum systems allowing to relate the flux produced by the Berry curvature \cite{Berry}, acting as the analogue of the magnetic field, with the first Chern number \cite{Chern} counting the number of monopoles \cite{Roushan,Boulder,Henriet,HH}. A quantized topological number equal to unity can be measured when driving from north to south pole \cite{Roushan,Boulder,Henriet,HH}. These developments show the realization of magnetic monopoles inside atomic, mesoscopic balls with a precise control on the parameters phase space and on the Hilbert space. These topological mesoscopic balls can also find practical applications, e.g. as a quantum dynamo effect, when rolling a spin-1/2 from north to south entangled to a set of harmonic oscillators forming a quantum universe \cite{Henriet,EphraimCyril}. Dirac magnetic monopoles are also realized in Bose-Einstein spinor condensates \cite{Ray} and analogues are found in condensed-matter systems such as spin ices \cite{Moessner,Bramwell}. The topological charge can be interpreted as a winding number on a unit sphere related to Skyrmions physics \cite{Nakahara,ZhangSkyrmions,Skyrme}. The same topological number \cite{Chern} can describe lattice models related to the quantum Hall effect, quantum anomalous Hall effect and Haldane model on the honeycomb lattice in two dimensions \cite{Haldane}. The quantum Hall effect was observed by K. von Klitzing, G. Dorda and M. Pepper \cite{Hall} in MOSFETs associated with theoretical developments \cite{Ando}, a century after the discovery of the Hall effect \cite{HallEdwin}. Such a topological system reveals a bulk-edge correspondence leading to a protected one-dimensional chiral flow around the sample with a quantized conductance $\frac{e^2}{h}$, where $e$ is the charge of an electron and $h$ the Planck constant \cite{Halperin,Buttiker}. Topological properties are measured from the quantum Hall conductivity \cite{Thouless,Simon} related to the dynamics of these edge states \cite{Hatsugai}, and also through a quantized circular dichroism of light \cite{Goldman,Hamburg,Klein}.

Berry curvatures can be measured in ultra-cold atoms through a correspondence with the Bloch sphere \cite{Weitenberg,Hauke}. The Karplus-Luttinger velocity has also engendered important developments since 1954 towards the understanding of anomalous Hall effect in materials \cite{KarplusLuttinger,Luttinger,Nozieres,Nagaosa}. The quantum anomalous Hall effect and Haldane model in two dimensions find applications in materials \cite{Liu}, cold atoms \cite{Jotzu} and light systems \cite{HaldaneRaghu,Joannopoulos,Ozawa,KLHlightnetworks}. Topological phenomena in the presence of interactions can yet reveal other intriguing phenomena such as the presence of fractional charges and statistics as observed in the fractional quantum Hall effect \cite{Stormer,Heiblum,Saminadayar,Kapfer,Bartolomei,TalKaryn} related to Laughlin phases \cite{Laughlin}. The sphere is also appropriate to describe topological aspects of Laughlin states \cite{Haldanesphere}. Geometrical foundations and Berry phase effects then play a key role on the understanding of topological properties of band theory \cite{Niu,Haldanegeometry,RMPColloquium,Book}. 
In this review, we will show that the formalism can define a common language or guiding principle that links classical electromagnetism, quantum physics and topological matter with crystalline structures through our recently developed geometrical approach or topometry \cite{Klein,HH,C2}. Here, we introduce smooth fields in Eq. (\ref{tildeA}) related to the vector potential in classical physics and the Berry connection (i.e. local gauge potential) in quantum mechanics (in Eq. (\ref{A})) allowing us to introduce local markers \cite{C2,OneHalfKLH} of global topological properties from the poles of the sphere down to specific points in the Brillouin zone of the honeycomb lattice i.e. the $K$, $K'$ and $M$ points \cite{graphene,Vozmediano}. The topological characterization at the poles then leads to a re-interpretation similarly as in cartesian coordinates. These vector fields will then unify the different Sections of this manuscript. 

In this task, we will present additional derivations and applications that we will find useful to highlight these findings. 

We present applications of this `topometry' for transport properties in time within a Newtonian approach in curved space related to the Parseval-Plancherel theorem in quantum mechanics \cite{HH} and also for the responses to circular polarizations of light locally in the reciprocal space of the lattice model \cite{Klein,C2}. The formalism reveals a relation between the response to circularly polarized light and the square of the global topological number at the poles of the sphere. We introduce a relation between smooth fields and Fourier series through a dynamical protocol to describe the topological response. This allows us to address the induction of topological protected energetics coming from the transversal kinetic energy as a result of the Dirac monopole. The robustness of the local topological responses persists even when we apply a smooth deformation of a sphere onto an ellipse and onto a cylinder where we can show analytically the presence of edge modes at the boundaries with the top and bottom disks.  Here, we elaborate on a relation between light response, quantum distance and metric \cite{Ryu,BlochMetric}. Related to quantum distance, the metric tensor has been recently measured within a superconducting qubit \cite{Tan}. Possible links between quantum distance, information geometry and quantum phase transitions have been addressed in the recent review \cite{Carollo}. Near the poles the metric is flat and can be interpreted as a vacuum for the gravitational field in the sense of Einstein field equation assuming a pure quantum state. The recent work \cite{BlochMetric} also identifies a possible relation between energy, entropy, Bloch bands and gravitational potential from the reciprocal space in the situation of mixed states.  The (global) topological characterization from the poles of the sphere is now encoded in the definition of smooth fields which imply that information on the topological charge is transported in a thin Dirac cylinder from the equatorial plane to each pole. We draw a parallel between light responses in the quantum Hall effect and in the topological Haldane model on the honeycomb lattice from the Dirac points \cite{C2}. A circularly polarized electric field on a sphere produces a topological phase either at the classical level through a skin effect or at the quantum level showing a way to induce a topological phase on a lattice through the light-matter interaction. Such protocols are commonly applied in cold atoms in optical lattices and topological light systems to implement artificial gauge fields within a Floquet approach \cite{Monika,GoldmanDalibard}. In this way graphene can also turn into a quantum anomalous Hall state through circularly polarized light \cite{McIver,Sato}.

Within this review, we also show that two interacting Bloch spheres  give rise to fractional $\frac{1}{2}$ topological numbers \cite{HH} in the presence of a Einstein-Podolsky-Rosen (EPR) entangled wavefunction or Bell pair \cite{EPR,Bell1964,Hagley,Aspect,Aspect2,Aspect3,Ekert} at one pole. We introduce a stable one-half topological number as the superposition of two geometries on each sphere, a halved surface radiating the flux associated to the Berry curvature and a halved surface participating in the spooky non-local correlations between the two spheres defining the quantum entangled geometry \cite{HH}. We elaborate on the mathematical relation between topological properties, quantum entangled wavefunction at one pole, and present several understandings of this phenomenon and show a correspondence towards Majorana fermions \cite{OneHalfKLH}. This is then similar to have two tori placed one on top of the other; when switching on the interaction, the topological response for each subsystem becomes equivalent to have a reduced (half) surface encircling a hole and the remaining surface participating in the quantum entanglement. This two-spheres' model  can in fact be engineered with existing capabilities in mesoscopic quantum circuits \cite{Roushan}. The $\frac{1}{2}$ topological number can be observed when driving in time from north to south pole \cite{HH}
and also from the response to a circularly polarized field at south pole \cite{OneHalfKLH}. This model links with merons or half Skyrmions occupying half a unit sphere \cite{Alfaro,Meron}. The identification of $\frac{1}{2}$ magnetic monopoles results in other recent developments through thoughts on the Berry phase and magnetic flux in the presence of Cooper pairs \cite{DeguchiFujikawa}. The presence of an edge in real space can similarly produce fractional Skyrmions  in magnetic systems \cite{fractionalSkyrmions,Hirosawa}. We present applications of the two spheres' model and fractional topological numbers for two-dimensional topological semi-metals, graphene and also bilayer honeycomb systems with Bernal stacking \cite{HH,bilayerQSH,Semimetal}. We develop the formalism further for topological coupled-planes models in a cube showing a relation between $\frac{1}{2}$ topological numbers, Ramanujan alternating infinite series and surface states of three-dimensional topological insulators \cite{SekineNomura}. One-half topological numbers of the two-spheres model also finds a correspondence in the physics of two-interacting topological superconducting wires \cite{Herviou,delPozo}. Solving the two-spheres' model at a pole with Majorana fermions, we show that an entangled wavefunction equivalently describes one `free' Majorana fermion on each sphere \cite{OneHalfKLH}. Finally, we elaborate on the smooth fields formalism applied to generalized resonating valence bond states \cite{HH}. It is also interesting to mention related geometrical developments in black holes physics \cite{Gibbons}. Through the presence of EPR pairs, the two-spheres' model may have applications for quantum cryptography and BB84 codes \cite{BennettBrassard}. The presence of different states at the two poles, pure state and EPR pair respectively, may allow for additional questions and tests between Bob and Alice for each pair. 

Topological insulators are phases of matter characterized through a $\mathbb{Z}_2$ topological number as a result of spin-orbit coupling \cite{RMPColloquium,QiZhang,Book}. In two dimensions, topological insulators equivalently refer to quantum spin Hall effect \cite{BernevigZhang,Murakami} and Kane-Mele model on the honeycomb lattice \cite{KaneMele1} where the $\mathbb{Z}_2$ topological order reveals a helical Luttinger liquid at the edges \cite{WuC}. In three dimensions, the system develops metallic surface states. Topological insulators find applications in two and three dimensions starting from Mercury \cite{Konig} and Bismuth materials \cite{HsiehTI,RMPColloquium} respectively. These states of matter have engendered interesting mathematical developments on symmetries related to the Pfaffian and Bloch theorem \cite{KaneMele2,FuKane,MooreBalents}. Through the topometry and a two-spheres' model, we show how we can measure the topological spin Chern number \cite{Sheng} at specific points in the Brillouin zone from the responses to circularly polarized light with a correspondence towards a $\mathbb{Z}_2$ spin pump \cite{C2}. We develop the geometrical approach in topological superconducting wires \cite{Kitaev} and $p+ip$ superconducting state on a square lattice \cite{ReadGreen} to reveal Majorana fermions \cite{Wilczekclass,WilczekMajorana} through the Bogoliubov de Gennes Hamiltonian \cite{Bogoliubov,deGennes,Tinkham} and Bardeen-Cooper-Schrieffer (BCS) theory \cite{Cooper,BardeenCooperSchrieffer}. Various approaches show the stability of topological phases towards weak interaction effects \cite{StephanReview} such as renormalization group methods \cite{KaneMele1}, perturbative, mean-field and gauge theories \cite{PesinBalents,Mott}. At the same time, strong interactions give rise to Mott physics  \cite{PesinBalents,Mott,Varney,WuQSH,Hohenadler,Zheng,Cocks}. Here, we present a way to include interaction effects within the topological (geometrical) description from a recently developed stochastic mean-field variational approach \cite{Klein,QSHstoch} with a simple understanding of the Mott transition line through the ground-state energy or free energy. This approach also allows us to include the light-matter interaction and evaluate the topological marker locally within the Brillouin zone \cite{Klein}. Strong interactions can also reveal other interesting states of matter such as topological Mott phases \cite{PesinBalents},  fractional topological insulators \cite{MaciejkoFiete} and fractionalized quantum spin Hall states. 

All these developments show that the topological characterization of interacting topological systems is an active field of interest, which will be addressed hereafter within this review. 

\subsection{Organization of the Review}

The organization of this review is as follows.

In the preliminaries of Sec. \ref{preliminaries}, we introduce the geometrical approach and the smooth fields in Eq. (\ref{tildeA}) from classical electrodynamics. We also show how a time-dependent circularly polarized electric field can induce a Dirac magnetic monopole. 

In Sec. \ref{quantumphysics}, we introduce geometrical aspects on the Bloch sphere of a spin-$\frac{1}{2}$ particle and show their relevance for the description of global topological properties locally from the poles of the sphere. We describe transport properties and energetics aspects, the correspondence with Fourier series and the responses to circularly polarized light from these smooth fields showing a topologically quantized photo-electric effect \cite{C2}. We show how circularly polarized light produces a topological Bloch sphere. By slightly transforming the sphere into a cylinder geometry reveals the occurrence of edge modes. The stability towards a dissipative environment (set of harmonic oscillators) is addressed related to quantum phase transitions, and a sphere coupled to  this ensemble of oscillators can also be useful as a quantum dynamo \cite{Henriet, EphraimCyril}. Then, we introduce the fractional entangled geometry \cite{HH} leading to $\frac{1}{2}$ topological numbers and the mesoscopic circuit implementation. We show a local relation between entangled wavefunction and fractional topological number in Eq. (\ref{topoquant}) \cite{OneHalfKLH}. The stability of the formalism towards a deformation of the surface is also discussed. 

In Sec. \ref{anomalous}, we apply the sphere approach to the Haldane model on the honeycomb lattice referring to quantum anomalous Hall effect discussing interaction and disorder effects with the stochastic variational approach \cite{Klein}. 

In Sec. \ref{Observables}, we show the link between smooth fields, quantum Hall conductivity, the responses to circularly polarized light and the ${\cal I}(\theta)$ function \cite{C2}, related to the quantum metric \cite{Ryu}, which reveals the square of the topological invariant at the poles of the spheres and equivalently at the Dirac points on the lattice. The light response at the $M$ point is also quantized in units of half of the (integer) topological invariant from symmetries applied to the tight-binding model. For a Haldane circuit quantum electrodynamics model on the honeycomb lattice, topological properties can be measured via the coupling to a local probe or long transmission line transporting the microwave AC signal \cite{JulianLight}. Then, we address the parallel with the light response in the quantum Hall regime of graphene \cite{Zhang,Novoselov} commenting on recent developments related to fractional quantum Hall physics. The formalism reveals the topological transition induced by a Semenoff mass \cite{Semenoff}. Related to the developed methodology, we describe the photovoltaic effect in graphene \cite{OkaAoki1,OkaAoki2,MoessnerCayssol} and the photogalvanic effect in Weyl semimetals in three dimensions \cite{Juan,Orenstein}. 

In Sec. \ref{quantumspinHall}, we develop the formalism towards the quantum spin Hall effect into two-dimensional topological insulators with $\mathbb{Z}_2$ symmetry showing a relation between the ${\cal I}(\theta)$ function through the light responses \cite{C2}, the topological spin Chern number \cite{Sheng} and the zeros of the Pfaffian as formulated by Kane and Mele \cite{KaneMele2}. The cylinder geometry establishes a correspondence to a spin pump. We also address interaction effects towards the Mott transition \cite{Mott,WuQSH,QSHstoch,Plekhanov}. 

In Sec. \ref{proximityeffect}, we present a topological proximity effect in bilayer systems between graphene and a topological material \cite{bilayerQSH}. Then, we elaborate on the correspondence between the half-topological number per sphere and a topological semimetal with applications in bilayer honeycomb systems \cite{HH} and also in one layer Fermi-liquid graphene model \cite{Semimetal}. We show that the smooth fields on the sphere at the poles can be measured at the Dirac points in the band theory.

In Sec. \ref{Planks}, we introduce a topological coupled-planes model in a cube and show a correspondence between the $\frac{1}{2}$ topological number and the Ramanujan infinite alternating series with applications in transport and circularly polarized light. This model also allows us to show a correspondence with the quantum Hall effect developing on the surface of three-dimensional topological insulators and axion electrodynamics \cite{Wilczek,QiZhang}. 

In Sec. \ref{further}, we develop the formalism for topological superconducting systems with a relation between the Nambu representation, the Anderson' pseudospin \cite{Anderson} and the Bloch sphere \cite{SatoAndo}. We propose an analogue of the ${\cal I}(\theta)$ function \cite{C2} and a protocol implementation related to the $\mathbb{Z}_2$ symmetry associated to the BCS theory of the Kitaev superconducting wire \cite{Kitaev}. We develop a Majorana fermions representation of the one-half topological number for two spheres and show that the $\frac{1}{2}$ topological number can also be defined as a gapless Majorana fermion at zero energy at one pole (on each sphere) \cite{OneHalfKLH}. We address a correspondence with the physics of superconducting wires \cite{Herviou,delPozo}. Through the topological $p+ip$ superconductor on the square lattice, we show a link between smooth fields and a topological marker introduced by Wang and Zhang through the Green's function at zero-frequency \cite{Wang}. 

In Sec. \ref{GRVBT}, we develop the geometrical approach and the formalism of smooth fields for small spin arrays showing the possibility of other fractional topological numbers \cite{HH}. 

In Sec. \ref{Summary}, we summarize the main findings related to this research.

In \ref{Berrycurvature}, we  present additional derivations showing a link between the Berry formalism, quantum distance, metric and the functions $\alpha(\theta)$ in Eq. (\ref{alpha}) and ${\cal I}(\theta)$ in Eq. (\ref{Itheta}) which are measurable 
through responses to circularly polarized fields on the Bloch sphere and on the lattice respectively \cite{C2}. 

In \ref{LaughlinQHResponse}, we introduce the Laughlin wavefunction related to geometry and Berry formalism and address a comparison with the situation of fractional topological state for the two-spheres' model.

In \ref{TopometryBandTheory}, we introduce the topometry formalism associated to energy bands of the topological lattice model. We show two approaches to measure locally the topological number for an energy band. 

In \ref{lightconductivity}, we develop the correspondence between light responses and photo-induced currents \cite{Klein}. 

In \ref{Halldrift}, we present a simple derivation of the quantized transport from the drift velocity related to light response in the quantum Hall regime of graphene.

In \ref{timereversal}, we introduce time-reversal symmetry from quantum mechanics to topological insulators. 

In \ref{GeometryCube}, we develop the geometry related to the Green and divergence theorem in the thermodynamical limit 
for an assembly of topological planks. 

In \ref{interactions}, we describe topological superconducting phase stability for one wire regarding interactions within the Luttinger liquid formalism \cite{HaldaneLuttinger}.

\section{Preliminaries}
\label{preliminaries}

\subsection{Magnetism and Vector Potential}
\label{potential}

Here, we discuss the vector potential solution(s) from classical electromagnetism in the presence of a radial magnetic field ${\bf B}=\bm{\nabla}\times{\bf A}=B{\bf e}_r$ with ${\bf e}_r$ being the radial unit vector.  We study the surface of a sphere with a fixed radius $r$ such that $B(r)=B$. These radial magnetic fields can be produced by a Dirac magnetic monopole at the origin via the equation $\bm{\nabla}\cdot{\bf B}=q_m\delta({\bf r})$ \cite{Dirac}.
Although this step may be perceived as a simple calculation related to electromagnetism, we find it useful to  precisely introduce Eq. (\ref{tildeA}) through vector fields $\tilde{\bf A}$ smoothly defined on the whole surface (that we have not seen introduced before). 
An analogous situation in quantum physics will be produced on the Bloch sphere of a spin-$\frac{1}{2}$ particle related to topological lattice models where the Dirac monopole will describe the topological charge associated to the model. 
Therefore, this Section aims at defining the topological response from the poles in a manner analogous to flat space. The smooth fields $\tilde{\bf A}$, defined on the whole surface, will form the foundations of the mathematical language together with the fields ${\bf A}'$ associated to Stokes' theorem. The form of the potential vector will link with the smooth fields and with topological properties of the system including quantum physics, the Berry connection \cite{Berry} is analogous to a momentum in quantum mechanics and therefore should have similar properties from symmetries as a classical vector potential. The discussion below can be adapted for an electric field produced by a charge $q_e$ at the origin introducing the vector ${\bf A}_e$ such that ${\bf E}=\bm{\nabla}\times{\bf A}_e=E{\bf e}_r$. 

In spherical coordinates, the radial component of the `nabla' (del) operator leads to the equation
\begin{equation}
\frac{1}{r\sin\theta}\left(\frac{\partial}{\partial\theta}(A_{\varphi}\sin\theta) - \frac{\partial A_{\theta}}{\partial_{\varphi}}\right) = B.
\end{equation}
Here, $\theta$ refers to the polar angle and $\varphi$ to the azimuthal angle. 
A solution of this equation in agreement with the Stokes theorem and the geometry admits the form $\frac{\partial A_{\theta}}{\partial\varphi}=0$; we can also set $A_{\theta}=0$ for this solution. Therefore, this leads to 
\begin{equation}
\label{differential}
\frac{\partial}{\partial\theta}(A_{\varphi}\sin\theta)=Br\sin\theta.
\end{equation}
To solve this equation, we can redefine 
\begin{equation}
\label{smoothclassical}
{A}'_{\varphi} = A_{\varphi}\sin\theta.
\end{equation}
From the geometrical foundations the ${\bf A}$ field (and ${\bf A}'$) should be a smooth, continuous function at each pole where we also have the prerequisite $A'_{\varphi}=0$.

Then, Eq. (\ref{differential}) simplifies to
\begin{equation}
\label{diff}
\partial {A}'_{\varphi} = -Br \partial (\cos\theta).
\end{equation}
Then, if we integrate this equation continuously from an angle $\theta=0$ up to $\theta$ then the solution at an angle $\theta=\pi$ would not satisfy ${A}'_{\varphi}=0$. Therefore, it is necessary at least to introduce two regions joining at an angle $\theta_c$; see the discussion in the book of Nakahara \cite{Nakahara}. We can integrate this equation on the north region (with a polar angle $\theta$ from $0$ to $\theta_c$) and on the south region (with a polar angle $\theta$ from $\theta_c$ to $\pi$). This gives rise to the equations \cite{Smooth}
\begin{eqnarray}
\label{vecpot}
{A}'_{\varphi}(\theta<\theta_c) &=& -Br(\cos\theta-1) = 2Br\sin^2\frac{\theta}{2} \\ \nonumber
{A}'_{\varphi}(\theta>\theta_c) &=& -Br(\cos\theta+1) = -2Br\cos^2\frac{\theta}{2}.
\end{eqnarray}
The symbols $\theta<\theta_c$ and $\theta>\theta_c$ in $A'$ can be equivalently understood as $\theta=\theta_c^-$ and $\theta=\theta_c^+$.
The field $A'$ is smoothly defined inside each region and it shows a discontinuity (jump) at the interface related to applicability of Stokes' theorem with a charge in the core of the surface.
Since $A'$ is defined as the difference of two potentials at different points it is defined in a gauge invariant way. Eqs. (\ref{vecpot}) are related to topological properties, the Aharonov-Bohm effect and Berry phase \cite{Smooth}. 
This way, we define solutions such that ${A}'_{\varphi}=0$ at the two poles. These fields $A'$ satisfy
\begin{equation}
\label{Acircles}
{A}'_{\varphi}(\theta<\theta_c) - {A}'_{\varphi}(\theta>\theta_c)=2Br.
\end{equation}

To describe topological properties of the surface from the poles we find it useful to introduce the field 
\begin{equation}
\label{tildeA}
\tilde{A}_{\varphi}(\theta)=-Br\cos\theta
\end{equation}
which has the property to be smooth on the whole surface. This leads to
\begin{eqnarray}
\label{tildefields}
{A}'_{\varphi}(\theta<\theta_c) &=& \tilde{A}_{\varphi}(\theta) -\tilde{A}_{\varphi}(0) \\ \nonumber
{A}'_{\varphi}(\theta>\theta_c) &=& \tilde{A}_{\varphi}(\theta) -\tilde{A}_{\varphi}(\pi),
\end{eqnarray}
such that
\begin{equation}
\label{polesA}
{A}'_{\varphi}(\theta<\theta_c) - {A}'_{\varphi}(\theta>\theta_c) = \tilde{A}_{\varphi}(\pi) - \tilde{A}_{\varphi}(0).
\end{equation}
From Eq. (\ref{Acircles}), the last identity measured from the poles is fixed to $2Br$. We emphasize that the $\tilde{A}_{\varphi}(\theta)$ function introduced in Eq. (\ref{tildefields}) is defined smoothly on the whole surface. This is an important property that we will precisely introduce for geometrical proofs related to quantum physics, lattice models and physical observables. This will play the role of the Berry connection field or gauge potential in the next Sections. This will also allow us to introduce a Dirac string or a thin cylinder in the topological quantum formalism and to show that topological properties can be revealed from the poles (only).

For our purposes, it is useful to introduce the Gauss law on the surface of the sphere $(S^2)$ through the smooth fields ${A}'_{\varphi}$ which will then allow for a simple re-formulation in a flat metric. On the one hand with the measure of an area in spherical coordinates 
$r^2\sin\theta d\theta d\varphi$, we have
\begin{equation}
\label{Phiclassical}
\Phi = Br^2\left(\int_0^{2\pi} d\varphi\right)\left(\int_0^{\pi} \sin\theta d\theta \right) = 4\pi r^2 B(r).
\end{equation}
From $\bm{\nabla}\cdot{\bf B}=0$ for $r\neq 0$ then we verify that $B(r)(4\pi r^2)$ is independent of the radius such that we can re-interpret $B=\frac{q_m}{2 r^2}=B(r)$ with the encircled magnetic charge $q_m=\frac{\Phi}{2\pi}$.
On the other hand, it is also useful to re-interpret Eq. (\ref{Phiclassical}) in flat space $(\theta,\varphi)$ with the area measure $r^2d\theta d\varphi$ through the introduction of $F_{\theta\varphi}=\frac{1}{r}\partial_{\theta}{A}'_{\varphi} = B\sin\theta$.
For a sphere with a radius $r$, then we identify
\begin{equation}
\label{Fclassical}
\Phi = r^2\left(\int_0^{2\pi} d\varphi \right)\left(\int_0^{\pi} F_{\theta\varphi} d\theta \right) = B(r)(4\pi r^2).
\end{equation}
This is equivalent to 
\begin{equation}
\label{Phi}
\Phi = 2\pi r^2\left(\int_0^{\theta_c^-} F_{\theta\varphi} d\theta + \int_{\theta_c^+}^{\pi} F_{\theta\varphi} d\theta \right).
\end{equation}
Therefore,
\begin{equation}
\label{Phi'}
\Phi = 2\pi r\left({A}'_{\varphi}(\theta_c^-) - {A}'_{\varphi}(0) + {A}'_{\varphi}(\pi) - {A}'_{\varphi}(\theta_c^+)\right). 
\end{equation}
The $A'$ fields are defined to be zero at the two poles, such that
\begin{equation}
\label{circles}
\Phi = 2\pi r\left({A}'_{\varphi}(\theta_c^-) - {A}'_{\varphi}(\theta_c^+)\right) = 2\pi r(\tilde{A}_{\varphi}(\pi) - \tilde{A}_{\varphi}(0)).
\end{equation}
The last identity measures the magnetic flux from the poles of the sphere. The last step in this equation reveals that topological properties can be measured from the poles (only) with a smooth field defined on the whole surface. This proof can be generalized to multiple topological spheres from geometry as in Sec. \ref{smooth}. 

This equation can also be re-written as
\begin{equation}
\label{tilde}
\Phi = \int_0^{\pi}\int_0^{2\pi} \tilde{F}_{\theta\varphi} d\theta d\varphi
\end{equation}
with $\tilde{F}_{\theta\varphi}=r\partial_{\theta}\tilde{A}_{\varphi}$, which will play a key role in our analysis hereafter.
We can also interpret Eq. (\ref{Fclassical}) as 
\begin{equation}
\label{curl}
\Phi = \iint_{S^2} {\bf F} \cdot d^2{\bf s}
\end{equation}
with $d^2{\bf s}=d\varphi d\theta {\bf e}_r$ and ${\bf F}=\bm{\nabla}\times {\bf A}'$ where $\bm{\nabla}=\left(\partial_{\theta},\partial_{\varphi}\right)$. As mentioned above, through the application of Gauss theorem, the double integral should be thought of as $\oiint=\iint_{S^2}$. In that sense, Eq. (\ref{circles}) can then be viewed as a consequence of the Stokes' theorem on each hemisphere where ${A}'_{\varphi}$ is independent of $\varphi$ (and $A'_{\theta}=0$ or $\partial_{\varphi} A'_{\theta}=0$). We underline here that Eq. (\ref{smoothclassical}) allows a re-interpretation of the curved space into a flat metric similarly as in cartesian coordinates. In spherical coordinates, we have the geodesic defined through $dl^2 = dx^2+dy^2+dz^2 = r^2 d\theta^2 + r^2\sin^2\theta d\varphi^2$. This is equivalent to define the (differential) vector
\begin{equation}
\label{geodesic}
d{\bf l} = r d\theta {\bf e}_{\theta} + r\sin\theta d\varphi {\bf e}_{\varphi},
\end{equation}
with ${\bf e}_{\theta}$ and ${\bf e}_{\varphi}$ being unit vectors along the polar and azimuthal angles directions. Therefore, for a sphere of radius unity, we have 
\begin{equation}
{\bf A}\cdot d{\bf l} = A_{\theta}d\theta + {A}'_{\varphi} d\varphi,
\end{equation}
which is another way to visualize the first equality in Eq. (\ref{circles}) through circles parallel to the equator.

As a remark, we observe that we also have
\begin{eqnarray}
A_{\varphi}(\theta<\theta_c) = Br\tan\frac{\theta}{2}
\end{eqnarray} 
and
\begin{eqnarray}
A_{\varphi}(\theta>\theta_c)=-\frac{Br}{\tan\frac{\theta}{2}}.
\end{eqnarray}
Then, we verify that these functions go to zero at the two poles. 

We show below that the fields ${A}'_{\varphi}$ and $\tilde{A}_{\varphi}$ in Eqs. (\ref{circles}) are interesting to understand physical properties of quantum and topological lattice models.

\subsection{Radial Magnetic Field from Time-Dependent Electric Field and Skin Effect}
\label{electricfield}

In my quest, I realize that a radial magnetic field can be produced in the vicinity of a metallic surface on a sphere through a time-dependent electric field, as a skin effect. 

With circular polarizations, waves will induce a time-dependent momentum boost for a charge on the surface parallel to the azimuthal angle. Here, the analysis of the charged particle and of the electromagnetism is purely classical assuming that the sphere is sufficiently macroscopic. This Section then shows that a `mean' (time-averaged) magnetic flux can be produced on the surface of the sphere through a Floquet perturbation periodic in time already at a classical level. These waves will also intervene in the quantum situation with a Bloch sphere when coupling an electric dipole to circularly polarized light or a magnetic spin-$\frac{1}{2}$ to a rotating magnetic field.

We introduce a circularly polarized wave propagating in $z$ direction from the point of view of Arago and Fresnel. We describe here the skin effect produced by an in-coming electromagnetic wave on a particle with charge $e$, mass $m$ and momentum ${\bf p}=m{\bf v}$. For simplicity, we study the response in the equatorial plane of the sphere where the interaction between matter and the electromagnetic wave is the most prominent meaning that this grows as the perimeter of the sphere related to the typical number of charged particles on the surface. From polar coordinates $(r,\varphi)$ in the equatorial plane defined through $z=0$, in the case of a right-moving (+)  or left-moving (-) wave moving with the rotating frame, this corresponds to set $\varphi=\mp \omega t$ respectively. 

A vector field associated to a right-moving and left-moving wave will be denoted ${\bf V}_{\pm}$ such that $+$
rotates clockwise (+) and $-$ anticlockwise (-). Then, we have the correspondence of vectors $-i e^{-i\omega t}({\bf e}_x \mp i {\bf e}_y) = \mp {\bf e}_{\varphi} - i {\bf e}_r$ for a right-moving (+) and left-moving wave (-) respectively. Here, ${\bf e}_x$ and ${\bf e}_y$ refer to unit vectors in cartesian coordinates in the plane whereas ${\bf e}_{\varphi}$ and ${\bf e}_{r}$ are unit vectors in polar coordinates in the equatorial plane.

A vector potential ${\bf A}_{\pm}(t)=A_0e^{-i\omega t}({\bf e}_x\mp i {\bf e}_y)$ for $z=0$ then corresponds to a static electric field in the moving frame ${\bf E}_{\pm}=-i E_0e^{-i\omega t}({\bf e}_x \mp i {\bf e}_y) = E_0(\mp {\bf e}_{\varphi} - i {\bf e}_r)$ with a real component along the azimuthal angle $\mp E_0 {\bf e}_{\varphi}$. Here, we suppose that on the surface of the sphere we have a time-dependent electric field 
\begin{equation}
{\bf E}(t)=E_0 e^{-i\omega t} e^{i k z}{\bf e}_{\varphi}
\end{equation}
where $E_0$ is fixed, corresponding to an $AC$ perturbation in time in the rotating frame. Within this Floquet perturbation we will then derive an effective real radial magnetic field from an average on a time period. We will analyse the response of a charge $e$ particle in the equatorial plane with $z\rightarrow 0$ in cylindrical coordinates $(r,\varphi,z)$ and the polar angle also corresponds to the azimuthal angle in spherical coordinates. In this specific application, in the Maxwell equations, the $\bm{\nabla}$ operator will be then developed in the same coordinates system. The modulated electric field in $z$ direction ${\bf E}(t)$ satisfies the one-dimensional d'Alembert equation along $z$ direction. 

The Newton's equation on a charge $e$ with $r=1$ on the surface of the sphere then results in
\begin{equation}
\dot{p}_{\varphi}(t) = eE_0 e^{-i\omega t} e^{i k z}.
\end{equation}
Integrating this equation
\begin{equation}
p_{\varphi}(t) = \frac{eE_0}{(-i\omega)}\left( e^{-i\omega t} -1 \right) e^{ikz}.
\end{equation}
By developing the response with $z\rightarrow 0$, then $p_{\varphi}(t)$ acquires both a real and imaginary parts 
\begin{equation}
\hbox{Re} p_{\varphi}(t) = \frac{e E_0}{\omega}\sin(\omega t)
\end{equation}
\begin{equation}
\hbox{Im} p_{\varphi}(t) = -\frac{e E_0}{\omega} 2\sin^2\frac{\omega t}{2},
\end{equation}
traducing an oscillating response. The real part produces a sinusoidal current associated to a charge $e$. The $\hbox{Im} p_{\varphi}(t)$ component is important to induce a real radial magnetic field when averaging on a Floquet time period. Since we introduce an incident wave propagating in $z$ direction, we can equally work in cylindrical coordinates $(r,\varphi,z)$. 

From Maxwell-Faraday equation $\bm{\nabla}\times {\bf E} = -\frac{\partial {\bf B}}{\partial t}$, we obtain an induced radial magnetic field on the surface 
\begin{equation}
B_r(t) = + \frac{ik}{e} p_{\varphi}(t).
\end{equation}
At time $t=0$, we fix $B_r=0$ to study the magnetic response in the presence of the light-matter interaction related to $p_{\varphi}(t)$ for $t>0$. Keeping the factor $e^{ikz}$, we identify a relation between $\bm{\nabla}\times{\bf B}$ and the moving particle 
\begin{equation}
\bm{\nabla}\times{\bm B} = -\frac{k^2}{e}p_{\varphi} {\bm e}_{\varphi}.
\end{equation}
Inserting the form of $p_{\varphi}(t)$, in the presence of the light-matter interaction then we can interpret Amp\` ere's law as:
\begin{equation}
\bm{\nabla}\times{\bf B} = \frac{1}{c^2}\frac{\partial {\bf E}}{\partial t} + \mu_0 \bar{\bf{J}}
\end{equation}
with 
\begin{equation}
\mu_0\bar{\bf{J}} = \frac{ik}{c}E_0 e^{ikz}{\bf {e}}_{\varphi}.
\end{equation}
This corresponds to a current contribution per unit area. Here, $c$ refers to the speed of light. 
The real part of $B$ then evolves as
\begin{equation}
\hbox{Re} B_r(t) = -\frac{k}{e}\hbox{Im} p_{\varphi}(t) = \frac{2 E_0 k}{\omega}\sin^2\frac{\omega t}{2}.
\end{equation}
The electric field is periodic in time with period $T=\frac{2\pi}{\omega}$. We define an effective averaged magnetic field on this time period
\begin{equation}
\bar{B}_r=\frac{1}{T}\int_0^T \hbox{Re} B_r(t) dt = \frac{E_0 k}{\omega} = \frac{E_0}{c}.
\end{equation}
In this way, we produce classically an effective magnetic flux on the surface of the sphere. 
The mean energy stored in this magnetic field $\frac{1}{2}\mu_0^{-1} \bar{B}_r \bar{B}_r = \frac{1}{2}\epsilon_0 |E_0|^2$ is then the symmetric entity of the electric energy density $\frac{1}{2}\epsilon_0|{\bf E}(t)|^2=\frac{1}{2}\epsilon_0 |E_0|^2$.
The energy in fluctuations related to $B_r B_r^*$ is associated to the mean kinetic energy of the particle(s). 

In addition to the radial magnetic field, the Maxwell-Faraday equation in cylindrical coordinates also produces a magnetic field along $z$ direction. Introducing the polar angle $z=\cos\theta$ on the unit sphere, we have
\begin{equation}
B_z = \frac{E_0}{i\omega}(e^{-i\omega t} - 1)e^{ik\cos\theta}.
\end{equation}
From the equatorial plane, then this gives rise to 
\begin{equation}
\hbox{Re} B_z = -\frac{E_0}{\omega}\sin(\omega t)-\frac{2kE_0}{\omega}\sin^2\frac{\omega t}{2}\cos\theta,
\end{equation}
producing a finite value
\begin{equation}
\bar{B}_z = -\frac{E_0}{c}\cos\theta
\end{equation}
if we average the response on a time period related to a loop of current in the equatorial plane. The averaged (sinusoidal) current on the loop is zero, but we identify a finite real part of the magnetic field along $z$ direction through the wave propagation factor $e^{i kz}=e^{ik\cos\theta}$ developed
from the equatorial plane (where the number of particles is related to the perimeter of the sphere). Since the measured magnetic flux is real, we can neglect the effect
of the imaginary component $\hbox{Im}B_z=\frac{2E_0}{\omega}\sin^2\frac{\omega t}{2}$. 

Rotating the sphere such that south pole becomes north pole and north pole becomes south pole, through the modification $\theta\rightarrow \theta+\pi$, we identify the magnetic field 
\begin{equation}
\bar{\bf B} = \hbox{Re}{\bf B} = \frac{E_0}{c}(\sin\theta\cos\varphi, \sin\theta\sin\varphi,\cos\theta),
\end{equation}
which precisely corresponds to a radial magnetic field on a unit sphere with a positive amplitude. This radial magnetic field then produces a flux on the surface according to Eq. (\ref{Fclassical}). It is in agreement with the form $\bar{\bf B}=\bar{B}{\bf e}_r$ with here
$\bar{B}=\frac{E_0}{c}$ as if we have fixed the radius to unity and $q_m=\frac{2E_0}{c}$. For a surface close to or at $r=1$, then from the calculation in cylindrical coordinates we can verify that $\bm{\nabla}\cdot \bar{{\bf B}}=0$ on the surface.

\section{Geometry and Topological Aspects in the Quantum}
\label{quantumphysics}

In this Section, we develop the formalism for quantum physics following our recent articles \cite{HH,C2,OneHalfKLH}. We show applications of this geometrical approach for quantum physics, referring to quantum topometry, with applications on quantum transport
and light-matter coupling. We introduce a relation between interacting Bloch spheres, fractional topological numbers and entangled wavefunctions in quantum mechanics \cite{HH,OneHalfKLH}.
We present additional relevant derivations in Secs. \ref{DynamicsSeries}, \ref{Geometry} and introduce a practical mesoscopic circuit in Sec. \ref{Mesoscopic} to realize the model of interacting Bloch spheres.

\subsection{Topological Spin-$\frac{1}{2}$}
\label{spin1/2}

In the presence of a time-dependent Hamiltonian, a plane wave or wavepacket may acquire a Berry phase in addition to the dynamical phase
\begin{equation}
\gamma_n =i\oint \langle \psi_n| \bm{\nabla}|\psi_n\rangle d{\bf R},
\end{equation}
with $n$ referring to an eigenstate and ${\bf R}$ parametrizing a surface as in \ref{Berrycurvature}.

Here, we introduce ${A}_{\varphi}$, the Berry connection or local gauge potential \cite{Berry} defined as 
\begin{equation}
\label{A}
A_{\varphi}=-i\langle \psi| \frac{\partial}{\partial\varphi} |\psi\rangle,
\end{equation}
related to a Berry phase $-\oint A_{\varphi} d\varphi$ when navigating in a closed loop in the equatorial plane on a unit sphere parametrized through the polar angle $\theta$ and azimuthal angle $\varphi$ (The $-$ sign is to have a precise link with the momentum or wave-vector when $\varphi$ refers to a position variable). 
The goal is to show that $A_{\varphi}$ has an interpretation of (local) topological marker associated to geometrical properties of a quantum system such as a spin-1/2.
Since $A_{\varphi}$ precisely corresponds to the smooth field $\tilde{A}_{\varphi}$ of the preceding section (related to the classical vector potential and $A'_{\varphi}$ functions) in Eq. (\ref{tilde}) this allows for a re-interpretation of the global topological number similarly as in flat space close to the poles \cite{HH,C2}. The Berry connection \cite{Berry} in (\ref{A}) is then introduced similarly as in flat space. This can be simply understood from the correspondence between spherical and cartesian coordinates $A_{\varphi}^s = -i\langle \psi| \frac{1}{\sin\theta}\frac{\partial}{\partial\varphi} |\psi\rangle = \frac{1}{\sin\theta}A_{\varphi}$. The Berry curvature acting as the magnetic field in classical physics satisfies the following form in spherical coordinates $F_{\theta\varphi}=\frac{1}{\sin\theta}
\frac{\partial}{\partial\theta}(\sin\theta A_{\varphi}^s) = \frac{1}{\sin\theta}\frac{\partial}{\partial\theta} A_{\varphi}$. This way, the flux integration of ${\bf F}$ on the two-dimensional curved surface is simply 
\begin{equation}
C = \frac{1}{2\pi}\oiint \frac{\partial}{\partial\theta} A_{\varphi}d\theta d\varphi.
\end{equation}
This equation is then identical as if we re-define $F_{\theta\varphi}=\partial_{\theta}A_{\varphi}$ with $d^2{\bf s}=d\theta d\varphi {\bf e}_r$, as we implicitly assume hereafter.
For another discussion on equivalence between cartesian versus spherical coordinates' representations, see Supplementary Information of Ref. \cite{Roushan}.  In the following, the $\bm{\nabla}$ operator will be then developed similarly as in cartesian space, $\bm{\nabla}=(\partial_{\theta},\partial_{\varphi})$. Since the vector potential plays a similar role as the momentum, here we show a simple correspondence between the classical and quantum formalisms on the sphere. The sphere here refers to the Bloch sphere of a spin-$\frac{1}{2}$ particle. It is interesting to anticipate that in the case where $A_{\varphi}$ would not depend on $\varphi$ then $C$ acquires a simple form $A_{\varphi}(\pi)-A_{\varphi}(0)$.

Suppose we start with a radial magnetic field in quantum mechanics such that the Hamiltonian reads 
\begin{equation}
{H}=-{\bf d}\cdot\mathbfit{\sigma}
\end{equation}
 with
\begin{equation}
\label{dvector}
{\bf d}(\varphi,\theta) = d(\cos\varphi\sin\theta,\sin\varphi\sin\theta,\cos\theta) = (d_x,d_y,d_z).
\end{equation}
The spin response will form an hedgehog structure. 
We define the Hilbert space with $\{|+\rangle_z; |-\rangle_z\}$ corresponding to the eigenstates in $z$ direction and we re-write these states in terms of two-dimensional orthogonal unit vectors. In this way, the eigenstates take the simple form 
\begin{equation}
\label{eigenstates}
|\psi_+\rangle =
\left(
\begin{array}{lcl}
\cos\frac{\theta}{2}e^{-i\frac{\varphi}{2}} \\
\sin\frac{\theta}{2} e^{i\frac{\varphi}{2}}
\end{array}
\right),
\hskip 0.25cm
|\psi_-\rangle =
\left(
\begin{array}{lcl}
-\sin\frac{\theta}{2} e^{-i\frac{\varphi}{2}} \\
\cos\frac{\theta}{2} e^{i\frac{\varphi}{2}}
\end{array}
\right).
\end{equation}
The eigenenergies are respectively $-|{\bf d}|$ and $+|{\bf d}|$ for the eigenstates $|\psi_+\rangle$ and $|\psi_-\rangle$.
Here, we represent the wave functions in a specific gauge $\varphi$-representation. The topological responses will be formulated from the geometry in a gauge-invariant way.
We study the topological response related to the lowest energy eigenstate $|\psi_+\rangle$. A similar calculation can be reproduced for the other eigenstate such that the two energy eigenstates will be characterized through opposite topological numbers (this can be visualized from the fact that the two eigenstates are related through the transformation $\theta\rightarrow \theta+\pi$).

\begin{figure}[t]
\begin{center}
\includegraphics[width=0.7\textwidth]{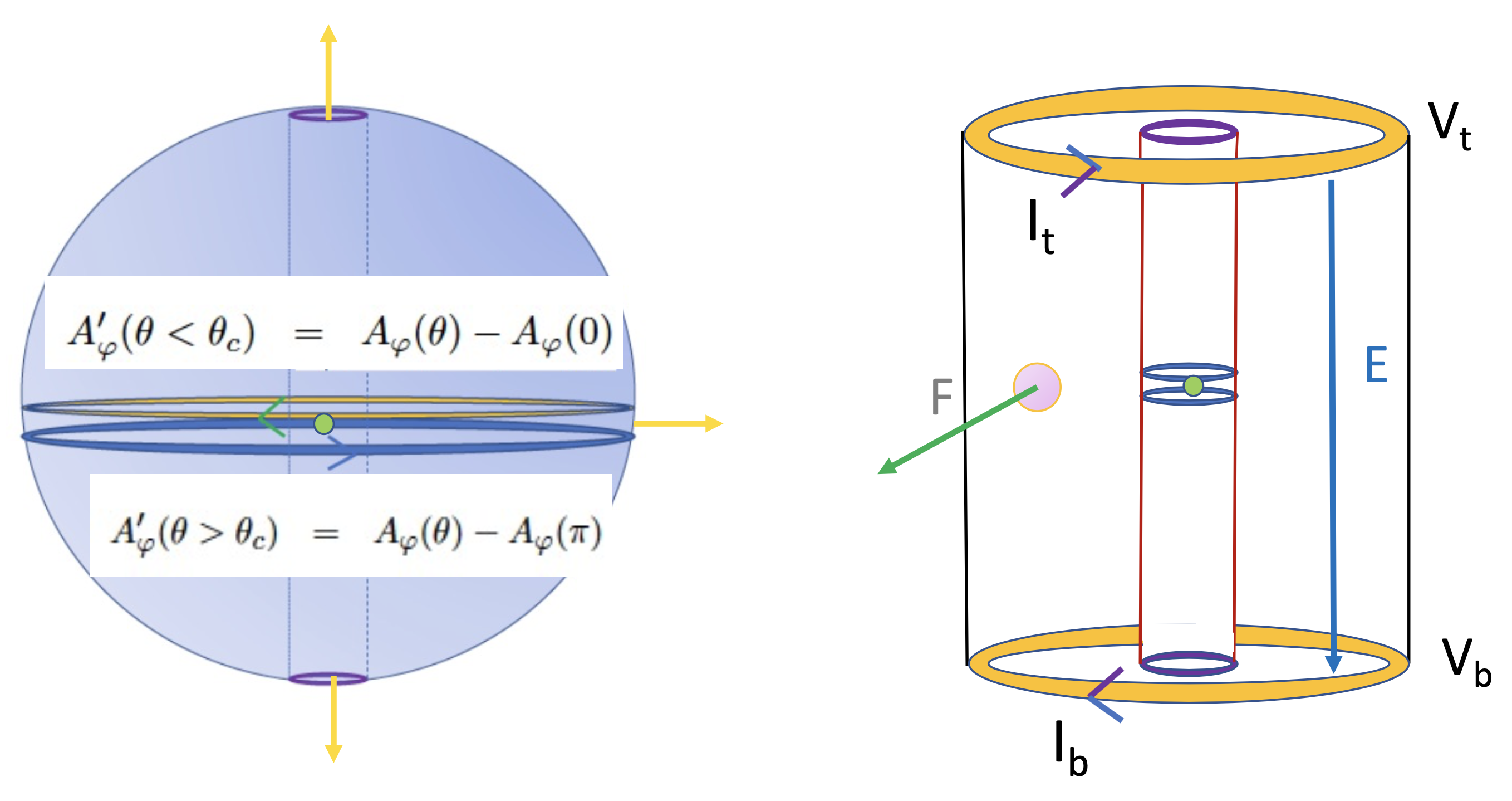}
\caption{(Left) Definition of smooth fields ${A}'_{\varphi}$ where the north and south hemispheres or regions meet at the angle $\theta_c$. The function $A_{\varphi}$ is smoothly defined on the whole surface. With the topometry analysis, we associate two thin cylinders inside the sphere joining each pole to the equatorial plane around the topological charge and transporting $A_{\varphi}(0)$ and $A_{\varphi}(\pi)$ respectively in the definitions of $A'_{\varphi}(\theta<\theta_c)$ and $A'_{\varphi}(\theta>\theta_c)$. At a pole, a purple circle can be adiabatically deformed through a small radius $r_c\rightarrow 0$ becoming a unit radius from Stokes' theorem since we can define $\bm{\nabla}\times {\mathbf{A}}=0$ outside the surface of the ball $S^2$. Therefore, we have ${A}_{\varphi}(0) =\lim_{\theta \rightarrow 0}{A}_{\varphi}(r_c,\theta,\varphi)=\lim_{\theta\rightarrow 0}{A}_{\varphi}(\theta)$ and similarly ${A}_{\varphi}(\pi) =\lim_{\theta \rightarrow \pi}{A}_{\varphi}(r_c,\theta,\varphi)=\lim_{\theta\rightarrow \pi}{A}_{\varphi}(\theta)$. The topological charge in green is surrounded by two circles in the equatorial plane of the sphere and the topological information is transported on a thin cylinder from the equator to a pole. (Right) Equivalent topological representation on the cylinder with a topological charge and a uniform Berry curvature ${\bf F}$. The cylinder geometry reveals two protected chiral edge modes with a quantized current analytically.}
\label{Edges.pdf}
\end{center}
\end{figure}

For the $|\psi_+\rangle$ eigenstate, 
\begin{equation}
\label{cosine}
A_{\varphi} = -\frac{\cos\theta}{2},
\end{equation} 
and $A_{\theta}=0$. The sphere as a fiber bundle then allows us to describe topological properties through one component of the vector ${\bm A}$ only leading to simple analytical calculations.
This will also justify the choice of a boundary (interface) defined parallel to the equatorial line in the application of Stokes' theorem with two regions or hemispheres. Since we can progressively move this interface close to the two poles then such a boundary can shrink to a point defining the pole or equivalently a small circle surrounding one of this pole. Related to definitions in Sec. \ref{potential}, we find it very useful to introduce $A'_{\varphi}(\theta<\theta_c)$ and $A'_{\varphi}(\theta>\theta_c)$ \cite{HH,C2} such that
\begin{eqnarray}
\label{smoothfields}
A'_{\varphi}(\theta<\theta_c) &=& A_{\varphi}(\theta) - A_{\varphi}(0) = \sin^2\frac{\theta}{2} \\ \nonumber
A'_{\varphi}(\theta>\theta_c) &=& A_{\varphi}(\theta) - A_{\varphi}(\pi) = -\cos^2\frac{\theta}{2}.
\end{eqnarray}
Here, $\theta_c$ refers to the angle of the boundary (interface). We remind that the symbols $\theta<\theta_c$ and $\theta>\theta_c$ in $A'$ refer equivalently to $\theta=\theta_c^-$ and $\theta=\theta_c^+$. These equations will also play an important role to relate with quantum transport
and light responses in the next sections.  The fields $A'$ are smoothly defined in each region and are defined to be gauge invariant. If we define the same $A'$ for any gauge representation, then one can also re-associate the same $A_{\varphi}(\theta)$ defined as in Eq. (\ref{smoothfields}).
The field $A_{\varphi}(\theta)$ is defined smoothly on the whole sphere and this equation also implies that the singularity at the position of the topological charge is transported in a thin Dirac cylinder (handle) from the center of the sphere to each pole (see Sec. \ref{smooth}). In the applicability of Stokes' theorem, Eqs. (\ref{smoothfields}) can be seen as if each hemisphere gives rise to two circles at the equator one linked to the field $A_{\varphi}(\theta)$ at radius $r=1$ and one linked to information at a pole $A_{\varphi}(0)$ or $A_{\varphi}(\pi)$ acting on a very small circle with infinitesimal radius linking each pole to the equator (see purple or violet circles in Fig. \ref{Edges.pdf}). The field $A'_{\varphi}$ is discontinuous at the interface related to topological properties from Stokes' theorem (Eq. (\ref{polesA})). 

These equations are indeed very similar to Eqs. (\ref{vecpot}) if we set $B=\frac{1}{2}$ for a sphere with $r=1$ and if we identify $\tilde{A}_{\varphi}(\theta)$ for the classical model precisely to $A_{\varphi}(\theta)$ for quantum physics. The Berry connection in Eq. (\ref{cosine}) corresponds precisely to the introduced classical field $\tilde{A}_{\varphi}$ in Eq. (\ref{tildefields}). 

In that case, Eq. (\ref{Fclassical}) then leads to
\begin{equation}
C = \frac{\Phi}{2\pi} = \frac{1}{2\pi} \left(\int_0^{2\pi} d\varphi \right)\left(\int_0^{\pi} F_{\theta\varphi} d\theta \right) = 1.
\label{formulaC}
\end{equation}
For the eigenstate $|\psi\rangle=|\psi_+\rangle$, the Berry curvature takes the form
\begin{equation}
F_{\theta\varphi}=\partial_{\theta}A'_{\varphi}=\frac{\sin\theta}{2}
\end{equation}
with ${\bf F}=\bm{\nabla}\times{\bf A}=\bm{\nabla}\times{\bf A}'$, which is a smooth (gauge-invariant) and continuous function on the whole surface of the sphere. 
The definition of $C$ is also in accordance with the first Chern topological number \cite{Chern}. 

From Eq. (\ref{Fclassical}), the definition of ${\bf A}'$ allows us to re-interpret the Gauss law as in flat space which will be useful for a simple identification between angles and distances in the plane when studying lattice models.
Similarly as the effect of a charge $e$ producing a radial electric field or a Dirac monopole producing a radial magnetic field, we can interpret $C=1$ as an effective topological charge encircled by the surface; see Fig. \ref{Edges.pdf}. We can also re-write the equation above locally at the poles \cite{HH}
\begin{equation}
\label{C}
C = \int_0^{\pi} \frac{\sin\theta}{2} d\theta = -\frac{1}{2}[\cos \theta ]_0^{\pi} = (A_{\varphi}(\pi) - A_{\varphi}(0)).
\end{equation}
From the Ehrenfest theorem, then we identify $\langle \psi_+ |\sigma_z |\psi_+\rangle = \langle \sigma_z(\theta)\rangle = \cos\theta$ \cite{Henriet} such that 
\begin{equation}
\label{polesC}
C = (A_{\varphi}(\pi) - A_{\varphi}(0)) = \frac{1}{2}\left(\langle\sigma_z(0)\rangle - \langle\sigma_z(\pi)\rangle\right).
\end{equation}
This sphere's model is realized in circuit quantum electrodynamics \cite{Roushan,Boulder} and can equally be realized in atomic physics. That way the global topological number \cite{Henriet,HH,C2} is described in terms of the Berry phases \cite{Berry} 
$\frac{1}{2\pi}\oint A_{\varphi}(\theta) d\varphi$, with a relative sign at the two poles for $\theta=0$ and $\theta=\pi$. A topological number $C=1$ can then be perceived as a total Berry phase of $2\pi$ from the poles.

The topological number can also be detected by rolling the spin adiabatically from north to south pole when changing linearly the polar angle with time $\theta=vt$. 
The topological protection is observed when a sphere turns into an ellipse \cite{Roushan}.
Then, we have 
\begin{equation}
C = -\frac{1}{2}\int_0^{\pi} \frac{\partial\langle\sigma_z\rangle}{\partial\theta} d\theta = -\frac{1}{2}\int_0^{T_{\pi}} \frac{\partial\langle\sigma_z\rangle}{\partial t} dt,
\label{drive}
\end{equation}
where $T_{\pi}=\frac{\pi}{v}$. Similarly as Eq. (\ref{Phi'}), Eq. (\ref{formulaC}) is equivalent to
\begin{equation}
\label{Cgeometry}
C = \left({A}'_{\varphi}(\theta_c^-) - {A}'_{\varphi}(0) + {A}'_{\varphi}(\pi) - {A}'_{\varphi}(\theta_c^+)\right). 
\end{equation}
Since $A'_{\varphi}(0)={A}'_{\varphi}(\pi)=0$ then this implies \cite{HH,C2}
\begin{eqnarray}
\label{CA'}
C &=&  A'_{\varphi}(\theta_c^-) - A'_{\varphi}(\theta_c^+) = 1 \nonumber \\
&=& A'_{\varphi}(\theta<\theta_c) - A'_{\varphi}(\theta>\theta_c).
\end{eqnarray}
This form of topological number, which can be understood from Stokes' theorem, then also justifies the re-interpretation of the curved space into an effective flat metric through the discussion around Eq. (\ref{geodesic}).

From shapes in mathematics, spheres are usually defined with an Euler characteristic $\chi=2$ and tori, donuts and cups with $\chi=0$ revealing the formula $\chi=2-2g$ for an orientable surface. Here, $g$ refers to the number of holes or handles. From this angle, it is then useful to introduce the definition \cite{Tan,Ma,Kolodrubetz}
\begin{equation}
\chi = \frac{1}{4\pi}\iint_{M} \sqrt{det g} d\mu d\nu = 2|C|
\end{equation}
with the quantum metric tensor $g_{\mu\nu}$ (with here $\mu,\nu=\theta,\varphi$) on the manifold introduced in \ref{Berrycurvature}. This number can be understood from the relation between curvature $R_{\varphi\varphi}$ along $d\varphi$ in spherical coordinates $\sin^2\theta$ and the metric component $g_{\varphi\varphi}=\frac{\sin^2\theta}{4}$ in \ref{Berrycurvature} related to the form of eigenstates. 
For one spin-$\frac{1}{2}$ within the topological phase, there is an interesting relation between quantum distance related to the eigenstates and the Berry curvature
\begin{equation}
\sqrt{det g}=\frac{\sin\theta}{4} = 2F_{\theta\varphi}.
\end{equation}

On the other hand, when a topological charge is introduced in the sphere, the topology of the sphere becomes also similar to that of the donut similarly as if $g$ can be identified to $|C|=1$, in agreement with the Poincar\' e-Hopf theorem.
From Stokes' theorem in Eq. (\ref{CA'}), in this way the topological number can also be viewed as two circles at the equator with radius unity (associated to the respective fields $A'$) encircling (surrounding) equally the topological charge, similarly as in the presence of vortices in a superfluid. The Euler characteristic of these two circles encircling the topological charge may be alternatively interpreted as $\chi_{circles}=0+0$. In the situation of a non-topological sphere, it is as if there is no hole or no singularity such that each circle can be transformed into a disk in this interpretation with $\chi_{disks}=1+1=2$, which is also in agreement with the Euler characteristic of a sphere. The disk means here that the encircled region has no charge or equivalently no singularity.  It is interesting to observe that Eq. (\ref{C}) leads to the same topological response as a one-dimensional Zak phase \cite{Zak} defined through a linear function along the polar angle
\begin{equation}
\label{Zak}
C = \frac{1}{\pi}\int_0^{\pi} |\bm{\nabla}\theta| d\theta = 1.
\end{equation}
It is interesting to mention here recent experimental developments with ultra cold atoms related to measurements on the Bloch sphere of spin observables and Zak phase associated to topological lattice models and Floquet engineering \cite{MunichRiceMele,Spielman}.

For a recent theoretical generalization of geometry and topology to multi-levels' systems, see \cite{CambridgeMulti}.

\subsection{Smooth Fields and Quantum TopoMetry}
\label{smooth}

Here, we show that the definition of $A'$ fields can be understood from quantum topometry on the Bloch sphere related to Fig. \ref{Edges.pdf}, elaborating on a geometrical proof with two boundaries per hemisphere \cite{HH}.
This proof is useful to show that the global topological number can be measured from the poles of the sphere (only) from a vector field defined smoothly on the whole sphere (see Eq. (\ref{cosine})). 
This proof will be applicable to one and also to multi-spheres systems. 

We define the smooth fields ${\bf A}'$ equal to zero at the poles, such that the topological charge can be defined as
\begin{equation}
C = \frac{1}{2\pi}\iint_{S^{2'}} \bm{\nabla}\times{\bf A}'\cdot  d^2{\bf s},
\label{topologicalnumber}
\end{equation}
when $S^{2'}$ corresponds to the slightly modified Riemann, Poincar\' e, Bloch sphere where we subtract the two poles (two points). Introducing the field ${\bf A}'$ leads to the definition of the area differential 
$d^2{\bf s}=d\varphi d\theta {\bf e}_r$ similarly as in Secs. \ref{potential} and \ref{spin1/2} with $\bm{\nabla}=(\partial_{\theta},\partial_{\varphi})$. We also have ${\bf F}=\bm{\nabla}\times{\bf A}'=\bm{\nabla}\times{\bf A}$ on $S^{2'}$ to validate $C$ as a topological number. This step then requires that ${\bf A}$ is continuous (smooth) at the two poles leading then to the azimuthal topological responses 
\begin{equation}
A_{\varphi}(0)=\lim_{\theta\rightarrow 0}{A}_{\varphi}(\theta)\hskip 0.2cm \hbox{and} \hskip 0.2cm A_{\varphi}(\pi)=\lim_{\theta\rightarrow \pi}{A}_{\varphi}(\theta). 
\end{equation}

If we define two regions, north and south hemispheres, meeting at the boundary angle $\theta_c$, then from Stokes' theorem applied on each region we have
\begin{equation}
\iint_{north'} \bm{\nabla}\times{\bf A} \cdot d^2{\bf s} = \int_0^{2\pi} \left(A_{N\varphi}(\theta,\varphi)-A_{\varphi}(0)\right) d\varphi.
\label{north'}
\end{equation}
To write this line, we take into account the information in caption of Fig. \ref{Edges.pdf}.
Here, $A_{N\varphi}=A_{\varphi}(\theta<\theta_c)$ refers to the azimuthal component of the Berry connection in the north region and $A_{\varphi}(0)=\lim_{\theta\rightarrow 0} A_{N\varphi}(\theta,\varphi)$.
Similarly, in south region, we have
\begin{equation}
\hskip -0.2cm \iint_{south'} \bm{\nabla}\times{\bf A} \cdot d^2{\bf s} = -\int_0^{2\pi} \left(A_{S\varphi}(\theta,\varphi)-A_{\varphi}(\pi)\right) d\varphi.
\label{south'}
\end{equation}
Here, $A_{S\varphi}=A_{\varphi}(\theta>\theta_c)$ refers to the azimuthal component of the Berry connection in south region and $A_{\varphi}(\pi)=\lim_{\theta\rightarrow \pi} A_{S\varphi}(\theta,\varphi)$. The relative $-$ sign between the two regions on the right-hand side comes
from the different orientations of the two surfaces. From these two equalities, we can define smooth fields on each region $A'_{\varphi}(\theta<\theta_c)=A_{N\varphi}(\theta,\varphi)-A_{\varphi}(0)$ and  $A'_{\varphi}(\theta>\theta_c)=A_{S\varphi}(\theta,\varphi)-A_{\varphi}(\pi)$. 
Here, $A_{S\varphi}(\theta,\varphi)$ means equivalently $A_{S\varphi}(\theta_c^+,\varphi)$ and $A_{N\varphi}(\theta,\varphi)$ means equivalently $A_{N\varphi}(\theta_c^-,\varphi)$.
Now, suppose we slightly move the boundary close to a pole, for instance setting $\theta_c\rightarrow 0^+$. Within our definitions, since ${\bf A}$ is smooth close to the pole this implies that $A_{S\varphi}(\theta\rightarrow 0^+)=A_{\varphi}(0)=A_{N\varphi}(\theta\rightarrow 0^+)$ and therefore that $C=A_{\varphi}(\pi)-A_{\varphi}(0)$ from Eq. (\ref{north'}) and (\ref{south'}). This means that the global topological Chern number can be defined locally from the poles through a gauge invariant response. 

Moving back slowly the boundary close to the equator through small portions, from the smoothness of all the fields in one hemisphere then we conclude that we can yet define $A_{S\varphi}(\theta,\varphi)=A_{N\varphi}(\theta,\varphi)=A_{\varphi}(\theta)$ for each interval such that the fields can be defined as $A'_{\varphi}(\theta<\theta_c)=A_{\varphi}(\theta)-A_{\varphi}(0)$ and $A'_{\varphi}(\theta>\theta_c)=A_{\varphi}(\theta)-A_{\varphi}(\pi)$ in accordance with definitions above (summing Eqs. (\ref{north'}) and (\ref{south'})) and with the classical definitions in Eqs. (\ref{vecpot}). Indeed, this is equivalent to say that the relation $C=A_{\varphi}(\pi)-A_{\varphi}(0)$ can be obained adding Eqs. (\ref{north'}) and (\ref{south'}) for any $\theta_c$, as $C$ is uniquely defined. 

For practical applications, $A'_{\varphi}(\theta>\theta_c)$ and $A'_{\varphi}(\theta<\theta_c)$ can be defined in a gauge invariant way; the Berry gauge field $A_{\varphi}$ can then be rephrased uniquely through Eqs. (\ref{smoothfields}) in a continuous and smooth manner on the whole sphere. In each region the field $A'_{\varphi}$ is also smoothly defined. Observe that we can equivalently transport the discontinuity of the function $A'_{\varphi}(\theta<\theta_c)-A'_{\varphi}(\theta>\theta_c)$ back to the poles which then leads to a local interpretation of the topological number in terms of $A_{\varphi}(\pi)-A_{\varphi}(0)$. 

Related to Fig. \ref{Edges.pdf}, we can precisely define two cylinders (handles) with a thin radius $r_c$ piercing or approaching the center of the sphere on each side of the boundary 
\begin{equation}
\oint {\bf A}'(\theta_c^{\pm})\cdot{\bf dl} = \oint {\bf A}(\theta_c^{\pm})\cdot{\bf dl} -\oint_{r=r_c} {\bf A}(pole^{S,N})\cdot{\bf dl},
\end{equation}
such that we effectively transport in a Dirac string of radius $r_c$ informations from a pole $S,N$ down to or up to an angle $\theta_c^{\pm}$. 
We emphasize here that since the global topological number can be defined locally from the poles, this implies that we may then slightly adjust a sphere onto an ellipse or a cylinder. In Sec. \ref{cylinderformalism}, we present an analytical understanding of smooth fields and the occurrence of edge modes on a cylinder geometry with uniform Berry curvatures. In Sec. \ref{Geometry}, we adiabatically reshape this geometry onto generalized cubic geometries or three-dimensional geometries with corners starting from superellipses in the plane.

Hereafter, we show that smooth fields provide important information related to physical observables associated to quantum transport and light-matter coupling.
In Table I, we review quantum symbols related to geometry and observables introduced in this Section (from one sphere). The function $\alpha(\theta)$ will be introduced in Sec. \ref{lightdipole} related to the light-matter interaction for one topological sphere.
The last line is related to Sec. \ref{lightKM} and the Pfaffian for two-dimensional topological insulators through the response to circularly polarized light.

\begin{table}[t]
\caption{Quantum Formalism and Symbols} 
\centering 
\begin{tabular}{||c|c||} 
\hline\hline 
Smooth Fields and Observables & \hskip 0.5cm Definitions \\  
\hline 
$A_{\varphi}$ & $-i\langle \psi| \partial_{\varphi} |\psi\rangle$ \\
$A'_{\varphi}(\theta<\theta_c)$ & $A_{\varphi}-A_{\varphi}(0)$  \\ 
$A'_{\varphi}(\theta>\theta_c)$ & $A_{\varphi}-A_{\varphi}(\pi)$  \\
$F_{\theta\varphi}=\partial_{\theta}A'_{\varphi}=\partial_{\theta}A_{\varphi}$ & $C=\int_0^{\pi} F_{\theta\varphi}d\theta$ \\
$C$ & $A_{\varphi}(\pi)-A_{\varphi}(0)$\\
$C$ & $A'_{\varphi}(\theta<\theta_c) - A'_{\varphi}(\theta>\theta_c)$ \\
$C$ & $\frac{1}{2}(\langle\sigma_z(0)\rangle - \langle \sigma_z(\pi)\rangle)$ \\
$J_{\perp}(\theta)$ & $\frac{e}{T}A'_{\varphi}(\theta<\theta_c)$ \\
$G$, $\sigma_{xy}$ & $\frac{q^2}{h}C$ \\
$\alpha(\theta)$ & $C^2 +2A'_{\varphi}(\theta<\theta_c)A'_{\varphi}(\theta>\theta_c)$  \\
$\alpha(0) = \alpha(\pi) = 2\alpha(\frac{\pi}{2})$ & $C^2$ \\
$g_{\mu\mu}$ & $\frac{1}{2}\frac{{\cal I}(0)}{m^2} =  \frac{1}{2}\frac{{\cal I}(\pi)}{m^2}$\\
$\alpha_{\uparrow}(\theta)+\alpha_{\downarrow}(\theta)$ & $|C_s|-(P({\bf k}))^2$ \\
\hline %
\end{tabular}
\label{tableI} %
\end{table}

\subsection{Transport from Newtonian Mechanics and Force, Quantum Mechanics and Parseval-Plancherel Theorem}
\label{ParsevalPlancherel}

This formalism is useful to study transport on the sphere due to an electric field ${\bf E}=E{\bf e}_{x_{\parallel}}=-\bm{\nabla} V$  where the unit vector ${\bf e}_{x_{\parallel}}$ refers to the direction of the electric field. 
To apply the Parseval-Plancherel theorem, we will define the wave-vector through $(k_{\parallel},k_{\perp})$ on the same sphere and for simplicity we choose the direction of 
${\bf e}_{x_{\parallel}}$ to be related to $k_{\parallel}$. Periodic boundary conditions for the wave-vector components will allow us to link with topological properties of lattice models from the reciprocal space.

Similar to the charged particle in a magnetic field, we will navigate from classical to quantum mechanics in a coherent way. We start from the Hamiltonian for a charge $q$, mass $m$ and spin-$\frac{1}{2}$ such that
the motion longitudinal to the direction of the electric field is associated to the polar angle direction:
\begin{equation}
H_{\parallel} = \frac{(\hbar k_{\parallel})^2}{2m} + qV - {\bf d}\cdot\mathbfit{\sigma}.
\end{equation}
We introduce the Planck constant $h$ such that $\hbar=\frac{h}{2\pi}$. To derive simple arguments, we start from the reciprocal space to define the unit sphere as $(k_{\parallel},k_{\perp})=(\theta,\varphi)$.  The analogy with a flat metric for the motion of the phase $\theta(t)$ can also be understood from the form of the geodesic on the sphere, see Eq. (\ref{geodesic}). For the motion side in the Hamiltonian, we should satisfy $\dot{x}_{\parallel} = \frac{p_{\parallel}}{m}=\partial_{p_{\parallel}}H$ and $\dot{p}_{\parallel} = - \partial_{x_{\parallel}}H$, to be consistent with the Newton equation $ma_{\parallel}=\dot{p}_{\parallel}=\hbar \dot{k}_{\parallel}=qE$ and the Coulomb force. Then, we obtain the equation
\begin{equation}
\label{distancetime}
\theta(t) = k_{\parallel}(t) = \frac{q}{\hbar} E t.
\end{equation}

This equation can also be understood from the Lagrangian form in spherical coordinates for a charged particle of mass $m$
\begin{equation}
\label{lagrangian}
L = \frac{1}{2}m\dot{\tilde{\theta}}^2 + \frac{1}{2}m\sin^2\tilde{\theta}\dot{\tilde{\varphi}}^2 - qV(\tilde{\theta})
\end{equation}
where $\tilde{\theta}=x_{\parallel}$ and $\tilde{\varphi}=x_{\perp}$ describe the polar angle and azimuthal angles of the unit sphere associated to the real space variables $(x_{\parallel},x_{\perp})$. From Euler-Lagrange equations, we have
\begin{equation}
\label{Lagrangian}
m\ddot{\tilde{\theta}} = qE + \sin\tilde{\theta}\cos\tilde{\theta} \dot{\tilde{\varphi}}^2= \hbar \dot{\theta},
\end{equation}
with the relation between potential and electric field $E=-\frac{dV(\tilde{\theta})}{d\tilde{\theta}}$. We start at the north pole where $\tilde{\theta}=0$ at time $t=0$. From the first term in Eq. (\ref{Lagrangian}), this gives rise to $m\dot{\tilde{\theta}} = qEt = \hbar \theta$ and to 
$\tilde{\theta} = \frac{1}{2m} qE t^2$. In a short-time $dt$, when writing the differentials, the term in $\sin\tilde{\theta}$ will only give negligible $dt^2$ corrections. For the transverse direction, the Euler-Lagrange equation in a short-time $dt$ leads to $\ddot{\tilde{\varphi}}=\dot{\varphi}=0$ or equivalently to 
$\dot{k}_{\perp}=0$. At initial time, we assume that $k_{\perp}=0$ which leads to $\dot{\tilde{\varphi}}=0$ in Eq. (\ref{Lagrangian}). In this way, the polar angle is defined as in Eq. (\ref{distancetime}) in agreement with the form of the geodesic in Eq. (\ref{geodesic}). 
In the spherical geometry related to the skin effect in Sec. \ref{electricfield}, including the effect of a gravitational potential will then result in a term similar to $qE$ in Eq. (\ref{Lagrangian}). 
The response of the charged particle is viewed as an analogue of a response to a gravitational potential. The gravitational potential may also have useful applications as an entanglement source \cite{UCL}.

Here, we study the perpendicular response to the electric field in a quantum mechanical form to show a quantized response induced by the topological properties coming from $- {\bf d}\cdot\mathbfit{\sigma}$ in the Hamiltonian. We derive the topological current in the transverse direction to the polar direction from the Parseval-Plancherel theorem. We define the averaged transversal current density as \cite{HH}
\begin{eqnarray}
J_{\perp} &=& \frac{q}{T}\int_0^T \frac{d\langle x_{\perp}\rangle}{dt} dt = \frac{q}{T}\left(\langle x_{\perp}\rangle(T)-\langle x_{\perp}\rangle(0)\right) \\ \nonumber
&=& \oint  \left(J_{\varphi}(\varphi,T)-J_{\varphi}(\varphi,0)\right)d\varphi,
\end{eqnarray}
such that the time $T$ in the protocol is related to the angle $\theta$ through the longitudinal dynamics $\theta(T)=\frac{q E T}{\hbar} = v^* T$ and $\varphi$ represents the wave-vector associated to the $x_{\perp}$ direction in real space.
For a fixed angle $\varphi$, then we identify
\begin{equation}
\label{Jtransverse}
J_{\varphi}(\varphi,\theta) = \frac{i q}{4\pi T}\left(\psi^{*}\frac{\partial}{\partial \varphi}\psi - \frac{\partial\psi^*}{\partial\varphi}\psi\right) = \frac{i q}{2\pi T}\psi^*\frac{\partial}{\partial \varphi}\psi,
\end{equation}
the factor $\frac{1}{4\pi}$ absorbs the normalization of the motion side in the wave-function such that within our definition $\langle \psi| \psi\rangle=1$ with $|\psi\rangle=|\psi_{motion}\rangle\otimes|\psi_+\rangle$. 
It should be noted that in the absence of the magnetic field, $J_{\perp}=0$ which is also in accordance with the classical equation of motion $\dot{k}_{\perp}=0$.
Furthermore, solutions of the quantum equation
\begin{equation}
i\hbar\frac{\partial}{\partial t}\psi(\varphi,t) = \frac{\hbar^2\varphi^2}{2m} \psi(\varphi,t),
\end{equation}
are $\psi(\varphi,t) = \frac{1}{\sqrt{4\pi}}e^{-i\frac{\hbar^2\varphi^2}{2m} t}$ which include solutions with $\pm \varphi$. If we average the two contributions for positive and negative $\varphi$ then this is equivalent to say that $k_{\perp}=0$ in the semi-classical approach. 
Therefore, in our situation, for $q=e>0$ such that $\theta>0$ through $J_{\perp}$ we probe the topological linear response to the electric field induced by the presence of the radial magnetic field on the sphere acting on the spin degrees of freedom. 
In this way, Eq. (\ref{Jtransverse}) turns into
\begin{eqnarray}
\label{Jperp}
|J_{\perp}(\theta)| = \frac{e}{2\pi T} \oint A'_{N\varphi}(\varphi,\theta)d\varphi &=& \frac{e}{T}A'_{\varphi}(\theta<\theta_c), \nonumber \\
\end{eqnarray}
where $A'_{N\varphi}$ is identical to the smooth field $A'_{\varphi}(\theta<\theta_c)$ defined in Eq. (\ref{smoothfields}). The particle starts at $t=0$ from the north pole and we assume that at time $T$ the particle remains inside the same north-N domain (to ensure an adiabatic protocol) such that we can adjust $\theta_c$ accordingly closer to the south pole. In fact, to obtain a gauge-invariant form, it is appropriate to fix here $\theta_c=\pi$ corresponding to the south pole. The typical time scale to reach the south pole is defined through $\theta(T)=\pi=\frac{q}{\hbar}ET$. When fixing $\theta_c=\pi$, this implies
$A'_{\varphi}(\theta_c^+)=0$ in Eq. (\ref{CA'}) and $A'_{\varphi}(\theta_c^-)=C$. 

It is interesting to observe that the relation between the transversal pumped current and the topological number is shown without any specific form of the wavefunction in the system such that this relation remains valid for multiple spheres. In that case, from Eq. (\ref{CA'}), we have $A'_{N\varphi}(\varphi,\theta\rightarrow\theta_c)=C$ and therefore the produced current density is \cite{HH}
\begin{equation}
\label{Jperp2}
|J_{\perp}(T)| = \frac{e}{T} C.
\end{equation}
The transversal pumped charge is $Q=J_{\perp}.T=eC$ and it is therefore quantized in units of $C$. This also means that if the particle acquires a macroscopic current 
\begin{equation}
\label{polarization}
|J_{\perp}|T=\frac{e\hbar (k_{\perp}T)}{m} = eC.
\end{equation}
The integrated current on the whole surface of the sphere reveals a quantization related to the presence of the topological charge. Similar to the Bohr quantization for the angular momentum, the transversal pumped response is quantized in units of $C$. In Sec. \ref{Observables}, we show a correspondence between this smooth-fields' formalism and general many-body theory relating the Karplus-Luttinger velocity \cite{KarplusLuttinger} and transport \cite{Thouless1983}. The formula $J_{\perp}=eC$ can be equivalently obtained by fixing the interface at any angle $\theta_c$ through Eq. (\ref{CA'}) via the transportation of a charge $e$ from north pole to this interface and simultaneously the transportation of a charge $-e$ from south pole to the same interface.

At this stage, one may question the feedback of the produced current $J_{\perp}$ onto the semiclassical analysis in Eq. (\ref{lagrangian}) which has resulted in the identification $\theta(T)=\pi=\frac{q}{\hbar}ET$.
From $A'_{\varphi}(\theta<\theta_c)=\sin^2\frac{\theta}{2}$, we observe that this gives rise to a small correction in  $\dot{\tilde{\varphi}}$ (being also related to $J_{\perp}$) proportional to $E^2 t$ at a time $t$ which means a (very small) correction of the order of $E^4 dt^2$ when evaluating the influence of $\sin\tilde{\theta}\cos\tilde{\theta} \dot{\tilde{\varphi}}^2$ on $\dot{\theta}$. Therefore, we can safely neglect this small effect in the evaluation of the angle $\theta(t)$ such that the typical time scale to reach the south pole is defined through $\theta(T)=\pi=\frac{q}{\hbar}ET$. 

\subsection{Dynamics, Protected Energetics and Fourier Series}
\label{DynamicsSeries}

Here, we provide additional information on the dynamics for a charge which navigates from north to south pole adiabatically (when adjusting appropriately $\theta_c=\pi$ such that $A'_{\varphi}$ and $A_{\varphi}$ remain smooth on the whole trajectory). 
The objective is to develop a correspondence with Fourier series and with the notion of protected topological energetics from the produced transverse kinetic energy of a particle. This simple approach, starting again from Newton equation, also verifies that we can produce an equivalent of the transverse topological current through the application of an electric field parallel to the equatorial line with an amplitude proportional to the electric field applied along the polar angle in Sec. \ref{ParsevalPlancherel}. 

To establish a link with Fourier series it is useful to evaluate the (mean) current 
\begin{equation}
|\bar{J}_{\perp}| = \frac{eC}{2T}.
\label{current}
\end{equation}
At a time $t$, we define the averaged pumped charge in the transversal direction when summing all the accumulated currents until angle $\theta$ associated to time $t=\frac{\hbar \theta}{e E}$
\begin{equation}
\bar{Q} = \frac{e}{t}\int_0^t \sin^2\frac{\theta}{2} d\theta.
\end{equation}
From the smooth fields, then we have the identity $\bar{Q}(T)= \frac{eC}{2}$ and this charge is accumulated in the time $T=\frac{\hbar\pi}{eE}$ such that $|\bar{J}_{\perp}|= \frac{eC}{2T}$.
The presence of the topological invariant $C$ in this identity can be precisely understood as follows. First, one may write down $\sin^2\frac{\theta}{2}$ as $C-\cos^2\frac{\theta}{2}$ and close to the north pole we also have the limit that $\cos^2\frac{\theta}{2}\rightarrow C$
such that $\cos^2\frac{\theta}{2}\rightarrow \frac{C}{2}(1+\cos\theta)$ on the trajectory. The averaged pumped charge is related to the encircled topological charge.
 In the sense of quantum mechanics, we can relate the current with mean velocity through $J = e \langle v_{\perp}\rangle$ with $\langle v_{\perp}\rangle = \frac{\hbar k_{\perp}}{m} = \frac{\hbar \varphi}{m}$. In this way, we have
a simple correspondence between the pumped current and transversal momentum. 
The quantity $J_{\perp}^2(\theta)$ is directly related to a gain in energy associated to the kinetic energy in the transversal direction produced by the topological phase and the electric field applied along the polar angle.
A particle (electron) navigating from north pole up to $\theta_c=\pi$ produces the transversal kinetic energy
\begin{equation}
\label{Ekin}
E_{kin} = \frac{1}{2m}(\hbar k_{\perp})^2 = \frac{m}{2T^2}C^2.
\end{equation}
We can also define an averaged kinetic energy
\begin{equation}
\bar{E}_{kin} = \frac{1}{\pi}\int_0^{\pi} \frac{m}{2}\frac{(eE)^2}{\hbar^2} \frac{\sin^4\frac{\theta}{2}}{\theta^2}d\theta \approx \frac{\pi m}{T^2}\frac{C^2}{8},
\end{equation}
which is slightly reduced but comparable to $E_{kin}$.

Now, we discuss a relation with Fourier series through the effect of an additional electric field applied parallel to the azimuthal angle. To reproduce the function $J_{\perp}(\theta)$ we suppose that this transversal electric field is proportional to the electric field applied along the polar direction. Then, we identify $\hbar\dot{\varphi} = \alpha' eE$ leading to $\dot{\varphi} = \alpha' \dot{\theta}$ and to a linear relation between the azimuthal angle and the polar angle $\varphi = \alpha' \theta$. Therefore,
\begin{equation}
{\cal J}_{\perp}(\theta) = |J_{\perp}(\theta)| = \frac{e\hbar}{m}\varphi = \frac{\alpha' e \hbar}{m} \theta = \alpha \theta,
\end{equation}
where we implicitly assume that $\alpha'>0$ and $\alpha>0$. This relation is applicable in the physical space $\theta\in[0;\pi]$ and we can in fact generalize this relation on the full cycle $\theta\in[0;2\pi]$ from the continuity of the current along any line parallel to the equator where ${\cal J}_{\perp}(\theta)$ is defined on $[0;\pi]$ and can be viewed as an even function on the interval $[0;2\pi]$. Then, we have 
\begin{eqnarray}
{\cal J}_{\perp}(\theta) &=& \alpha \theta \hskip 0.2cm \hbox{for} \hskip 0.2cm \theta\in [0;\pi] \nonumber \\ 
{\cal J}_{\perp}(\theta) &=& -\alpha\theta + \alpha 2\pi  \hskip 0.2cm \hbox{for} \hskip 0.2cm  \theta\in [\pi;2\pi]. 
\end{eqnarray}
We obtain the same decomposition of ${\cal J}_{\perp}(\theta)$ by viewing the charge $-e$ navigating
from south to north pole if we redefine the angle $\theta$ such that $\theta=\pi$ at $t=0$ and $\theta=0$ at $vt=\pi$.
If we visualize this function as a periodic function of period $2\pi$ then this corresponds to a triangular periodic function which can
be decomposed as a Fourier series. Through the correspondence of averaged current 
\begin{equation}
\frac{1}{2\pi}\int_0^{2\pi} {\cal J}_{\perp}(\theta) d\theta = \frac{1}{\pi}\int_0^{\pi} {\cal J}_{\perp}(\theta) d\theta = \frac{e C}{2{T}} = \bar{{\cal J}}_{\perp}
\end{equation}
we obtain
\begin{equation}
\alpha = \frac{eC}{\pi {T}}.
\end{equation}

To calculate simply the Fourier coefficients, it is useful to redefine $\theta' = \theta-\pi$ such that ${\cal J}_{\perp}(\theta)={\cal J}_{\perp}(\theta'+\pi)=f(\theta')$ with $f(\theta') = \alpha(\theta'+\pi)$ for $\theta'\in [-\pi;0]$ and $f(\theta') = \alpha (-\theta' +\pi)$ for $\theta'\in [0;\pi]$ such that $f(-\theta')=f(\theta')$.
In this way, the function can be visualized through the Fourier decomposition
$$
f(\theta') = f_0 +\sum_{n=1}^{+\infty} a_n \cos(n \theta'),
$$
which corresponds to
\begin{equation}
{\cal J}_{\perp}(\theta) = f_0 +\sum_{n=1}^{+\infty} a_n (-1)^n \cos(n \theta) = \frac{e}{T} A'_{\varphi}(\theta<\theta_c).
\end{equation}
Here, $f_0 = \bar{J}_{\perp}=\frac{eC}{2T}$. To determine the Fourier coefficients $a_n$, we can write down the boundary conditions ${\cal J}_{\perp}(0)={\cal J}_{\perp}(2\pi)=0$ and ${\cal J}_{\perp}(\pi)=\alpha\pi=\frac{e}{T}C$.
The equation ${\cal J}_{\perp}(0)={\cal J}_{\perp}(2\pi)=0$ comes from the definition of the function ${\cal J}_{\perp}(\theta)$ which also implies that the function $A'_{\varphi}$ is defined to be zero at the poles.
From the Riemann and Hurwitz-Zeta functions, the solution takes the form $a_{2n}=0$ and 
\begin{equation}
a_{2n+1} = \frac{8}{\pi^2} \frac{1}{(2n+1)^2} f_0.
\end{equation}

By developing the series to order $a_1$, we have
\begin{equation}
{\cal J}_{\perp}(\theta) = \frac{eC}{2T}\left(1-\frac{8}{\pi^2}\cos\theta\right)
\end{equation}
which is in fact very close to the result produced by the radial magnetic field in Eq. (\ref{Jperp})
\begin{equation}
{\cal J}_{\perp}(\theta) = \frac{e}{2T}(1-\cos\theta).
\end{equation}
This function indeed evolves linearly in a (large) region around the equatorial plane.
We can also evaluate the gain in transverse energetics related to the Fourier series. The mean energy when averaging on $\theta\in [0;2\pi]$ is then related to the Parseval-Plancherel theorem and is equivalently related to the interval $\theta\in [0;\pi]$
\begin{eqnarray}
\hskip -0.5cm \frac{1}{\pi}\int_0^{\pi} \frac{m}{2e^2}{\cal J}_{\perp}^2(\theta) d\theta &=&
\frac{m}{2e^2}\left(f_0^2 +\frac{1}{2}\sum_{n=1}^{+\infty} a_n^2\right) \nonumber \\
&=& \frac{mC^2}{6{T}^2}.
\end{eqnarray}

In Sec. \ref{curvature}, we will justify this correspondence further from the fact that the topological state on the sphere can be described through the Karplus-Luttinger velocity \cite{KarplusLuttinger} producing indeed a transversal velocity which is proportional to the longitudinal electric field.

\subsection{Cylinder Geometry with Uniform Berry Curvatures}
\label{cylinderformalism}

Here, we show the application of geometry on a cylinder geometry (see Fig. \ref{Edges.pdf}). The cylinder will allow us to define edge modes localized around the top disk and the bottom disk related to the topological number $C$ of a topological dispersive Bloch  band \cite{C2}. The topological charge is present in the core of the cylinder.

We define the surface of the (large) vertical cylinder $2\pi H$, with $H$ the height, to be precisely the same as the surface of the sphere $4\pi$ with the radius $r=1$. This leads to fix the height of the cylinder as $H=2$. Within this definition, in polar coordinates we have ${\bf F}=F{\bf e}_r$ with $F=F_{\theta\varphi}(\theta=\frac{\pi}{2})=\frac{1}{2}$ from Sec. \ref{spin1/2} such that the flux on the surface or equivalently the topological charge is equal to $C=\frac{\Phi}{2\pi}=1$ on the vertical region of the cylinder. This solution admits a similar form of $A_{\varphi}$ than on the unit sphere. Indeed, in cylindrical coordinates \cite{C2,OneHalfKLH}
\begin{equation}
F(\varphi,z)=\partial_{\varphi}A_z-\partial_z A_{\varphi}=\frac{1}{2}
\end{equation}
 allows us to write $A_{\varphi}=-\frac{z}{2}$ with the identification $z=\cos\theta$ on the sphere for the same $r=1$ (in cylindrical and spherical coordinates). Therefore, this solution on the cylinder reveals the smooth fields of Sec. \ref{smooth}. From topological properties, the sphere and the (vertical region of the) cylinder are then identical from the forms of $C$ and from the smooth fields on the surface. These arguments then agree with the fact that $A_{\varphi}=cst$ at the boundary disks with $z=\frac{H}{2}=+1$ at the north disk and $z=-\frac{H}{2}=-1$ at the bottom disk resulting in $A_{\varphi}(0^+)=-\frac{1}{2}$ and $A_{\varphi}(\pi^-)=+\frac{1}{2}$ from the eigenstates in Eq. (\ref{eigenstates}). On the top and bottom surfaces ${\bf A}=cst$ such that the integral $\oint {\bf A}\cdot {\bf e}_{\varphi} d\varphi$ is identical along a small purple circle associated with the sphere and along a golden circle related to the cylinder geometry in Fig. \ref{Edges.pdf}. All the outgoing flux from the surface comes from the vertical region, similarly as if the cylinder is `open', therefore we can describe analytically the presence of edge modes coming from the vertical region at the boundaries with top and bottom disks.
 
We apply the protocol of Sec. \ref{ParsevalPlancherel} with an electrical field on the sphere acting on a charge $q$ through the Coulomb force along the direction ${\bf e}_{\theta}$ related to the polar angle. Since the smooth fields are identical in spherical and cylindrical coordinates, Eq. (\ref{Jperp}) is also valid in cylindrical coordinates. In the correspondence, we assume then that for the same electric field and the same time $T=\frac{h}{2q E}$, the particle reaches the south pole on the sphere and the bottom of the cylinder. For the physical correspondence below, we simply use the fact that Eq. (\ref{Jperp}) and Eq. (\ref{Jperp2}) are equivalent in both geometries, corresponding to the current produced in the perpendicular direction along the ${\bf e}_{\varphi}$ unit vector in the same time $T$. It is also correct that we must satisfy the linear relation between time and distance from Eq. (\ref{distancetime}). In the cylindrical geometry this corresponds to a non-linear evolution between distance and time such that 
\begin{equation}
d(\arccos(z))=-\frac{dz}{\sqrt{1-z^2}}=\frac{qE}{\hbar}dt. 
\end{equation}
In this way, one can formulate a corresponding (Lorentz) force $qE\sqrt{1-z^2}$ which is reduced at $z=+1$ and $z=-1$ due to the curvature effects on the sphere close to the poles. We can now measure the conductance located at the two edges on the cylinder introducing a difference of potential such that $H.E=(V_t-V_b)$ and we suppose that the potential is fixed on
the top and bottom disks. 

Using the form of the final time $T$ for a charge $q$, $T=\frac{h}{2q E}$, then we have
\begin{equation}
| J_{\perp} | = \frac{q^2}{h}2EC = \frac{q^2}{h} C(V_t-V_b) = (I_t - I_b).
\end{equation}
We can now visualize $I_b$ and $I_t$ as two edge currents associated with a quantized conductance 
\begin{equation}
\label{conductance}
G = \frac{dI}{dV} = \frac{q^2}{h}C.
\end{equation}
Due to the applied electric field, a charge $e$ goes in one direction and a charge $-e$ in opposite direction justifying the two quantized conductances at the two edges in this protocol. 
Therefore, $C=1$ measures one edge mode (located at the boundaries of the top and bottom disks) in agreement with the Landauer-B\" uttiker formula \cite{Landauer,Buttiker,Imry}. Similarly as the quantum Hall effect and Laughlin cylinder \cite{LaughlinQHE,Fabre,LiHall}, the smooth fields here allow us to reveal analytically the formation of edge modes at the boundary with the top and bottom disks. 

From Stokes' theorem on the vertical surface of the cylinder $\Sigma$, we also have
\begin{equation}
\frac{e^2}{h}\iint_{\Sigma} F(\varphi,z) d\varphi dz =\frac{e^2}{h}(A_{\varphi}(\pi)-A_{\varphi}(0)) =\frac{e^2}{h}C.
\end{equation}
The left-hand side can then be interpreted as a quantum Hall response (for an electron).

In Sec. \ref{Geometry}, we show a correspondence towards cube or rectangle geometries and also generalized boxes with a superellipse form in the plane. 

\subsection{Topological Response from Circularly Polarized Light: Quantized and Protected Photo-Electric Effect}
\label{lightdipole}

Here, we show that coupling an electric dipole to circularly polarized light gives rise to a quantized response at the poles of the sphere related to the square of the topological invariant \cite{C2}. Through this $C^2$, there is then a relation between the topological transversal current produced when driving from north to south poles in Eq. (\ref{Ekin}) and the responses to circularly polarized light. 

From definitions in Sec. \ref{electricfield}, we describe the effect of a time-dependent electric field 
\begin{equation}
{\bf E}_{\pm}=E_0 e^{-i\omega t} e^{ikz}({\bf e}_x \mp i {\bf e}_y)
\end{equation}
in the Jones formalism, producing an interaction with an atom or electric dipole located in the  $z=0$ plane \cite{Klein}
\begin{equation}
\label{energyshift}
\delta H_{\pm} = E_0 e^{\pm i\omega t} |a\rangle \langle b| +h.c. = E_0 e^{\pm i\omega t} \sigma^+ +h.c.
\end{equation}
Here, $|a\rangle$ and $|b\rangle$ refer to two discrete energy levels which can also be identified to the $|+\rangle_z$ and $|-\rangle_z$ states of a spin-$\frac{1}{2}$ particle and the light field is assumed to be classical here. Similarly, this Hamiltonian corresponds to a rotating magnetic field on a 
spin-$\frac{1}{2}$ particle similarly as in the nuclear magnetic resonance. Suppose the dipole or spin-$\frac{1}{2}$ is in the topological phase described through a radial magnetic field as in Sec. \ref{spin1/2}. Below, we describe the possibility of a topologically quantized photo-electric effect
from the poles of the sphere in the situation of resonance. The resonance situation is obtained from the transformation on states 
\begin{eqnarray}
|b\rangle &=& e^{\mp i \frac{\omega t}{2}}|b'\rangle \\ \nonumber
|a\rangle &=& e^{\pm i \frac{\omega t}{2}}|a'\rangle
\end{eqnarray}
 such that $E_b-E_a=\pm \hbar\omega$ for the $(\pm)$ polarization. Since we will evaluate real transition rates for the dipole induced by the right-handed circular polarization of light for instance, we can assume here that $E_0$ is real. When using this formalism to describe lattice models, the vector potential itself will couple to the pseudo-spin.

Then, we can equivalently re-write $\delta H_{\pm}$ in the basis of the eigenstates $|\psi_+\rangle$ and $|\psi_-\rangle$ introduced in Sec. \ref{spin1/2}  \cite{Klein}
\begin{eqnarray}
\label{deltaH}
\delta H_{\pm} &=& E_0 \sin\theta \cos(\varphi \pm \omega t)(|\psi_+\rangle \langle \psi_+|  - |\psi_-\rangle \langle \psi_-| )  \\ \nonumber
&-& E_0 e^{\pm i\omega t} e^{i\varphi} A'_{\varphi}(\theta>\theta_c)|\psi_+\rangle \langle \psi_-| +h.c. \\ \nonumber
&-& E_0 e^{\mp i\omega t} e^{-i\varphi} A'_{\varphi}(\theta<\theta_c) |\psi_+\rangle \langle \psi_-| +h.c.
\end{eqnarray}
The $+$ $(-)$ sign refers to the right-handed (left-handed) circularly polarized light. 
The smooth fields are identical to the ones in Eq. (\ref{smoothfields}).  The ground state is described through $|\psi_+\rangle$ with a topological number $C=1$. Here, we show a simple relation between the topological properties of the dipole or spin-$\frac{1}{2}$ and the response to this time-dependent electric field or magnetic field. 

For this purpose, we can apply the Fermi golden's rule approach and calculate the transition rate:
\begin{equation}
\Gamma_{\pm}(\omega)=\frac{2\pi}{\hbar}|\langle \psi_+ | \delta H_{\pm} |\psi_-\rangle |^2 \delta(E_b - E_a \mp\hbar\omega).
\label{rates}
\end{equation}
Due to the radial structure of the magnetic field on the surface of the sphere, the north pole will prominently interact with the right-handed wave such that $E_b-E_a=E_- - E_+ = +\hbar\omega>0$
and the south pole will prominently couple with the left-handed wave such that $E_b - E_a = E_+ - E_- = -\hbar\omega$. 
From the geometrical definitions above, then we obtain the time-independent result
\begin{equation}
\Gamma_{\pm}(\omega) = \frac{2\pi}{\hbar}E_0^2{\alpha}(\theta) \delta(E_b - E_a \mp\hbar\omega),
\end{equation}
with the function \cite{C2}
\begin{equation}
\label{alpha}
\alpha(\theta) = \left(\cos^4\frac{\theta}{2} +\sin^4\frac{\theta}{2}\right).
\end{equation}
The introduction of smooth fields $A'_{\varphi}$ allows then a simple interpretation of this function in terms of a topological response. Indeed, from the square of Eq. (\ref{CA'}) we have
\begin{equation}
\label{polarizationstheta}
\alpha(\theta) = C^2 +2A'_{\varphi}(\theta<\theta_c)A'_{\varphi}(\theta>\theta_c).
\end{equation}
Here, $C^2$ precisely refers to the square of the topological invariant defined through Stokes' theorem. Measuring the response at specific angles $\theta$ reproduces the global topological number $C=1$. For instance one can select the angles close to north and south poles such that $\theta\rightarrow 0$ and $\theta\rightarrow \pi$ respectively:
\begin{equation}
\label{alpha}
\alpha(0) = \alpha(\pi) = C^2.
\end{equation}
To the best of of our knowledge, this formula was not introduced earlier in the literature and the relation between the transition rates and $\alpha(\theta)$ gives a nice application to the functions $A'_{\varphi}(\theta<\theta_c)$ and $A'_{\varphi}(\theta>\theta_c)$.
In \ref{Berrycurvature}, we show that $\alpha(\theta)$ and the transition rates are also related to a new geometrical function and to the second-order correction in energy due to the light-matter interaction. In \ref{Berrycurvature}, we also show a link between the $\alpha(\theta)$ function and quantum metric related to the eigenstates in Eq. (\ref{eigenstates}).  We emphasize here that Eqs. (\ref{rates}) and (\ref{alpha}) are specific to the topological phase, a specific light polarization interacting with one pole only reflecting then the direction of the radial field structure. Transition towards the non-topological phase(s) will be addressed e.g. in Sec. \ref{Semenoff}.

It should be mentioned here that in the situation of a time-dependent linearly polarized perturbation, Eq. (\ref{polarizationstheta}) would acquire a dependence on the azimuthal angle for a fixed value of $\theta$, yet at the two poles the results in Eq. (\ref{alpha}) would be identical.

It is also interesting to discuss the response in the equatorial plane similarly as for the classical analysis of Sec. \ref{preliminaries}. In this situation, the structure of the eigenstates allow each circularly polarized light to interact with the system. The response in the equatorial plane can be written as
\begin{equation}
\label{onehalf}
\alpha\left(\frac{\pi}{2}\right) =  C^2 - \frac{1}{2}
\end{equation}
for each light polarization equally. This formula (\ref{alpha}) then can also be reformulated as
\begin{equation}
\alpha(\theta) = |\langle \psi_+ | \sigma_x | \psi_-\rangle|^2 + |\langle \psi_+ | \sigma_y | \psi_-\rangle|^2.
\end{equation}
The key point here is that $C^2$ can be measured from the evolution in time of the lowest energy-band population \cite{C2}, as developed in Sec. \ref{light} related to topological lattice models.  
The  response will take a similar form as for nuclear magnetic resonance with the topological response $\alpha(\theta)$ entering as a prefactor. A quantized photo-electric effect can then be observed close to the poles of the sphere and a one-half response is revealed at the equator for one circular polarization of light. This half response can be understood from the correspondence with the quantum metric; see Eq. (\ref{metrictheta}) in \ref{Berrycurvature}. 

This describes a topological protection of the photo-electric effect due to the presence of $C^2$ which may find practical applications related to the production of electronic quasiparticles and current through
a light field. In Sec. \ref{Semenoff}, we justify that topological properties close to the poles remain identical within the topological phase when including a uniform field $M\sigma_z$. 
Another interesting remark here is that this optical analogy to nuclear magnetic resonance may find practical applications in (medical) imaging. 

Related to this analysis, we also mention that recently a four-dimensional scaling for the atomic (dipole) polarisability \cite{Szabo} has been emphasized which then reveals that some interesting formulae may yet be identified in quantum physics.

\subsection{Topological Phase from Circularly Polarized Light}
\label{polarizationlight}

In fact, Eq. (\ref{energyshift}) also allows us to show that the time-dependent electric field is able to induce a topological phase for the electric dipole or spin-$\frac{1}{2}$ similarly as the classical analysis of Sec. \ref{electricfield}. 

Applying the unitary transformation $U_{\pm}(t) = e^{\mp i\frac{\omega t}{2}\sigma_z}$, the Hamiltonian in the rotating frame takes the form \cite{C2}
\begin{equation}
\delta \tilde{H}_{\pm} = U_{\pm} \delta H_{\pm} U_{\pm}^{-1} \pm \frac{\hbar\omega}{2}\sigma_z,
\end{equation}
with 
\begin{equation}
U_{\pm} \delta H_{\pm} U_{\pm}^{-1}  = E_0\sigma_x.
\end{equation}
The symbol $\pm$ refers to each circular polarization, right-handed $+$ and left-handed $-$ respectively related to ${\bf E}_{\pm}=\mp E_0{\bf e}_{\varphi}$ as defined in Sec. \ref{electricfield}. For the dipole (spin) Hamiltonian, we can also reach a time-independent form if we adjust the azimuthal angle as $\varphi = \mp \omega t$ and $U_{\pm} (- d_x\sigma_x - d_y\sigma_y) U_{\pm}^{-1} = -d\sin\theta \sigma_x$. If we include the presence of a topological term $-d_z\sigma_z$ with $d_z=d\cos\theta$ then we verify that a resonance can be obtained such that $\pm \frac{\hbar\omega}{2} = \pm d$ corresponding to the right-handed wave interacting with north pole and the left-handed wave interacting with south pole.  We can indeed adjust the angle $\varphi$ preserving the topological properties of the system since they only depend on the variable $\theta$. Assuming $E_0\ll d$, then the $E_0\sigma_x$ term will not modify the spin polarizations at the poles such that the topological number remains unity. This also justifies the perturbative analysis of \ref{lightdipole}. The term $E_0\sigma_x$ is precisely equivalent to the terms proportional to $|\psi_+\rangle\langle \psi_-|$ in Eq. (\ref{deltaH}) allowing to mediate inter-band transitions. 

From the rotating frame, we can verify that $C^2$ can be measured from the evolution in time of the lowest-energy band population \cite{C2}.

Suppose now that $d_z=0$ and that we modify the ${\bf d}$ vector acting on the Bloch sphere such that close to the poles
\begin{equation}
{\bf d} = d(\sin\theta \cos\varphi, \sin\theta \sin (\zeta \varphi),0),
\end{equation}
with $\zeta=1$ close to the north pole and $\zeta=-1$ close to the south pole. This situation will precisely correspond to the case of graphene discussed in Sec. \ref{spherelattice}. We discuss here an implementation of a light-induced topological phase on the Bloch sphere through the analogy with a radial magnetic field. It is useful to introduce the rotating frame through the transformation $U_{\pm}(t)=e^{\mp i \frac{\omega t}{2}\sigma_z}$ such that $U_{\pm} (- d_x\sigma_x - d_y\sigma_y) U_{\pm}^{-1} = -d\sin\theta e^{-i\zeta\varphi} e^{\mp i \omega t}\sigma^+ +h.c.$. The azimuthal angle then effectively rotates in different directions at the two poles. For a specific light polarization, then we can reach the rotating frame fixing $\zeta\varphi = \mp\omega t$ as in the preceding equation. The light-matter interaction turns into $E_0\sigma^+ e^{\pm i \omega t} e^{\mp i\omega t}+h.c.$. From the rotating frame, a resonance condition can be reached with the two poles of the sphere if the right-handed light polarization interacts with the north pole and the left-handed light polarization with the south pole. Physically, this implementation may be realized with a dipole or effective
spin $\frac{1}{2}$ interacting with an electric field in the plane at $z=0$ with two circular components such as ${\bf E}={\bf E}_+ + {\bf E}_- = E_0 e^{-i k z} e^{-i\omega t} ({\bf e}_x + i {\bf e}_y) + E_0 e^{ikz} e^{-i\omega t}({\bf e}_x - i {\bf e}_y)$. Here, we assume two symmetric beams propagating in opposite vertical direction and meeting on the plane $z=0$ such that the form of the electric field at $z=0$ precisely corresponds to the definitions in Sec. \ref{lightdipole}. 

In the vicinity of $\theta=0$ and $\pi$ the time-independent model in the rotated frame then reads
\begin{eqnarray}
\label{showmatrix}
H_{eff} =
\begin{pmatrix}
\frac{\hbar\omega}{2}\cos\theta & E_0 - d\sin\theta\\
E_0 - d\sin\theta &  -\frac{\hbar\omega}{2}\cos\theta \\
 \end{pmatrix}.\quad 
 \end{eqnarray}
 We observe that the induced dynamical term along one diagonal of the matrix is equivalent to a $d_z$ term. In the rotated basis, the effective ${\bf d}_{eff}$ vector acting on the spin in the vicinity of the poles can be written as
 \begin{equation}
 \label{E0field}
 {\bf d}_{eff} = \left((\tilde{d}+d)\sin\theta,0, -\frac{\hbar \omega}{2}\cos\theta\right),
 \end{equation}
 where $\tilde{d}\sin\theta=-E_0$. This produces an ellipse in the space of parameters. The induced gap between the bands $|\psi_+\rangle$ and $|\psi_-\rangle$ tends to $\hbar\omega$ at the poles if $E_0\ll \hbar\omega$.
 A similar conclusion can be reached in the $|\psi_+\rangle$ and $|\psi_-\rangle$ eigenstates representation
 such that
 \begin{equation}
\pm (|a\rangle\langle a|-|b\rangle\langle b|) =  \pm\cos\theta(|\psi_+\rangle \langle \psi_+| - |\psi_-\rangle \langle \psi_-|).
 \end{equation}
 Therefore, for the $+$ polarization, at the north pole the lowest-energy eigenstate is $|\psi_-\rangle=|b\rangle$ and at south pole, for the $-$ polarization, it becomes $|\psi_-\rangle=|a\rangle$ similarly as in the presence of a radial magnetic field. 
 
 In Ref. \cite{MoessnerCayssol}, a similar situation was discussed through two light polarizations with a vector potential ${\bf A}=-A_0(\sin(\omega t){\bf e}_x +\sin(\omega t -\phi){\bf e}_y)$ and $E_0=A_0\omega$ using
 a high-frequency Magnus development within the Floquet formalism \cite{GoldmanDalibard}.  The case $\phi=\frac{\pi}{2}$ can be precisely visualized as the superposition of two circular right- and left-handed polarizations.
For $e A_0 v_F \ll \hbar\omega$, this gives rise to a topological $d_z$ term of the form $-\frac{(e v_F A_0)^2}{\hbar\omega}(\sigma_z\zeta)$ close to the poles of the sphere. This
particular form of circularly polarized light field(s) leads to the induction of a topological semimetal in graphene \cite{Semimetal}.

\subsection{Dissipation and Radiation}

The topological number is generally a robust quantity towards perturbations. Here, we show the stability of the topological number towards dissipation effects coming from a bath or from an ensemble of bosons $a_i$ (harmonic modes) describing a resistance through a long transmission line \cite{Leggett,Weiss}. The Hamiltonian becomes
\begin{equation}
H = -{\bf d}\cdot \mathbfit{\sigma} +\sum_i \hbar\omega_i a^{\dagger}_i a_i +\sigma_z \sum_i \lambda_i(a_i+a^{\dagger}_i).
\end{equation}
We implicitly suppose that the ${\bf d}$ vector plays the role of a radial magnetic field. Fixing the azimuthal angle $\varphi=0$, then this model is equivalent to a spin-boson model characterized through the spectral function $J(\omega) = \pi\sum_i \lambda_i^2 \delta(\omega-\omega_i)$. A situation of particular interest is for ohmic dissipation where $J(\omega)=2\pi\alpha\omega e^{-\omega/\omega_c}$. As soon as $\alpha\neq 0$, there is an energy coupling between the system (spin) and the reservoir. We address the quantum limit at low temperature. The spin and reservoir become entangled characterized through the entropy $E= - p_+ \ln p_+ - p_- \ln p_-$ with $p_{\pm} = (1\pm \sqrt{\langle \sigma_x\rangle^2 +\langle \sigma_z\rangle^2})/2$. The Hamiltonian
is symmetric under the transformation $\sigma_y\rightarrow -\sigma_y$ implying $\langle \sigma_y\rangle=0$.

To show the stability of the topological number with respect towards $\alpha$, we can evaluate the ground state energy through Bethe Ansatz \cite{Ponomarenko} or perturbatively in the transverse field \cite{CedraschiButtiker,LeHur}. Using the identifications close to the north pole $h=-d_z=-d\cos\theta>0$ (with for instance $d<0$ and $\theta\rightarrow 0$, $h\ll \omega_c$) and $\Delta=d\sin\theta$ then the spin magnetization can be evaluated from the ground state energy of the total system $\langle \sigma_z\rangle = -\frac{\partial {\cal E}_g}{\partial h}$. This parameter $h$ is precisely introduced in Ref. \cite{LeHur} and 
should be distinguished from the Planck constant. This results in 
\begin{equation}
\langle \sigma_z\rangle = 1 - \frac{1}{2}\left(\frac{\Delta}{\omega_c}\right)^2 (1-2\alpha)\Gamma(1-2\alpha)\left(\frac{h}{\omega_c}\right)^{2\alpha-2}.
\end{equation}
To obtain the result at south pole this is equivalent to modify $h\rightarrow -h$ or $\sigma_z\rightarrow -\sigma_z$. We observe that as long as $\alpha<1$, the magnetizations at the poles remain indeed identical and stable. From Eq. (\ref{polesC}), then
this shows that the topological number remains identical. The point $\alpha=1$ means a Berezinsky-Kosterlitz-Thouless quantum phase transition \cite{Berezinskii,Berezinskii2,KT} where the topological number jumps to zero. This transition is also analogous to the quantum phase transition in the Kondo model \cite{Hewson}. Another way to visualize this instability is through the calculation of the Berry curvature $F_{\theta\varphi}$ which also shows a similar instability at $\alpha=1$ \cite{Henriet}. For $\alpha<1$, the global topological number remains quantized to unity. 

From entanglement entropy in the equatorial plane corresponding to $h\rightarrow 0$, $E$ reaches a maximum at $\alpha=\frac{1}{2}$ such that the topological number 
remains quantized even in the situation where $\alpha>\frac{1}{2}$ as long as $\alpha<1$, referring to the strong-coupling regime \cite{LeHur}. Related to the entanglement entropy $E$, we can also associate an effective
temperature induced by the coupling between spin and bath \cite{Williams}. Here, it is judicious to mention progress in measuring the entanglement entropy in many-body systems \cite{Harvard,Neill,Satoor}.
Rolling the spin on the sphere in the presence of a cavity or a bath can also be useful for practical (energy) applications as a quantum dynamo \cite{Henriet,EphraimCyril}. 
The produced coherent field in the bath shows a relation with the square of the topological number, dynamically, through the stocked energy.
A bath may be thought of as a resistance or long transmission line in a mesoscopic circuit \cite{CedraschiButtiker,LeHurBath} or a Bose-Einstein condensate \cite{Recati,Orth}.
A similar effect can be observed with a cavity or quantum electrodynamics circuit.

\subsection{Fractional Topological Entangled Geometry}
\label{fractionaltopology}

In the recent work \cite{HH}, we introduced the possibility of fractional topological entangled geometry from the curved space. Before elaborating on the model and on the proofs, we emphasize on the main properties related to this fractional topological entangled geometry. The occurrence of one-half topological numbers on the sphere defines the occurrence of quantum entanglement  at one pole \cite{OneHalfKLH}. Here, each sphere presents a coherent superposition of two (halved) surfaces, one participating in the flux production of the Berry curvature on the Riemann surface and one participating in the non-local spooky quantum correlations. The half flux quantization on each sphere is meaningful, stable and occurs as a result of the EPR pair formation similar to the half-flux quantum in a superconductor $\Phi_s$ produced by Cooper pairs in comparison to the metallic (normal) regime, $\Phi_s=\frac{\Phi_0}{2}$ with $\Phi_0=\frac{h}{e}$ \cite{Imry}. The correspondence is also similar to have two tori placed one on top of the other; the topological response for each subsystem becomes equivalent to having a reduced surface encircling a hole and the remaining surface participating in the quantum entanglement \cite{HH}.

A model to obtain fractional topological numbers can be formulated through two spins-$\frac{1}{2}$ as \cite{HH}
\begin{equation}
{H} =-({\bf d}_1\cdot{\mathbfit{\sigma}}_1 + {\bf{d}}_2\cdot{\mathbfit{\sigma}}_2) + r f(\theta){\mathbfit{\sigma}}_{1z}{\mathbfit{\sigma}}_{2z}. \label{eq:H}
\end{equation}
The magnetic field ${\bf d}_i$ acts on the same sphere parameterized by $(\theta,\varphi)$ and may be distorted along the $\hat{z}$ direction with the addition of the uniform field $M_i$ according to: 
\begin{equation}
\label{fieldsphere}
{\bf{d}}_i=(d\sin\theta\cos\varphi,d\sin\theta\sin\varphi,d\cos\theta+M_i),
\end{equation}
for $i=1,2$. Below, we analyze the properties of this model such that we have a tensor product state at one pole or a pure state (for instance, the north pole) and an entangled state at the other pole (for instance, at south pole) implying a symmetry between $1$ and $2$ in the Hamiltonian 
with ${\bf d}_1={\bf d}_2$. We will remind how to adjust the parameters to realize these prerequisites in the second paragraph below Eq. (\ref{J}) \cite{HH}. From the smooth fields and the general definitions in Sec. \ref{smooth} we define $C_j=A_{j\varphi}(\pi) - A_{j\varphi}(0)$ for each spin $i=1,2$ such that the Nabla or del operator is defined on each sphere locally through $\partial_{j\varphi}$ or equivalently $\partial_{1\varphi}\otimes \mathbb{I}$ and $\mathbb{I}\otimes \partial_{2\varphi}$ for the two spheres. 

The eigenstates of such an Hamiltonian can be written as 
\begin{eqnarray}
\label{wavefunction}
|\psi\rangle=\sum_{kl}c_{kl}(\theta)|\Phi_k(\varphi)\rangle_1|\Phi_l(\varphi)\rangle_2,
\end{eqnarray}
with $k,l=\pm$ and $|\Phi_+(\varphi)\rangle_j$ and $|\Phi_-(\varphi)\rangle_j$, defining the Hilbert space, can be chosen related to $|+\rangle_z$ and $|-\rangle_z$ introduced earlier in Sec. \ref{spin1/2} modulo a $\varphi$ phase corresponding to a particular gauge choice in the wavefunction. Here, we define 
$|\Phi_k(\varphi)\rangle_1|\Phi_l(\varphi)\rangle_2 = |\Phi_k(\varphi)\rangle_1\otimes|\Phi_l(\varphi)\rangle_2$. An important deduction from Sec. \ref{smooth} is that we can choose the same $\varphi$-gauge for the eigenstates in the north and south regions when defining the smooth fields $A'_{j\varphi}(\theta<\theta_c)$ and $A'_{j\varphi}(\theta>\theta_c)$ which have resulted in the equality $C_j=A_{j\varphi}(\pi) - A_{j\varphi}(0)$. This implies that we can introduce the same form of Hilbert space with the same phases-definitions in Eq. (\ref{wavefunction}) for all values of $\theta$. Since we can adiabatically tune the parameter $r$ this implies that we can also fix the same ``gauge'' for the definitions of $|\Phi_k(\varphi)\rangle_1$ and $|\Phi_l(\varphi)\rangle_2$ for all the phase diagram in the parameters space. We can then assume that the topological responses at the poles should be independent of $\varphi$ as all azimuthal angles are equivalent, $c_{kl}=c_{kl}(\theta)$.

Suppose we adjust the interaction $r$ such that the ground state evolves from a product state or pure state 
\begin{equation}
|\psi(0)\rangle = |\Phi_+\rangle_1|\Phi_+\rangle_2 = |\Phi_+\rangle_1\otimes |\Phi_+\rangle_2
\end{equation}
 at $\theta=0$ to an entangled state at $\theta=\pi$ represented by a Einstein-Podolsky-Rosen (EPR) pair \cite{EPR} or Bell pair \cite{Bell1964,Hagley,Aspect}
 \begin{equation}
 |\psi(\pi)\rangle=\frac{1}{\sqrt{2}}(|\Phi_+\rangle_1|\Phi_-\rangle_2+|\Phi_-\rangle_1|\Phi_+\rangle_2).
 \end{equation}
 Within our deductions, we can also incorporate the fact that when $r\rightarrow 0$ at south pole the corresponding wavefunction is $|\Phi_-\rangle_1 |\Phi_-\rangle_2$ (modulo a global phase independent of $\varphi$). 
In this way, we obtain \cite{OneHalfKLH}
\begin{equation}
\label{eq2}
A_{j\varphi}(\pi) = -i\langle \psi(\pi) | \partial_{j\varphi} |\psi(\pi)\rangle = \frac{A_{j\varphi}(0)}{2} + \frac{A^{r=0}_{j\varphi}(\pi)}{2},
\end{equation}
where $A^{r=0}_{j\varphi}(\pi)$ corresponds to the Berry curvature evaluated on $|\Phi_-\rangle_1 |\Phi_-\rangle_2$ equivalent to the state at south pole for $r=0$ (assuming that $M<d$).
Now, we can use the fact that for $r=0$, each sphere is in a topological phase such that
\begin{equation}
\label{eq1}
A^{r=0}_{j\varphi}(\pi) - A_{j\varphi}(0) = q =1
\end{equation}
and $q$ corresponds to the encircled topological charge on each sphere normalized to the charge $e$. We have $A_{j\varphi}(0)=A^{r=0}_{j\varphi}(0)=A^{r\neq 0}_{j\varphi}(0)$ since the wavefunction at the north pole remains identical if $r=0$ or $r\neq 0$. Plugging Eq. (\ref{eq1}) into (\ref{eq2}) then results in the identity 
\begin{equation}
\label{Aj}
A_{j\varphi}(\pi) - A_{j\varphi}(0) = q\frac{1}{2} = C_j,
\end{equation}
for the situation with an entangled wave-function at south pole. In this formula, the factor $\frac{1}{2}$ is related to the probability to be in a given quantum state $|\Phi_+\rangle_j$ or $|\Phi_-\rangle_j$ for the sub-system $j$ at south
pole, therefore it hides information on the entangled wavefunction. Through Sec. \ref{smooth}, we can also reformulate this equation as
\begin{equation}
\label{q2}
\frac{1}{2\pi}\iint_{S^2}\bm{\nabla}_j\times{\bf A}_j \cdot d^2{\bf s} = \frac{q}{2},
\end{equation}
with $d^2{\bf s}=d\theta d\varphi {\bf e}_r$, $\bm{\nabla}_1=\bm{\nabla}_1\otimes\mathbb{I}$ and $\bm{\nabla}_2=\mathbb{I}\otimes\bm{\nabla}_2$, revealing the half-flux quantization for a sphere.
The half surface does not necessarily correspond to a specific hemisphere of the two-spheres' model, yet the topological response of each sphere is equivalent to a half surface radiating the Berry curvature.

Measuring the quantum distance in \ref{Berrycurvature} through a circularly polarized field at south pole of a sphere reveals the same information as for one sphere at $r=0$ at the equator \cite{OneHalfKLH}. This is then similar as if we set 
$A'_{j\varphi}(\theta>\theta_c)=A_{j\varphi}(\frac{\pi}{2})-A_{j\varphi}(\pi)=0$ and $A'_{j\varphi}(\theta<\theta_c)=A_{j\varphi}(\frac{\pi}{2})-A_{j\varphi}(0)=q\frac{1}{2}$ from Stokes' theorem. It is then similar as if one circle attached to one hemisphere
encircles a topological charge $q$ similar to a half Skyrmion \cite{Meron}. In this way, this fractional topological number can also be viewed as the Euler characteristic related to a circle on top of a disk (describing a mirror) resulting in $\chi=0+1=1$ (which may be re-interpreted as $\chi=2-2g_{eff}$ with $g_{eff}=\frac{1}{2}$ on a unit sphere). Two-dimensional black-holes in a Schwarzschild space-time metric can be similarly described by an Euler characteristic $\chi=1$ as a result of a boundary for the system \cite{Gibbons}. It is interesting to observe developments relating the Euler characteristic of black holes with a topological interpretation of the Hawking-Bekenstein radiation \cite{Robson}. We emphasize here developments motivated by gauge fields, Aharanov-Bohm effect and Berry phase on the geometrical description of half monopoles \cite{DeguchiFujikawa}. 

A relation with the spin magnetization at the poles can be written similarly as in Eq. (\ref{polesC}). Indeed, we can re-write Eqs. (\ref{eq2}) and (\ref{eq1}) as $A_{j\varphi}(\pi) - A_{j\varphi}(0) = \frac{1}{4}(\langle \sigma^{r=0}_{jz}(0)\rangle
- \langle \sigma^{r=0}_{jz}(\pi)\rangle) = \frac{1}{2}\langle \sigma_{jz}^{r=0}(0)\rangle = \frac{1}{2}\langle \sigma_{jz}(0)\rangle$. Since we have $\langle \sigma_{jz}(\pi)\rangle = 0$  for the situation with the entangled wavefunction at south pole then we verify
\begin{equation}
\label{Cjspin}
C_ j = \frac{q}{2} = \frac{1}{2}(\langle \sigma_{jz}(0) \rangle - \langle \sigma_{jz}(\pi) \rangle).
\end{equation}
Eq. (\ref{Cjspin}) is important as it also represents a physical observable of the situation with $\frac{1}{2}$-topology. It can be verified when rolling the spin from north to south pole along a meridian line adiabatically. The above equations can also be re-written as $C_1=C_2=\frac{q}{2}$ and $C_1+C_2=q=\frac{1}{2}(\langle S_z(0)\rangle -\langle S_z(\pi)\rangle)$ where
$S_z=\sigma_{1z}+\sigma_{2z}$. The total Chern number can then be measured in the triplet sub-space of the two spins model where the wavefunctions at north and south poles find representations in the $S_z=+1$ and $S_z=0$ sectors. The correspondence between Eqs. (\ref{eq1}) and (\ref{Cjspin})
can also be understood from the relation
\begin{eqnarray} 
A_{1\varphi}(\theta) &=& \sum_l |c_{+l}(\theta)|^2 A_{1\varphi}(0) + \sum_l |c_{-l}(\theta)|^2 A_{1\varphi}^{r=0}(\pi), \nonumber \\
&=& - \frac{1}{2}\langle \sigma_{1z}(\theta)\rangle,
\end{eqnarray}
with $A_{j\varphi}(0)=-\frac{1}{2}$ and $A^{r=0}_{j\varphi}(\pi)=\frac{1}{2}$ if $q=1$. This allows for a local interpretation between $A_{j\varphi}$ and a physical observable. Topological properties are defined accordingly in a gauge-invariant way through differences of two fields.
Fixing ${\bf d}_1={\bf d}_2$ in the Hamiltonian then leads to $c_{+-}=c_{-+}$ such that $A_{1\varphi}=A_{2\varphi}$ and $\sigma_{1z}=\sigma_{2z}$.

Here, we formulate local relations between correlation functions and topological properties \cite{OneHalfKLH}. At north pole, on the one hand for the situation of fractional entangled geometry we have the identities $\langle \sigma_{1z}(0)\sigma_{2z}(0)\rangle = |c_{++}(0)|^2=\langle \sigma_{jz}(0)\rangle^2 = (2C_j)^2=q^2$ with $q=1$. This means that the information on the topological charge can propagate in a thin cylinder from the equatorial plane to the north pole. At south pole, we can also write down $\langle \sigma_{1z}(\pi)\sigma_{2z}(\pi)\rangle = |c_{++}(\pi)|^2 + |c_{--}(\pi)|^2 - |c_{+-}(\pi)|^2 - |c_{-+}(\pi)|^2=1-2(|c_{+-}(\pi)|^2 + |c_{-+}(\pi)|^2)$ equivalent to the Bell correlation function \cite{Aspect} if we measure the two sub-systems in $z$ direction. In the second equality, we have invoked the normalization of the wavefunction at south pole. Now, using the fact that for the entangled wavefunction at south pole we have $c_{++}(\pi)=c_{--}(\pi)=0$ then the normalization of the wavefunction at north and south poles also implies $|c_{++}(0)|^2 = |c_{+-}(\pi)|^2 + |c_{-+}(\pi)|^2$ which 
results in the local identities at south pole
\begin{equation}
\label{correlation}
\langle \sigma_{1z}(\pi)\sigma_{2z}(\pi)\rangle = -(2C_j)^2=1 - 2(2C_j)^2 =-1.
\end{equation}
There is then a link between the entangled wavefunction and the $\frac{1}{2}$ topological number.
It is important to emphasize here that to show this equality we have used the formula $C_j = A_{j\varphi}(\pi) - A_{j\varphi}(0) =  \frac{1}{2}\langle \sigma_{jz}(0)\rangle$ which is only applicable in the situation where we have an entangled state around south pole implying $C_j=\frac{1}{2}$. 
For an angle $\theta$, we have a local relation between geometry, local spin observables and correlation functions. Indeed, $A'_{\varphi}(\theta<\theta_c) = -\frac{1}{2}\langle \sigma_{jz}(\theta)\rangle +\frac{1}{2} = |c_{--}(\theta)|^2 + |c_{+-}(\theta)|^2$.
We also have $A'_{\varphi}(\theta>\theta_c) = -\frac{1}{2}\langle \sigma_{jz}(\theta)\rangle = -\frac{1}{2}|c_{++}(\theta)|^2 + \frac{1}{2}|c_{--}(\theta)|^2$ and $\langle \sigma_{1z}(\theta)\sigma_{2z}(\theta)\rangle = |c_{++}(\theta)|^2 + |c_{--}(\theta)|^2 -2|c_{+-}(\theta)|^2$.
This way, for $C_j=\frac{1}{2}$, we also have
\begin{equation}
2|c_{+-}(\theta)|^2 = C_j -\frac{1}{2}\langle \sigma_{1z}(\theta)\sigma_{2z}(\theta)\rangle.
\end{equation}

We can also introduce a relation with an observable defining entanglement such as the bipartite fluctuations \cite{Song}. For many-body systems such as free fermions, we have shown a precise equality between many-body entanglement and the full counting statistics associated to an observable such as charge or spin magnetization (along $z$) \cite{FluctuationRelation}.  For the EPR pair or Bell pair, from an exact series relating entanglement entropies from the full counting statistics associated to the charge or spin magnetization along $z$ direction in a bipartite partitioning, this requires in principle a finite number of charge cumulants to reproduce the $\ln 2$ entropy. On the other hand, the second cumulant defined as bipartite fluctuations and corresponding here to the variance of the spin magnetization for one spin at south pole already contains the same information as the entanglement entropy \cite{WittenCourse} in terms of probabilities to be in the $\sigma_{jz}=+1$ or $\sigma_{jz}=-1$ compared to the measure at north pole referring to a pure state along the same axis. The variance on a spin magnetization measure reveals the information
\begin{eqnarray}
F_1 &=& \langle \psi (\pi)| \sigma_{1z}^2\otimes \mathbb{I} |\psi(\pi)\rangle - \langle \psi(\pi) | \sigma_{1z}\otimes \mathbb{I} |\psi(\pi)\rangle^2, \nonumber \\
F_2 &=& \langle \psi(\pi)| \mathbb{I}\otimes \sigma_{2z}^2 |\psi(\pi)\rangle - \langle \psi(\pi) | \mathbb{I}\otimes \sigma_{2z} |\psi(\pi)\rangle^2.
\end{eqnarray}
These bipartite fluctuations are defined positively through a measure along $z$ direction for each sub-system and are symmetric $F_1=F_2$. 
For the case of Bell or EPR pair at south pole, $F_1=F_2$ takes the maximum value $1$. The first term is $1$ since the matrix is equivalent to the identity matrix.
Then, we have the simple relation
\begin{equation}
\label{topoquant}
F_1=F_2=|\langle \sigma_{1z}(\pi)\sigma_{2z}(\pi)\rangle| = (2C_j)^2=1.
\end{equation}
Here, $F_1=F_2=1$ reveals the quantum uncertainty as a result of the formation of the entangled wavefunction through $\langle \psi(\pi) | \sigma_{jz}|\psi(\pi)\rangle=0$. For a pure state occurring at north pole for $r=0$, we have $F_1=F_2=1-|c_{++}(0)|^2=0$ 
and at south pole $F_{j} = |c_{++}(0)|^2 = (2C_j)^2=q^2=1$. Measuring $C_j=\frac{1}{2}$ then may be defined as a measure of entanglement in the situation of spheres with radial magnetic fields. We insist here on the fact that $|\langle \sigma_{1x}(\pi)\sigma_{2x}(\pi)\rangle|$ is also equal to $1$ for the Bell or EPR pair at $\theta=\pi$. If we define the bipartite fluctuations $F_i$ along the $x$ axis, then they would equally give $F_i=1$ at the two poles. At $\theta=\pi$, this reveals entanglement and at $\theta=0$ this would reveal the lack of information if we measure the pure state (polarized along $z$ axis) along $x$ direction instead.  Similar to probabilities in the entropy, bipartite fluctuations must be defined along $z$ direction with the reference of a pure state polarized along $z$ direction (at $\theta=0$). Here, $F_i$ is also equal to $1$ at the equator but then drops for $\theta>\frac{\pi}{2}$ and becomes equal to $1$ again in the close vicinity of $\theta=\pi$ when the EPR pair forms.
Fluctuations and compressibility theorems were introduced by J. Bell related to the physics of superconductors in 1963 \cite{Bell1963}. Quantum entanglement also plays a crucial role in properties of solids through correlation functions \cite{Brukner}.

We can then interpret $C_j=\frac{1}{2}$ from geometrical and transport properties \cite{OneHalfKLH}. When the boundary is at $\theta_c=0^+$, from Sec. \ref{smooth} we obtain the equation
\begin{equation}
\label{transportCj}
- \oint A'_{j\varphi}(0^+) d\varphi = 2\pi C_j.
\end{equation}
Here, we use the correspondence for $\theta_c=0$: $A'_{j\varphi}(0^-)=A'_{j\varphi}(\theta<\theta_c)=A_{j\varphi}(0)-A_{j\varphi}(0)=0$ and $A'_{j\varphi}(0^+) = A'_{j\varphi}(\theta>\theta_c)=A_{j\varphi}(0)-A_{j\varphi}(\pi)=-C_j$. In this way, each sphere now presents a total $\pi$ Berry phase transported around the north pole of the sphere. These equations are also equivalent to say that in this case the Dirac string is equivalent to a pair of $\pi$ winding numbers, related to Fig. \ref{Edges.pdf}. The total system presents a Berry phase of $2\pi$ equivalent to a Chern number $C_1+C_2=1$. Now, we can re-interpret the left-hand side of Eq. (\ref{transportCj}) in the sense of transport on a cylinder geometry. Eq. (\ref{transportCj}) is equivalent to a current moving along the edge defined here for a unit time
\begin{equation}
\label{J}
J_{\perp}^j = \frac{e}{2\pi}\oint \psi^*(0^+) i \partial_{j\varphi} \psi(0^+)d\varphi = eC_j.
\end{equation}
Here, $\psi(0^+)$ refers to the two-particles' wavefunction at the top of the cylinder. This gives rise to a half current compared to one sphere. 

The $\frac{1}{2}$ factor multiplying the charge $e$ may also be understood as follows, as a result of entanglement. The charge at $\theta_c=0^+$ should be in the appropriate quantum state $|\Phi_+\rangle$ for each sphere or each cylinder. In Sec. \ref{ParsevalPlancherel}, we have introduced the Newton equation and de Broglie principle which implies to follow the motion of a particle with charge $e$ and with a fixed spin quantum number along the path to define the pumped current in a transverse direction for an angle $\theta=\pi\rightarrow \theta_c=0^+$. Therefore, to transmit a charge $e$ in this protocol from south pole along the path this effectively corresponds to project $|\psi(\pi)\rangle$ onto either $\frac{1}{\sqrt{2}}|\Phi_+\rangle_1|\Phi_-\rangle_2$ or $\frac{1}{\sqrt{2}}|\Phi_-\rangle_1|\Phi_+\rangle_2$ with a probability $\frac{1}{2}$. The key point of the gauge-invariant argument in Eq. (\ref{J}) is that whatever the choice of the projected state, the produced edge current at the north pole will be halved compared to that of one sphere. Similarly as the interpretation of the topological number as $q\frac{1}{2}$, Eq. (\ref{J}) can be thought of as $e\frac{1}{2}$ where $\frac{1}{2}$ refers to the entangled wavefunction and projection protocols required by the measure. A similar interpretation justifies $C_j=\frac{1}{2}$ in Eq. (\ref{conductance}) at the edges on the cylinder leading to a halved quantum Hall response. This projection measure on one sphere at south pole can be understood through a half Berry phase. This is then another mechanism leading to fractional 
transport, compared to Laughlin states for fractional quantum Hall systems; see \Ref{LaughlinQHResponse}. 

\begin{figure}[t]
\begin{center}
\includegraphics[width=0.6\textwidth]{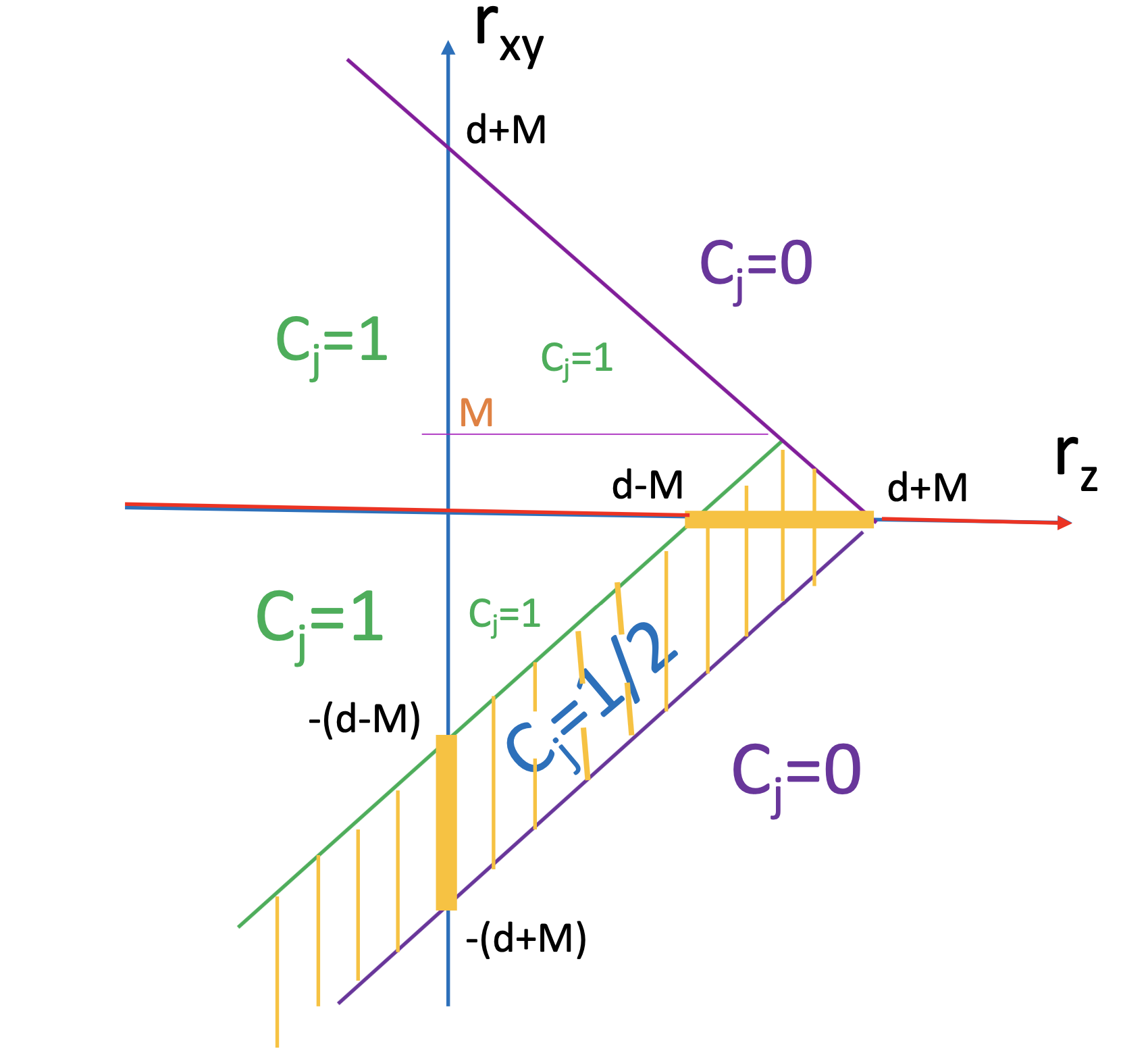} 
\caption{Phase diagram of two spheres with an Ising interaction $r_z=r$ and an interaction $r_{xy}$ in the $xy$ plane when we respect the $\mathbb{Z}_2$ symmetry $1\leftrightarrow 2$ implying that $M_1=M_2=M$. Three phases occur with $C_j=1$, $C_j=0$ and $C_j=\frac{1}{2}$.}
\label{Phasesrxyrz}
\end{center}
\end{figure}

Here, we emphasize on the parameters related to Eq. (\ref{eq:H}) to realize this fractional topological entangled geometry.  
This situation can be precisely realized through a fine-tuning of the interaction $r$ for the situation of $\mathbb{Z}_2$ symmetry characterizing the $1\leftrightarrow 2$ symmetry in the presence of a global field $M_1=M_2=M$ \cite{HH}. The ground state at north pole with $\theta=0$ is $|\Phi_{+}\rangle_1|\Phi_{+}\rangle_2$ provided that $r f(0)<d+M$. At south pole with $\theta=\pi$, the ground state is $|\Phi_-\rangle_1|\Phi_-\rangle_2$ for $rf(\pi)<d-M$, but it is degenerate between the anti-aligned configurations for $rf(\pi)>d-M$. In that case, the presence of the transverse fields in the Hamiltonian along the path from north to south poles will then produce the analogue of resonating valence bonds. Indeed, from the south pole, the second-order perturbation theory induces an effective Ising coupling similar to 
\begin{equation}
\label{HamiltonianIsing}
H_{eff} = - \frac{d^2 \sin^2\theta}{r}\sigma_{1x}\sigma_{2x}
\end{equation}
(assuming here $r\gg d-M$) which will then effectively favor the $S_z=0$ sector of the triplet state. The curved space here is important on the one hand to ensure that the two classical antiferromagnetic states are degenerate at south pole and on the other hand to induce locally entanglement around the south pole. As a result, we obtain half-integer Chern numbers. Furthermore, for the simple constant interaction $f(\theta)=1$, this occurs within the range 
\begin{equation}
\label{HM}
d-M<r<d+M.
\end{equation}
At south pole, the states $\frac{1}{\sqrt{2}}(|\Phi_+\rangle_1|\Phi_-\rangle_2 + |\Phi_-\rangle_1|\Phi_+\rangle_2)$ and   $\frac{1}{\sqrt{2}}(|\Phi_+\rangle_1|\Phi_-\rangle_2 - |\Phi_-\rangle_1|\Phi_+\rangle_2)$ have energies
$-r -\frac{d^2\sin^2\theta}{r}$ and $-r + \frac{d^2\sin^2\theta}{r}$.  Driving from north to south pole allows then for the measure of the topological number of each sphere according to Eq. (\ref{drive}) and (\ref{Cjspin}). Adjusting the angle $\theta=vt$ in time is equivalent to apply an electric field along the polar direction as  in Sec. \ref{ParsevalPlancherel}. In addition, we have verified Eq. (\ref{Aj}) numerically which reveals a clear signature of the $q\frac{1}{2}$ topological number from the poles in particular for the limit where $M\rightarrow 0^+$ with $r\rightarrow q$. In that case, the information at the poles is yet meaningful and in fact characterizes the topological properties of the system.

Including a ferromagnetic $xy$ interaction $r_{xy}<0$ between spins $2 r_{xy}(\sigma_1^{+}\sigma_2^- + \sigma_1^{-}\sigma_2^+)$ enhances the energy gap between these two states and therefore reinforces the occurrence of such a phenomenon.  At north pole, the $r_{xy}$ interaction does not change the ground state to first order in perturbation theory.  This ensures a certain stability of the phase with $C_j=\frac{1}{2}$ on the two spheres which is also important for practical applications. It is interesting to observe that even in the situation where $r_z=0$ then the fractional
topological state $C_j=\frac{1}{2}$ can yet occur when $-d-M<r_{xy}<-|d-M|$ \cite{HH}. The phase diagram in the plane $(r_{xy},r_z)$ then shows a stable and prominent dashed region with $C_j=\frac{1}{2}$, see Fig. \ref{Phasesrxyrz}. In particular, as long as we respect the $\mathbb{Z}_2$ $1 \leftrightarrow 2$ symmetry and Eq. (\ref{HM}) is satisfied then the $C_j=\frac{1}{2}$ number occurs.

\begin{table}[t]
\caption{Symbols related to the $C_j=\frac{1}{2}$ fractional phase} 
\centering 
\begin{tabular}{||c|c||} 
\hline\hline 
Smooth Fields and Observables & \hskip 0.5cm Definitions \\  
\hline 
$C_j$ & $\frac{1}{2\pi}\iint \bm{\nabla}\times{\bf A}'_j\cdot  d^2{\bf s}=\frac{q}{2}=\frac{1}{2}$\\
$C_j$ & $A_{j\varphi}(\pi)-A_{j\varphi}(0)=\frac{q}{2}=\frac{1}{2}$ \\
$C_j$ & $\frac{1}{2}\left(\langle \sigma_{jz}(0)\rangle-\langle \sigma_{jz}(\pi)\rangle\right)=\frac{q}{2}$\\ 
$\langle \sigma_{1z}(\pi)\sigma_{2z}(\pi)\rangle=\langle 2i\alpha_2\eta_1\rangle$ & $-(2C_j)^2=-1$ \\
$\chi_j$ & $2-2C_j=0+1$ \\
$G_j$, $\sigma_{xy}^j$ & $\frac{1}{2}\frac{e^2}{h}$ \\
$-\oint A'_{j\varphi}(0^+) d\varphi$ & $2\pi C_j$ \\
\hline %
\end{tabular}
\label{tableII} %
\end{table}

The proofs of Eqs. (\ref{Aj}) and (\ref{q2}) can be generalized to three spheres and multiple spheres simply from the fact that we assume a resonating valence state at one pole. We give further detail in Sec. \ref{GRVBT}
to show the usefulness and simplicity of the formalism related to a special class of wavefunctions allowing resonance between all possible states with one bound state forming a domain wall in a classical antiferromagnet at south pole as a result of frustration in a ring geometry with a odd number of spins.  For instance, for three spheres with ground state wave functions $|\psi(0)\rangle = \Pi_{i=1}^3 |\Phi_+\rangle_i$ and $|\psi(\pi)\rangle=\frac{1}{\sqrt{3}}(|\Phi_+\rangle_1|\Phi_-\rangle_2|\Phi_-\rangle_3 + |\Phi_-\rangle_1|\Phi_+\rangle_2|\Phi_-\rangle_3+|\Phi_-\rangle_1|\Phi_-\rangle_2|\Phi_+\rangle_3)$,
we obtain $C_j = \frac{2}{3}q$.  Each sphere is yet in a quantum superposition of a topological surface participating in the flux production of the Berry curvature and in a quantum entangled region. These generalized fractions are verified numerically when driving from north to south poles through Eq. (\ref{Cjspin}) \cite{HH}.

In Table \ref{tableII}, we summarize quantum symbols related to the $C_j=\frac{1}{2}$ phase including a correspondence towards Majorana fermions through $\langle \sigma_{1z}(\pi)\sigma_{2z}(\pi)\rangle=\langle 2i\alpha_2\eta_1\rangle$ as developed in Sec. \ref{pwavewire}.

\subsection{Mesoscopic Engineering of $C_j=\frac{1}{2}$}
\label{Mesoscopic}

There are certainly several ways to realize these two spins-$\frac{1}{2}$ models in atomic and mesoscopic systems. The topological number(s) for two spheres has been measured in Ref. \cite{Roushan} through mesoscopic quantum circuits.  In this system, the authors have built a two-spins' model coupled through an $xy$ interaction from g-mons qubits and the $M$ parameter was activated on one spin only such that the fractional topological state $C_j=\frac{1}{2}$ remains to be revealed. This existing platform shows that it is possible to observe the fractional topological state with current technology assuming that a term $-M\sigma_{jz}$ acts symmetrically on the two spheres \cite{HH}. Here, we propose an alternative double-dot device to engineer the $C_j=\frac{1}{2}$ fractional topological state, from charge qubits. An antiferromagnetic $r_z$ interaction and a ferromagnetic $r_{xy}$ interaction may be both engineered. Each dot is a superconducting charge qubit such that $\sigma_{iz}$ measures locally the charge on each individual island. We assume that the coherence time of the system is sufficiently long such that we can measure the charge response (activating the angular parameter $\theta=\omega t\in [0;\pi]$ on the sphere) in real time.  For an introduction on charge qubits, see \cite{SPEC,Nakamura,Yale}. 

The system is built from two superconducting islands or charge qubits \cite{Heij}. The precise mapping between spin-$\frac{1}{2}$ and the charge operator is such that $\sigma_{jz} = 2\hat{q}_j -1$ with $\hat{q}_j=0$ and $1$ corresponding to the presence of zero or one additional Cooper pair on each island compared to the charge neutrality situation. This situation can be achieved by varying a (global) gate voltage capacitively coupled to each island in the plane. This corresponds to engineer the parameter $M$ symmetrically on the two islands in Eq. (\ref{fieldsphere}). The presence of a Coulomb interaction $E_c(\hat{q}_j - n)^2$, with $\hat{q}_j$ counting the number of additional Cooper pairs and $2 E_c n=V_g$ on each island, is essential to invoke the charge-spin correspondence. The system will function close to a charge degeneracy point with $n=\frac{1}{2}+\delta n$ on each dot and $E_c \delta n=M$ such that $\delta n\ll 1$ to have $0$ or $1$ additional Cooper pairs. The charging energy $E_c$ is then the dominant energy scale for the mesoscopic system implying also that $k_B T\ll E_c$
with $T$ the temperature and $k_B$ the Boltzmann constant. Below, we will fix $M$ to satisfy the prerequisite shown in Eq. (\ref{HM}) and realize the one-half topological number on each sphere. The term $-d\cos\theta\sigma_{jz}$ can be implemented as an additional AC gate voltage $-V_0\cos(\omega t)$ to the DC gate voltage or $M$ parameter such that $d=\frac{V_0}{2}$ and the polar angle on the Bloch sphere then reads $\theta=\omega t$. For $M=d=0$, the charge states $0$ and $1$ are perfectly degenerate on each island. 

To implement the transverse field acting on each island, we can proceed similarly as for superconducting charge qubits coupled to a superconducting reservoir (from the top) via Josephson junctions (usually sketched as  ``crosses''). The transfer of one Cooper pair from the reservoir
to each island is implemented through a term $\Delta \sigma_{j}^{+}+h.c.$ \cite{Yale}. Formally here we should interpret $\Delta=\langle b\rangle$ where $b$ corresponds to transmit one Cooper pair from the reservoir into an island. The subtle step here is to implement
the time-dependent prefactor $e^{i\omega t}$ such that the transverse field also depends on the angle $\theta$ playing the role of the polar angle on the sphere. The Cooper pairs in the superfluid reservoir are described by the Hamiltonian
\begin{equation}
H_{SF}= (V(t)+E_0)b^{\dagger} b + w \sum_{i} (b\sigma_{j}^{+}+h.c.)
\end{equation}
 with $E_0$ the energy for the superfluid fraction, $V(t)$ is a potential term produced from a battery attached to the superconducting reservoir and $w$ represents the hopping of Cooper pairs from the superconducting substrate to each island. In the literature \cite{Yale}, it is common to modify $b$ by $\langle b\rangle$ in the coupling term $w$. From the equations of motion similarly as in quantum mechanics, the potential $V(t)$ can be absorbed as a time-dependent phase shift for the bosons or Cooper pairs describing the Cooper pairs $\langle b\rangle(t) = \langle b\rangle e^{\frac{i}{\hbar}\int_0^{t} V(t')dt'}$. The key point will be to tune $V(t)=V=\hbar\omega$ such that the coupling between the superfluid reservoir and each island takes the required form $\tilde{w} e^{i\omega t}\sigma_{i}^{+}+h.c.$ with $\tilde{w}=w\langle b\rangle$.
Adjusting $\tilde{\omega}=d$ realizes Eq. (\ref{fieldsphere}) with a particular choice of azimuthal angle $\varphi=\frac{\pi}{4}$. To measure the topological number when driving from north to south pole, any meridian line is equivalent. Deviating from $\tilde{\omega}=d$ smoothly modifies the sphere into an ellipse preserving the same topological properties from the poles.

Here, we address the implementation of the Ising interaction $r\sigma_{1z}\sigma_{2z}$ with $r>0$ and of a ferromagnetic $XY$ interaction $r_{xy}\sigma_{1}^+\sigma_2^- +h.c.$ with $r_{xy}<0$, which will also require to adjust the parameter $M$.
We assume that the capacitive coupling is prominent compared to the tunnel coupling between quantum dots. The mutual capacitance between islands gives rise to a term proportional to $r(\sigma_{1z}+1)(\sigma_{2z}+1)$. In addition to the required Ising interaction, we observe the occurrence of an additional $r\sigma_{jz}$ term which may be absorbed in the gate voltage such that $M\rightarrow M'=(M-r)$. At the north pole, the additional gate potential term $-\sum_{j=1,2} M'\sigma_{jz}$ helps to stabilize the ground state $(1,1)$ if $M'>0$; a state $(q_1,q_2)$ counts the number of charges on the dot $1$ and on the dot $2$ with respect to the electrostatic charge neutrality situation. To realize the fractional topological state then we must also adjust $d-M'<r<d+M'$ with $d-M'>0$ to ensure a Dirac monopole in each sphere. 
To stabilize a dominant ferromagnetic interaction $r_{xy}$ between islands we add a cavity, for instance, inductively or capacitively coupled to the two dots (in the plane) through a Jaynes-Cummings Hamiltonian \cite{JaynesCummings} $H_{JC} = \hbar\omega_a a^{\dagger} a + g\sum_{i=1,2} a^{\dagger}\sigma_{j}^{-} +h.c.$. The fact that the two islands couple to the same cavity mode produces a ferromagnetic $XY$ interaction which can be visualized by completing the square such that $a\rightarrow \tilde{a}=a+ \frac{g}{\hbar\omega_a}\sum_{j=1,2}\sigma_{i}^-$ in the limit where $g\ll \hbar\omega$ then maintaining the commutation relations for the harmonic oscillator.
Then, 
\begin{equation}
H_{JC} = \hbar\omega_a \tilde{a}^{\dagger}\tilde{a} - \frac{g^2}{\hbar \omega_a} \sigma_1^{+}\sigma_2^{-}+h.c.. 
\end{equation}
Therefore, the coupling $- \frac{g^2}{\hbar \omega_a}$ gives rise to a ferromagnetic $XY$ interaction that would help stabilizing the fractional topological state with $C_j=\frac{1}{2}$, the wavefunction $\frac{1}{\sqrt{2}}(|10\rangle + |01\rangle)$ being stabilized around the south pole. 
Here, we define $|q_1 q_2\rangle = |q_1\rangle |q_2\rangle = |q_1\rangle\otimes |q_2\rangle$. The sign of $r_{xy}$ is important to decrease the energy of the entangled state at south pole. This induced coupling from a cavity and the occurrence of the eigenstate $\frac{1}{\sqrt{2}}(|10\rangle + |01\rangle)$ is observed in mesoscopic circuits \cite{Majer}.

The detection of the topological number for an island requires to measure the evolution of the (average) charge in real time according to Eq. (\ref{Cjspin}). The charge response to an AC gate voltage has engendered recent technological progress \cite{Filippone}.
We also mention here recent efforts in double-dot graphene systems to tune the system from
the $(1,1)$ state to the charge degeneracy region comprising the $(1,0)$ and $(0,1)$ states by increasing the power of a microwave cavity \cite{Deng}. 

These facts together with the recent progress realized in Ref. \cite{Roushan} suggest that it is possible to observe
the fractional topological numbers, address the relation to entanglement properties and also engineer entangled states through smooth gates in real time (here activated by the linear evolution in time of  $\theta=\omega t$). 

\subsection{Robustness of Geometry through Superellipses}
\label{Geometry}

Here, we emphasize on the stability of the geometry from the poles through the formula $C=A_{\varphi}(\pi)-A_{\varphi}(0)$. The geometry can be adjusted from a Gabriel Lam\' e curve or superellipse
in the plane. Let us proceed similarly as for the cylinder geometry in Sec. \ref{cylinderformalism} and use the dressed coordinates $(r,\varphi,z)$ such that $z=\cos\theta$ with $\theta$ referring to the polar angle in spherical coordinates and $\varphi$ represents the angle in the plane. We can adjust the applied magnetic field as
\begin{equation}
{\bf d} = d(R^{-\frac{1}{n}}|x|^{\frac{1}{n}}, R^{-\frac{1}{n}}|y|^{\frac{1}{n}}, \cos\theta).
\end{equation}
In the equatorial plane, the parameters' space is defined as
\begin{equation}
d_x^2+d_y^2 = d^2 R^{-\frac{2}{n}} (|x|^{\frac{2}{n}} + |y|^{\frac{2}{n}}).
\end{equation}
Suppose that we study trajectories for the spin-$\frac{1}{2}$ with constant energy implying that $d_x^2+d_y^2=d^2$ then this leads to
the equation of a superellipse of the form
\begin{equation}
\label{superellipse}
|x|^\frac{2}{n} + |y|^\frac{2}{n} = R^{\frac{2}{n}}.
\end{equation}
This is the equation of a Lam\' e curve or superellipse which can now be solved through (for $n$ even)
\begin{eqnarray}
x &=& R\left(\cos\varphi\right)^n \hbox{sgn}(\cos\varphi) \\ \nonumber
y &=& R\left(\sin\varphi\right)^n \hbox{sgn}(\sin\varphi).
\end{eqnarray}
and $x=R(\cos\varphi)^n$, $y=R(\sin\varphi)^n$ for $n$ odd.

The case $n=1$ with $R=1$ corresponds to the unit sphere with the equation of  a circle in the equatorial plane.
Starting from the equatorial plane we build a surface delimited by the equation in (\ref{superellipse}) and the height $H=2$. Then, $z=\frac{H}{2}=+1$ corresponds to $\theta=0$ and $z=-\frac{H}{2}=-1$ corresponds to $\theta=\pi$. The cylinder geometry of Sec. \ref{cylinderformalism} corresponds to the special case $n=1$ where the structure in the plane is a circle of unit radius $R=1$.  We can identify the Berry curvature $F(\varphi,z)=-\partial_z A_{\varphi} = \frac{1}{2}$ implying that $A_{\varphi}=-\frac{z}{2}$ with $z=\cos\theta$ from the sphere. The Berry curvature is oriented in the plane such that only the vertical region of the surface participates in the topological properties. At the top surface we have $A_{\varphi}(0^+)=-\frac{1}{2}$ and at the bottom surface we have $A_{\varphi}(\pi^-)=\frac{1}{2}$ such that the Berry curvature is zero similar to an open cylinder geometry allowing then to describe the formation of edges at the top and bottom of the vertical region. 

We can adjust the parameter $R$ for a given value of $n$ to reproduce $\frac{C}{2\pi}=1$ as in Sec. \ref{cylinderformalism}. This is precisely the identity to obtain a topological charge unity in the core of the geometry.
For $n=2$, the system acquires the same topology as a cube with a topological charge inside. In that case, we have the identification $4\sqrt{2}R=2\pi$ when adjusting the length $\oint \sqrt{dx^2+dy^2}$ of the edge of the rhombus in the plane to the perimeter of the circle forming the structure of the cylinder geometry for $n=1$. For $n=3$, then we have an astroid in the plane and yet we can adjust $3R=2\pi$ to obtain a topological charge inside the three-dimensional geometry. 

An important deduction from Sec. \ref{cylinderformalism} is that we can apply a difference of potentials $V_t-V_b$ such that within the same travelling time $T=\frac{h}{2qE}$ for a charge to navigate from north to south pole
we will observe the same quantized conductance $G=\frac{q^2}{h}C$ associated to the two edge currents according to Eq. (\ref{conductance}). A similar observation is applicable for the quantum Hall response on the cylinder.

Also, since the vertical direction is $z=\cos\theta$ in all these geometries similarly as for the
unit sphere we can then include interactions between two of those geometries and from the top and the bottom regions realize similarly a fractional topological state as in Sec. \ref{fractionaltopology}. 

\section{Quantum Anomalous Hall Effect}
\label{anomalous}

Here, we show the applicability of this formalism to topological lattice models from the honeycomb lattice. We begin the discussion with a situation in the presence of dispersive topological Bloch energy bands, the Haldane model \cite{Haldane}, referring to the quantum anomalous Hall effect. A quantum Hall effect driven by a uniform magnetic field, initially observed in typical MOSFET structures \cite{Hall}, can equally arise on the honeycomb lattice \cite{Zhang,Novoselov} and will be addressed in the next Section related to the light-matter coupling and quantum transport. These models belong to class $A$ in classification tables following E. Cartan notations; see for instance Refs. \cite{BernevigNeupert,RyuTable,KitaevBottTable,AltlandZirnbauer,Zirnbauer1,Zirnbauer2}. 

\begin{figure}[ht]\centering
\includegraphics[width=0.3\textwidth]{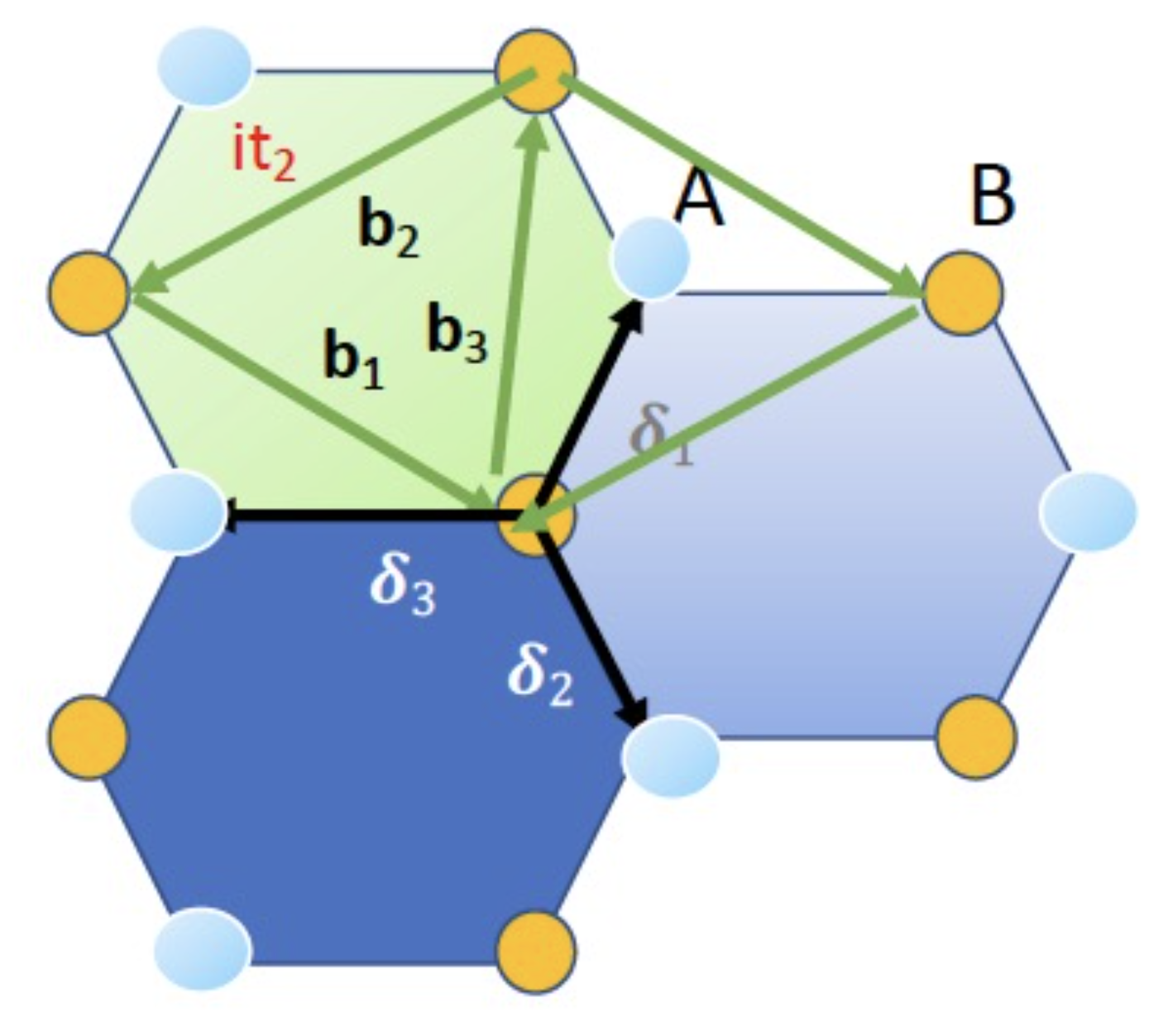} 
\includegraphics[width=0.3\textwidth]{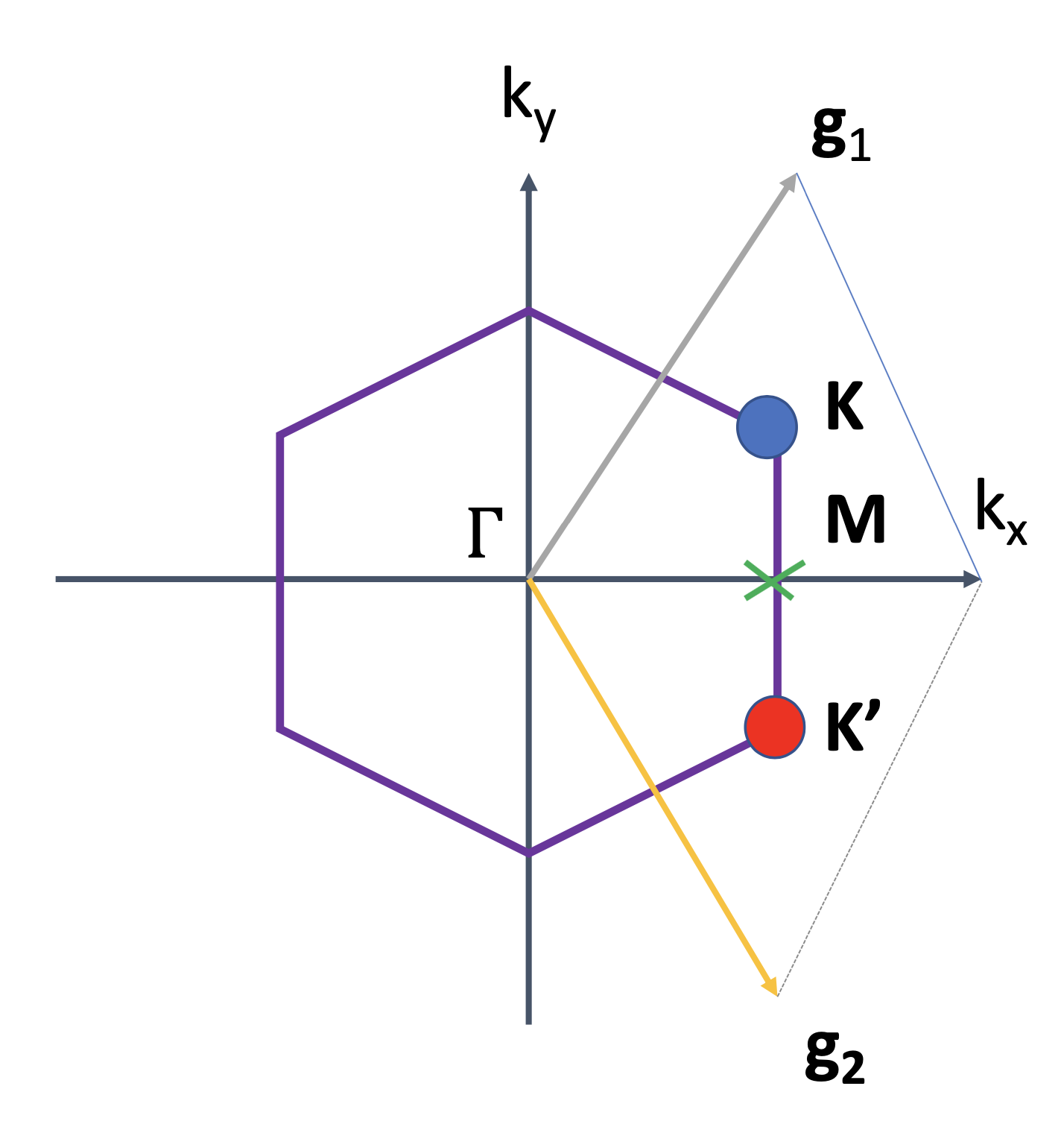} 
\caption{(Left) Honeycomb lattice in real space with the two sublattices $A$ and $B$ and two Bravais lattice vectors ${\bf b}_1$ and $-{\bf b}_2$ defining a triangular lattice. The vectors $\mathbfit{\delta}_i$ characterize the hopping of particles on nearest-neighbors' sites giving rise to the Dirac cones in the energy spectrum. The Haldane model also includes the $it_2$ hopping term (with $t_2\in \mathbb{R}$) between the second nearest neighbors. (Right) Brillouin zone. Here, we have represented one $M$ (high-symmetry) point. A high-symmetry point in the Brillouin zone satisfies the translation symmetry with respect to a reciprocal lattice vector $-{\bf \Gamma}_i = {\bf \Gamma}_i +{\bf G}$ with ${\bf G}=n_1{\bf g}_1 +n_2{\bf g}_2$ and $(n_1,n_2)\in\mathbb{Z}$.}
\label{graphenefig}
\end{figure}

\subsection{Honeycomb lattice, Graphene and Realization of a Topological Phase}
\label{spherelattice}

The orbital $2p_z$ of the carbon atom leads effectively to a one-band model on the honeycomb lattice as a result of the $sp^2$ hybridization. The system is half-filled if the chemical potential is  zero. Its band structure was calculated by P. R. Wallace in 1947 \cite{Wallace} related to three-dimensional graphite. The physics of one graphene plane (layer) has attracted tremendous attention in the community \cite{graphene}. The honeycomb lattice is interesting on its own due to the relation with the Dirac equation. The graphene is a semimetal presenting two inequivalent $K$ and $K'$ Dirac points in its Brillouin zone
at zero energy. Introducing $K=(\frac{2\pi}{3a},\frac{2\pi}{3\sqrt{3}a})$ and $K'=(\frac{2\pi}{3a},-\frac{2\pi}{3\sqrt{3}a})$ with $a$ the lattice spacing in real space, the Hamiltonian of the tight-binding model can be written similarly as a spin-$\frac{1}{2}$ particle such that $H=\sum_{{\bf p}}\Psi^{\dagger}({\bf p})H({\bf p})\Psi({\bf p})$ with $\Psi({\bf p})=(c_{A{\bf p}},c_{B{\bf p}})$ related to electron operators acting on sublattice $A$ or $B$. 

Close to the two Dirac points, the model takes a linear form
\begin{eqnarray}
\label{Kspectrum}
H({\bf p}) &=& \hbar v_F(p_x\sigma_x + \zeta p_y\sigma_y),
\end{eqnarray}
with $\sigma_x$ and $\sigma_y$ referring to Pauli matrices acting on the Hilbert space formed by a dipole $|A,B\rangle$ between two nearest neighbors. The Pauli matrix $\sigma_z$ plays the role of a pseudo-spin magnetization measuring the relative occupancy on each sublattice. Here, we introduce ${\bf p}=(p_x,p_y)$ as a wavevector measured from each Dirac point such that the Fermi velocity reads $v_F=\frac{3}{2\hbar}ta$ with $t$ the hopping amplitude between nearest neighbors defined through the vectors $\mathbfit{\delta}_j$ equivalently written in terms of the Bravais (triangular) lattice vectors $\mathbfit{\delta}_1-\mathbfit{\delta}_3=-{\bf b}_2$ and $\mathbfit{\delta}_2-\mathbfit{\delta}_3={\bf b}_1$ of the triangular lattice. The ${\bf b}_j$ vectors on Fig. \ref{graphenefig} are then defined as ${\bf b}_1 = \frac{a}{2}(3,-\sqrt{3})$, ${\bf b}_2 = -\frac{a}{2}(3,\sqrt{3})$ and ${\bf b}_3 = (0,\sqrt{3}a)$. The sum on ${\bf b}_j$ includes 6 second nearest-neighbors which can be added two by two similarly to the square lattice. Here and hereafter, we will apply the definition that $\zeta=\pm$ at the $K$ and $K'$ points respectively. The energy spectrum is linear $E({\bf p})=\pm \hbar v_F|{\bf p}|$ close to a Dirac point and through the identification $p_x=|{\bf p}|\cos\tilde{\varphi}$ and $p_y=|{\bf p}|\sin\tilde{\varphi}$ this gives rise to a cone structure centered around $K$ and $K'$.

The presence of a linear $-i\hbar\bm{\nabla}$ operator implies the presence of positive and negative energy bands referring to occupied particle band and empty hole bands for the ground state at zero temperature at half-filling and linking matter with anti-matter. This is an application of the Dirac equation (in two dimensions) 
\begin{equation}
i\hbar\gamma^{\mu}\partial_{\mu}\psi-mc\psi=0, 
\end{equation}
where Clifford matrices, usually called $\gamma^{\mu}$, are Pauli matrices. Here, we have $\gamma_1=\sigma_x$ and $\gamma_2=\sigma_y$. The particles  --- for instance, the electrons in graphene --- play the role of massless Weyl particles (with mass $m=0$) and $v_F$ can be seen as the speed of light $c$ (being 300 times smaller for graphene). When $m=0$, the two Dirac points show opposite $\pm \pi$ Berry phases \cite{graphene}. 

Here, we introduce a class of topological models where the system becomes an insulator, with the opening of an energy gap through the occurrence of a term $-m\zeta \sigma_z$, as in Sec. \ref{polarizationlight}. The energy spectrum at the two Dirac points turns into $E=\pm\sqrt{(\hbar v_F)^2|{\bf p}|^2+m^2}$ such that the gap $m$ can also be identified to a mass in the relativistic sense defining $m=m^*c^2$ with here $c=v_F$. 

\begin{figure}[ht]\centering
\includegraphics[width=0.5\textwidth]{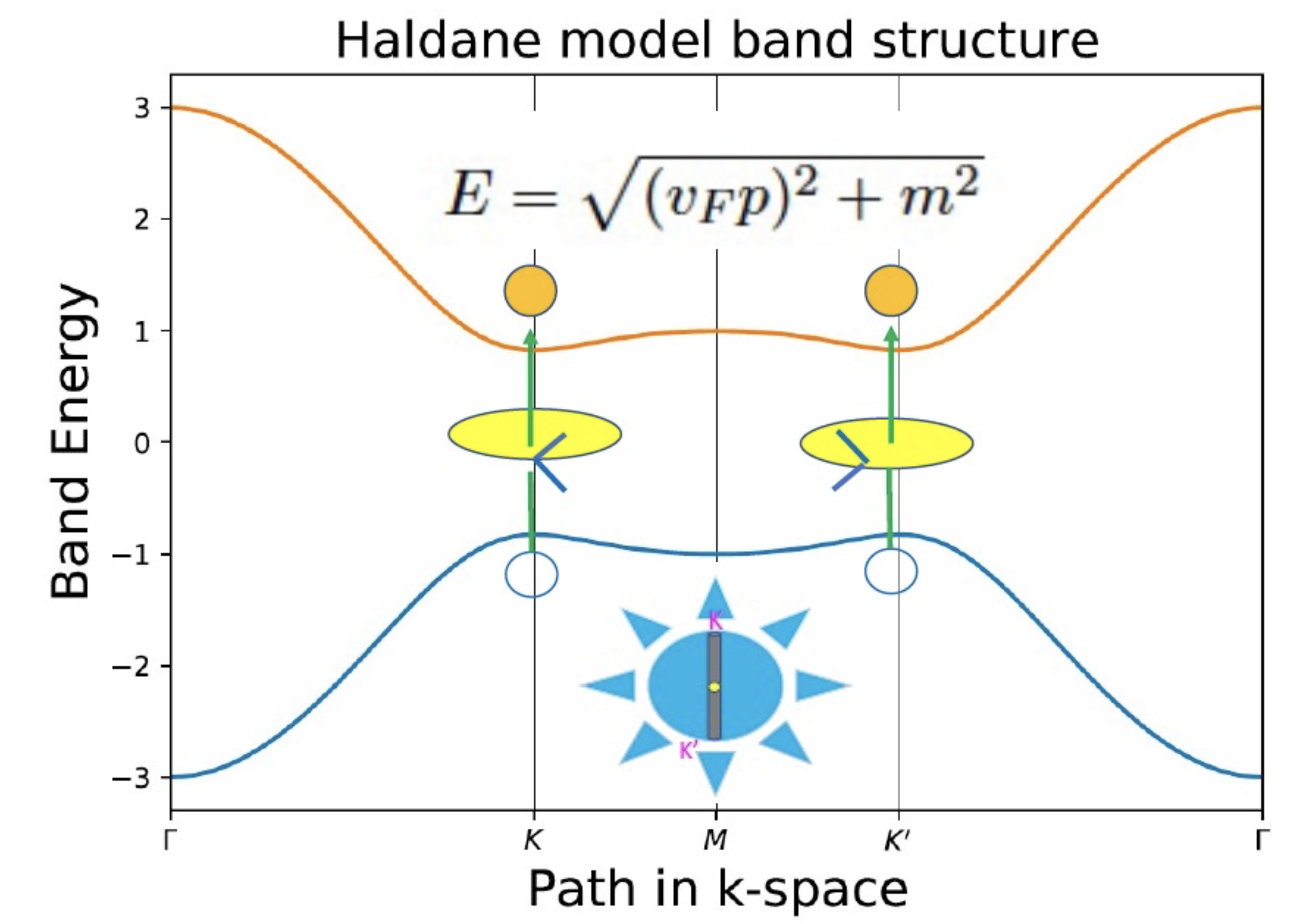} 
\hskip 0.5cm
\includegraphics[width=0.45\textwidth]{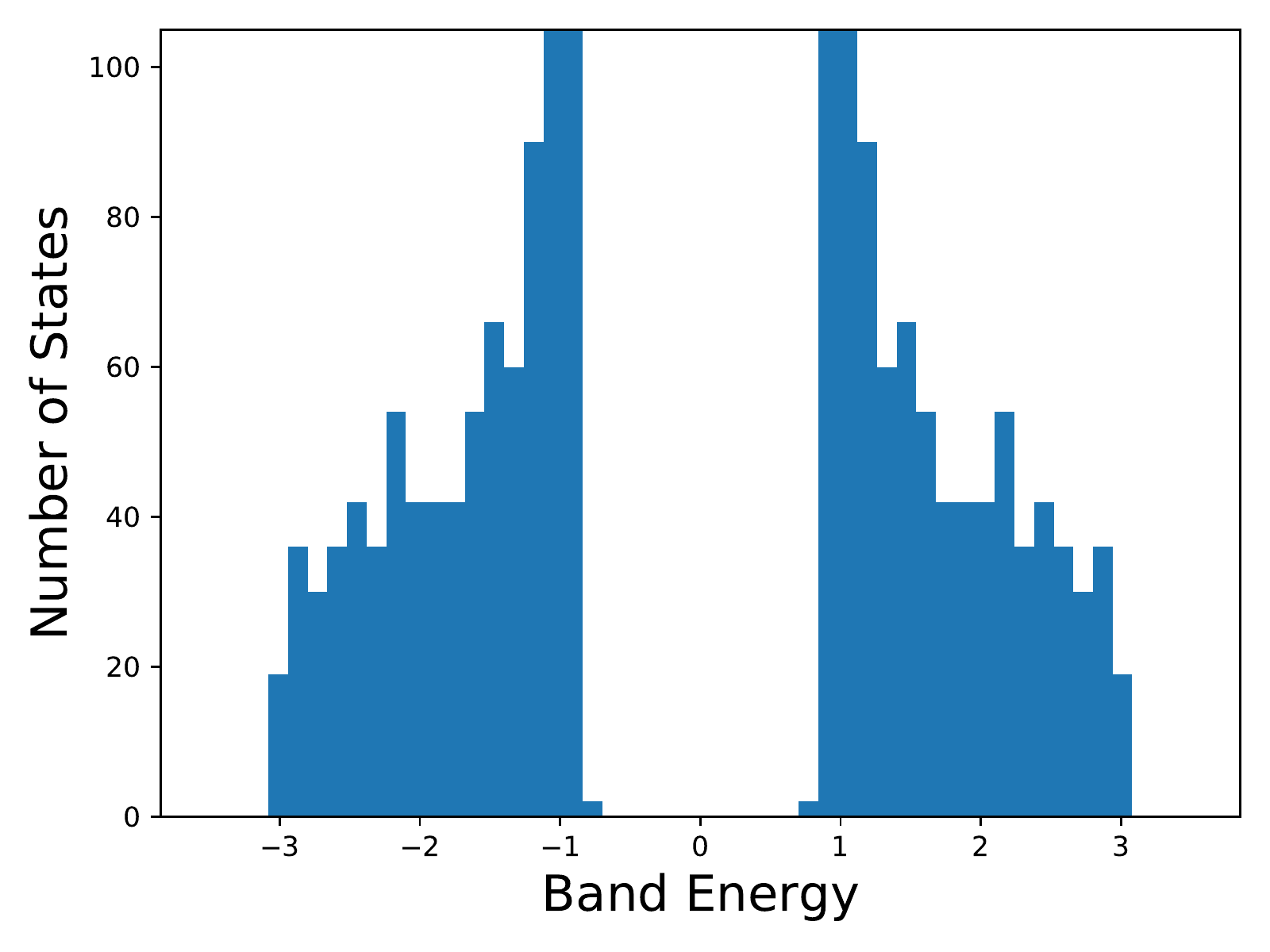}
\caption{(Top) Energy spectrum for the Haldane model with $t_2=0.15$ and $t=1$ (with $\hbar=1$). We also present the protocol showing the effect of each circular polarization of light at the two Dirac points related to Sec. \ref{light}. We analyze the topological and physical responses through the Bloch sphere. (Bottom) Density of states showing the importance of states close to $E=\pm m$ at the Dirac points for these parameters.}
\label{Haldanespectrum.pdf}
\end{figure}

In this way, we observe a simple identification with the Bloch sphere formalism and
radial magnetic field
\begin{eqnarray}
\label{correspondence}
&& -d(\cos\varphi\sin\theta, \sin\varphi \sin\theta, \cos\theta) = \\ \nonumber
&& (\hbar v_F|{\bf p}|\cos\tilde{\varphi},\hbar v_F|{\bf p}|\sin(\zeta\tilde{\varphi}),-\zeta m).
\end{eqnarray}
Choosing $d,m>0$ and $d=m$, we can then identify the $K$ Dirac point at north pole of the sphere with $\theta=0$ and the $K'$ Dirac point at south pole with $\theta=\pi$. Close to the north pole or $K$ point, we can now relate the polar angle associated to the cone geometry in the energy band structure $\tilde{\varphi}$ with the azimuthal angle $\varphi$ of the sphere, such that $\tilde{\varphi}=\varphi\pm \pi$. 
The Dirac cone at the $K$ point is now centered at the north pole with 
\begin{equation}
\label{tan}
\tan\theta = \frac{\hbar v_F|{\bf p}|}{m}. 
\end{equation}
Close to south pole corresponding to the $K'$ point, since $\zeta$ flips its sign this requires to modify $\varphi\rightarrow -\varphi$ and we also have $m\rightarrow -m$. 
In this way, this tight-binding model is equivalent to the topological Bloch sphere with an effective radial magnetic field and the `mass' term $-m\zeta\sigma_z$. The key point in the formalism is that the smooth fields defined in Sec. \ref{smooth} are invariant under the modification $\varphi\rightarrow -\varphi$. This allows us to conclude that this insulator is described from its ground state (through $|\psi_+\rangle$ on the sphere) by a topological invariant $C=+1$. 

Here, it is important to emphasize that the situation of graphene corresponding to $m=t_2=0$ is special since $\tan\theta\rightarrow \pm\infty$ for any infinitesimal $v_F|{\bf p}|$ around the two Dirac points. Then we will have formally the two Dirac cones in the equatorial plane (with $C=0$).

\subsection{Haldane Model and Quantum TopoMetry}
\label{topometrylattice}

Here, we relate the physics associated to the term $-\zeta m \sigma_z$ with the Haldane model \cite{Haldane}. 

We introduce the second-nearest-neighbor term $t_2 e^{i\phi}$ from the definitions of Fig. \ref{graphenefig} (with $t_2$ real and to illustrate simply the effect we fix $\phi=\frac{\pi}{2}$) such that on a green triangle in Fig. \ref{graphenefig} the phase
accumulated is non-zero. The important point here is to have a complex phase attached to the $t_2$ hopping term. In this discussion, particles or electrons are assumed to be spin-polarized.

If we invert the direction on a path, then we should modify $\phi\rightarrow -\phi$. If a wave performs a closed path on a honeycomb cell or ring A-B-A-B-A-B-A, since the Peierls phase associated with the nearest neighbor term is zero assuming $t\in\mathbb{R}$, we conclude that the total phase acquired is zero. From Stokes theorem, on a honeycomb cell we infer that $\iint {\bf B}\cdot {\bf n} d^2 s =0$ with ${\bf B}$ the induced magnetic field from the vector potential. We are in a case where on each triangular lattice formed by each sub-lattice the magnetic flux $\frac{3\pi}{2}$ (defined for simplicity in unit of the phase $\frac{\pi}{2}$) is staggered. Therefore, we can check that on a honeycomb ring on the lattice, the total magnetic flux is $\frac{3\pi}{2}-\frac{3\pi}{2}=0$. Locally, the magnetic fluxes on a triangle can yet induce a topological phase.
For $\phi=\frac{\pi}{2}$, this gives rise to an additional term in the Hamiltonian
\begin{equation}
\label{t2term}
H_{t_2}({\bf k}) = H_{t_2}^A  + H_{t_2}^B =  -\sum_{\bf k}\sum_{{\bf b}_j} 2 t_2 \sin({\bf k}\cdot{\bf b}_j) \sigma_z.
\end{equation}
The phase $\phi\neq 0$ is important in this proof to produce a odd function under the parity transformation ${\bf k}\rightarrow -{\bf k}$ and going from $B$ to $A$ is equivalent to modify ${\bf b}_j\rightarrow -{\bf b}_j$. 
For the general situation of arbitrary values of $\phi$, this results in an additional $\sin \phi$ prefactor in Eq. (\ref{t2term}).
We can write the Haldane model as an effective spin-$\frac{1}{2}$ model 
\begin{eqnarray}
\label{dvectorzone}
{\bf d}({\bf k}) = \left(t\sum_{\mathbfit{\delta}_i} \cos({\bf k}\cdot \mathbfit{\delta}_i), t\sum_{\mathbfit{\delta}_i}\sin({\bf k}\cdot \mathbfit{\delta}_i), 2 t_2 \sum_{{\bf b}_j} \sin({\bf k}\cdot{\bf b}_j)\right).
\label{dvector}
\end{eqnarray}
Each value of ${\bf k}$ corresponds to a point on the sphere $(\theta,\varphi)$. Using the form of ${\bf b}_j$ vectors, then we identify
\begin{equation}
\label{hz}
d_z({\bf K}) = 2t_2 \sum_{{\bf b}_j} \sin({\bf K}\cdot {\bf b_j}) = 3\sqrt{3}t_2 =m.
\end{equation}

Each term gives the same contribution since $\sin({\bf K}\cdot{\bf b}_1)=\sin(\frac{2\pi}{3})=\frac{\sqrt{3}}{2}$, $\sin(K_y b_3)=\sin(\frac{2\pi}{3})$ and $\sin({\bf K}\cdot{\bf b}_2)=\sin(-\frac{4\pi}{3})=\frac{\sqrt{3}}{2}$. All the $K'$ points in the Brillouin zone have the same properties. From the form of $d_z$, changing ${\bf k}\rightarrow -{\bf k}$, then we find $d_z({\bf K}')=-m$. We verify this specific form from the identity $\sin(K_j' b_j)=-\sin(K_j b_j)$. The $d_z$ term then is equivalent to $\zeta m$ close to the two Dirac points with $\zeta=\pm 1$ at the $K$ and $K'$ points respectively. From the preceding Sec. \ref{spherelattice}, this allows us to conclude that the $d_z$ term opens a gap at the Fermi energy and that the lowest-energy band is characterized through a topological number $C=+1$. Also, through Eq. (\ref{polesC}) the topological number can be visualized as the addition of the Berry phases around the Dirac points. On the lattice, $(2\pi)(A_{\varphi}(\pi)-A_{\varphi}(0))$ precisely corresponds to the addition of Berry phases at the two Dirac points. The relative minus sign between $A_{\varphi}(\pi)$ and $A_{\varphi}(0)$ takes into account that close to the south pole we have redefined $\varphi\rightarrow -\varphi$ according to the discussion around Eq. (\ref{correspondence}).

We emphasize here that the poles of the sphere play a special role in the proof of Sec. \ref{smooth}. Indeed, we have defined the topological number on $S^{2'}$ subtracting the two poles on the surface of the sphere where ${\bf A}'=0$. Also, ${\bf A}$ is uniquely defined at these points and smooth (if we fix a $\varphi$-representation of the Hilbert space) through the identification 
\begin{equation}
C = \frac{1}{2\pi}\iint_{S^{2'}} \bm{\nabla}\times{\bf A}'\cdot  d^2{\bf s} = \frac{1}{2\pi}\iint_{S^{2'}} \bm{\nabla}\times{\bf A}\cdot  d^2{\bf s}.
\end{equation}
This has precisely resulted in the identification $C=A_{\varphi}(\pi)-A_{\varphi}(0)$, which is then defined to be gauge invariant from the formulation of Stokes' theorem, when moving progressively the boundary towards one of the two poles.  For a comparison, in graphene the sum of the Berry phases around the two Dirac points is zero \cite{graphene,bilayerQSH}. The peculiarity of the present approach when applying Stokes' theorem with two domains is that, compared to previous approaches \cite{Kohmoto}, the Berry field ${\bf A}$ is continuous at the interface and the discontinuity is absorbed in the definition of ${\bf A}'$. In \ref{TopometryBandTheory}, we introduce the quantum topometry formalism within the band theory and elaborate on the relations between the $M$ point and Dirac points with the sphere.

\begin{figure}[ht]\centering
\includegraphics[width=0.45\textwidth]{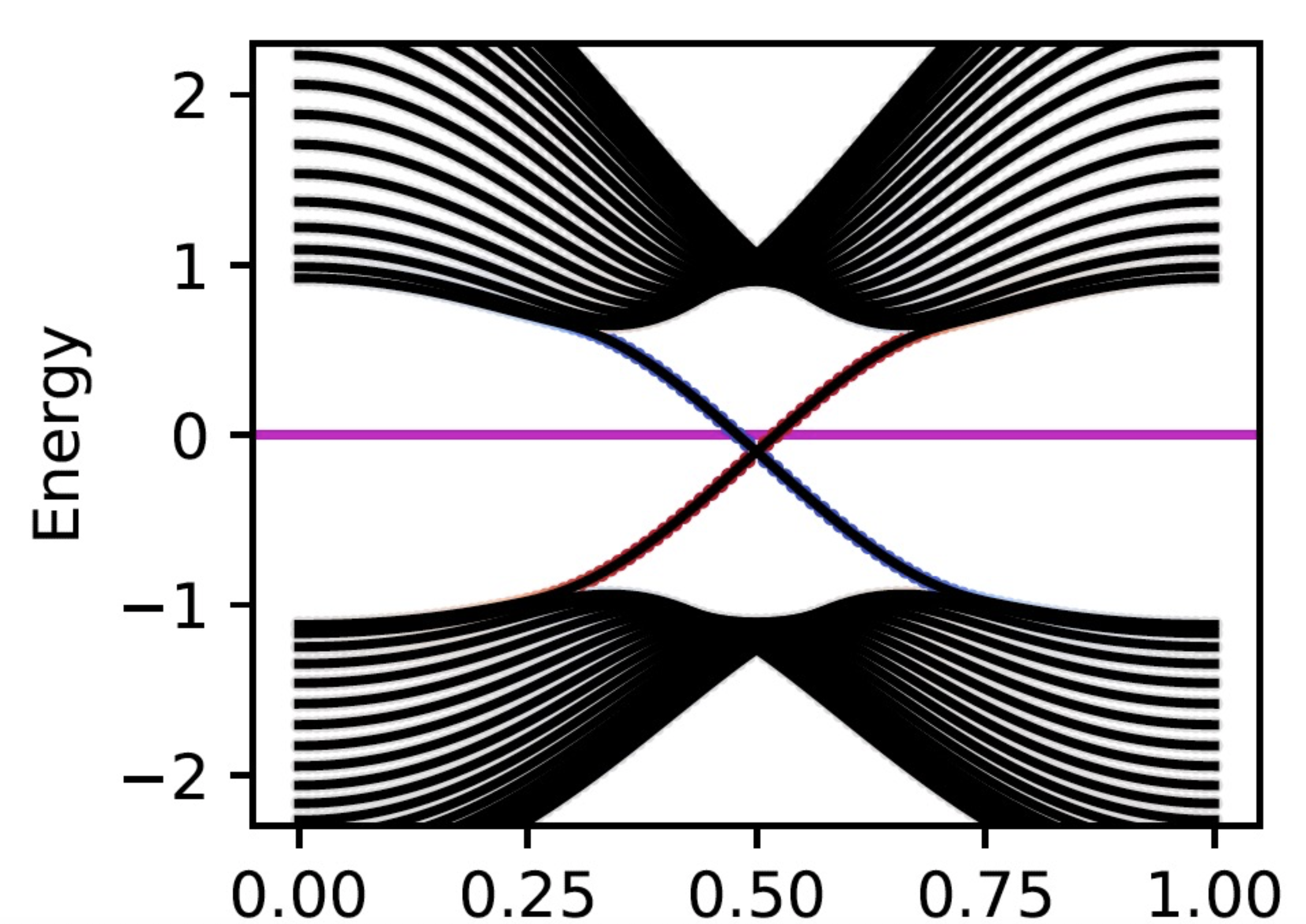}
\includegraphics[width=0.45\textwidth]{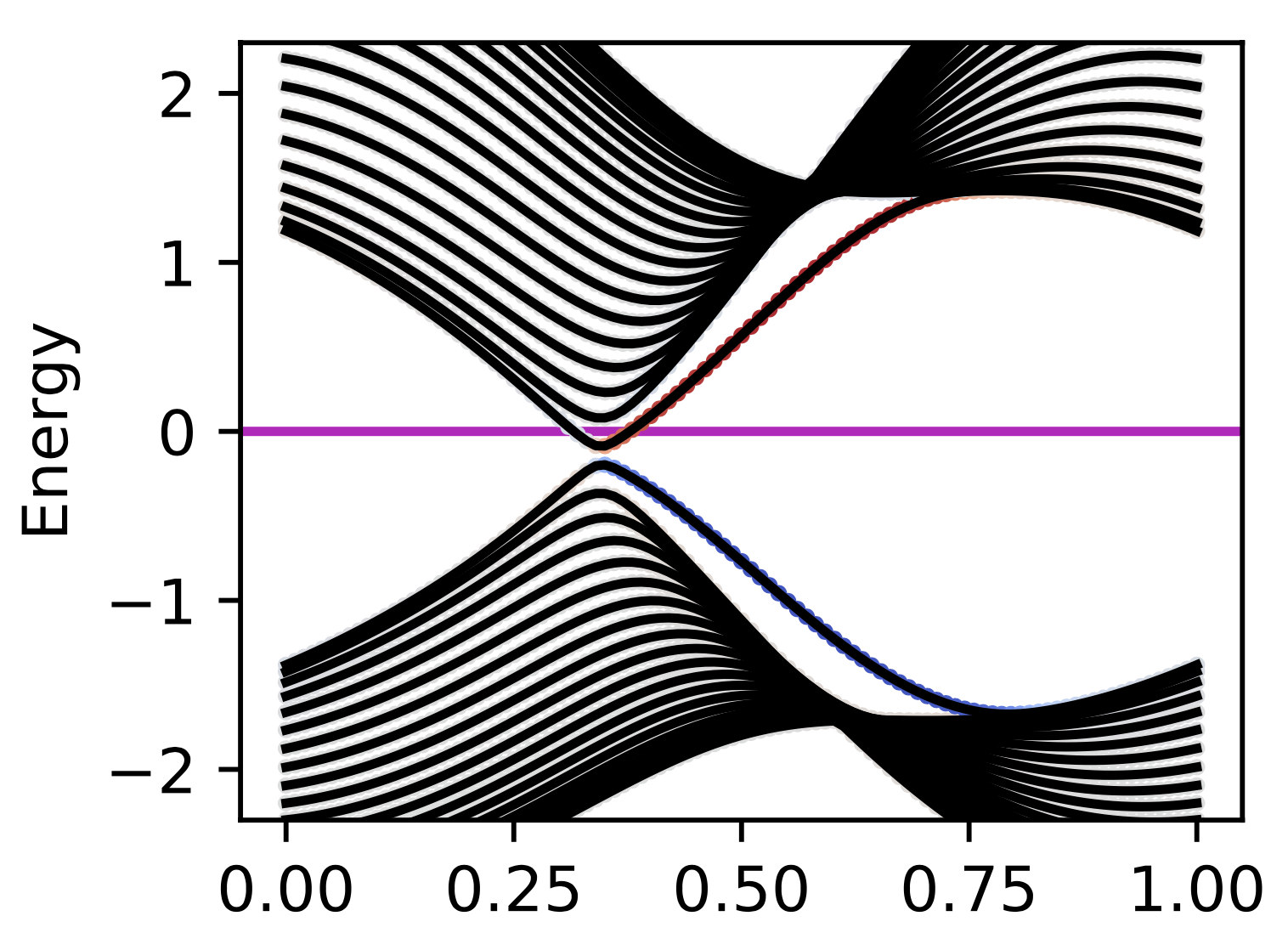}
\caption{(Top) Topological properties of the Haldane model are revealed when solving the energy band structure on a cylinder or geometry presenting edges. The number of discretized points in the two directions is 100 and 30. The parameters are identical to those in Fig. \ref{Haldanespectrum.pdf}. The system has a chiral zero-energy state localized at the edges in the topological phase associated to $C=1$. On the cylinder geometry, the zero-energy modes at the two edges turn with opposite velocities resulting in two branches in the band structure crossing the (pink) Fermi energy.
(Bottom) If $|M|=|m|$ the gap closes at one Dirac point only leading to $C=\frac{1}{2}$ (finite size effects result in an infinitesimal band gap at this point). The Semenoff mass is included through $+M\sigma_z$ in the Hamiltonian (If $|M|>m$, then $C=0$ and a gap re-opens at the two Dirac points).}
\label{Haldanespectrum2.pdf}
\end{figure}

In Fig. \ref{Haldanespectrum.pdf}, we show the energy band structure for the specific situation with $t=1$, $t_2=0.15$ and $m=0.779423$.  The density of states is important close to the Dirac points showing that the description close to the poles of the sphere is particularly meaningful for this range of $t_2$ values. As shown in Fig. \ref{Haldanespectrum2.pdf}, the system shows one chiral edge mode at the edges of the sample similar to the quantum Hall effect. On the cylinder geometry, the colors blue and red are associated to the edge modes in the proximity of top and bottom disks. In Sec. \ref{Paritysymmetry}, we will show that the topological number can also be measured from the $M$ point between $K$ and $K'$ in the Brillouin zone via the light-matter interaction through the properties of  the tight-binding model. This allows us to verify that we remain within the same topological phase for larger $t_2$ where the spectrum flattens around the Dirac points. 

The letter $M$ represents either a high-symmetry point in the Brillouin zone or a term $M\sigma_z$ in the Hamiltonian as introduced in Eq. (\ref{fieldsphere}).
If we include a Semenoff mass $M\sigma_z$ corresponding to a staggered potential on the lattice \cite{Semenoff}, then one can close effectively the gap in the band structure at one Dirac point only when $m=M$ (see Fig. \ref{Haldanespectrum2.pdf}) and induce a transition towards a charge density wave with only one sublattice occupied in real space. This topological transition can also be interpreted from the fact that the topological charge leaks out from the sphere in the sense of the Poincar\' e-Hopf theorem. The topological number is described through a step Heaviside function leading to $C=\pm\frac{1}{2}$ at the transition from Eq. (\ref{polesC}) and from the fact that one pole of the sphere shows $\langle \sigma_z\rangle=0$ and the other pole $\langle \sigma_z\rangle = \pm 1$. This also leads to $C=\frac{1}{2}=A_{\varphi}(\pi)-A_{\varphi}(0)$ such that the sum of the two Berry phases encircling the two Dirac points is equal to $\pi$ (in a symmetric gauge). This fact can be verified numerically. This conclusion will be developed in Sec. \ref{Semenoff} from the smooth fields. The topological charge is in a superposition of `leak-in' and `leak-out' related to the sphere. Experiments on circuit quantum electrodynamics measure $C\approx 0.4$ at the topological transition related to the step function profile \cite{Roushan}. For the Haldane model, the occurrence of such a topological semi-metal occurs at the phase transition only when tuning $M$. In Secs. \ref{bilayer} and \ref{semimetalclass}, related to the two-spheres' model, we will show the possibility of topological nodal ring semi-metals where one-half topological numbers defined locally from the Dirac points, can be stabilized in a finite region of the phase space. 

The Haldane model \cite{Haldane} becomes a standard model as it is realized in quantum materials \cite{Liu}, cold atoms \cite{Hamburg,Jotzu,Monika}, light systems \cite{HaldaneRaghu,Joannopoulos,Ozawa,KLHlightnetworks} and when shining circularly polarized light on graphene \cite{McIver} such that one can adjust the phase associated to the $t_2$ parameter in several platforms.  Applying the results of Sec. \ref{polarizationlight}, the present formalism reveals that circular polarizations of light can turn graphene into an Haldane topological insulator (see Sec. \ref{graphenelight}). The quantum anomalous Hall effect can also be realized on the Kagome lattice as a result of artificial gauge fields for instance produced by Josephson junctions circulators \cite{Koch,KagomeAlex}. Progress in realizing these gauge fields locally in circuit quantum electrodynamics circuits have been recently achieved \cite{SantaBarbaraChiral}.

From the quantum field theory perspective, coupling the electromagnetic fields to Dirac fermions can also reveal a correspondence towards Chern-Simons theory and a Green's function approach \cite{Yakovenko}.
This is related to general questions on parity anomaly \cite{Redlich,NiemiSemenoff}.  It is relevant to mention early efforts on $\hbox{Pb-Te}$ semi-conductors related to
applications of the parity anomaly in condensed-matter systems \cite{Fradkin}. 

\subsection{Interaction Effects and Mott Transition}
\label{Mott}

Interactions can also mediate charge density wave or Mott transitions when the interaction strength becomes significant or comparable to the energy band gap. 
Here, we show that the geometrical formalism allows us to include interaction effects from the reciprocal space of the lattice model \cite{Klein} and through an analogy with the phase transition induced by a Semenoff mass we provide a simple estimate for the transition line. This momentum space approach then allows for a simple approach to describe topological properties from the poles of the sphere or from the Dirac points in the presence of interactions. A similar momentum-space representation of interactions
has allowed to describe Weyl Mott insulators \cite{NagaosaInteraction}.

For spin-polarized electrons, the dominant interaction takes the form
\begin{equation}
\label{interaction}
H_V = V\sum_{i,p} \hat{n}_i \hat{n}_{i+p}
\end{equation}
with $i\in\hbox{sublattice}(A,B)$, $\hat{n}_i=c^{\dagger}_i c_i$ and $p$ sums on the 3 nearest neighbors from the other `color' or sublattice. To englobe interaction effects, we introduce the second quantization representation with creation and annihilation operators on each sublattice. We begin with a simple mean-field approach such that 
\begin{eqnarray}
\hskip -0.5cm H_V = V \sum_{i,p} [-(\phi_0 +\phi_z)c^{\dagger}_{i+p}c_{i+p} - (\phi_0-\phi_z)c^{\dagger}_i c_i && \\ \nonumber
+ c^{\dagger}_i c_{i+p}(\phi_x-i\phi_y) + c^{\dagger}_{i+p} c_{i}(\phi_x+i\phi_y) && \\ \nonumber
-\left(\phi_0^2 - \phi_z^2 -\phi_x^2 - \phi_y^2\right)]. &&
\end{eqnarray}
Minimizing the ground-state energy is equivalent to introduce the definitions
\begin{equation}
\label{phir}
\phi_r = - \frac{1}{2}\langle \Psi_i^{\dagger} \mathbfit{\sigma} \Psi_i \rangle
\end{equation}
with $\Psi_i = (c_i , c_{i+p})$, $\sigma_r$ are the Pauli matrices with $\sigma_0=\mathbb{I}$ the $2\times 2$ identity matrix. We can then introduce the effect of interactions from the reciprocal space which results in the $2\times 2$ matrix \cite{Klein}
\begin{equation}
H(\bm{k}) = \begin{pmatrix}
                              \gamma({\bm k}) & -g({\bm k}) \\
                              -g^*({\bm k}) & - \gamma({\bm k})
 			  \end{pmatrix} \quad
   \end{equation}
with
\begin{eqnarray}
\label{parameters}
\gamma({\bf k}) &=& 3V\phi_z -2t_2\sum_{\bf p} \sin({\bf k}\cdot{\bf b}_p) \\ \nonumber
g({\bf k}) &=& (t_1-V(\phi_x+i\phi_y))\cdot\left(\sum_{\bf p} \cos({\bf k}\cdot\mathbfit{\delta}_p) - i\sin({\bf k}\cdot\mathbfit{\delta}_p)\right).
\end{eqnarray}

From Secs. \ref{spherelattice} and \ref{topometrylattice}, using the mapping with the Bloch sphere close to the Dirac points, we identify
\begin{eqnarray}
\cos\theta({\bf p}) &=& \frac{1}{\epsilon({\bf p})}(\zeta d_z({\bf p}) -3V\phi_z) \\ \nonumber
\sin\theta({\bf p}) &=& \frac{1}{\epsilon({\bf p})}(\hbar v_F -\frac{3}{2}Va(\phi_x + i\phi_y))|{\bf p}|,
\end{eqnarray}
with $\epsilon({\bf p})=\sqrt{|g({\bf p})|^2+\gamma^2({\bf p})}$, $\zeta=\pm$ at the $K$ and $K'$ Dirac points and ${\bf p}$ corresponds to a small wave-vector displacement from a Dirac point. 
This gives rise to
\begin{equation}
\langle c^{\dagger}_{A{\bf p}}c_{A{\bf p}}\rangle = \frac{1}{2} +\frac{1}{2\epsilon({\bf p})}\left(\zeta d_z({\bf p}) -3V\phi_z\right),
\end{equation}
and
\begin{equation}
\langle c^{\dagger}_{B{\bf p}}c_{B{\bf p}}\rangle = \frac{1}{2}-\frac{1}{2\epsilon({\bf p})}\left(\zeta d_z({\bf p}) -3V\phi_z\right).
\end{equation}
Therefore, this leads to the simple identifications
\begin{eqnarray}
\langle c^{\dagger}_{A{\bf p}}c_{A{\bf p}}\rangle &=& \frac{1}{2} +\frac{1}{2}\hbox{sgn}(\zeta d_z({\bf p}) -3V\phi_z)\\ \nonumber 
\langle c^{\dagger}_{B{\bf p}}c_{B{\bf p}}\rangle &=& \frac{1}{2} -\frac{1}{2}\hbox{sgn}(\zeta d_z({\bf p}) -3V\phi_z).
\end{eqnarray}

From the analogy with the effect of a Semenoff mass discussed in Fig. \ref{Haldanespectrum2.pdf}, a quantum phase transition implies that the energy gap is reduced to zero at one Dirac point such that $\langle c^{\dagger}_{A{\bf K}}c_{A{\bf K}}\rangle = \langle c^{\dagger}_{B{\bf K}}c_{B{\bf K}}\rangle = \frac{1}{2}$ corresponding to $d_z=3\sqrt{3}t_2 = -3V\phi_z$ whereas at the other Dirac point the dipole is yet polarized in the ground state so that $\langle c^{\dagger}_{B {\bf K}'} c_{B{\bf K}'}\rangle =1$. If we approximate the Fourier transform of the charge densities as the average responses at the two Dirac points, from Eq. (\ref{phir}) we obtain a jump of $\phi_z$ from zero to $\phi_z\sim -\frac{1}{4}$ for a given bond in real space. This analytical argument supports the occurrence of a first order charge density wave or Mott transition induced by the nearest-neighbors' interaction which was first reported via Exact Diagonalization (ED) in the literature \cite{Varney}. This roughly leads to the estimate $V_c\sim 4\sqrt{3}t_2$. The linear increase of $V_c$ is also in agreement with a numerical self-consistent solution of the coupled equations and with the Density Matrix Renormalization Group (DMRG) results \cite{Klein}. From the definition of the topological number written in terms of $\langle \sigma_z\rangle=\cos\theta$ at the two poles of the sphere, we deduce that $C$ will also jump from $1$ to $0$ at the transition in accordance with numerical results \cite{Klein}.

It is also useful to evaluate energetics from the ground state. Using the Hellmann-Feynman theorem such that $-2\phi_z=\langle c_i^{\dagger} c_i - c_{i+p}^{\dagger} c_{i+p}\rangle$. In this calculation, we suppose a specific sublattice $i$ which then contributes to $1/2$ of the bond-energy in the definition of the interaction energy (\ref{interaction}). This results in the correspondence  $-2\phi_z=\frac{1}{6N}\sum_{\bf k} \frac{\partial E_{gs}}{\partial (V\phi_z)}$ with the ground-state energy $E_{gs}=-\sum_{\bf k}\epsilon({\bf k})$ and $N$ the number of unit cells $\{A;B\}$.  This gives rise
to 
\begin{equation}
\frac{4}{3V}=\frac{1}{N}\sum_{\bf k}\frac{1}{\epsilon({\bf k})}.
\end{equation}
In the limit $t_2\rightarrow 0$, a numerical evaluation of this equation leads to $V_c\sim\frac{4}{3}t_1$. This value tends to agree with results from exact diagonalization leading to $V_c\sim 1.38t_1$ \cite{Capponi}. 
Assuming that the energy spectrum is smoothly varying with ${\bf k}$ and approximating $\epsilon({\bf k})\sim m=3\sqrt{3}t_2$ as in Fig. \ref{Haldanespectrum.pdf}, then we also verify from this equation $V_c\sim 4\sqrt{3}t_2$. 
A careful numerical analysis of the Ginzburg-Landau functional shows a relation with a $\phi^6$-theory \cite{Klein}. 

The possibility of stabilizing a quantum anomalous Hall phase from interactions in two dimensions is questioned in the literature in relation with topological Mott insulators (which are predicted in three dimensions \cite{PesinBalents} or in multi-flavors systems \cite{Mott,Kallin}) through the introduction of a second-nearest-neighbor or a long-range interaction at a mean-field level \cite{Honerkamp,RKKY}. Present numerical calculations such as exact diagonalization do not confirm this possibility \cite{Capponi}. This remains as an open question. A topological Mott insulator i.e. a quantum anomalous Hall state in a correlated limit, has been shown in a twisted bilayer system through DMRG \cite{Vafek}. A Hund's coupling between local magnetic impurities and conduction electrons can stabilize a quantized Hall conductivity in the presence of magnetism as recently observed in Kagome materials \cite{Guguchia,Felser,LegendreLeHur}. The presence of localized magnetic impurities has also be shown to produce topological Kondo physics \cite{Dzero}. 

The discussion above is specific to fermions. A similar discussion can be addressed for bosons including the possibility of superfluidity. The specific form of the honeycomb lattice can allow an additional chiral superfluid phase with condensation of the bosons close to the Dirac points as in a Fulde-Ferrell-Larkin-Ovchnnikov (FFLO) superconducting phase \cite{FuldeFerrell,Larkin}. The Mott phase can reveal topological particle-hole pairs excitations in this case which have been identified through Dynamical Mean-Field Theory, ED and also Cluster Perturbation Theory within a Random Phase Approximation. The phase diagram of the interacting bosonic Haldane model can be found in Ref. \cite{Vasic}.

\subsection{Stochastic Topological Response and Disorder}

The mean-field approach can be reformulated as a path integral approach \cite{Schulz} and the variational principle through the identification \cite{Klein}
\begin{equation}
e^{\frac{V}{8}(c_i^{\dagger}\sigma_r c_{i+p})^2} = \int D\phi e^{-2V\left(\phi_{r}^{i+\frac{p}{2}}\right)^2+V\left(\phi_{r}^{i+\frac{p}{2}}\right)(c^{\dagger}_i \sigma_r c_i)}.
\end{equation}
Here, $\phi_r=\phi_{r}^{i+\frac{p}{2}}$ is centered in the middle of a bond formed by the two sites $i$ and $i+p$. Within this approach, the stochastic variables $\phi_r$ can be thought of as classical static variables and the sampling ranges from $-\infty$ to $+\infty$.
Even though one can redefine the prefactor of each term, their relative weight is fixed according to the variational principle $\frac{\partial S}{\partial \phi_r}=0$, which here reproduces Eq. (\ref{phir}). This is equivalent in the reciprocal space to an action in imaginary time 
\begin{eqnarray}
\label{action}
S &=& \int_0^{\beta} d\tau \sum_{\bf k} \Psi_{\bf k}(\partial_{\tau}-{\bf d}\cdot\mathbfit{\sigma})\Psi_{\bf k} \\ \nonumber 
&+&\sum_{\bf k,q,p} \Psi_{\bf q}^{\dagger} h_V({\bf k},{\bf q},{\bf p}) \Psi_{\bf k} + \sum_{{\bf k},r} 6V|\phi_r^{\bf k}|^2. 
\end{eqnarray}
For simplicity, we use the same symbols for the electron fields or creation/annihilation operators in second quantization and the Grassmann variables defining the path integral. 
To characterize the ground state, we can use the Fourier transform and perform a development of Eq. (\ref{action}) around $\omega\rightarrow 0$ and in the long-wave length limit ${\bf k}\rightarrow {\bf q}$ corresponding to a momentum transfer ${\bf p}\rightarrow {\bf 0}$ for the stochastic variables. The interacting part of the Hamiltonian $h_V$ becomes quite simple such that we obtain a mean-field Hamiltonian of the form $H_{mf}({\bf k})$. The stochastic variational approach is also in agreement with a Ginzburg-Landau analysis \cite{Klein}. Compared to quantum Monte-Carlo which shows a sign problem for the interacting Haldane model, the present approach allows us to have a good understanding of interactions effects from the reciprocal space to evaluate ground state observables. 

The formalism is very useful to incorporate disorder effects \cite{Klein}. We can assume a disordered interaction $V$ with Gaussian fluctuations defining $\tilde{v}=(\tilde{V}-V)/V$. From statistical physics, we can then define the disordered averaged magnetization
\begin{equation}
\langle \sigma_z\rangle = \int_{-\infty}^{+\infty} d\tilde{v} P(\tilde{v})\langle \sigma_z(\tilde{v})\rangle
\end{equation}
with
\begin{equation}
P(\tilde{v}) = \frac{1}{\sqrt{2\pi\xi(V)}} e^{-\frac{1}{2}\tilde{v}^2\xi^{-1}(V)},
\end{equation}
and with the normalization factor $\xi(V)=1/(12V)$. This implies that we can also define a disordered-averaged topological number
\begin{eqnarray}
\langle C\rangle &=& \frac{1}{2}\int_{-\infty}^{+\infty} d\tilde{v} P(\tilde{v})(\langle \sigma_z(0,\tilde{v})-\langle \sigma_z(\pi,\tilde{v})\rangle) \\ \nonumber
&=& \int_{-\infty}^{+\infty} d\tilde{v} P(\tilde{v})C(\tilde{v}).
\end{eqnarray}
The key point here is that the matrix $H_{mf}({\bf k})$ is symmetric under the variable $V$ and $\phi_r$. For the calculation of $\langle \sigma_z\rangle$ only the variable $\phi_z=\phi$ will enter in the calculation of $\langle C\rangle$ such that we
can equivalently write it as a stochastic topological number
\begin{equation}
\langle C \rangle = \int_{-\infty}^{+\infty} d\phi P(\phi) C(\phi)
\end{equation}
with the identification $\tilde{v}=(\phi-\phi_{mf}^z)$ and $\phi_{mf}^z$ corresponds to the ground-state value of the stochastic variable such that $\phi_{mf}^z=0$ if we start with $V<V_c$. For a given value of $V$, then we identify a critical value of $\phi$ due to
the Gaussian fluctuations such that $3V|\phi_c|=3\sqrt{3}t_2$. Therefore, averaging on samples with different disorder configurations then leads to
\begin{equation}
\langle C\rangle \approx 1- e^{-\frac{(2m)^2}{(k_B T_{eff})^2}} \hskip 0.2cm \hbox{and} \hskip 0.2cm k_B T_{eff}\propto \sqrt{V}. 
\end{equation}
Corrections to the topological number from fluctuations are driven by values of $\tilde{v}=\phi\approx |\phi_c|$. This shows that fluctuations induced by the disorder can play a similar role as temperature (heating) \cite{Rivas} or driving effects mediating inter-band transitions \cite{MunichWilczekZee}. In particular, the correction to the topological number takes a similar form as the probability to cross into the upper energy band in a dynamical Landau-Zener-Majorana protocol \cite{HH}. In the present case, the correction of $\langle C\rangle$ from unity is slightly different from an Arrhenius law at finite temperature due to the Gaussian form of the disorder distribution. Averaging the topological response on stochastic variables can then be an efficient way to measure fluctuations effects from the ground state. Experiments in cold atoms allow to measure inter-band transitions in the topological response \cite{MunichWilczekZee}. 

Recently, a numerical study of the interacting Haldane model with disorder was performed using ED and also DMRG \cite{Yi}. This shows that the transition from topological to Mott phase becomes continuous when increasing
the disorder strength. This analysis also finds that the averaged topological number becomes different from the one in the presence of disorder within the (interacting) topological phase. In addition, the Anderson topological insulating phase induced by disorder \cite{ATI,Beenakker} is also robust towards interaction effects.

\section{Observables from Geometry and Local Markers}
\label{Observables}

Here, we describe the usefulness of the formalism to access analytically observables, specifically transport properties and responses to circularly polarized light, within the topological phase of the lattice model \cite{Klein,C2}.
The topological responses may be measured locally from the Dirac points corresponding to the poles of the sphere and also from the $M$ point in the Brillouin zone. Then, we introduce local Chern markers in the reciprocal space.
Recent works have introduced local Chern markers in real space \cite{BiancoResta,Cambridge}. 

Then, we will develop the methodology to study the response to circularly polarized light in the quantum Hall regime on the honeycomb lattice, the photovoltaic
effect in graphene and the light response of Weyl semimetals. 

\subsection{Berry Curvature and Conductivity}
\label{curvature}

We begin with Eq. (\ref{polarization}) where the induced charge polarization reads
\begin{equation}
\label{DeltaP}
\Delta P = eC = \int_0^T dt j(t).
\end{equation}
Here, $j(t)=J_{\perp}$ corresponds to the measured transverse current in the protocol when driving a charge $e$ particle from north to south pole as a result of a longitudinal electric field ${\bf E}=E{\bf e}_{\theta}$. From Eq. (\ref{drive}), we have a correspondence between
the transverse current density and pseudo-spin observable. We can write general relations and build a link with the Karplus and Luttinger velocity \cite{KarplusLuttinger} (defined hereafter in Eq. (\ref{velocity})). We define the vector associated to the Chern number
\begin{equation}
\label{pumpingC}
{\bf C} = \frac{1}{2\pi}\iint d{\bf k}\times {\bf F},
\end{equation}
where ${\bf F}=\bm{\nabla}\times {\bf A}$ parallel to the normal vector to the surface (here the sphere) and ${\bf C}$ has the direction of the induced perpendicular current. Here, we use the identification $d{\bf k}=d\varphi d\theta {\bf e}_{\theta}$. 
We emphasize here that the correspondence $\bm{\nabla}\times {\bf A}=\bm{\nabla}\times {\bf A}'$ justifies the fact to write $C$ as in flat space allowing then a simple adaptation of the formalism in the plane.
We will not use the specific periodicity of the boundary conditions (such that the proof can also be adapted for a graphene plane defined with the appropriate Brillouin zone). In this dynamical protocol, the Berry curvature depends on the component of the wave-vector parallel to the electric field as in Sec. \ref{ParsevalPlancherel} and ${\bf F}={\bf F}(t)$. From Newton's equation on a charge $q=e$, we have 
\begin{equation}
\label{chargeE}
\hbar\dot{\bf k}=e{\bf E}.
\end{equation}
We can then integrate on $\varphi \in[0;2\pi]$ such that
\begin{equation}
{\bf C} = \int dt \frac{e}{\hbar} {\bf E}\times {\bf F}.
\end{equation}
These relations lead to the anomalous velocity or anomalous current density
\begin{equation}
{\bf j}({\bf k})=\frac{e^2}{\hbar} {\bf E}\times {\bf F}.
\end{equation}
The measured current density is perpendicular to the electric field referring to the topological response here. The density of particles in the reciprocal space is $n=1$, therefore the anomalous velocity on the lattice participating in the topological properties is 
\begin{equation}
\label{velocity}
{\bf v} = \frac{e}{\hbar} {\bf E}\times {\bf F}.
\end{equation}
The possibility of anomalous Hall currents in crystals was for instance reviewed in Refs. \cite{Nagaosa,Liu}. Progress in ultracold atoms allows to measure $C$ with a high accuracy from a semiclassical analysis of wavepackets
and Karplus-Luttinger velocity \cite{MunichKarplusLuttinger}. 
The total current density on the lattice is
\begin{equation}
{\bf j} = \iint \frac{dk_x d k_y}{(2\pi)^2} {\bf j}({\bf k}).
\end{equation}
The two directions of the Brillouin zone are defined (symmetrically) in an identical manner. Integrating the current on ${\bf k}$ and playing with cross product rules leads to
\begin{equation}
|{\bf j}| = \frac{e^2}{h}\iint |(d {\bf k}\times {\bf F})\cdot {\bf E}| = \frac{e^2}{h} C | {\bf E}|,
\label{xyconductivity}
\end{equation}
therefore to
\begin{equation}
\label{transportxy}
\sigma_{xy} = \frac{e^2}{h}C.
\end{equation}
Since we also have the local formulation of the global topological number $C = (A_{\varphi}(\pi) - A_{\varphi}(0))$ from Sec. \ref{smooth} this implies that the quantum Hall conductivity (usually defined with an integration on the Brillouin zone) can be measured from the Dirac points only.  Eq. (\ref{transportxy}) is in agreement with the general analysis of edges on the cylinder geometry of Sec. \ref{cylinderformalism}.

From the Karplus-Luttinger velocity in Eq. (\ref{velocity}), we can also relate results of Sec. \ref{ParsevalPlancherel} with general relations of quantum many-body physics from the current density in a one-dimensional pump geometry \cite{Thouless1983}
\begin{equation}
{\bf j} = e\int \frac{d k}{2\pi} {\bf v}(k)
\end{equation}
with $k=k_{\parallel}$ referring to the wave-vector parallel to the motion of the particle. For the sphere we have the identification between time and angle $\theta$ from Newton mechanics, de Broglie principle and Coulomb force, $\theta=eEt/\hbar$. The Karplus-Luttinger velocity reads 
$|{\bf v}| = \frac{e}{\hbar}E |F_{\theta\varphi}|$. Therefore, this gives rise to
\begin{equation}
\Delta P = e\oint \frac{d\varphi}{2\pi} A'_{\varphi}(\theta<\theta_c),
\end{equation}
when integrating on the polar angle from $0$ to an angle $\theta$, in agreement with Eq. (\ref{Jperp}). We emphasize here that this equation is applicable for one and also for interacting spheres leading then to a one-half transverse pumped charge in the situation of Sec. \ref{fractionaltopology}. In Secs. \ref{quantumspinHall} and \ref{topomatter}, we address applications of this formalism to the quantum spin Hall effect and multi-planes systems.

The semiclassical approach is generally judicious to verify certain laws of quantum transport which can also be obtained using many-body physics and a Green's function approach. 
To link with the next Section, it is useful to relate the quantum Hall conductivity \cite{Thouless} with the Thouless-Kohmoto-Nightingale-Nijs formula $\sigma_{xy}=\lim_{\omega\rightarrow 0}\sigma_{xy}(\omega)$ from the Kubo formula \cite{Shen}.
Then, $\sigma_{xy}(\omega)$ can be re-written as $\frac{e^2}{\hbar}\Pi_{xy}$ where
\begin{eqnarray}
\label{xy}
\Pi_{xy} &=& \frac{1}{N}\sum_{{\bf k},n,m} \hbox{Im}(\langle {\bf k}, n | \partial_{k_x} H | {\bf k}, m\rangle \\ \nonumber
&\times& \langle {\bf k}, m | \partial_{k_y} H | {\bf k}, n\rangle)\frac{f_{{\bf k},n} - f_{{\bf k},m}}{(E_n({\bf k}) - E_{m}({\bf k}))^2}.
\end{eqnarray}
For the honeycomb lattice, $N$ means the number of unit cells $\{A;B\}$ (with a normalized lattice spacing equal to one) and the current density is written through the general definition of the velocity $\frac{1}{\hbar}\partial_{k_i} H$ and $i=x,y$. We have introduced a general notation of energy band eigenstates $|{\bf k},n\rangle$ defined for a given wavevector ${\bf k}$. At zero temperature, the Fermi functions related to the lower and upper bands satisfy respectively $f_{{\bf k},n}=1$ and  $f_{{\bf k},m}=0$ or vice-versa.

\subsection{Local and Global Quantized Topological Responses}

Here, we show that the quantum Hall conductivity is related to the Berry curvatures at the Dirac points (only) \cite{C2}. The idea is simply to swap the indices
from $F_{\theta\varphi}$ on the sphere to $F_{p_x p_y}$ on the lattice with definitions of Sec. \ref{spherelattice}.  Again, we will work closely to the two Dirac points of the energy band structure or close to the poles of the sphere such that we can use the Dirac equation. 

From Eq. (\ref{Kspectrum}), we have the important equalities
\begin{equation}
\label{swap}
\partial_{p_x}{H}= \frac{\partial{H}}{\partial p_x} = \hbar v_F\sigma_x\ \hbox{and}\ \partial_{\zeta p_y}{H}=\frac{\partial{H}}{\partial (\zeta p_y)} = \hbar v_F\sigma_y,
 \end{equation} with $\zeta=\pm 1$ at the $K$ and $K'$ Dirac points. Now, we can use the Pauli matrix representations of the pseudo-spin operators and apply them on the two eigenstates $|\psi_+\rangle$ and $|\psi_-\rangle$ on the sphere. From general properties of the Berry curvatures (see \ref{Berrycurvature}), we can equivalently write
 \begin{equation}
\label{F0}
F_{p_x p_y}(\theta) =  i\frac{(\langle \psi_-|\partial_{p_x}{H}|\psi_+\rangle\langle \psi_+|\partial_{p_y}{H}|\psi_-\rangle-(p_x\leftrightarrow p_y))}{(E_--E_+)^2},
\end{equation}
with $|\psi_+\rangle$ on the sphere being the lowest-energy state and $E_- - E_+ = 2m$ at the Dirac points for the topological lattice model. This formula allows a direct link with the general formulation of quantum Hall conductivity (\ref{xy}).
For a relation between polarization, transport and Wannier centers on the lattice see Ref. \cite{Vanderbilt}.
Using Eqs. (\ref{eigenstates}), close to the north pole of the sphere or the $K$ Dirac point with $\theta\rightarrow 0$, then we derive \cite{C2,Meron}
 \begin{equation}
 \label{FBerry}
F_{p_y p_x}(\theta) = -F_{p_x p_y}(\theta)=\frac{(\hbar v_F)^2}{2 d^2} \cos\theta.
\end{equation}
This identity supposes that $t_2\neq 0$ on the lattice or that the energy spectrum has a gap between the ground state and excited state. This form is gauge invariant and is also invariant under the change $\varphi\rightarrow -\varphi$ from 
$\zeta=-1$ in the energy band structure close to the $K'$ Dirac point. Close to the $K'$ point with $\theta+\pi\rightarrow \pi$, then we also
have 
\begin{equation}
F_{-p_y p_x}(\theta+\pi)= \frac{(\hbar v_F)^2}{2 d^2} \cos(\theta+\pi)= - F_{p_y p_x}(\theta+\pi).
\end{equation}
From Eq. (\ref{polesC}), we can then formulate a relation to the topological properties \cite{C2}:
\begin{eqnarray}
\label{F}
\left(F_{p_y p_x}(0) \pm F_{\pm p_y p_x}(\pi)\right) = C \frac{(\hbar v_F)^2}{m^2}.
\end{eqnarray}
Eq. (\ref{F}) is valid assuming that the Dirac approximation of the energy spectrum is correct. From Sec. \ref{Mott}, we deduce that it remains correct in the presence of interactions as long as we stay within the same topological phase and at the Mott transition it will jump to zero. This formula may be verified within current technology in ultra-cold atoms in optical lattices. We can measure the topological number $C$ from the Dirac points only either through Eq. (\ref{polesC}) or through Eq. (\ref{F}). 
This local description from the Dirac points will be useful to  study the magnetoelectric effect in topological insulators \cite{Morimoto,SekineNomura} in Sec. \ref{3DQHE}. 

We can verify that Eq. (\ref{FBerry}) agrees with general lattice relations for the conductivity related to Eq. (\ref{xy}). Indeed, from the form of the Hamiltonian $H=-{\bf d}\cdot \mathbfit{\sigma}$ we have the identification
$\partial_{k_x} H = \partial_{k_x} \sum_{i=x,y,z} d_i \sigma_i$. Now, using the form of ${\bf d}$ derived from the Dirac points in Eq. (\ref{correspondence}) this leads to 
\begin{eqnarray}
\sigma_{xy} &=& \frac{e^2}{\hbar}\sum_{{\bf k},m,n}  \frac{\partial_{k_x} d_x \partial_{k_y} d_y}{4d({\bf k})^2} \\ \nonumber
&\times&\hbox{Im}(\langle {\bf k},n| \sigma_x | {\bf k},m\rangle\langle {\bf k},m| \sigma_y| {\bf k},n\rangle).
\end{eqnarray}
From Eqs. (\ref{F0}) and (\ref{FBerry}), we identify
\begin{equation}
\label{Im}
\hbox{Im}(\langle {\bf k},n| \sigma_x | {\bf k},m\rangle\langle {\bf k},m| \sigma_y| {\bf k},n\rangle) = 2\cos\theta = 2\frac{d_z({\bf k})}{d({\bf k})}
\end{equation}
such that the conductivity can be equivalently written in terms of the ${\bf d}$ vector \cite{Volovik}
\begin{eqnarray}
\sigma_{xy} = \frac{e^2}{2N\hbar} \sum_{\bf k} \frac{(\partial_{k_x} d_x)(\partial_{k_y} d_y) d_z}{d({\bf k})^3}.
\end{eqnarray}
Equivalently, we have
\begin{equation}
\label{dvectorsigma}
\sigma_{xy} = \frac{e^2}{h}\iint \frac{d^2{\bf k}}{4\pi} \frac{(\partial_{k_x} d_x)(\partial_{k_y} d_y) d_z}{d({\bf k})^3}.
\end{equation}
For a relation with the Green's function approach, see \cite{Volovik,Ishikawa}. This can be equally written in terms of the normalized ${\bf n}=\frac{\bf d}{|{\bf d}|}$ vector such that
\begin{equation}
\label{nvectorsigma}
\sigma_{xy} = \frac{e^2}{h}\iint \frac{d^2{\bf k}}{4\pi} (\partial_{k_x} n_x).(\partial_{k_y} n_y)n_z.
\end{equation}
Then, 
\begin{equation}
\label{W}
W= \iint \frac{d^2{\bf k}}{4\pi} (\partial_{k_x} {\bf n})\times (\partial_{k_y} {\bf n}) \cdot {\bf n}
\end{equation}
can be precisely visualized as a winding number on the unit sphere \cite{QiZhang,Volovik} and has a similar interpretation in quantum Hall ferromagnets \cite{Girvin} related to Skyrmion physics \cite{Skyrme,NagaosaTokura} where we turn the momentum into real space variables. A simple correspondence between the spin-$\frac{1}{2}$ Hamiltonian and the winding number $W$ is formulated in the book of Nakahara, see Eq. (10.163) \cite{Nakahara}. This is also a direct measure of the topological charge related to this effective Dirac monopole. This formula also shows that the norm of the ${\bf d}$ vector can be effectively transformed locally as unity which then ensures the correspondence between the plane and the band structure with ${\bf d}={\bf d}({\bf k})$ and ${\bf d}={\bf d}(\theta,\varphi)=d{\bf e}_r$ being described through the same normalized ${\bf n}$ vector.

It is relevant to emphasize here a recent direct experimental determination of the topological winding number through polarized X-ray scattering in Skyrmions materials \cite{ZhangSkyrmions}.
In the case of a meron \cite{Meron}, this is similar as a half Skyrmion on half of the sphere similarly as for the two entangled spheres of Sec. \ref{fractionaltopology}.
From an historical perspective, the meron was first predicted related to the Yang-Mills equation and the possibility of $\frac{1}{2}$ instantons \cite{Alfaro}. In the two spheres' model, the $\frac{q}{2}$ topological numbers are intrinsically protected from the form and nature
of the entangled wavefunction at one pole, as shown through Eq. (\ref{Cjspin}). It is then interesting to comment that the fractional quantum Hall conductivity for the two-spheres' model related to Sec. \ref{cylinderformalism}, \ref{fractionaltopology} and \ref{Geometry} can be obtained integrating Eq. (\ref{W}) on one half of the surface.

In the presence of interactions, the characterization of topological properties from momentum space \cite{Volovik,Gurarie} presents certain advantages such that in the presence of quasiparticle and quasihole excitations, the pole of the single-particle Green's function leads to the robustness of the quantized topological response through the Ishikawa-Matsuyama formula \cite{Ishikawa}. The momentum representation of Sec. \ref{Mott} allows then a simple analytical stochastic variational approach to describe the Mott transition on the sphere. Also, the sphere elegantly reveals topological properties of quantum Hall states \cite{Haldanesphere,Papic,Fluctuations}. Dynamical mean-field theory methods \cite{DMFT} are also very efficient to evaluate topological properties \cite{WuQSH,KimKrempa,Cocks,Vasic,Plekhanov,Julian}. Another interesting approach in the strong-coupling regime is the cluster perturbation theory combined with a Random Phase Approximation which also allows for the evaluation of the quasiparticle spectral function. For the interacting bosonic Haldane model, this has allowed to reveal topological excitations at finite frequency on top of the insulating ground state \cite{Vasic}. Quantum Monte Carlo methods may show a signal problem in the interacting Haldane model, yet important progress has been realized in particular related to the Kane-Mele-Hubbard model discussed hereafter \cite{Hohenadler,Meng,ZCWei}.

It is also important to remind here that Niu {\it et al.} \cite{NiuThoulessWu} have introduced the technique of twisted boundary conditions such that the topological first Chern number can be rephrased in terms of twisted phases in the parameter space for two-dimensional insulators. This approach links the bulk topological number with the properties of the edge states \cite{QiWuZhang}. 

\subsection{Effect of a Semenoff Mass}
\label{Semenoff}

The sphere formalism allows us to include a Semenoff mass \cite{Semenoff}. We describe the situation as in Fig. \ref{Haldanespectrum2.pdf} where the gap closes at the $K$ point only  with a term $+M\sigma_z$ in the Hamiltonian.
In the vicinity of the poles, we can yet write down eigenstates in the same form as in Eq. (\ref{eigenstates}). Close to the north pole of the sphere corresponding to the $K$ point on the lattice, we identify
\begin{equation}
\cos(\theta\rightarrow 0) = \frac{\tilde{d}_z(0)}{\sqrt{\tilde{d}_z(0)^2 +(\hbar v_F)^2 |{\bf p}|^2}}
\end{equation}
with
\begin{equation}
\tilde{d}_z(0)=d_z-M=m-M.
\end{equation}
This way, close to the north pole, we have
\begin{equation}
{A}_{\varphi}(\theta\rightarrow 0,\tilde{d}_z(0))=-\frac{\cos(\theta\rightarrow 0)}{2}.
\end{equation}

Similarly, close to the south pole of the sphere corresponding to the $K'$ point on the lattice, we identify
\begin{equation}
\cos(\theta\rightarrow \pi) = \frac{\tilde{d}_z(\pi)}{\sqrt{\tilde{d}_z(\pi)^2 +(\hbar v_F)^2 |{\bf p}|^2}}
\end{equation}
with
\begin{equation}
\tilde{d}_z(\pi)=-d_z-M=-m-M.
\end{equation}
Close to south pole, we have
\begin{equation}
{A}_{\varphi}(\theta\rightarrow \pi,\tilde{d}_z(\pi))=-\frac{\cos(\theta\rightarrow \pi)}{2}.
\end{equation}
Precisely at the poles we have $\sin\theta\rightarrow 0$ corresponding to $v_F|{\bf p}|\rightarrow 0$. Then, as long as $M<m$ meaning that we are within the same topological phase, we verify ${A}_{\varphi}(\pi)-{A}_{\varphi}(0)=C=+1$, with ${A}_{\varphi}(0)=-\frac{1}{2}$ and ${A}_{\varphi}(\pi)=+\frac{1}{2}$ through the eigenstates of Eq. (\ref{eigenstates}). This shows the perfect quantization of the topological number as long as we stay within the same topological phase (assuming a very clean sample).
The equation relating to the conductivity in Eq. (\ref{F}) is also modified as
\begin{equation}
\frac{\tilde{d}_z^2(0)}{(\hbar v_F)^2}\tilde{F}_{p_y p_x}(0) \mp \frac{\tilde{d}_z^2(\pi)}{(\hbar v_F)^2}\tilde{F}_{p_y\mp p_x}(\pi) =C,
\end{equation}
where $\tilde{F}$ takes a similar form as $F$ when adjusting the $d_z$ component with $\tilde{d}_z$ such that
\begin{equation}
\frac{\tilde{d}_z^2(0)}{(\hbar v_F)^2}\tilde{F}_{p_y p_x}(0)  = \frac{C}{2}\hbox{sgn}(m-M).
\end{equation}
This term remains identical to $F_{p_y p_x}(0)$ as long as we remain within the topological phase with $M<m$. Similarly, at the $K'$ Dirac point, we identify
\begin{equation}
\frac{\tilde{d}_z^2(\pi)}{(\hbar v_F)^2}\tilde{F}_{p_y p_x}(\pi)  = \frac{C}{2}\hbox{sgn}(m+M).
\end{equation}
This also leads to the important equality:
\begin{equation}
A_{\varphi}(\pi)-A_{\varphi}(0) = \frac{C}{2}\left(\hbox{sgn}(\tilde{d}_z(\pi)) - \hbox{sgn}(\tilde{d}_z(0))\right).
\end{equation}
As long as the gap between the topological valence and conduction bands does not close then the winding number $W$ will remain quantized, which can also
be deduced from $\langle \sigma_z\rangle$ at the poles in Eq. (\ref{polesC}) \cite{Orsaytheory,CayssolReview}, and similarly for the quantum Hall conductivity. We observe that the quantum phase transition with one Dirac point yet engenders recent interest in the formulation of topological invariants \cite{Verresen} related to Chern-Simons theory with half-integer prefactors. Within the present description, the topological transition at $m=M$  corresponds to ${A}_{\varphi}(\theta=\pi)-{A}_{\varphi}(\theta=0)=\frac{1}{2}$ through the limit $v_F|{\bf p}|=0^+$. This confirms that $C=\frac{1}{2}$ when $m=M$ and that there is also a jump of the topological number.  Above the transition, we also have $A_{\varphi}(\theta=0)=A_{\varphi}(\theta=\pi)=\frac{1}{2}$ and $C=0$.

An equivalent approach is to say that the effect of the term $M\sigma_z$ is to introduce a `dressed' angle such that $\tan\tilde{\theta}= \frac{\sin\theta}{\cos\theta-\frac{M}{d}}$. As long as $M<d$, $\tilde{\theta}\in [0;\pi]$. When $M=d$, we have
$\tilde{\theta}\in ]\frac{\pi}{2};{\pi}]$ and for $M>d$, it is as if the system navigates around the pole $\tilde{\theta}=\pi$ such that $A_{\varphi}(\theta=0)=A_{\varphi}(\theta=\pi)$.

In Sec. \ref{light}, we show that these relations are useful to make a link between the quantum Hall conductivity on the lattice and response to circular polarizations of light from the Dirac points only. The formalism is similar to that of a spin-$\frac{1}{2}$ atom and therefore it allows simple derivations regarding the light-matter interaction building an analogy with nuclear magnetic resonance. 

\subsection{Quantized Photo-Electric Response and ${\cal I}(\theta)$ function}
\label{light}

Recent works, related to circular dichroism of light and experiments, have introduced the photo-induced current on the whole Brillouin zone and a relation with the quantum Hall conductivity is elegantly built from the Kubo formula \cite{Goldman,Hamburg}. Within our formalism, we will address locally the light response from the Dirac points and show a quantized photo-electric effect related to $C^2$ in Eq. (\ref{alpha}) following our recent progress \cite{C2,Klein}. The word local here refers to specific points in the Brillouin zone. We also show that $C^2$ is measurable from the evolution in time associated to the lowest band population \cite{C2}. In our description, the dipole description is  built from the pseudo-spin $\frac{1}{2}$ associated to the two sublattices $A$ and $B$ of the honeycomb lattice whereas in Ref. \cite{Goldman}, the dipole is measured with the position operator.  

The wave propagates in $z$ direction and we suppose circular polarizations of the vector potential. Within the Dirac equation, the light-matter interaction can be written through a term ${\bf A}\cdot \mathbfit{\sigma}$ \cite{Klein}.

First, we can use the results of Sec. \ref{lightdipole}. From Eqs. (\ref{alpha}), we deduce that the response to circularly polarized light is quantized at the two Dirac points and reveal $C^2$
if we satisfy the proper energy conservation. In fact, $C^2$ in Eq. (\ref{alpha}) acquires another physical interpretation on the lattice close to the Dirac points through \cite{C2}
\begin{equation}
\label{alphalight}
\alpha(\theta) = \frac{{\cal I}(\theta)}{2(\hbar v_F)^2},
\end{equation}
where
\begin{eqnarray}
\label{Itheta}
{\cal I}(\theta) &=& \left\langle \psi_+ \left|\frac{\partial {H}}{\partial p_x} \right|\psi_-\right\rangle \left\langle \psi_- \left|\frac{\partial {H}}{\partial p_x} \right|\psi_+\right\rangle \nonumber \\
&+&  \left\langle \psi_+ \left|\frac{\partial {H}}{\partial p_y} \right|\psi_-\right\rangle \left\langle \psi_- \left|\frac{\partial {H}}{\partial p_y} \right|\psi_+\right\rangle \nonumber \\
&=& 2(\hbar v_F)^2\left(\cos^4\frac{\theta}{2} +\sin^4\frac{\theta}{2}\right).
\end{eqnarray}
It is perhaps relevant to emphasize here the usefulness of the introduced smooth fields in Eq. (\ref{smoothfields}) to relate locally ${\cal I}(\theta)$ with the topological properties. In particular, in Ref. \cite{Klein}, the relation between light response and the topological invariant was revealed when acting through small portions (slices) on the Bloch sphere from the equatorial plane onto the poles, in agreement with the form of the photo-induced currents. From the smooth fields in Eq. (\ref{CA'}), we observe that ${\cal I}(\theta)$ directly measures the square of the topological invariant $C^2$ at the poles of the sphere \cite{C2}. To the best of our knowledge, the relation between the ${\cal I}(\theta)$ function and the topological properties from a local interpretation of the cosine and sine functions of the spin-$\frac{1}{2}$ eigenstates (surprisingly) was not mentioned before in the literature.

In \ref{Berrycurvature}, we show that $C^2$ is related to the quantum distance. In this Appendix, we also address the metric, the ${\cal I}(\theta)$ function and light. It is interesting to emphasize the recent relation between quantum metric from the reciprocal space and a gravitational approach through the Einstein Field Equation \cite{BlochMetric}. 

From energy conservation, the $K$ Dirac point will interact with the right-handed polarization described through the vector potential ${\bf A}_+=A_0 e^{-i\omega t} e^{+ikz}({\bf e}_x - i {\bf e}_y)$ and the $K'$ Dirac point will interact with 
the left-handed polarization described through ${\bf A}_-=A_0 e^{-i\omega t} e^{ikz}({\bf e}_x + i {\bf e}_y)$.  Through the correspondence between momentum and vector potential, the light-matter coupling can be described on the Bloch sphere through the vector potential itself. The system is in the plane $z=0$. Fixing $A_0<0$, then the electric field takes the form ${\bf E}_{\pm}=\mp \omega |A_0| {\bf e}_{\varphi}$ referring to the right-handed and left-handed polarizations (with definitions of Sec. \ref{electricfield}).

Here, we show the relation with the evolution in time of the lowest-band population due to the light-matter coupling at the Dirac points. In terms of smooth fields, to address the physics at the $K$ Dirac point we place the interface $\theta_c$ close to the north pole 
where  $-{A}'_{\varphi}(\theta>\theta_c)=-{A}'_{\varphi}(\theta=0^+)=C$. From Sec. \ref{lightdipole}, at the north pole, we can write the light-matter coupling with the right-handed $(+)$ polarization, as 
\begin{equation}
\label{evolve}
\delta{H}_+ = A_0 e^{i\omega t} e^{-i\varphi} (-A_{\varphi}'(\theta>\theta_c)) |\psi_-\rangle \langle \psi_+| +h.c.
\end{equation}
and $-{A}'_{\varphi}(\theta>\theta_c)=C$. Here, we must be careful with the correspondence with the honeycomb lattice as formulated in Sec. \ref{spherelattice}. Indeed, the polar angle in the reciprocal space of the tight-binding model takes the form $\tilde{\varphi}=\varphi\pm \pi$ and also we must re-adjust $\varphi\rightarrow -\varphi$ between the north and south poles. At the north pole, the additional phase $\pm \pi$ is equivalent for instance to change ${\bf e}_x\rightarrow -{\bf e}_x$ and therefore is equivalent to turn a right-moving wave
into a left-moving wave. To compensate for this lattice effect compared to the Bloch sphere description of Sec. \ref{lightdipole}, we can then change $\omega\rightarrow -\omega$ in Eq. (\ref{deltaH}) which results in Eq. (\ref{evolve}).  At south pole, since we modify
$\varphi\rightarrow -\varphi$ to validate the sphere-lattice correspondence, then we do not need to modify $\omega\rightarrow -\omega$ in Eq. (\ref{deltaH}). Indeed, $\varphi\rightarrow -\varphi$ is equivalent to swap back the direction of the wave. 

Then, developing the evolution operator in time to first order in $\delta{\cal H}_+$, we have 
\begin{eqnarray}
 |\psi_{+}(t)\rangle &=& e^{\frac{i}{\hbar}m t}|\psi_{+}(0)\rangle \\ \nonumber
 & - & \frac{e^{\frac{i}{\hbar}m t}}{\hbar\tilde{\omega}} A_0 e^{i\varphi} {A}'_{\varphi}(\theta>\theta_c)\left(e^{i\tilde{\omega}t} -1\right)|\psi_-(0)\rangle,
\end{eqnarray}
with $\tilde{\omega}=\omega-2m/\hbar$ and $|\psi_-(0)\rangle=|B\rangle$. 
Here, we have selected the right-handed polarisation term $\delta {H}_+$ because in the limit $t\rightarrow +\infty$ we verify that it satisfies the energy conservation in agreement with the Fermi golden rule approach. In this way, we obtain the transition probability ${\cal P}(\tilde{\omega},t)= |\langle B | \psi_{+}(t)\rangle|^2$ to reach the upper energy band \cite{C2}
\begin{equation}
\label{short}
{\cal P}(\tilde{\omega},t) = \frac{4 A_0^2}{(\hbar\tilde{\omega})^2} ({A}'_{\varphi}(\theta>\theta_c))^2 \sin^2\left(\frac{1}{2}\tilde{\omega}t\right).
\end{equation}
Here, $|B\rangle$ represents the upper energy state at the north pole or the $K$ Dirac point in the topological band structure. This formula is reminiscent of the nuclear magnetic resonance inter-band transition formula where we identify an additional geometrical factor encoding the topological properties from the radial magnetic field. Inter-band transition probabilities in time can be measured with current technology, and for instance, in ultra-cold atoms \cite{MunichWilczekZee}.
From Fourier transform, the signal will reveal two $\delta$ peaks which in nuclear magnetic resonance find applications for imaging. At short times, we obtain
\begin{equation}
{\cal P}(dt^2) = \frac{A_0^2}{\hbar^2} C^2 dt^2,
\end{equation}
such that
\begin{equation}
\label{density}
\frac{dN_+}{dt^2} =  - \frac{A_0^2}{\hbar^2} C^2.
\end{equation}
Here, $N_+(t) = |\langle \psi_+(t) |\psi_+(t)\rangle|^2= N_+(0) - {\cal P}(t) = 1-{\cal P}(t)$ describes the normalized number of particles in the lowest band at time $t$. This equation shows locally the relation with the smooth fields and the global topological invariant $C^2$. In fact, if we select the resonance frequency $\tilde{\omega}\rightarrow 0$, then the relation
\begin{equation}
{\cal P}(t) \sim \frac{A_0^2}{\hbar^2} C^2 t^2,
\end{equation}
can be measured for long(er) times. 

To describe the physics at the $K'$ point, within the geometrical approach we can move the interface  $\theta_c\rightarrow \pi$, such that we have the identification $C={A}'_{\varphi}(\theta<\theta_c)={A}'_{\varphi}(\pi^-)$ for the left-handed polarisation. Using the identity between eigenstates $|\psi_+(\theta=0)\rangle = -|\psi_-(\theta=\pi)\rangle$ and $|\psi_-(0)\rangle = |\psi_+(\theta=\pi)\rangle$, then we obtain
\begin{eqnarray}
\delta{H}_- &=& A_0 e^{i\omega t} e^{-i\varphi}(-C)|\psi_+(\pi)\rangle \langle \psi_-(\pi)| +h.c.  \\ \nonumber
&=& A_0 e^{i\omega t} e^{-i\varphi}C|\psi_-(0)\rangle \langle \psi_+(0)| +h.c.
\end{eqnarray}
We obtain a similar formula as in Eq. (\ref{evolve}) close to the $K$ point. From the time evolution of the population in the lower band (or equivalently upper band), the two light polarizations play a symmetric role one at a specific Dirac point $K$ or $K'$. In \ref{lightconductivity}, we verify that the photo-induced currents measure $|C|=C^2$ from the poles of the sphere. This reveals circular dichroism of light \cite{Goldman} referring to an induced current from the left-handed polarization being different compared to to the induced current from the right-handed polarization. In the present situation, the difference between the two responses is topologically quantized. 

This \ref{lightconductivity} also shows why in the calculation of the photo-induced currents it is identical to integrate the light response on all the wavevectors in the Brillouin zone, as in Ref. \cite{Goldman}, or to consider the Dirac points only for the analysis of the light response \cite{Klein,C2}.  When approaching a quantum phase transition such as the Mott transition in Sec. \ref{Mott}, from the analogy between the stochastic approach and a Semenoff mass, we can re-write the results to observe a jump of the topological number through the light response similarly \cite{Klein}. When crossing the transition, a specific light polarization may interact with the two Dirac points. 

Topological properties of the lattice can also be revealed from coupling a circuit quantum electrodynamics honeycomb array to a local microscope (probe) selecting the frequency of the incoming $AC$ signal to measure the physics at the Dirac points \cite{JulianLight}. In this case, it is then possible to resolve the topological information from the Dirac points, without the presence of (circular) polarizations for the light field in the probe, from the energy conservation. 

\subsection{Parity Symmetry and Light at the $M$ point}
\label{Paritysymmetry}

Here, we show useful relations from the lattice at the high-symmetry $M$ point in Fig. \ref{graphenefig} related to the light response. The superposition of the two circularly polarized lights at the $K$ and $K'$ Dirac points is equivalent to
a linearly polarized wave.

 If we use the Bravais lattice vectors ${\bf u}_1=-{\bf b}_2=\frac{a}{2}(3,\sqrt{3})$ and ${\bf u}_2 = {\bf b}_1 = \frac{a}{2}(3,-\sqrt{3})$, we can write the graphene Hamiltonian at the $M$ point in the form
\begin{equation}
{H}(M) =  w\sigma^+ +h.c.,
\end{equation}
with 
\begin{equation}
\label{w}
w = -t\left(1+\sum_{i=1}^2 e^{-i {\bf k}\cdot{\bf u}_i}\right),
\end{equation}
and $k^M_x=\frac{2\pi}{3a}$, $k^M_y=0$. We can justify the choice of local gauge in Eq. (\ref{w}) as follows. Within our definition of the Brillouin zone, at this $M$ point, since $k_y=0$ the Hamiltonian should be invariant under the symmetry $k_y\rightarrow -k_y$ which implies that the term $d_2\sigma_y$ in the formulation of Fu and Kane \cite{FuKane} should be defined to be zero. The Hamiltonian at this $M$ point should be equivalent to $d_1\sigma_x=d_1\hat{P}$, with $d_1=\frac{1}{2}\left(w+w^*\right)$ and with $\hat{P}$ defined to be the parity operator defined in a middle of a bond in a unit cell in real space, corresponding then to interchange $A\leftrightarrow B$ sublattices through the transformation $x\rightarrow -x$ or $k_x\rightarrow -k_x$. This $M$ point in the middle of $K$ and $K'$ is in fact special since $\hbox{sgn}(d_1)=-1$ within our definitions of $w$ whereas at the other high symmetry points, we find $\hbox{sgn}(d_1)=+1$. These definitions are also in agreement with the fact that the light-matter response is invariant under $\varphi\rightarrow -\varphi$. For graphene, the latter also allows us to verify the group velocity for the lowest-energy (upper-energy) band $v=\pm\frac{1}{\hbar}\frac{\partial w}{\partial k_y} = \mp t\sqrt{3}a\sin\left(k_y \frac{a}{2}\sqrt{3}\right)$ with $K_y = \frac{2\pi}{3\sqrt{3}a}$ and $K_y' = - \frac{2\pi}{3\sqrt{3}a}$. This gives rise accordingly to $v(K)=\mp v_F$ and $v(K')=\pm v_F$ and since the sine function is odd there is no integrated current or charge accumulation within the Brillouin zone (in the absence of electric field).

This results in
\begin{eqnarray}
\frac{\partial{w}}{\partial k_x} = -i t (2u_x)\hbox{sgn}(d_1) = (3ita),
\end{eqnarray}
with the identification $u_x=u_{1x}=u_{2x}$. Here, $\hbox{sgn}(d_1)=-1$ reveals that the eigenvalue of $\sigma_x$ or parity operator on the lattice takes a negative value at this specific point, as in the definition of Fu and Kane.
In a similar way, 
\begin{eqnarray}
\frac{\partial{w}}{\partial k_y} = 0.
\end{eqnarray}
Then, we verify the same form as the one obtained above within the Dirac approximation
\begin{eqnarray}
\hskip -0.7cm \frac{1}{(\hbar v_F)^2}\left\langle \psi_+ \left|\frac{\partial H}{\partial k_x} \right|\psi_-\right\rangle \left\langle \psi_- \left|\frac{\partial H}{\partial k_x} \right|\psi_+\right\rangle = 4\cos^4\frac{\theta}{2}.
\end{eqnarray}
Here, we take into account the $\delta(E_b-E_a\mp \hbar\omega)$ function in Eq. (\ref{rates})  such that either $w\sigma^+$ or $w\sigma^-$ contributes for a specified light polarization. 
Since the sine and cosine functions are equal at $\theta=\frac{\pi}{2}$, it allows us to verify that this quantity at the $M$ point is also equal to ${\cal I}(\theta)$. At the $M$ point for $\theta=\frac{\pi}{2}$, then we have
\begin{eqnarray}
\left\langle \psi_+ \left|\frac{\partial H}{\partial k_x} \right|\psi_-\right\rangle \left\langle \psi_- \left|\frac{\partial H}{\partial k_x} \right|\psi_+\right\rangle = {\cal I}(M),
\end{eqnarray}
for all values of $t_2$. This results in \cite{C2}
\begin{equation}
\label{resonancemiddle}
{\cal I}(M) = \frac{{\cal I}(0)}{2} = \frac{{\cal I}(\pi)}{2}
\end{equation}
for the light response, in agreement with Eq. (\ref{onehalf}). 

These equations are also in agreement with the fact that the addition of electric fields around the two Dirac points produces an electric field along $x$ direction at the $M$ point, similarly as a linearly polarized wave:
$$
{\bf E} = {\bf E}_+ + {\bf E}_- = 2 e^{i\frac{\pi}{2}}A_0 \omega e^{-i\omega t} {\bf e}_x.
$$
For a specified light polarization, from geometry, the response of the system will be halved compared to the Dirac points because $\frac{\partial w}{\partial k_y}=0$. 

\subsection{Quantum Hall Effect and Light}

Here, we compare topological properties of the sphere model with $C=1$, the Haldane model and the quantum Hall effect with a uniform magnetic field in $z$ direction. 

The effect of the uniform magnetic field is now directly included within the Dirac formalism on the honeycomb lattice. The validity of the Dirac approximation can be verified from the lattice and from the Azbel-Harper-Hofstadter Hamiltonian \cite{Azbel,Harper,Hofstadter} including Peierls phases associated to the magnetic field \cite{graphene}. The magnetic field in $z$ direction is described (similarly as the light-matter coupling) through the vector potential in a one-dimensional gauge ${\bf B}=B{\bf e}_z = (-Ay,0,0)$. We can absorb the vector potential in the $2\times 2$ matrix associated to Eq. (\ref{Kspectrum}). The wavefunction of the system can be written as $\Phi({\bf r}) = e^{ikx} \Phi(y)$, where $\Phi(y)$ with ${\bf r}=(x,y)$ is associated to the spinor $|\Phi_A(y),\Phi_B(y)\rangle$ such that we have a plane wave solution along $x$ direction.

It is judicious to introduce the dimensionless position operator $\hat{r} = -\frac{y}{l_B}+k l_B$ with the momentum operator $-i \hbar\partial_r = l_B(i\hbar\partial_y)$ such that $[\hat{r},-i\hbar\partial_r] = i\hbar$. The cyclotron length takes the usual form
$l_B = \sqrt{\frac{\hbar}{q B}}$ with $q>0$. Introducing 
the normalized ladder operator 
\begin{eqnarray}
\label{O}
{\cal O} &=& \frac{1}{\sqrt{2}}\left(\hat{r} +\partial_r\right) ={\cal O}_K = {\cal O}^{\dagger}_{K'} \\
{\cal O}^{\dagger} &=& \frac{1}{\sqrt{2}}\left(\hat{r} - \partial_r\right) = {\cal O}^{\dagger}_K = {\cal O}_{K'}
\end{eqnarray}
such that $[{\cal O},{\cal O}^{\dagger}]=1$, the Hamiltonian takes the form
\begin{equation}
H = \hbar \omega_c^* \left( 
\begin{matrix}
0 & {\cal O}^{\dagger} \\
{\cal O} & 0
\end{matrix}
\right),
\end{equation}
with the cyclotron frequency $\omega_c^*=\sqrt{2}\frac{v_F}{l_B}$. Introducing the operator $\hat{N}={\cal O}^{\dagger}{\cal O}$ such that $[H,\hat{N}]=0$, the energy eigenvalues read
\begin{equation}
E = \pm \hbar\omega_c^* \sqrt{N}.
\end{equation}
The energy eigenvalues at the two Dirac points become quantized in units of $\hbar\omega_c^*$ as mentioned by J. W. McClure in 1956 \cite{McClure}.  The main difference compared to the solution of the Schr\" odinger equation is that at $N=0$, then $E=0$. The spectrum is doubly degenerate since the $K$ and $K'$ points give identical solutions. For $N=0$, we have a zero-energy mode shared between the $K$ and $K'$ Dirac points. The ground state at the $K$ point is projected on $\Phi_A$ and satisfies ${\cal O}\Phi_A=0$ and equivalently $\Phi_B=0$. Due to the inversion of ${\cal O}(K)$ and ${\cal O}^{\dagger}(K')$, at the $K'$ point, the ground state is projected on $\Phi_B$, similarly as for the Haldane model. The Haldane model can be viewed as a truncation to two levels $N=0$ and $N=1$ of this model on a sphere with the transverse quantized current of Eq. (\ref{polarization}) corresponding to the current pumped at south pole on the sphere or equivalently at $K'$ within the Brillouin zone. 

Since we have positive and negative energy plateaus, this implies electron and hole conductivity plateaus.
The formation of Landau energy levels in graphene is observed in various experiments \cite{Zhang,Novoselov}. We present a simple derivation of the quantum Hall response related to the drift velocity in \ref{Halldrift}. 
Here, $\sigma_{xy}$ takes the form of $\pm 2(2N+1)\frac{e^2}{h}$ associated to the filled Landau energy levels. For $N=0$, the factor $2$ comes from the presence of two Dirac energy ladders at the $K$ and $K'$ points. 
The formation of the quantum Hall effect in graphene occurs at relatively large temperatures comparable to room temperatures. Heat transport is also measured related to the integer and fractional quantum Hall phases in graphene \cite{LeBreton}.

The light response in quantum Hall systems can be studied using an analogy with spin-$\frac{1}{2}$ particles \cite{NathanNigel}. For the specific situation of the quantum Hall effect in graphene, we show below that the effect of circular polarizations of light can be described using the same formalism as in Sec. \ref{light} allowing a correspondence with the Haldane model if we analyse specifically the induced transitions between energy levels $N=0$ and $N=1^+$.
Including the vector potential due to the light field results in the two equations
\begin{eqnarray}
\left(\hbar\omega_c^*{\cal O}+\hbar\omega_c^*(A_0 l_B) e^{\mp i\omega t}\right)\Phi_A &=& i\hbar\frac{d}{dt}\Phi_B \\ \nonumber
\left(\hbar\omega_c^*{\cal O}^{\dagger} +\hbar\omega_c^*(A_0 l_B) e^{\pm i\omega t}\right)\Phi_B &=& i\hbar\frac{d}{dt}\Phi_A.
\end{eqnarray}
The signs in these two equations refer to the two light polarizations, right- and left-handed, respectively.
The corrections in energy occur to second-order in $A_0$. This can be seen simply as the light-matter coupling modifies 
$$
{\cal O}\rightarrow {\cal O}+(A_0 l_B)e^{\mp i\omega t},
$$
such that corrections in energy occur through ${\cal O}^{\dagger}{\cal O}=N+(A_0 l_B)^2$, and therefore to second-order in $A_0 l_B$. This implies that similarly as for the Haldane model, to calculate the first-order correction in $A_0 l_B$ of
the eigenstates, we can replace on the right-hand side of these equations $\Phi_{\pm}(t)=\Phi e^{\mp i\sqrt{N}\hbar\omega_c^* t}$, with the unmodified (`bare') energies for the quantum Hall system such that
\begin{eqnarray}
\left(\hbar\omega_c^*{\cal O}+\hbar\omega_c^*(A_0 l_B) e^{\mp i\omega t}\right)\Phi_A &=& \pm\sqrt{N}\hbar\omega_c^* \Phi_B \\ \nonumber
\left(\hbar\omega_c^*{\cal O}^{\dagger} +\hbar\omega_c^*(A_0 l_B) e^{\pm i\omega t}\right)\Phi_B &=& \pm\sqrt{N}\hbar\omega_c^*\Phi_A.
\end{eqnarray}
We assume here that we start with the Landau level $N=0$ filled and describe transitions towards the $+1$ state through absorption of $\hbar\omega$. For the plateau at $N=0$, then 
\begin{eqnarray}
\left(\hbar\omega_c^*{\cal O}+\hbar\omega_c^*(A_0 l_B) e^{\mp i\omega t}\right)\Phi_A &=& 0 \\ \nonumber
\left(\hbar\omega_c^*{\cal O}^{\dagger} +\hbar\omega_c^*(A_0 l_B) e^{\pm i\omega t}\right)\Phi_B &=& 0.
\end{eqnarray}
A natural ansatz here is $\Phi_B=0$ and 
\begin{equation}
\tilde{\Phi}_A(0)=\Phi_A(0) - f_A\Phi_A(1)e^{-i\omega_c^*t}.
\end{equation}
Since we have fixed the energies to zero, this implies that $f_A$ here should be time-independent such that $\dot{f}_A$ does not provide a contribution in $A_0$. Selecting the right-handed polarization when we study transitions $0\rightarrow 1+$, for $\omega=\omega_c^*$ then this leads to $f_A=(A_0 l_B)$. This solution requires the synchronization of cyclotron orbits with the circular light polarizations. The probability to reach the upper band at short times $t\ll 1/\omega_c^*$ is $(A_0 l_B)^2 (\omega_c^* t)^2$. 

Compared to the Haldane model, the prefactor $C^2=1$ coming from smooth fields becomes similarly $1$ in this case if we define the dimensionless $\tilde{A_0}=A_0 l_B$ and the dimensionless time unit $\omega_c^* t$. The $1$ counts the number of Dirac points or zero-energy mode involved for a given light polarization similarly as the quantum Hall plateau present between the $N=0$ and $1+$ Landau levels. Taking into account the degeneracy of a Landau level, the response will be multiplied by $\nu_{max}=\frac{\Phi}{\Phi_0}$ with the magnetic flux $\Phi=BS$, $S$ the area of the plane and  $\Phi_0$ the flux quantum. We obtain a similar effect at the $K'$ Dirac point. This is equivalent to modify ${\cal O}\rightarrow {\cal O}^{\dagger}$ in the equations changing the role of $\Phi_A$ and $\Phi_B$ similarly as in the Haldane model, and to modify the light polarization $+\rightarrow -$. If we add the effect of the two light polarizations then similarly as the quantum Hall conductivity we measure the fact that a zero-energy state can be equally distributed at the $K$ or $K'$ Dirac point. 

The fractional quantum Hall effect (FQHE) \cite{Laughlin} is also observed in graphene as an effect of interactions and for instance a quantum plateau at $\nu=\frac{1}{3}$ is clearly identified \cite{Andrei,Bolotin}. The observation of the fractional quantum Hall state has also been shown
in a graphene electron-hole bilayer \cite{MITQHE}. We mention here recent theoretical efforts to detect the FQHE with circular dichroism via ED on a Laughlin state of bosons at $\nu=\frac{1}{2}$ \cite{CecileNathan}. There is then an interesting link between the many-body topological number and the circular dichroic signal. 

Quantum Hall states have been generalized at a level of a ladder or an assembly of wires \cite{TeoKane,Kanewires}, in three-dimensional models \cite{Halperin3D,Montambaux,Berneviggraphite} and also in four-dimensional systems \cite{Price}, such that the light response in these systems could be studied further through the smooth fields. The quantum Hall effect has been recently observed in a ladder in a strongly-correlated regime \cite{Zhou}, in coupled planes and three dimensions \cite{Li3D} and also in four dimensions \cite{Munich4D}. The bosonic Laughlin state at $\nu=\frac{1}{2}$ can be realized in a ladder \cite{PetrescuLeHur,PetrescuPiraud,Taddia,Mazza} through Thouless pump measures, edge properties and quantum information tools towards coupled-ladders geometries \cite{FanKaryn} showing a relation with high-$T_c$ cuprates, Andreev and Mott physics \cite{KarynMaurice}. 

In Sec. \ref{3DQHE}, we relate the quantum Hall effect on top and bottom surfaces of topological insulators and the spheres model with $\frac{1}{2}$  topological number.

\subsection{Photovoltaic Hall Effect in graphene}
\label{graphenelight}

Recent experiments on quantum anomalous Hall effect induced in graphene through circular polarizations of light show a growing signal when increasing the laser drive pulse fluence (intensity)\cite{McIver}. Motivated by these results,  we describe the effect of circularly polarized light on graphene when $t_2=m=0$. 

A quantum Hall response can be measured when applying an additional fixed DC electric field ${\bf E}$. This effect was introduced as a photovoltaic Hall effect in graphene \cite{OkaAoki1,OkaAoki2} and means that if we produce a topological phase either through the Floquet theory within a Magnus high-frequency expansion \cite{MoessnerCayssol} or within the rotating frame, then we can effectively induce a topological phase. The response may be calculated through a Floquet formalism combined with the Keldysh approach \cite{OkaAoki1,OkaAoki2,Balseiro}. A similar Floquet protocol is applied to produce the quantum anomalous Hall effect in ultra-cold atoms \cite{Jotzu,Kitagawa,Weitenberg,Hauke} and in circuit quantum electrodynamics architectures \cite{FloquetQAH}. Theoretical calculations then verify that the quantum Hall conductivity scales as the light intensity \cite{OkaAoki1,OkaAoki2}. 

Here, we analyze the formation of the light-induced topological phase from the geometry and results of Sec. \ref{polarizationlight} fixing the resonance condition with both Dirac points in the rotating frame. The objective is to evaluate transport properties from the geometry and results of Sec. \ref{ParsevalPlancherel}. Within this protocol, this is achieved through the superposed effect of the left-handed and right-handed light polarizations. The light-induced topological phase from Sec. \ref{polarizationlight} corresponds to a Haldane model with a sign change of the $d_z$ term at the two Dirac points producing an effective radial magnetic field acting on the sub-lattice Hilbert space of graphene. 

This allows for a simple $2\times 2$ matrix approach that can be then combined with the geometry
to evaluate transport properties. Including the light-matter coupling from the `resonant' rotating frame is equivalent to have the effective Hamiltonian close to the poles of the sphere \ref{polarizationlight}
\begin{eqnarray}
H_{eff} =
\begin{pmatrix}
\frac{\hbar\omega}{2}\cos\theta & A_0 + \hbar v_F|{\bf p}| \\
A_0 + \hbar v_F|{\bf p}| & -\frac{\hbar\omega}{2}\cos\theta \\
 \end{pmatrix}.\quad 
 \end{eqnarray}
 Here, $|{\bf p}|$ measures a wave-vector deviation from one of the two Dirac points. 
 Within the Dirac approximation, the vector potential $A_0=\frac{E_0}{\omega}$ couples directly to the pseudo-spin $\bm{\sigma}$. The off-diagonal term $A_0$ will then slightly modify the form of the effective angle $\theta$ at the poles of the sphere.
The eigenstate associated with the lowest eigen-energy $-d=-\sqrt{\left(\frac{\hbar\omega}{2}\right)^2 + (A_0+\hbar v_F|{\bf p}|)^2}$ takes the form
\begin{equation}
|\psi_-\rangle = - \sin\left(\frac{\tilde{A}_0}{\hbar\omega}\right)|a'\rangle + \cos \left(\frac{\tilde{A}_0}{\hbar\omega}\right)|b'\rangle 
\end{equation}
with $|a'\rangle$ and $|b'\rangle$ introduced in Sec. \ref{polarizationlight} and $\tilde{A}_0=A_0+\hbar v_F|{\bf p}|$. From the point of view of the rotated frame, the effective azimuthal angle is zero.
Going back to the original frame, this is identical to have
\begin{equation}
\label{psi-}
|\psi_-\rangle = - \sin\left(\frac{\tilde{A}_0}{\hbar\omega}\right)e^{-i\frac{\omega t}{2}} |a\rangle + \cos\left(\frac{\tilde{A}_0}{\hbar\omega}\right)e^{i\frac{\omega t}{2}} |b\rangle.
\end{equation}
For simplicity, we have omitted the global phase factor of the eigenstate coming from the evolution in time. On the Bloch sphere, this is indeed similar to have an effective polar angle `boost' $\theta=\frac{2\tilde{A}_0}{\hbar\omega}$ and effective azimuthal angle $\varphi=\omega t$. 
From the results of Sec. \ref{ParsevalPlancherel}, we deduce that the light-matter coupling itself will then produce a transverse pumped charge $e\sin^2\left(\frac{\theta}{2}\right)\sim e\left(\frac{\tilde{A}_0}{\hbar\omega}\right)^2$ on the Bloch sphere from north pole proportional to the light intensity.

A similar protocol with circularly polarized light may be useful to induce a topological nodal ring semimetal in graphene discussed in Sec. \ref{semimetalclass} \cite{Semimetal}.

\subsection{Weyl Semimetals and Light}

Weyl and Dirac semimetals analogues of graphene in three dimensions have attracted attention recently and several reviews have been written on the subject \cite{SekineNomura,Armitage,Rao}. A Weyl semimetal presents points in its Brillouin zone with a linear energy dispersion similarly to graphene and it develops interesting topological properties giving rise to surface states as a result of the chiral anomaly. The anomalous Hall conductivity is finite and becomes quantized when we have an energy difference between two Weyl points or the two valleys \cite{BurkovBalents,Haldanesemimetal,Wan,Juan}. 
Here, the objective is to discuss the response to circularly light polarized from Sec.\ref{graphenelight} on graphene. Weyl semimetals are described through the Hamiltonian $H_{0}({\bf k}) = \pm\hbar v_F {\bf k}\cdot \bm{\sigma}$ where  
$\zeta=\pm$ for each valley and $\bm{\sigma}$ acts on the space of Pauli matrices. The energy spectrum presents branches described through $\pm \hbar v_F\sqrt{k_x^2+k_y^2+k_z^2}$. We describe the induction of an anomalous Hall effect from the coupling to circularly polarized light which gives rise in a valley to a term $-{\bf b}\cdot \bm{\sigma}$ with ${\bf b}$ a radial magnetic field defined according to Eq. (\ref{E0field}). We suppose an energy difference $\mu_5 \tau_z=\mu_5\zeta$ between the left and right valleys such that we can select the frequency of the wave and study the response at one Weyl point.

The action of such systems takes the form
\begin{eqnarray}
S &=& \int dt d^3 k \psi^{\dagger}i(\partial_t - ie A_0)\psi \\ \nonumber
  &-& \psi^{\dagger}(H_0({\bf k}) - {\bf b}\cdot \bm{\sigma} - \mu_5\tau_z)\psi.
\end{eqnarray}
Through an infinitesimal gauge transformation related to the Fujikawa method \cite{Fujikawa}, then a Weyl semimetal gives rise to a so-called ${\bf E}\cdot{\bf B}$ term in the quantum electrodynamics of the specific term
$\frac{\alpha}{4\pi^2}\int dt d^3 r \theta({\bf r},t) {\bf E}\cdot {\bf B}$ with the fine-structure constant $\alpha=\frac{e^2}{(\hbar c)}=\frac{1}{137}$. In the presence of the $\bf{b}$ term, the $\theta$ parameter takes the form $2({\bf b}\cdot {\bf r}-\mu_5 t)$ and should be distinguished from the polar angle. For this aspect, we can refer for instance to the detailed recent review of Sekine and Nomura \cite{SekineNomura}. This axion type quantum electrodynamics \cite{Wilczek} has attracted attention related to the understanding of cosmology, quantum chromodynamics, strings and dark matter \cite{SvrcekWitten,PreskillWiseWilczek,PecceiQuinn,darkmatterreview}. For a recent review with applications in condensed-matter systems, see Ref. \cite{Nenno}. 

This also produces interesting transport properties and in particular gives rise to the circular photogalvanic effect in the presence of the light-matter coupling as shown by Juan {\it et al.} \cite{Juan}. The presence of the ${\bf b}$ term, acting as an effective magnetic field in the sub-space of $\bm{\sigma}$, produces a current density as \cite{SekineNomura}
\begin{equation}
\label{nabla}
{\bf j}({\bf r},t) = \frac{e^2}{4\pi^2\hbar} \bm{\nabla}\theta\times {\bf E}.
\end{equation}
Here, light also produces a ${\bf b}$ term from the rotating frame similarly as for the chiral magnetic effect.
Similarly as in Sec. \ref{graphenelight}, we study the photo-induced current as a function of the light field amplitude in one valley, which is allowed as a result of the $\mu_5$ term. 
From the rotating frame in Sec. \ref{polarizationlight} in the presence of one polarization $\pm$ for circular light interacting with one valley in the reciprocal space, we have $\bm{\nabla}\theta = 2eA_0{\bf e}_r=2e\frac{E_0}{\omega}{\bf e}_r$ with ${\bf e}_r\sim {\bf e}_x$ and ${\bf E}_{\pm}=\mp E_0 {\bf e}_{\varphi}\sim\mp E_0 {\bf e}_y$ in the $xy$ plane. The factor $e$ in $2eE_0{\bf e}_r$ comes from the fact that the dipole operator is defined through a unit charge in Eq. (\ref{energyshift}). Eq. (\ref{nabla}) then produces a current perpendicular to the plane with a universally quantized response written in terms of the light intensity with a relative $\mp$ sign 
for the two light polarizations. From the original frame, the calculation of  \cite{Juan} also reveals that the universal prefactor hides the chiral anomaly. 

Various experiments have been performed to observe this effect and we mention here applications in $\hbox{RhSi}$ Weyl semimetals \cite{Orenstein}. Recent experiments report a spatially dispersive circular photogalvanic effect in \hbox{MoTe$_2$} and 
\hbox{Mo$_{0.9}$W$_{0.1}$Te$_2$} Weyl semi-metals \cite{Ji}.

\section{Quantum Spin Hall Effect and Two Spheres}
\label{quantumspinHall}

Here, we introduce the formalism related to two-dimensional topological insulators and the quantum spin Hall effect \cite{Book,BernevigZhang,KaneMele1}. We show \cite{C2} that the $\mathbb{Z}_2$ spin Chern number \cite{Sheng} can be measured locally in the reciprocal space from light related to the zeros of the Pfaffian \cite{KaneMele1}. We also elaborate on interaction effects and the Mott transition \cite{Mott,WuQSH,QSHstoch,Plekhanov}.

\subsection{Two Spheres and Two Planes}

Here, we generalize the formalism and introduce the Berry connection for multiple spheres
\begin{equation}
A_{j\nu}({\bf R}) = -i\langle \psi |\partial_{j\nu} |\psi\rangle,
\end{equation}
where $j$ refers to a sphere, $\partial_{j\nu}=\frac{\partial}{\partial_{j{\nu}}}$ to $\partial_{j\theta}$ and $\partial_{j\varphi}$. We assume a general form for the wave-function $|\psi\rangle$ of the system.
We introduce the fields ${\bf A}'_j$ (allowing a re-interpretation in a flat space) such that 
\begin{equation}
\bm{\nabla}_j\times {\bf A}_j=\bm{\nabla}_j\times {\bf A}'_j={\bf F}_j
\end{equation}
where $\bm{\nabla}_1$ (written in terms of $\partial_{1\nu}$) is equivalently written as $\bm{\nabla}_1\otimes\mathbb{I}$ and $\bm{\nabla}_2$ is equivalently written as $\mathbb{I}\otimes\bm{\nabla}_2$.
Similarly as in Eq. (\ref{smoothfields}), we introduce
 \begin{eqnarray}
 A'_{i\varphi}(\theta<\theta_c) &=& A_{i\varphi}(\theta) - A_{i\varphi}(0) \nonumber \\
 A'_{i\varphi}(\theta>\theta_c) &=& A_{i\varphi}(\theta) - A_{i\varphi}(\pi). 
 \end{eqnarray}
If the spheres are subject to the same radial magnetic field ${\bf d}$ giving rise to the Hamiltonian
 \begin{equation}
 H=\sum_i H_i = -\sum_i {\bf d}_i\cdot\mathbfit{\sigma}_i
\end{equation} 
then we can measure the same topological number 
\begin{equation}
\label{topological}
C_i = \frac{1}{2\pi}\iint_{S^2} {\bf F}_i\cdot d^2 {\bf s} = 1
\end{equation}
defined on each sphere with $d^2{\bf s}=d\theta d\varphi$ and with the Berry curvature 
\begin{equation}
{\bf F}_i=F_i {\bf e}_r = F_i = \partial_{i\theta} A'_{i\varphi}- \partial_{i\varphi} A'_{i\theta}. 
\end{equation}
On each sphere, from the results in Sec. \ref{smooth} we also have the correspondence
\begin{equation}
C_i = A_{i\varphi}(\pi) - A_{i\varphi}(0) = A'_{i\varphi}(\theta<\theta_c) - A'_{i\varphi}(\theta>\theta_c).
\label{Ci}
\end{equation}
In particular, the topological information can yet be resolved from the poles.

Two  spheres can for instance describe two graphene planes described by a Haldane model resulting in a total topological number $C_{tot}=\sum_{i=1}^2 C_i = 2$.

\subsection{Quantum Spin Hall Effect}

Here, we introduce the Kane-Mele model on the honeycomb lattice, where we include the spin dynamics \cite{KaneMele1}. In graphene, the spin-orbit coupling is usually small, yet the model finds various applications for instance related to Mercury HgTe/CdTe materials and the Bernevig-Hughes-Zhang model  \cite{Konig,BernevigZhang,Book}, Bismuth thin films \cite{Wurzburgfilms}, and to three dimensional Bismuth materials, introducing a class called topological insulators \cite{Konig,RMPColloquium,Murakami}. We mention here the synthesis of monolayer 1T'-WTe$_2$ with a large bandgap in a robust two-dimensional materials family of transition metal dichalcogenides \cite{Te2}. A quantum spin Hall insulator has also been revealed in 1T'-WSe$_2$ \cite{Crommie}. The model is also realized with other platforms such as light \cite{Ozawa,Rechtsman} and progress is also on-going in ultra-cold atoms \cite{Monika,Ketterle}. Topological insulators belong to the AII class in the tables \cite{BernevigNeupert}.

The spin-orbit coupling can be seen as an atomic spin-orbit interaction $L_z s_z$ with $s_z$ measuring the spin-polarization of a spin-$\frac{1}{2}$ electron $\uparrow$ or $\downarrow$ and $L_z$ referring to the $z$ component of the angular momentum. The angular momentum is proportional to the momentum which produces a $t_2$ term with an imaginary $i$ prefactor in real space similar to the Haldane model. The spin-orbit coupling then produces a term (for each spin polarization) of the form \cite{KaneMele1} 
\begin{equation}
H_{t_2}^{KM} = - h_z\sigma_z\otimes s_z,
\end{equation}
with $h_z({\bf K}')=-h_z({\bf K})$. The Pauli matrix $\mathbfit{\sigma}$ acts on the sublattice space and the Pauli matrix $\mathbfit{s}$ acts on the spin space. We have implicitly assumed the second quantization structure $H=\sum_{\bf k}\Psi^{\dagger}({\bf k})H_{t_2}^{KM}\Psi({\bf k})$ where $\Psi({\bf k})=(c_{A{\bf k}\uparrow}, c_{B{\bf k}\uparrow}, c_{A{\bf k}\downarrow}, c_{B{\bf k}\downarrow})$.
The parameter $h_z$ is related to $t_2$ similarly as the term $d_z$ in Eq. (\ref{dvector}). 

\begin{center}
\begin{figure}[ht]
\hskip 1cm
\includegraphics[width=0.8\textwidth]{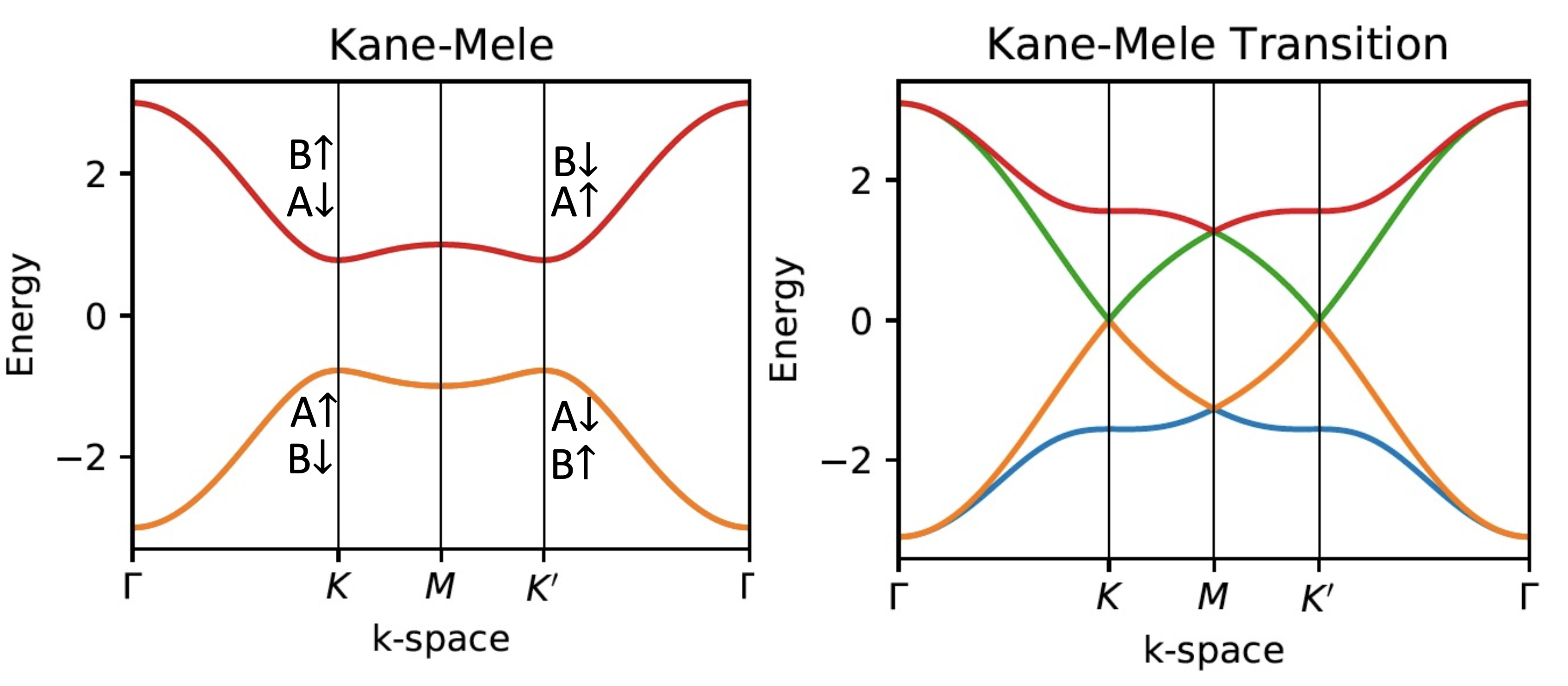}
\caption{(Left) Band structure of the Kane-Mele model with $t_2=0.15$ and $t=1$ showing a double (spin) degeneracy associated to the $\mathbb{Z}_2$ symmetry. The double degeneracy corresponds to simultaneously change $\sigma_z\rightarrow -\sigma_z$ and $s_z\rightarrow -s_z$
such that the states $A_{\uparrow}$ $(A_{\downarrow})$ and $B_{\downarrow}$ $(B_{\uparrow})$ have the same energy. (Right) Transition when including a Semenoff mass $+M\sigma_z$ for $M=3\sqrt{3}t_2=0.779423...$. At the $K$ point, the states $A_{\uparrow}$ and $B_{\uparrow}$ meet at $E=0$ and similarly at the $K'$ point the states $A_{\downarrow}$ and $B_{\downarrow}$ meet at $E=0$.}
\label{KaneMeleSpectrum}
\end{figure}
\end{center}

Similarly as an Ising interaction, this term shows a $\mathbb{Z}_2$ symmetry corresponding here to simultaneously change $\sigma_z\rightarrow -\sigma_z$ and $s_z\rightarrow -s_z$. At the $K$ point, the states $|A\rangle\otimes|\uparrow\rangle$ and  $|B\rangle\otimes|\downarrow\rangle$ are degenerate in energy associated to the energy $-m$, and the states $|B\rangle\otimes|\uparrow\rangle$ and  $|A\rangle\otimes|\downarrow\rangle$ are degenerate in energy associated to the energy $+m$. At the other Dirac point, the energies will evolve according to $m\rightarrow -m$. In Fig. \ref{KaneMeleSpectrum},
 we show the energy spectrum for the Kane-Mele model, including also a Semenoff mass $M$ corresponding to an energy difference between $A$ and $B$ sublattices. At the transition, here the gap is closing at the two Dirac points simultaneously.
 
 To describe the $\mathbb{Z}_2$ order associated to the Kane-Mele model it is useful to introduce the Dirac algebra representation \cite{KaneMele2}
 \begin{equation}
 \label{classification}
 H({\bf k}) = d_1({\bf k})\Gamma_1+d_{12}({\bf k})\Gamma_{12} +d_{15}({\bf k})\Gamma_{15}.
 \end{equation}
 Here, $\Gamma_1=\sigma_x\otimes\mathbb{I}$, $\Gamma_{12}=-\sigma_y\otimes\mathbb{I}$ and $\Gamma_{15}=\sigma_z\otimes s_z$. The $2\times 2$ matrix $\mathbfit{\sigma}$ acts on the sublattice subspace and the $2\times 2$ matrix $\bf{s}$ acts on the spin polarization subspace.
 The first two terms are then related to the graphene Hamiltonian and the $d_{15}=-m\zeta=-h_z$ term describes the mass structure in this model.
 Within the linear spectrum approximation close to a Dirac point, we have $d_1(\zeta K)=v_F|{\bf p}|\cos\tilde{\varphi}$, $d_{12}(\zeta K)=v_F|{\bf p}|\sin(\zeta\tilde{\varphi})$ with ${\bf p}$ measuring deviations from a Dirac point. 
  
This model is invariant under time-reversal symmetry; changing time $t\rightarrow -t$ can be seen as changing the wave-vector ${\bf k}\rightarrow -{\bf k}$ and also the spin polarization $\uparrow \rightarrow -\downarrow$ of a particle (see \ref{timereversal} for a simple description of this symmetry). This symmetry has deep consequences on physics and in particular implies the existence of two counter-propagating edge modes at the edges of the sample. As already mentioned in Sec. \ref{Paritysymmetry}, the characterization of the parity symmetry also plays an important role in the physical responses. Within the choice of our Brillouin zone, the parity transformation takes the form $\hat{P}=\sigma_x\otimes \mathbb{I}$ such that $[H({\bf k}),\hat{P}]=0$ at the $M$ point between $K$ and $K'$. 

Therefore, following Fu and Kane \cite{FuKane}, since $H({\bf k})$ commutes with both the time-reversal symmetry and this parity symmetry, we obtain useful information from the Pfaffian and from symmetries allowing to define a $\mathbb{Z}_2$ topological invariant from the four high-symmetric points in the Brillouin zone, the three $M$ points in the middle of the 3 pairs of $K$ and $K'$ points and the $\Gamma$ point in the center in Fig. \ref{graphenefig}. This invariant reads \cite{FuKane}
 \begin{equation}
 (-1)^{\nu} = \prod_{i=1}^4 \delta_i
 \end{equation} 
 with the product running on these four points and the function $\delta=-\hbox{sgn}(d_1)$ defined in Sec. \ref{Paritysymmetry}. In the topological phase, $\nu=1$ and for a non-topological band insulator, $\nu=0$. Generalizations to three-dimensional topological insulators exist \cite{MooreBalents,FuKaneMele,Teo,Roy}. 
 A quantum field theory description is developed in Ref. \cite{Qi}. The quantum spin Hall phase is also generalized on the square \cite{Cocks} and Kagome lattice \cite{Franz} where the physics of on-site potentials and interactions can also be addressed through various analytical and numerical methods \cite{IrakliJulian}. 
 
\subsection{$\mathbb{Z}_2$ number, Pfaffian, Light and Spin Pump}
\label{lightKM}

 To describe physical observables, here we apply the correspondence with the spheres' model. 

This model can be seen as two spheres described by radial magnetic fields such that ${H}_{\uparrow}({\bf k})=-{\bf d}_{\uparrow}({\bf k})\cdot\mathbfit{\sigma}_{\uparrow}$  and ${H}_{\downarrow}({\bf k})={\bf d}_{\downarrow}({\bf k})\cdot\mathbfit{\sigma}_{\downarrow}$ and $d_{\uparrow x}=d_{\downarrow x}=d_1$, $d_{\uparrow y}=d_{\downarrow y}=d_{12}$, $d_{\uparrow z}=\zeta m$ and $d_{\downarrow z}=d_{\uparrow z}$. 
We keep the same definition as before such that $\zeta=\pm$ at the $K$ and $K'$ Dirac points respectively.

\begin{center}
\begin{figure}[ht]
\includegraphics[width=0.9\textwidth]{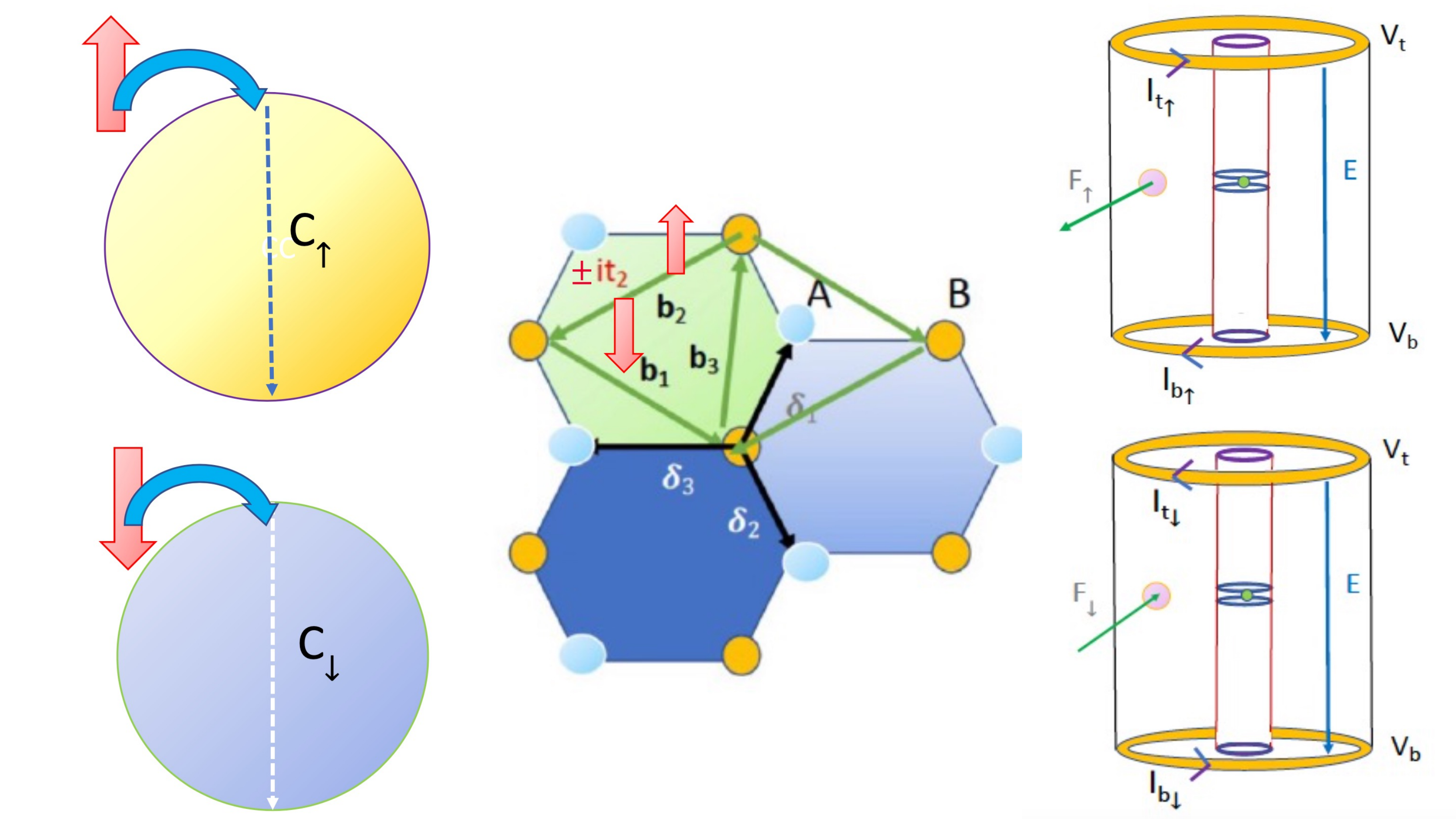}
\caption{(Left) Two Spheres' model describing the Kane-Mele model with a topological $\mathbb{Z}_2$ spin Chern number $C_s=C_{\uparrow}-C_{\downarrow}=\pm 2$. (Center) Lattice representation including a spin-orbit interaction on second nearest-neighboring sites represented through the imaginary hopping terms $\pm it_2$ in real space for an electron with spin polarization $\uparrow$ and $\downarrow$ respectively. (Right) Cylinder representation showing that the edge structure is characterized  by a $\uparrow$ particle moving in one direction and a $\downarrow$ particle moving in the other direction.}
\label{KaneMele}
\end{figure}
\end{center}

Related to Eq. (\ref{correspondence}), within our definitions, the polar angle around a Dirac point in the recripocal space is related to the Bloch sphere as $\tilde{\varphi}_\uparrow=\varphi\pm \pi$ and $\tilde{\varphi}_\downarrow=\varphi$. Going from sphere $1=\uparrow$ to sphere $2=\downarrow$ is equivalent to change the role of the lower and upper energy eigenstates in Eq. (\ref{eigenstates}) and to adjust the topological number accordingly as $C_{\uparrow}=+1$ and $C_{\downarrow}=-1$.  This modifies ${A}'_{\varphi}(\theta>\theta_c)\rightarrow - {A}'_{\varphi}(\theta>\theta_c)=\cos^2\frac{\theta}{2}$ and ${A}'_{\varphi}(\theta<\theta_c)\rightarrow -{A}'_{\varphi}(\theta<\theta_c)=-\sin^2\frac{\theta}{2}$ for sphere $2$. The quantum Hall conductivity is zero in this case since $\sum_{i=1}^2 C_i=0$. In Fig. \ref{KaneMele}, at the edges of the cylinder, a $\uparrow$ particle moves in one direction whereas the $\downarrow$ particle moves in the counter direction with $I_{b\uparrow}=-I_{b\downarrow}$ and similarly $I_{t\uparrow}=-I_{t\downarrow}$. Therefore, one can introduce the spin Chern number \cite{Sheng}
\begin{equation}
C_s = C_{\uparrow} - C_{\downarrow} = \pm 2,
\end{equation}
as a $\mathbb{Z}_2$ formulation of the topological invariant. The $\mathbb{Z}_2$ number is defined modulo $\pm $ related to the $1\leftrightarrow 2$ or $\uparrow\leftrightarrow \downarrow$ symmetry of the system or structure of smooth fields.  

Above, we have implicitly assumed a structure with symmetric topological masses $m_{\uparrow}=-m_{\downarrow}$. The situation with asymmetric masses can also be realized, such as in bilayer systems from a topological proximity effect in graphene \cite{bilayerQSH}. In this case, it is yet possible to observe a topological spin Chern number equal to $2$. The $4\times 4$ matrix description allows us to describe the occurrence of a $\mathbb{Z}_2$ topological spin Chern number for the two lowest bands \cite{bilayerQSH} following the same approach as in \ref{TopometryBandTheory}. We will discuss specifically the situation of asymmetric masses in Sec. \ref{hop} related to an induced topological proximity effect in graphene. The quantum spin Hall phase is also stable towards  other forms of anisotropic spin-orbit couplings at weak interactions \cite{Shitade,ShitadeLiu,Thomale}.

This topological number $C_s$ can be measured directly when driving from north to south pole on each sphere simultaneously or through circular polarizations of light resolved at the Dirac points. 

If we write generally the light-matter coupling as in Eq. (\ref{evolve}) for both spin polarizations, then from the energy conservation, the $+$ light polarization will promote interband transitions at the $K$ point for the $\uparrow$ sphere and at the $K'$ point for the $\downarrow$ sphere. Therefore, detecting the effect of the $+$ (right-handed) and $-$ (left-handed) light polarizations in the inter-band transition probabilities, we can measure $C_s=C_{\uparrow}^2+C_{\downarrow}^2=|C_{\uparrow}|+|C_{\downarrow}|$ from the Dirac points. The additivity of the light responses for the two spin polarizations or the occurrence of the spin topological number $C_s$ can also be understood from the photo-currents \cite{C2}. On the lattice model, the topological information can yet be resolved from the $M$ point in the Brillouin zone (see Sec. \ref{Paritysymmetry}).

The analysis of the light responses leads to an interesting relation with the nuclear magnetic resonance \cite{C2}. For the Kane-Mele model, adding the responses for the two spin polarizations around the two Dirac points this also establishes a relation to the Pfaffian $P({\bf k})$ \cite{C2}
\begin{equation}
\label{Pf}
\alpha_{\uparrow}(\theta) + \alpha_{\downarrow}(\theta) = |C_s| - \left(\hbox{P}({\bf k})\right)^2
\end{equation}
with 
\begin{equation}
\hbox{Pf}_{ij}=\epsilon_{ij}P({\bf k})=\langle u_i({\bf k})| U |u_j({\bf k})\rangle=\hbox{Pf}_{ji}
\end{equation}
and $(i,j)=(\uparrow,\downarrow)$ where the time-reversal operator is introduced as $U$ in \ref{timereversal}. The function $\alpha_i(\theta)$ corresponds to the light response 
for each spin polarization generalizing Eq. (\ref{alphalight}). 

Eq. (\ref{Pf}) is shown in Table \ref{tableI}. 

Here, we define eigenstates related to the two-lowest filled energy bands on the lattice and show a relation between $P({\bf k})$ and the sphere angles. Close to the $K$ point, an eigenstate related to Eq. (\ref{classification}) and spin polarization $\uparrow$ with energy $E$ can be re-written from the lattice as
\begin{eqnarray}
\hskip -0.2cm |u_{\uparrow}({\bf K})\rangle=\frac{1}{\sqrt{(E+m)^2 +v_F^2 |{\bf p}|^2}}
\left(
\begin{matrix}
v_F|{\bf p}| \\
(E+m) e^{i\tilde{\varphi}}
\end{matrix}
\right) &&
\end{eqnarray}
with ${\bf k} = {\bf K}+{\bf p}$. For an eigenstate with energy $E$ and spin polarization $\downarrow$, assuming symmetric masses, we have
\begin{eqnarray}
\hskip -0.4cm |u_{\downarrow}({\bf K})\rangle =\frac{1}{\sqrt{(E-m)^2 +v_F^2 |{\bf p}|^2}}
\left(
\begin{matrix}
v_F|{\bf p}| \\
(E-m) e^{i\tilde{\varphi}}
\end{matrix}
\right). && 
\end{eqnarray}
In this way, 
\begin{equation}
Pf_{\uparrow\downarrow}= \langle u_{\uparrow}({\bf p}) | u_{\uparrow}(-{\bf p})\rangle^*. 
\end{equation}
Going from $K$ to $K'$ corresponds to modify $m\rightarrow -m$ and $\tilde{\varphi}\rightarrow -\tilde{\varphi}$. Inserting the eigen-energies $E\rightarrow \pm \sqrt{m^2+v_F^2|{\bf p}|^2}$, then we obtain
\begin{equation}
Pf_{\uparrow\downarrow} = \frac{v_F |{\bf p}|}{m} = Pf_{\downarrow\uparrow}.
\end{equation}
For the case of symmetric masses, the Pfaffian satisfies
\begin{equation}
\label{Pkmetric}
P({\bf k}) =\frac{v_F|{\bf{p}}|}{m}\approx \sin\theta.
\end{equation}
This result can also be verified on the sphere. 

The perfect quantized light response at the Dirac points or poles of the sphere corresponds to the zeros of the Pfaffian. The light response at the Dirac points measures $|C_s|$. This argument is in fact yet valid in the presence of additional perturbations such as a mass asymmetry or a Rashba spin-orbit interaction $\alpha (\mathbfit{s}\times {\bf p})\cdot \bf{e}_z$ \cite{KaneMele2,Sheng} (with ${\bf p}$ the momentum and $\bf{e}_z$ a unit vector in $z$-direction) as long as the properties of the smooth fields at the poles of the sphere are unchanged and the single-particle gap is not closing. For asymmetric masses, we can adjust one sphere smoothly into an ellipse showing the robustness of $C_s$. 

\vskip -0.2cm
We can also formulate a relation between the light response and a spin pump analysis in the cylinder geometry. We have the correspondence from the sphere
\begin{equation}
\sigma_{\uparrow z} - \sigma_{\downarrow z} = (\hat{n}^A_{\uparrow} - \hat{n}^A_{\downarrow}) - (\hat{n}^B_{\uparrow} - \hat{n}^B_{\downarrow}) = s_{Az} - s_{Bz},
\end{equation}
where we introduce the projector $\hat{n}^i_{j}$ with $i=A,B$ and $j=\uparrow,\downarrow$ related to the two spheres. On the left-hand side, we generalize the pseudo-spin magnetization $\sigma_z$ of \ref{lightconductivity} for each sphere or each spin polarization. On the right-hand side, we have the spin magnetization resolved on a sublattice. In this way, the topological spin Chern number $C_s=C_1-C_2$ is equal to
\begin{equation}
C_s = \frac{\langle s_{Az}(0)\rangle -\langle s_{Bz}(0)\rangle -\langle s_{Az}(\pi)\rangle +\langle s_{Bz}(\pi)\rangle}{2}.
\end{equation}
Using the structure of the eigenstates at the two Dirac points in Fig. \ref{KaneMeleSpectrum}, we can re-write $C_s$ in terms of the spin magnetization on one sublattice (for instance, $A$)  as
\begin{equation}
\label{Cs}
C_s = \langle s_z(0)\rangle - \langle s_z(\pi)\rangle = -\int_0^{\frac{\pi}{v}} \frac{\partial \langle s_z\rangle}{\partial t} dt.
\end{equation}
In this sense, the topological spin Chern number is related to the transport of the spin magnetization from north to south pole on the sphere in a pump geometry. Now, we can equivalently link this analysis to the cylinder geometry of Sec. \ref{cylinderformalism} to reveal
the spin structure for the edge states. For the Kane-Mele model, due to the structure of the smooth fields, we have now two cylinders such that ${\bf F}_1={\bf F}_{\uparrow}={\bf F}$ and ${\bf F}_2={\bf F}_{\downarrow}=-{\bf F}$. 

To activate the spin pump we apply an electric field ${\bf E}$ parallel to the polar angle, from north to south pole on the sphere, acting on a charge $q$ such that from Newton equation $\theta(t)=vt$ with $v=\frac{q E}{\hbar}$ in Eq. (\ref{Cs}). From the Parseval-Plancherel theorem in Sec. \ref{ParsevalPlancherel}, this produces transverse currents on the two spheres related to the smooth fields $J_{\perp}^1=J_{\perp}(\theta)=\frac{q}{t}A'_{\varphi}(\theta<\theta_c)$ and $J_{\perp}^2=-J_{\perp}(\theta)$. To link transport with the light response, we navigate such that $\theta\in[0;\pi]$ in a time $T=\frac{h}{2qE}$ producing a spin current 
\begin{equation}
\label{Cscylinder}
J_{\perp}^1-J_{\perp}^2=\frac{2q^2}{h}C_s E. 
\end{equation}
The factor $2$ specifies that a charge $-q$ also navigates in opposite direction. On the cylinder, we have the same spin current from the smooth fields identification. 

If we introduce a voltage drop on the cylinders $EH=(V_t-V_b)$ similarly as in Sec. \ref{cylinderformalism} we verify the formation of edge modes at the boundaries with the disks, 
\begin{equation}
J_{\perp}^1-J_{\perp}^2=G_s(V_t-V_b)\ \hbox{and}\ G_s=\frac{q^2}{h}C_s.
\end{equation}
The analysis at the edges of the cylinder geometry then reveals that the light response resolved at the two Dirac points is related to a spin conductance measurement through $C_s$. 

In the presence of a Zeeman effect polarizing electrons, a quantum anomalous Hall effect can then be obtained as experimentally observed with magnetic dopants in $\hbox{HgTe}$ \cite{Budewitz} and Bismuth thin films \cite{Chang}.

\subsection{Interaction Effects}
\label{MottKM}

The stability of the quantum spin Hall phase towards interactions can be shown in various ways from renormalization group arguments \cite{KaneMele1}, gauge theories and simply mean-field theories \cite{Mott}. Here, we show that the mean-field theory can be developed in a controllable variational stochastic way
to reproduce the Mott transition line in one equation analytically in agreement with Cluster Dynamical Mean-Field Theory (CDMFT) \cite{WuQSH} and Quantum Monte-Carlo \cite{Hohenadler,Meng,ZCWei}. The key point is a proper derivation of the logical theoretical steps. Similarly as in Sec. \ref{Mott}, it is useful to first write down a mean-field Hamiltonian of the Hubbard interaction $H_U = U\hat{n}_{i\uparrow}\hat{n}_{i\downarrow}$ as
\begin{eqnarray}
H_U &=& -U\sum_i (\phi_0+\phi_z)c^{\dagger}_{i\downarrow}c_{i\downarrow} - U \sum_i (\phi_0 - \phi_z)c^{\dagger}_{i\uparrow}c_{i\uparrow}  \nonumber \\
&-& U\sum_i (\phi_0^2-\phi_z^2) \nonumber \\
&+& U\sum_i (\phi_x-i\phi_y) c^{\dagger}_{i\uparrow}c_{i\downarrow} + U\sum_i (\phi_x+i\phi_y)c^{\dagger}_{i\downarrow}c_{i\uparrow}\nonumber \\
&+& U\sum_i (\phi_x^2+\phi_y^2).
\label{interactionsPhi}
\end{eqnarray}
The Mott transition in the Kane-Mele-Hubbard model corresponds in fact to a magnetic or N\' eel transition in the $XY$ plane \cite{Mott} since the energy band gap in the electron spectral function calculated from CDMFT does not reduce to zero \cite{WuQSH}. The system evolves adiabatically from a band insulator into a Mott insulator. In the quantum spin Hall phase, from the wave-function at $U=0$, the spin-spin correlation functions decay vary rapidly with the distance similarly as in a gapped quantum spin liquid phase \cite{Mott}. Here, we introduce the magnetic channels $S_r= c^{\dagger}\sigma_r c$
with $r=x,y,z$ \cite{QSHstoch} and to minimize the interaction energy we verify the solution $\phi_r = -\frac{1}{2}\langle S_r\rangle$.  

At half-filling, we also identify $\phi_0 = - \frac{1}{2}\langle c^{\dagger}_{i\uparrow}c_{i\uparrow} +  c^{\dagger}_{i\downarrow}c_{i\downarrow}\rangle=-\frac{1}{2}$
and this works equally for the quantum spin Hall phase and the Mott phase. 

Similar to the interacting Haldane model in Sec. \ref{Mott}, the equality $\phi_r = -\frac{1}{2}\langle S_r\rangle$ can be implemented from a path-integral approach re-writing the interaction as a spin interaction $U\sum_i \hat{n}_{i\uparrow}\hat{n}_{i\downarrow} = U\sum_{i,r} \eta_r S_{ir} S_{ir}$. Since for $t_2\rightarrow 0$, the magnetic ground state would equally prefer to occur in any direction this implies to select $\eta_x=\eta_y=\eta_z$ which then leads to the specific choice $\eta_x=\eta_y=\eta_z=-\frac{1}{8}$ and $\eta_0=\frac{1}{8}$ \cite{QSHstoch}. In this way, we have the precise correspondence
\begin{equation}
H_U = \frac{U}{8}\sum_{i,r=x,y,z} {S}_{ir}\cdot {S}_{ir} +\frac{U}{4}\sum_i (\hat{n}_{i\uparrow} + \hat{n}_{i\downarrow}).
\end{equation}
We can then introduce the stochastic variables $\phi_i$ in a path-integral manner such that they satisfy the minimum action principle. To evaluate ground-state properties, we can then Fourier transform the interaction terms assuming the long wavelength and zero-frequency limit corresponding to
the limit of `static' and uniform Gaussian stochastic variables such that the participation function becomes
\begin{eqnarray}
Z \sim \int \Pi_{{\bf k},r=x,y,z} D\phi_{r} \int D\psi^{\dagger}_{\bf k} \psi_{\bf k} e^{-S_{KM}[\psi^{\dagger}_{\bf k},\psi_{\bf k}]} && \nonumber \\ 
\times e^{\frac{U}{2}\int_0^{\beta} d\tau \phi_{r}\phi_{r} +\phi_{r}S_{{\bf k}r}} &.&
\end{eqnarray}
Here, $S_{KM}$ corresponds to the action of the Kane-Mele model. The last term can then be absorbed in the Kane-Mele Hamiltonian such that we obtain the (degenerate) eigenenergies which depend on the stochastic variables $\phi_x$ and $\phi_y$,
\begin{equation}
\epsilon^{\pm}({\bf k}) = \frac{U}{2}\pm \sqrt{\epsilon({\bf k})^2 + \left(\frac{U}{2}\right)^2(\phi_x\phi_x+ \phi_y\phi_y)},
\end{equation}
and $\epsilon({\bf k)}$ the eigen-energies of the Haldane model with $V=0$ in Sec. \ref{Mott}. We assume here for simplicity that $\phi_z=0$ for the Mott insulating phase which can be justified from a strong-coupling effective theory when $t_2\neq 0$ \cite{Mott}. 

The Mott transition can be identified from various ways \cite{QSHstoch} such as the Green's function approach and free energy. Here, we show that the Hellmann-Feynman theorem gives a similar and simple answer for the ground-state energetics 
similarly as in Sec. \ref{Mott} for the Haldane model. The ground state energy takes the simple form $E_{gs}=2\sum_{\bf k} \epsilon^-({\bf k})$ where the factor $2$ accounts for the spin-degeneracy. Then, minimizing $E_{gs}$ with respect to $\phi_x$ gives the following equation
\begin{equation}
\frac{\partial E_{gs}}{\partial \phi_x}= -\frac{U^2}{2}\sum_{\bf k} \frac{\phi_x}{\epsilon({\bf k})},
\end{equation}
assuming here $\phi_x\rightarrow 0^+$  close to the quantum phase transition. From the definition of the mean-field theory we also have $\frac{\partial E_{gs}}{\partial \phi_x} = U\sum_i \langle c^{\dagger}_{i\uparrow}c_{i\downarrow}+c^{\dagger}_{i\downarrow}c_{i\uparrow}\rangle =
-2UN\phi_x$ where $N$ represents the number of unit cells (or half the number of total sites) such that we obtain the equation for the transition line
\begin{equation}
\frac{1}{U_c} = \frac{1}{4N}\sum_{\bf k} \frac{1}{\epsilon({\bf k})}.
\end{equation}
This equation is in good agreement \cite{QSHstoch} with numerical methods and also in the limit of $t_2\rightarrow 0$ where it is usually difficult to perform analytical calculations beyond the so-called large-$N$ method \cite{Herbut} where $N$ refers to the number of flavors or species or similarly the $3-\epsilon$ expansion \cite{Mottepsilon}. We identify $U_c(t_2\rightarrow 0)\sim 4t$. Within the present approach, a magnetic transition for the channel $\phi_z$ would occur at larger $U$ values when $t_2\neq 0$. For $U<U_c$, then $\phi_x=0$ which shows that the smooth field description of the quantum spin Hall phase remains quasi-identical as the situation for $U=0$. The variational approach also allows a control on the fluctuations from the calculation of the polarization bubbles \cite{QSHstoch}.  When adding a small mass term in the theory at the Dirac points, the fluctuations in the infra-red become reduced compared to graphene \cite{polarizationgraphene} since we should satisfy $\hbar\omega>\sqrt{(\hbar v_F q)^2+(2m)^2}$. The one-loop contribution to the polarizability $\Pi({\bf q},\omega)=i\frac{e^2}{8}\frac{{\bf q}^2}{\sqrt{v_F^2 |{\bf q}|^2 -\omega^2}}$ becomes regularized by $2m$ when $q=|{\bf q}|\rightarrow 0$. 

From the numerical sense, adding a small term opening a gap is then interesting to control the infra-red divergences of the fluctuations. For the typical situation of graphene with $t_2=0$, the Mott transition is only quantitatively tractable through the large $N$ method with $N$
referring to the number of flavors associated to fermions leading to $U_c\sim \frac{v_F}{2N}$ in this case \cite{Herbut}. This situation can be for instance achieved from a Hofstadter model on the square lattice where tuning magnetic fluxes allows to vary the number $q$ of Dirac 
points leading to $U_c\sim q^{-2}$ \cite{Cocks}. The mean-field variational stochastic approach was recently generalized to the Kagome lattice \cite{Julian}. 

The edge theory of the quantum spin Hall phase corresponds to the helical Luttinger liquid theory revealing the two counter-propagating modes in a quantum field theory way \cite{WuC}. The helical structure of the edge states can be addressed experimentally \cite{Weber}.
At the Mott transition, the presence of a magnetic order in the $XY$ plane then produces a massive theory for the helical liquid description at the edges reflecting the breaking of the time-reversal symmetry \cite{WuQSH}.  
From the edge theory on the cylinder in Eq. (\ref{Cscylinder}), we infer that $C_s=0$ at the transition. This approach is justified from the reciprocal space when the Mott transition develops a magnetic order and can then be combined with the smooth field description.

Gauge theories also allow the identification of topological Mott phases described by a quantum spin Hall effect for the spin degrees of freedom \cite{Kallin,Mott,PesinBalents}. However, the analysis of gauge fluctuations following Polyakov \cite{Polyakov} must be taken with care in two dimensions of space such that this requires additional flavors or spin species to stabilize this phase of matter. Such a topological Mott phase has also been identified in three dimensions related to iridates' materials \cite{PesinBalents}. For two-dimensional iridates, the Mott phase is described by spin textures \cite{ShitadeLiu, Thomale}. Furthermore, it is relevant to mention the emergence of a chiral spin state for the Kane-Mele-Hubbard model for bosons identified through DMFT, ED and quantum field theory \cite{Plekhanov}. 

\section{Topological Proximity Effects}
\label{proximityeffect}

In this Section, we show that a topological state on the honeycomb lattice may be induced through proximity effect leading to a possible $\mathbb{Z}_2$ topological classification for asymmetric masses \cite{bilayerQSH}. We comment on the role of inter-planes hopping and Coulomb interaction. We present applications of the two-spheres' model in a bilayer model \cite{HH} and also in a monolayer graphene \cite{Semimetal}. This allows us to identify a protected topological semimetal on the honeycomb lattice with Fermi liquid properties. 

\subsection{Induction of a Topological State in Graphene from Interlayer Hopping}
\label{hop}

Topological insulating states induced from proximity effect have attracted attention these last years theoretically \cite{Hsieh,Hofstetter} and experimentally \cite{AndoProximity}. Here, we introduce a $\mathbb{Z}_2$ topological proximity effect when coupling a graphene plane to a thin material described by a topological Haldane model forming then another plane \cite{bilayerQSH}. A small hopping term between planes induces a topological phase in graphene to second-order in perturbation theory. We mention here recent specific progress in engineering materials with graphene to achieve such a topological proximity effect \cite{QSHgraphene,Tiwari}.

The tunneling Hamiltonian takes the form
\begin{equation}
{H}_t = \int \frac{d{\bf k}}{(2\pi^2)} \sum_{\alpha\beta} \left(r c^{\dagger}_{g\alpha} \mathbb{I} c_{h\alpha} + \gamma({\bf k}) c^{\dagger}_{g\alpha} \sigma^{+}_{\alpha\beta} c_{h\beta} + h.c.\right).
\end{equation}
Here, $c^{\dagger}_{g}$ and $c^{\dagger}_h$ refer to electron (creation) operators in graphene and Haldane systems and the Pauli matrices $\bm\sigma$ act on the sub-lattice subspace (A,B) common for the two systems. Here, $r$ involves coupling between the same
sublattice $A$ or $B$ in the two planes. 

The topological proximity induced by a weak-$r$ value can be understood as follows.  A particle starts from graphene in sub-lattice $A$, then hops onto the same sub-lattice in the Haldane layer, and after the action of the second nearest-neighbor tunneling term $t_2$ giving a phase $+\phi$, the particle goes back in the graphene lattice producing an effective $t_2^{eff}$ term in the graphene layer proportional to $-\frac{|r|^2}{|d_z^h|^2} d_z^h\sigma_z$ in Eq. (\ref{hz}). As described below, the $-$ sign results from the second-order in perturbation theory, and should also reveal that for the B sub-lattice the perturbation theory gives an opposite sign because of the nature of the $t_2$ term in the Haldane layer. Assuming a $AA$ and $BB$ stacking between planes, in this case the effect of the $\gamma({\bf k})$ term in Eq. (\ref{parameters}) coupling here different sublattices is negligible in the proximity effect. This results from the fact that the sum of the three vectors ${\bf b}_i$ gives zero in Fig. \ref{graphenefig} when summing the effect of the different sites involved in the $\gamma({\bf k})$ term in the Haldane system. The proximity effect then can be simply described through a constant $r$ coupling between the two thin materials.

Here, we describe the effect in terms of two simple mathematical approaches (adapted from Supplemental Material in \cite{bilayerQSH}). 
For $r\ll t_2$, the effective Hamiltonian in the graphene layer takes the form
\begin{equation}
{H}_{eff}^g = P H_g P + P H_t (1-P) \frac{1}{(E-H)} (1-P) H_t P,
\end{equation}
where $H_g$ refers to the graphene lattice Hamiltonian.
Here, the projector $P$ acts on the sub-space where the lowest band of the Haldane system is completely filled linked to the ground state $|GS\rangle$, and the projector $(1-P)$ produces virtually a quasiparticle in the upper Haldane Bloch band. In this sense, the second term in $H_{eff}^g$ can be understood as follows, after introducing the notation ${c^{\dagger}}^u_{h\alpha}({\bf k}) | GS\rangle$ which creates a quasiparticle in the upper Haldane band and refers to the $(1-P)$ projector :
\begin{eqnarray}
\hskip -0.5cm - \frac{|r|^2}{|d_z^h({\bf k})|} \langle GS | c^{\dagger}_{g\alpha}({\bf k})c_{h\alpha}^u({\bf k}){c^{\dagger}}^{u}_{h\alpha}({\bf k}) |GS\rangle \frac{d^h_z({\bf k})\sigma_z}{| d^h_z({\bf k}) |} \\ \nonumber
\times\langle GS | c_{h\beta}^u({\bf k}) {c^{\dagger}}^u_{h\beta}({\bf k}) c_{g\beta}({\bf k}) |GS \rangle.
\end{eqnarray}
This gives rise to an effective Hamiltonian in the graphene system
\begin{equation}
{H}_{eff}^g = P H_g P + P \frac{-|r|^2}{|d_z^h({\bf k})|^2} c^{\dagger}_{g\alpha}({\bf k}) d_z^h({\bf k}) \sigma_z c_{g\beta}({\bf k}) P.
\end{equation}
The term $d_z^h({\bf k})\sigma_z$ reveals the phase accumulated for a particle in sub-lattice $A$ or/and sub-lattice $B$ when travelling in the Haldane
layer. At weak $r$-coupling, the induced $d_z$ term is of the form $-|r|^2/(27t_2^2 \sin^2\phi)d_z^h\left({\bf k}\right)\sigma_z $ in Eq. (\ref{hz}) which changes its sign at the two Dirac points, in the two valleys. In the analysis above, the system is spin-polarized as a result of Zeeman effect. 

As an application of quantum field theory, this result can also be verified through a path integral approach and simple transformations on matrices. The induced term in graphene can be obtained defining  $\zeta_h({\bf k}) = (c_{hA}({\bf k}), c_{hB}({\bf k}))$ and $\bar{\zeta}_h({\bf k}) = (c^{\dagger}_{hA}({\bf k}), c^{\dagger}_{hB}({\bf k}))$ for the partition function describing the Haldane system. The induced term in graphene can be obtained similarly as with classical numbers, when completing the `square' to obtain a Gaussian integral. It is appropriate to introduce the matrix
$M({\bf k})={\bf d}^h({\bf k})\cdot\bm\sigma$ and its inverse such that $M^{-1}({\bf k})\cdot M({\bf k})=M({\bf k})M^{-1}({\bf k})=\mathbb{I}$. In the weak-coupling limit $r\ll |{\bf d}^h({\bf k})|$, we can then redefine the dressed (fermionic) operator such that
\begin{equation}
({\zeta}_h^*({\bf k}))^T \approx (\zeta_h({\bf k}))^T + r M^{-1}({\bf k})(\zeta_g({\bf k}))^T
\end{equation}
and
\begin{equation}
\bar{\zeta}_h^*({\bf k}) \approx \bar{\zeta}_h({\bf k}) + r^* \bar{\zeta}_g({\bf k})M^{-1}({\bf k}).
\end{equation}
After completing the square, this gives rise to the following induced term in the partition function of graphene
\begin{eqnarray}
\hskip -0.5cm \int {\cal D}\zeta^*_h({\bf k}) {\cal D}\bar{\zeta}_h^*({\bf k}) e^{-\int_0^{\beta} d\tau \int \frac{d^2 k}{2\pi^2} \bar{\zeta}_h^*({\bf k})[\partial_{\tau} + M({\bf k})](\zeta_h^*({\bf k}))^T} && \\ \nonumber
\times e^{+\int_0^{\beta} d\tau \int \frac{d^2 k}{2\pi^2} |r|^2 \bar{\zeta}_g({\bf k}) M^{-1} (\zeta_g({\bf k}))^T} &.&
\end{eqnarray}
The induced term in graphene is only correct for time scales sufficiently long such that the gap in the Haldane layer has been formed. We neglect the dynamical effect in $\omega_n |r|^2/|{\bf d}^h({\bf k})|^2$ where $\omega_n$ are Matsubara frequencies related to real frequencies through the Wick rotation 
$\omega_n\rightarrow \omega+i0^+$, since we study ground-state properties. The effective Hamiltonian and the induced gap in graphene are in agreement with the perturbation theory. 

We can also re-interpret this result as an effective Ising coupling induced between two spins-1/2 in ${\bf k}$-space. If we introduce two distinct Pauli matrices $\bm\sigma_g$ and $\bm\sigma_h$ to describe the Hamiltonians of each system, the induced term in the graphene layer can be identified as 
\begin{equation}
- \frac{|r|^2}{|d_z^h({\bf k})|^2}d_z^h({\bf k}) \sigma_{gz} \rightarrow \frac{|r|^2}{|d_z^h({\bf k})|} \sigma_{gz}({\bf k})\sigma_{hz}({\bf k})
\end{equation}
The induced pseudo-magnetic field in the $z$ direction in the graphene layer for a given ${\bf k}$ wave-vector depends on the direction of the pseudo-spin (position of the particle in $A$ or $B$ sublattice) in the Haldane layer. 

Here, we discuss the applicability of the topological $\mathbb{Z}_2$ number in the situation of asymmetric masses related to the occurrence of this topological proximity effect. In the classification of Kane and Mele in Eq. (\ref{classification}) \cite{KaneMele2}, a mass asymmetry $\delta m=m_2-m_1$ gives rise to a perturbation of the form 
\begin{equation}
\delta {H}=d_2\sigma_z\otimes\mathbb{I}+\tilde{d}_{15} \sigma_z\otimes s_z=d_2\Gamma_2+\tilde{d}_{15}\Gamma_{15},
\end{equation}
in Eq. (\ref{classification}) where $d_2=\zeta\frac{\delta m}{2}$ and $\tilde{d}_{15}=-\zeta\frac{\delta m}{2}$ where $\zeta=\pm$ at $K$ and $K'$. In this way, the mass asymmetry is equivalent to the effect of a global staggering potential $d_2$ on the lattice which produces a topological mass $d_{15}+\tilde{d}_{15}=-\zeta\frac{1}{2}(m_1+m_2)$ (or equivalently
with $d_{15}=-\zeta m_1$) in the $\Gamma_{15}$ term. The quantum spin Hall effect is known to be stable as long as $|d_2|<|d_{15}+\tilde{d}_{15}|$ which means here $|m_1-m_2|<(m_1+m_2)$. This emphasizes the robustness of the $\mathbb{Z}_2$ topological phase even if one topological mass is small(er), as in the bilayer model described above. This is also in agreement with Fig. \ref{KaneMele} and with the fact that the topological number $C_s=C_1-C_2$ would keep the same form when driving from north to south pole (see Eq. (\ref{polesC})). As long as the directions of the effective magnetic fields at the poles do not change sign then the $C_s$ Chern number remains identical. 

However, the bulk-edge correspondence must be discussed with care in the topological proximity effect since a $r$ tunnel coupling at the edges would open a gap to first order in perturbation theory. The bulk proximity effect is revealed to second-order in $r$. Therefore, to stabilize two counter propagating modes at the edges in Fig. \ref{KaneMele} this requires to have two cylinders of different lengths, as verified numerically \cite{bilayerQSH}. 

\subsection{Coulomb Interaction Between two Planes}
\label{Coulomb}

Here, we describe interaction effects from an analogy with the Ising spin interaction, which will be useful to identify the possibility of a mapping towards the two spheres' model with $C_j=\frac{1}{2}$
described in Sec. \ref{fractionaltopology}. Each plane is described by a topological Haldane model with identical topological mass terms. 

We introduce the Coulomb interaction between the two planes. For a model of two spheres or two planes $1$ and $2$, we can project the interaction on the lowest filled band(s) to write an effective interaction at the Dirac points. For this purpose, we introduce the form of the projectors
\begin{equation}
\hat{n}_1^i = \frac{1}{2}\left(1\pm\sigma_{1z}\right)
\end{equation}
which act respectively in the Hilbert space $\Psi_1^{\dagger}=(c_{1A}^{\dagger},c_{1B}^{\dagger})$ 
and similarly for plane $2$
\begin{equation}
\hat{n}_2^i =\frac{1}{2}\left(1\pm\sigma_{2z}\right),
\end{equation}
Now, we use the structure of the eigenstates at the Dirac points such that for the two planes described by a Haldane model, the dominant interaction at the $K$ point is of the form $H_{Int}({\bf K})=\lambda \hat{n}_1^A \hat{n}_2^A$.
Similarly, close to the $K'$ point within the lowest band, the main interaction channel is of the form $H_{Int}({\bf K}')=\lambda \hat{n}_1^B \hat{n}_2^B$. These two local interactions at the Dirac points can be re-written as
\begin{eqnarray}
\label{Ising}
H_{Int}^1 = \frac{\lambda}{4}\sigma_{1z}\sigma_{2z} +\frac{\lambda}{4} +\frac{\lambda}{4}\zeta(\sigma_{1z}+\sigma_{2z}).
\end{eqnarray}
The Coulomb interaction between planes gives rise to an antiferromagnetic Ising interaction between the pseudo-spins, measuring the relative occupancy on each sublattice $A$ or $B$, 
and to a renormalization of the mass structure at the Dirac points. One can add the radial magnetic field acting on the two spheres and show that as long as $\lambda<2m$ the pseudo-spin polarization at the poles $\langle \sigma_{iz}\rangle$ remains identical such that from Eq. (\ref{polesC}), the total topological number $C_1+C_2$ remains equal to $2$.
We can also include long-range interactions in the reciprocal space between Dirac points. This results in an additional term
\begin{equation}
H_{int}^2 =\lambda'\left(\hat{n}_1^A({\bf K})\hat{n}_2^B({\bf K}') +\hat{n}_1^B({\bf K}')\hat{n}_2^A({\bf K})\right),
\end{equation} 
that we take as constant for simplicity. 
This increases the energy of the lowest energy states at the poles 
\begin{equation}
E_{AA}({\bf K})+E_{BB}({\bf K}')=(-4d+2\lambda+2\lambda').
\end{equation}
From the ground state, this is in fact identical to define a ``dressed'' interaction $\lambda_{eff}=\lambda+\lambda'$ in Eq. (\ref{Ising}). A similar argument can be made for the Kane-Mele model. This derivation shows that as long as the interaction
$\lambda_{eff}$ is smaller than the band gap, the topological phase is stable, as also discussed in Secs. \ref{Mott} and \ref{MottKM}.

When including Semenoff masses $M_1$ and $M_2$ in the two spheres we can obtain fractional geometry for relatively weak interactions as described in Sec. \ref{fractionaltopology}. 
The term $M_1=M_2$ on the lattice corresponds to a global staggered potential or modulated potential. This Section opens some perspective on realizing fractional topological numbers from the Coulomb interaction
and an analogy between charge and spin. In Sec. \ref{further}, we show a correspondence between the two spheres' model and two wires coupled through a Coulomb interaction \ref{further}.

\subsection{Topological Semimetal in a Bilayer System}
\label{bilayer}

Bilayer systems in graphene with Bernal stacking \cite{graphene,McCann} and also Moir\' e pattern \cite{AndreiMacDonald} have attracted a lot of attention these last years related to the quest of novel phases of matter \cite{Yazdanimagic,Louk,Herrero}. Topological semimetals in three dimensions have also attracted a growing attention these last years experimentally \cite{Bian,YanFelser,GuoDresden} and theoretically \cite{Burkovsemimetal,Ezawa,FangFu,YangNagaosa}. In 2015 Young and Kane have suggested the possibility of a two-dimensional Dirac semimetal on the square lattice \cite{YoungKane} with time-reversal symmetry.

Here, we show the possibility to realize a topological nodal ring semimetal in a bilayer system of two planes coupled through a hopping term $r$ as described in Sec. \ref{hop} \cite{HH}. This situation can be realized in optical lattices \cite{bilayerQSH}. Regarding quantum materials, one may design appropriate platforms to observe this topological semimetal. In Sec. \ref{semimetalclass}, we show a precise application of a (protected) topological nodal ring semimetal with one layer graphene \cite{Semimetal}. We also present a relation between this topological model and the two spheres' model with $C=\frac{1}{2}$ per sphere or per plane in Sec. \ref{topomatter}.
We mention here the very recent theoretical work reporting also the possibility of quantum anomalous semimetals linked to fractional topological numbers \cite{Bo}.

 \begin{center}
\begin{figure}[t]
\hskip 2cm
\includegraphics[width=0.7\textwidth]{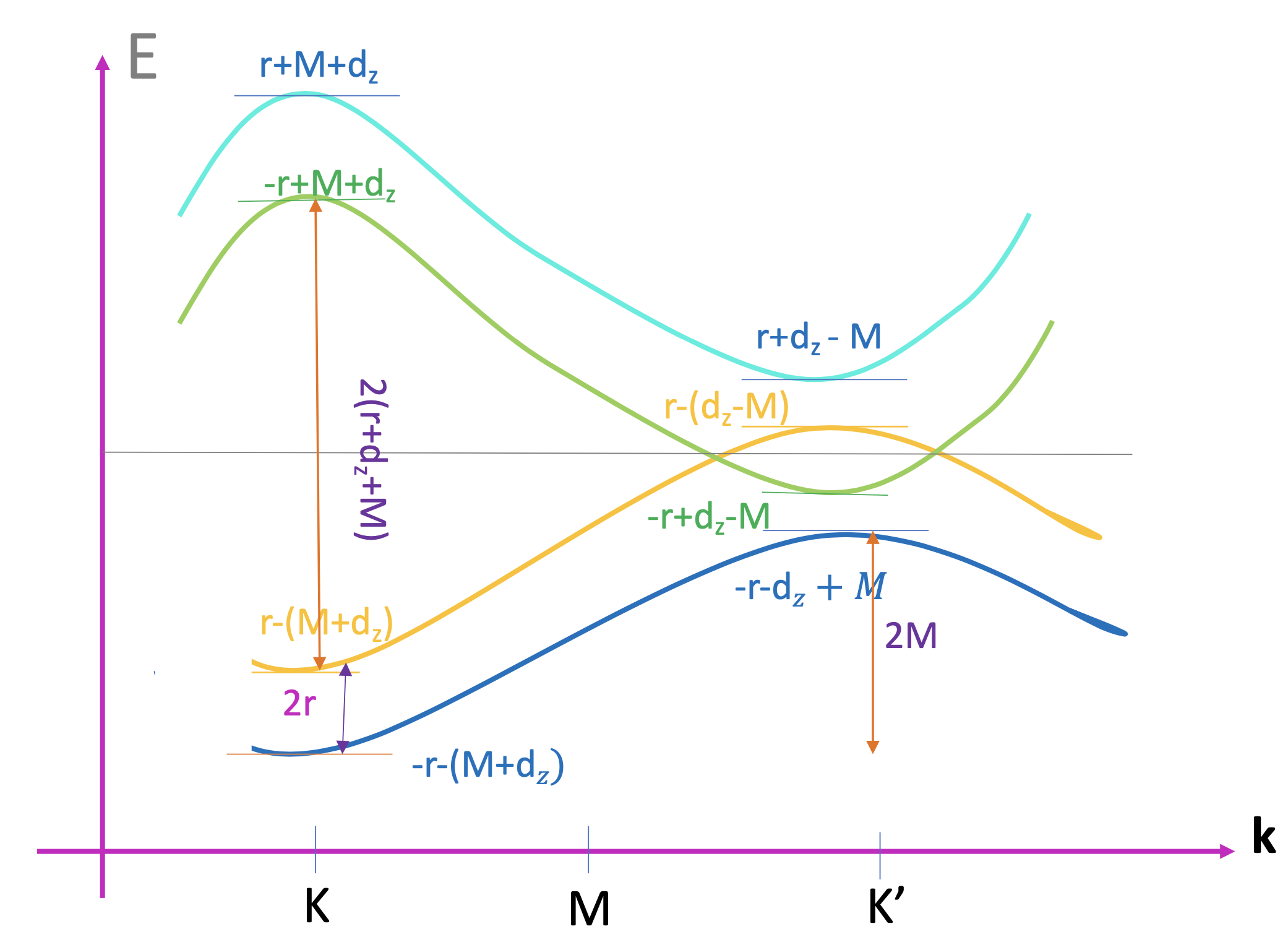}
\caption{Band Structure of the four-band model in the topological semimetal phase for $M$ of the order of $d_z$ with $0<d_z-M<r<d_z+M$ showing the Berry phases at the $K$ and $K'$ points.}
\label{4bandstructure}
\end{figure}
\end{center}

\vskip -1cm
The matrix model in the Hilbert space basis $(c^{\dagger}_{A1}, c^{\dagger}_{B1}, c^{\dagger}_{A2}, c^{\dagger}_{B2})$ takes the form
\begin{equation}
\label{semimetalmatrix}
H(\bm{k}) = \begin{pmatrix}
                           \zeta d_z +M & d_x-id_y & r & 0 \\
                            d_x+id_y & - \zeta d_z - M & 0 & r \\
                            r & 0 & M+\zeta d_z & d_x - i d_y \\
                            0 & r & d_x+i d_y & -\zeta d_z - M
 			  \end{pmatrix} \quad .
   \end{equation}
   Here, the components $d_x$ and $d_y$ describe the graphene physics in each plane according to Sec. \ref{spherelattice} and $\zeta=\pm$ at the $K$ and $K'$ Dirac points, respectively. We assume that  $d_z=m>0$ and $M>0$ with $M<d_z$. 
   The hopping term between planes is identical to that in Sec. \ref{hop}.
   
   We diagonalize the matrix and verify the four energy eigenstates close to the $K$ point \cite{HH}:
   \begin{eqnarray}
   E_1({\bf K}) &=& -r -\sqrt{d_x^2 + d_y^2 + (d_z+M)^2} \\ \nonumber
   E_2({\bf K}) &=& r - \sqrt{d_x^2+d_y^2 + (d_z+M)^2} \\ \nonumber
   E_3({\bf K}) &=& -r + \sqrt{d_x^2 + d_y^2 + (d_z+M)^2} \\ \nonumber
   E_4({\bf K}) &=& r + \sqrt{d_x^2+d_y^2 + (d_z+M)^2}.
   \label{Eeigenvalues}
   \end{eqnarray}
   The energies are structured such that $E_1$ corresponds to the lowest eigenenergy and $E_4$ the largest eigenenergy; see Fig. \ref{4bandstructure}. The corresponding eigenstates at the $K$ Dirac point are of the form $\psi_1=\frac{1}{\sqrt{2}}(0,-1,0,1)$, $\psi_2=\frac{1}{\sqrt{2}}(0,1,0,1)$, $\psi_3=\frac{1}{\sqrt{2}}(-1,0,1,0)$ and $\psi_4=\frac{1}{\sqrt{2}}(1,0,1,0)$. Precisely at the $K$ point, we have $d_x=d_y=0$ such that $E_1$ and $E_2$ correspond to the two lowest filled energy eigenstates at half-filling as long as $r<(d_z+M)$. In this case, the ground state corresponds to the two-particles wavefunction $\psi_1 \psi_2=e^{i\pi} c^{\dagger}_{B1} c^{\dagger}_{B2}|0\rangle$ with $|0\rangle$ referring to the vacuum state at wave-vector ${\bf K}$ precisely, the states below in energy being filled. 

  Similarly at the $K'$ point, changing $d_z\rightarrow -d_z$, the eigenenergies read:
   \begin{eqnarray}
   E_1' &=& -r -\sqrt{d_x^2 + d_y^2 + (d_z-M)^2} = E_1({\bf K}') \\ \nonumber
   E_2' &=& r - \sqrt{d_x^2+d_y^2 + (d_z-M)^2} = E_3({\bf K}') \\ \nonumber
   E_3' &=& -r + \sqrt{d_x^2 + d_y^2 + (d_z-M)^2} = E_2({\bf K}')\\ \nonumber
   E_4' &=& r + \sqrt{d_x^2+d_y^2 + (d_z-M)^2} = E_4({\bf K}').
   \end{eqnarray}
   The energy bands $E'_i$ refer to the same energy bands as defined at the $K$ point if we simply change $d_z\rightarrow -d_z$. The related eigenstates are of the form $\psi'_1 = \frac{1}{\sqrt{2}}(-1,0,1,0)$, $\psi'_2=\frac{1}{\sqrt{2}}(1,0,1,0)$, $\psi'_3=\frac{1}{\sqrt{2}}(0,-1,0,1)$ and $\psi'_4=\frac{1}{\sqrt{2}}(0,1,0,1)$. To realize a topological semi metal, the important prerequisite is to select accordingly $0<d_z-M<r<d_z+M$ similarly as in Eq. (\ref{HM}) for the two spheres. Indeed, in that case $E'_1=E_1({\bf K}')$ yet corresponds to the lowest-energy band, but then we also have an inversion between the bands $2'$ and $3'$ in this situation such that we can redefine $E'_3=E_2({\bf K}')$ and $E'_2=E_3({\bf K}')$. Within our definitions, energies are classified such that $E_1<E_2<E_3<E_4$ in Fig. \ref{4bandstructure}. At the $K'$ Dirac point there is a gap from the band $E'_3=E_2({\bf K}')$ to the Fermi energy.
 The occurence of a nodal ring in the region of the $K'$ Dirac point can be seen from the fact that  $E_2=E_3=0$ producing
 \begin{equation}
   v_F^2|{\bf p}|^2 = r^2 - (d_z-M)^2,
 \end{equation}  
 and a crossing effect around the $K'$ point. In this equation, the angle $\varphi$ can vary from $0$ to $2\pi$ which then defines a circle equation in the plane.
  At $K'$, the ground state corresponds to the two-particles wavefunction with one particle in band of energy $E_1({\bf K}')$ and one particle in band of energy $E'_3({\bf K}')=E_2({\bf K})$ such that the two-particles wavefunction reads $|\psi_g\rangle = \psi_1' \psi_3' |0\rangle = \frac{1}{2}(-c_{A1}^{\dagger} + c_{A2}^{\dagger})(-c_{B1}^{\dagger} + c_{B2}^{\dagger})|0\rangle$. It can be re-written as \cite{HH}
 \begin{equation}
 \label{psig}
 |\psi_g\rangle = \frac{1}{2}(c^{\dagger}_{A1} c^{\dagger}_{B1} - c^{\dagger}_{A1} c_{B2}^{\dagger} - c^{\dagger}_{A2} c^{\dagger}_{B1} + c^{\dagger}_{A2} c^{\dagger}_{B2})|0\rangle.
 \end{equation}
 
The topological nature of the system is revealed from the presence of one edge mode at zero energy in the band structure \cite{HH}. The total topological number $1$ is in agreement with the occurrence of one edge mode in the system which has $50\%$ chances to occupy each plane. The stable co-existence of the nodal semi-metallic ring and of the edge mode for $d_z-M<r<d_z+M$ can be understood from the fact that scattering events involving the $K$ and $K'$ regions would correspond to a wavelength that is not commensurate with the lattice spacing.  

Including a hopping term $\gamma({\bf k})$ between the planes, as in Sec. \ref{hop}, does not modify the wavefunction at the $K$ and $K'$ Dirac points which then reveals the stability of the nodal ring semi-metal towards this perturbation. 

 At $r=d_z-M$ we observe that the nodal ring shrinks as $E_2({\bf K}')=E_3({\bf K}')=0$ in that case. This implies the presence of an insulating phase with total topological number $2$ when $r<d_z-M$. For $r<d_z-M$, the physics is similar to that at $r=0$ showing two planes (or equivalently the two lowest bands) with the same topological number $1$.  For $r>d_z+M$, the two lowest bands show $+1$ and $-1$ topological numbers as in the Kane-Mele model with asymmetric topological masses \cite{bilayerQSH}
 giving rise effectively to an effective topological number equal to $0$ in each plane. 
 
 Now, we describe the properties of the semimetal related to the two spheres' model with $C_j=\frac{1}{2}$.   
 
  \subsection{$C_j=\frac{1}{2}$: Topological Properties of Bilayer model and Light}
 \label{topomatter}
 
To show the relevance of fractional topological numbers for this situation, in this section we present precise relations \cite{OneHalfKLH} between information on geometry (locally) from the poles of the sphere and within the energy bands at $K$ and $K'$, starting from Sec. \ref{fractionaltopology}. 
The specific energy of a particle is hidden hereafter and the precise detail of the band structure corresponds to adjust the shape of each fiber such that the radius of each sphere reproduces locally $|{\bf d}_i({\bf k})|$ for each plane. Below, we zoom on this geometrical response close to $K$ and $K'$ introducing a local topological marker from each side of the band-crossing effect.

The two-particles ground state wavefunction at $K$ is
\begin{equation}
|GS\rangle = |\psi_1(K)\rangle |\psi_2(K)\rangle = e^{i\pi} |\Phi_-\rangle_1 |\Phi_-\rangle_2= e^{i\pi} c^{\dagger}_{B1} c^{\dagger}_{B2} |0\rangle.
\end{equation}
This is equivalent to
\begin{equation}
 \sum_{j=planes 1,2} A_{j\varphi}(K) = \sum_{j=spheres 1,2} A_{j\varphi}(\theta=0).
\end{equation}
We have the precise correspondence
\begin{equation}
A_{j\varphi} = - i_K\langle GS| \partial_{\varphi}^j |GS\rangle_K,
\end{equation}
with $j$ referring equally to plane (sphere), particle or energy band in this equation. The function $A_{j\varphi}$ is smoothly defined around the $K$ point.
At $K'$, the two-particles ground-state wavefunction can be re-writen as
\begin{equation}
|GS\rangle_{K'} = \frac{1}{\sqrt{2}}|GS\rangle_{\theta=\pi^-} + \frac{1}{2} \left(c^{\dagger}_{A1} c^{\dagger}_{B1} + c^{\dagger}_{A2} c^{\dagger}_{B2}\right)|0\rangle.
\end{equation}
Here, the first term corresponds to an entangled state on the 2 spheres' model 
\begin{equation}
|GS\rangle_{\theta=\pi^-} = \frac{-1}{\sqrt{2}}(c^{\dagger}_{A1}c^{\dagger}_{B2} + c^{\dagger}_{A2}c^{\dagger}_{B1})|0\rangle=\frac{-1}{\sqrt{2}}(|\Phi_+\rangle_1^1 |\Phi_-\rangle_2^2 - |\Phi_-\rangle_1^1 |\Phi_+\rangle_2^2).
\end{equation}
The lowerscript refers to the sphere index and the upperscript refers to the particle index. If we measure the Berry phase in the sphere or particle basis, it gives the same information at $K'$. 
In the 2-spheres' model, to allow for a particle to navigate from north to south, this requires the two-spins' state to reside within the triplet state. For the analysis of the topological semimetal, we can e.g. redefine $|\Phi_+\rangle_2^2\rightarrow - |\Phi_+\rangle_2^2$
such that $|GS\rangle_{\theta=\pi^-}$ corresponds to the triplet state with $S_z=0$. This allows for a precise analogy with the two-spheres' model as formulated in Sec. \ref{fractionaltopology}.

We have the important equality \cite{OneHalfKLH}
\begin{equation}
A_{j\varphi}(\pi) = - i\langle GS(\theta=\pi^-)| \partial_{j\varphi} | GS(\theta=\pi^-)\rangle = \frac{A_{j\varphi}(0)}{2} + \frac{A_{j\varphi}^{r=0}(\pi)}{2}.
\end{equation}
In this equation, $j$ refers to a sphere corresponding then to a plane. We can then use from Sec. \ref{fractionaltopology}
\begin{equation}
A_{j\varphi}^{r=0}(\pi) - A_{j\varphi}(0) = q = -1.
\end{equation}
Therefore
\begin{equation}
A_{j\varphi}(\pi) = - i\langle GS(\theta=\pi^-)| \partial_{j\varphi} | GS(\theta=\pi^-)\rangle = A_{j\varphi}(0) +q\frac{1}{2}.
\end{equation}
Related to the first term in $|GS\rangle_{K'}$
\begin{equation}
\frac{1}{2}A_{j\varphi}(\pi) = - \frac{i}{2}\langle GS(\theta=\pi^-)| \partial_{j\varphi} | GS(\theta=\pi^-)\rangle = \frac{1}{2}A_{j\varphi}(0) +q\frac{1}{4}.
\end{equation}
Now, if we define the second term in $|GS\rangle_{K'}$ as
\begin{equation}
(*) = \frac{1}{2} \left(c^{\dagger}_{A1} c^{\dagger}_{B1} + c^{\dagger}_{A2} c^{\dagger}_{B2}\right)|0\rangle,
\end{equation}
we can write it down as
\begin{equation}
(*) = \frac{1}{2}(|\Phi_+\rangle_1^1 |\Phi_-\rangle_1^2 + |\Phi_-\rangle_2^1 |\Phi_+\rangle_2^2).
\end{equation}
For this term, the two particles occupy the same sphere. If we do the measure on a sphere from Sec. \ref{fractionaltopology}
\begin{equation}
- i \langle (*) | \partial_{j\varphi} | (*)\rangle = \frac{1}{4} A_{j\varphi}(0) + \frac{1}{4} A_{j\varphi}^{r=0}(\pi).
\end{equation}
We also have 
\begin{equation}
- i \langle (*) | \partial_{j\varphi} | (*)\rangle = \frac{1}{2} A_{j\varphi}(0) + \frac{1}{4}q .
\end{equation}
Therefore, we obtain the nice identity for a plane $j$
\begin{equation}
\label{semimetaltopo}
A_{j\varphi}(K') = A_{j\varphi}(0) +\frac{1}{2}q = A_{j\varphi}(K) +\frac{1}{2}q
\end{equation}
The function $A_{j\varphi}(K')$ is smoothly defined around the wavevector $K'$ such that the difference $A_{j\varphi}(K')-A_{j\varphi}(K)$ measured on each side of the band-crossing region can be introduced as a local topological marker. We can then introduce $A_{j\varphi}(K')-A_{j\varphi}(K)$ as a measure of a $\frac{1}{2}q$ topological invariant on a plane. This is in agreement with the edge structure on each plane showing 50\%-50\% probability for an injected particle \cite{HH}.

The quantum Hall response of the lowest band gives rise to $q \frac{e^2}{h}$, corresponding to add the topological responses of the two planes $j=1,2$ in Eq. (\ref{semimetaltopo}) 
\begin{equation}
\label{edgesplanes}
\sum_{j=1,2} (A_{j\varphi}(K') - A_{j\varphi}(K)) = q = \int_K^{K'} F_{k_x k_y}^{(1)} d k_y,
\end{equation}
where the summation on the left-hand side acts on the plane' subspace and $F_{k_x k_y}^{(1)}$ refers to the Berry curvature of the first energy band with the definition of the Brillouin zone $k_x=\varphi$ and $\theta=\frac{\pi}{2} - \frac{k_y 3\sqrt{3}a}{4}$, such that here $F_{k_x k_y}^{(1)}= F_{\theta\varphi}^{(1)}$. This quantum Hall response is equally distributed in each plane providing a bulk transport interpretation to the $q\frac{1}{2}$ number. In each plane, the left-hand side of Eq. (\ref{edgesplanes}) gives rise to edge modes on each cylinder
related to a plane $j$, which then provide an edge interpretation to the local topological marker.

The band $2$ is characterized through two domains which can be identified to the north and south regions of a sphere if we introduce $\theta_c$ as the crossing point between bands $2$ and $3$ along the path in momentum space. We can integrate the Berry curvature related to band $2$ for the two domains $\theta\in [0;\theta_c^-[$ and $\theta\in ]\theta_c^+;\pi]$. Since the Berry phases at $K$ and $K'$ are equal for band $2$, this results in an additional `interface' term, which is now a bulk contribution in momentum space, corresponding
to $A_{j\varphi}^{r=0}(\tilde{\theta}_c)=\frac{\cos\tilde{\theta}_c}{2}$ for a plane $j$. It is useful to introduce the dressed angle on the sphere including the `mass' $M$ defined as $\tan\tilde{\theta}=\sin\theta/(\cos\theta+\frac{M}{d_z})$ introduced in Sec. \ref{Semenoff}. This term gives rise to an additional transverse current located at the interface in momentum space. Solutions of $A_{j\varphi}^{r=0}(\tilde{\theta}_c)=0$ are particularly interesting because they preserve the topological information. They correspond to $\cos\theta_c+\frac{M}{d_z}=0$ such that $\tilde{\theta}_c=\frac{\pi}{2}$. If $\theta_c\rightarrow \pi$, this requires to adjust $M\rightarrow d_z^-$. The limit $M\rightarrow 0$ then would correspond to fix $r\rightarrow \sqrt{d_x^2+d_y^2}$ on the sphere, such that the nodal ring reaches the high-symmetric $M$ point for $\theta_c=\frac{\pi}{2}$.  If we now set $\theta_c\rightarrow 0$, due to the mass inversion $d_z\rightarrow -d_z$ between the $K$ and $K'$ points this leads to $M\rightarrow -d_z^+$. In this case, we also have the physical interpretation of a half-Skyrmion from the point of view that if we add the contributions from bands $1$ and $2$ for the integrals of the Berry curvatures, this reveals the charge $q$ in the domain $[0;\theta_c[$ and $0$ in the domain $]\theta_c;\pi]$. Therefore, the quantum Hall response per plane $q\frac{1}{2}\frac{e^2}{h}$ also measures the topological information of a half-Skyrmion. In this sense, fixing the interface (bulk) contribution to zero is a way to reveal a half-Skyrmion. Usually the presence of a Fermi surface alter the quantization of topological properties \cite{Haldane2004,KagomeAlex}. Here, since the crossing region is reduced to a point along the path then it is possible to obtain a purely topological response per plane. The $q\frac{1}{2}$ response in each plane is a consequence of the inversion symmetry between planes.

In the two-spheres' model of Sec. \ref{fractionaltopology}, measuring the quantum distance through a circularly polarized field acting on one sphere at south pole can reveal the fractional topological number \cite{OneHalfKLH}. 
Here, we discuss the light response for the topological semimetal, related to Sec. \ref{light}, and present an alternative understanding of the topological response $q\frac{1}{2}$ per plane. For simplicity, we suppose an incoming wave along $z$ direction perpendicular to the system.  We assume an identical light-matter coupling for the two planes or two spin polarizations which can be re-written in terms of the eigenstates for the bilayer (2-planes) system similarly as in Sec. \ref{lightdipole}. At the $K$ Dirac point, since the two-filled bands forming the ground state are related to the eigenenergies $E_1$ and $E_2$ then the light-matter coupling for a right-handed circularly polarized light reads
 \begin{equation}
 \label{light1/2}
 \delta {H}_{+} = A_0 e^{i\omega t}(|\psi_3\rangle \langle \psi_1| + |\psi_4\rangle\langle \psi_2|) +h.c.
 \end{equation}
 A similar representation is obtained in second quantization identifying $|\psi_ i \rangle = \psi^{\dagger}_i$ and $\langle \psi_i |=\psi_i$ with $i=1,2,3,4$ referring to the band energy similarly as in Sec. \ref{bilayer}. Then, adapting the calculation of Eq. (\ref{short}) in time we find a similar result for transition probabilities with $\tilde{\omega}=\omega-E_3+E_1$ or $\tilde{\omega}=\omega-E_4+E_2$. The energies at the $K$ Dirac point are shown in Eqs. (\ref{Eeigenvalues}). For inter-band transitions between bands $1$ and $3$ the prefactor can be interpreted as $C^2=|C|=q^2=1$ with $C$ referring to the topological number of band $1$. Transitions between bands $2$ and $4$ reveal a similar topological prefactor at $K$ point which is in agreement with the fact that bands $1$ and $2$ are characterized through the same Berry phase. Selecting the light frequency of one polarization we can then mediate transitions from $1\rightarrow 3$ or $2\rightarrow 4$. We can proceed in the same way at the $K'$ Dirac point. Writing the operator $|a_1\rangle\langle b_1| + |a_2\rangle\langle b_2|$ in terms of the eigenstates $\psi_i'$ we find that since the energy bands $E_1'$ and $E_3'$ form the ground-state and the two bands at energies $E_2'$ and $E_4'$ are empty then the light-matter cannot generate inter-band transitions at the $K'$ Dirac point as a result of inversion between bands $2$ and $3$, independently of the choice of the light polarization. This is similar as if $C^2=|C|=0$ in the response or equivalently $A^{(1)}(K')+A^{(2)}(K')=0$.  Averaging the light signal on the two regions then reproduces the superposition of two topologically distinct regions in the reciprocal space in accordance with a topological number $|C|=\frac{1}{2}(1+0)$ for a plane.  The light response measures locally the same information as for the situation of a half-Skyrmion in the quantum Hall response.
 
 The light response and the quantum Hall conductivity or transport then reveal complementary information on the topological entangled system's nature. 
 
 In the next section, we address the situation where the two planes would correspond to the spin polarizations of an electron $|+\rangle_z;|-\rangle_z$ in graphene \cite{Semimetal}. The $r$ term will correspond to an in-plane magnetic field such that measuring the quantum Hall responses in the basis of spin polarizations $|+\rangle_x;|-\rangle_x$ is also justified in this case. Within the band theory, we can describe transitions from a quantum Hall to a quantum spin Hall effect through the topological semimetal \cite{SariahKarynNew}. In this formulation, the domain of band $2$ related to $K$ is polarized along $|+\rangle_x$ direction and the domain of band $2$ related to $K'$ is polarized along $|-\rangle_x$ direction.

 \subsection{Topological Protected Fermi liquid, Classification and Realization in one Layer Graphene}
 \label{semimetalclass}
 
 Here, we elaborate on the properties of this nodal ring semimetal with $\mathbb{Z}_2$ symmetry re-introducing the two spins-$\frac{1}{2}$ matrices of the Kane-Mele and two-planes' models such that $\mathbfit{\sigma}$ acts on the sublattice sub-space and $\bf{s}$ on the flavor, spin or plane sub-space. This representation gives further insight on the realization of this state of matter in a mono-layer graphene system respecting the $\mathbb{Z}_2$ symmetry. The system is stable towards interaction and disorder effects and can be viewed as a symmetry-protected topological state \cite{Semimetal}. 
  
The model reads $H=\sum_{\bf k} \psi^{\dagger}({\bf k}) {H}({\bf k})\psi({\bf k})$ with $\psi({\bf k})=(c_{A{\bf k}\uparrow},c_{B{\bf k}\uparrow},c_{A{\bf k}\downarrow},c_{B{\bf k}\downarrow})$ and
\begin{eqnarray}
\label{Hmodel}
\hskip -0.5cm H({\bf k}) &=& (\zeta d_z + M)\sigma_z\otimes \mathbb{I} + d_1 \sigma_x\otimes \mathbb{I} + d_{12}\sigma_y\otimes \mathbb{I} \\ \nonumber
&+& r\mathbb{I}\otimes s_x.
\end{eqnarray}
Here, $d_1$ and $d_{12}$ correspond to the graphene Hamiltonian as defined in \ref{timereversal}, $\zeta d_z$ to the topological 
term in each plane induced through circularly polarized light such that $\zeta=\pm 1$ at the two Dirac points; see Sec. \ref{anomalous}. 
We observe that $[H,\mathbb{I}\otimes s_x]=0$ such that we can classify the eigenstates in terms of $|\psi_{\pm}\rangle$ in Eq. (\ref{eigenstates}) associated to the radial magnetic field for one sphere or one plane and the eigenstates of $s_x=\pm $ for the planes or layers.

An important aspect to realize this model is the form of the $M\sigma_z\otimes \mathbb{I}$ potential term. Physically, suppose general potential terms $V_{1}^A, V_{1}^B, V_2^A, V_2^B$ acting on the two planes and resolved on a given sublattice, then we have
the general relation
\begin{eqnarray}
&&V_1^A \hat{n}_1^A + V_1^B \hat{n}_1^B + V_2^A \hat{n}_2^A +V_2^B \hat{n}_2^B \\ \nonumber
&=& \frac{1}{4} \left( (V_1^A +V_1^B) - (V_2^A +V_2^B)\right)\mathbb{I}\otimes s_z \\ \nonumber
&+& \frac{1}{4} \left( (V_1^A-V_1^B) + (V_2^A-V_2^B)\right)\sigma_z\otimes \mathbb{I} \\ \nonumber
&+& \frac{1}{4} \left( (V_1^A-V_1^B) - (V_2^A-V_2^B)\right)\sigma_z\otimes s_z.
\end{eqnarray}
This equality can be understood from $\hat{n}_j^i = \hat{P}_i\otimes \hat{P}_j$ with $\hat{P}_i$ the projector on sublattice $i$ and $\hat{P}_j$  the projector in plane $j$. There is also an additional term proportional to $\mathbb{I}\otimes\mathbb{I}$ which shifts
the energy scale. Therefore, to realize the $\mathbb{Z}_2$ symmetry corresponding to invert the two flavors or two planes in the Hamiltonian, this requires to have $V_{1}^A = V_2^A$ and $V_{1}^B = V_2^B$. In this sense, there is no potential difference
 in the transverse direction to the planes. We observe that this also requires $V^A \neq V^B$. The $\mathbb{Z}_2$ symmetry of the nodal ring semimetal
implies the form $M\sigma_z\otimes \mathbb{I}$ of the potential term which is essential to realize a topological semimetal (if the symmetry $1\leftrightarrow 2$ is not respected in the form of the potentials, then a disorder term $\sigma_z\otimes s_z$ would be a relevant perturbation). 

To satisfy the perfect equalities $V_{1}^A = V_2^A$ and $V_{1}^B = V_2^B$ and the $\mathbb{Z}_2$ symmetry,
the model then can be implemented with one layer (thin plane) of graphene \cite{Semimetal}. Suppose a graphene plane with spins-$\frac{1}{2}$ electrons such that $\bf{s}$ acts on the physical spin space. The two planes then mean the two spin polarizations of an electron along $z$ direction with $\uparrow$ and $\downarrow$ referring to $s_z=\pm 1$ respectively. Applying a modulated potential term in the plane then would satisfy $V_{1}^A = V_2^A$ and $V_{1}^B = V_2^B$ in this situation. This way, the $r$ term can also be implemented as a Zeeman effect from a tunable magnetic field along $x$ direction. Tuning the magnetic field in the plane then allows us to satisfy $d_z-M<r<d_z+M$ and therefore to realize the topological semimetal. Applying an in-pane magnetic field is a key feature of the model.

The term $\sigma_z\otimes\mathbb{I}$ can be precisely engineered from a modulated potential applied in the graphene plane such that $V_A\neq V_B$ or also through
 interaction effects with a honeycomb layer in a Mott phase or CDW phase \cite{Klein}. The particles of the CDW system sit either on $A$ or $B$ sites and act as a potential on the graphene lattice. A Coulomb interaction reads $V\left(\hat{n}_{CDW}^A\hat{n}_g^A + \hat{n}_{CDW}^B\hat{n}_g^B\right)$ with $\langle \hat{n}_{CDW}^A \rangle =1$ and $\langle \hat{n}_{CDW}^B \rangle = 0$, and therefore we have $V\hat{n}_g^A = \frac{V}{2}(1+\sigma_z\otimes \mathbb{I})$. In this case, we would assume that $r$ is sufficiently strong such that the spins of the particles are polarized along $x$ direction and such that the two spin polarizations 
 $\uparrow$ or  $\downarrow$ see the same potential on each site. We have the identification $M=\frac{V}{2}$ with the prerequisite $d_z-M<r<d_z+M$ to realize the topological nodal ring semimetal.

Then, the $\zeta d_z \sigma_z \otimes \mathbb{I}$ term can be implemented \cite{Semimetal} though circularly polarized light for the two spin states $\uparrow$ and $\downarrow$ equally through the protocol(s) of Sec. \ref{polarizationlight} and Sec. \ref{graphenelight} which is realizable with current technology in graphene \cite{McIver}. A similar protocol can be implemented with atoms in optical lattices \cite{Jotzu}.

Within this representation, we have \cite{Semimetal}
\begin{equation}
H = {\bf d}\cdot\mathbfit{\sigma}\otimes\mathbb{I}+ r\mathbb{I}\otimes s_x
\end{equation}
and 
\begin{eqnarray}
H^2 = \left(|{\bf d}|^2 +r^2\right)\mathbb{I}\otimes\mathbb{I}
+2r {\bf d}\cdot\mathbfit{\sigma}\otimes s_x
\end{eqnarray}
with the form of ${\bf d}$ vector
\begin{equation}
{\bf d} = (d_1,d_{12},(\zeta d_z+M))
\end{equation}
such that $|{\bf d}|^2={\bf d}\cdot {\bf d} = d_1^2 + d_{12}^2 + (\zeta d_z+M)^2$. Close to the two Dirac points $d_1 = v_F|{\bf p}| \cos\tilde{\varphi}$, $d_{12}=v_F |{\bf p}|\sin\tilde{\varphi}$ such that $d_1^2 + d_{12}^2 = v_F^2 |{\bf p}|^2$. 
To classify eigenstates we can then introduce $|\psi_+\rangle$ and $|\psi_-\rangle$ as defined in Eq. (\ref{eigenstates}) corresponding here to energies $\pm |{\bf d}|$ and $|+\rangle_x$, $|-\rangle_x$ corresponding to spin eigenvalues $s_x=\pm 1$. 
The lowest and top energy levels correspond to $|\psi_-\rangle \otimes |-\rangle_x$ and $|\psi_+\rangle \otimes |+\rangle_x$ such that $2r {\bf d}\cdot\mathbfit{\sigma}\otimes s_x = 2r {\bf d}\cdot\mathbfit{\sigma}$ when acting on
these two states. Therefore, the energies of these two states satisfy 
\begin{equation}
E^2 = (r+ |{\bf d}|)^2,
\end{equation}
with respectively
\begin{equation}
E_1 = -(r+|{\bf d}|)
\end{equation}
and
\begin{equation}
E_4 = (r+|{\bf d}|).
\end{equation}
The energies at the two Dirac points are different due to $\zeta=\pm 1$. 

Now, we study the occurrence of the semi-metal through the two middle or intermediate bands which correspond to the two eigenstates $|\psi_-\rangle \otimes |+\rangle_x$ and $|\psi_+\rangle \otimes |-\rangle_x$ such that  $2r {\bf d}\cdot\mathbfit{\sigma}\otimes s_x =-2r {\bf d}\cdot\mathbfit{\sigma}$. 
These two energy states are then described through
\begin{equation}
E^2 = (-r+ |{\bf d}|)^2.
\end{equation}
To have a semimetal this requires these two bands to meet when $E^2=0$ implying the general relation $r=|{\bf d}|=\sqrt{d_1^2+d_{12}^2+(\zeta d_z+M)^2}$. The formation of the semimetal implies to adjust $r$ such that $d_z-M <  r <d_z+M$ close to the Dirac points which can be satisfied only if $\zeta=-1$. There is no inversion symmetry ${\bf k}\rightarrow -{\bf k}$. The $\mathbb{Z}_2$ symmetry leads to a pair of degenerate bands at $E^2=0$ with a double-orthogonality structure in the crossing eigenstates \cite{Semimetal} protecting the system  (see Fig. \ref{4bandstructure}). 

At $K'$ point, the eigenstate of the energy band $E_2'=E_3({\bf K}')=r-|{\bf d}|$ reads $|\psi_-\rangle\otimes|+\rangle_x$ and the eigenstate of the energy band $E_3'=E_2=|{\bf d}|-r$ reads $|\psi_+\rangle\otimes|-\rangle_x$. 
The ground state at $K'$ Dirac point corresponds to the two-particles' wavefunction such that the two lowest-energy bands are occupied. Therefore, this corresponds to
the (2-particles) ground-state wavefunction
\begin{equation}
|\psi_g\rangle = \left(|\psi_-\rangle\otimes|-\rangle_x\right)\left(|\psi_+\rangle\otimes|-\rangle_x\right),
\end{equation}
which corresponds to another way to write Eq. (\ref{psig}). 
Setting $\theta=\pi$ in Eq. (\ref{eigenstates}), then $|\psi_+\rangle$ corresponds to a particle in sublattice $B$ and $|\psi_-\rangle$ a particle in sublattice $A$. In the planes' or spin polarizations subspace, 
$|-\rangle_x = \frac{1}{\sqrt{2}}(|+\rangle_z - |-\rangle_z) = \frac{1}{\sqrt{2}}(|1\rangle - |2\rangle)$ corresponding to a quantum superposition. For simplicity, we have fixed the 
relative phases entering as a gauge choice to zero. 
  
The stability of the semimetal towards interaction effects in graphene can be verified at a mean-field level and also perturbatively. From Eq. (\ref{interactionsPhi}), we observe that the $U\phi_x$ term can be absorbed in the $r$ term corresponding to a transverse magnetic field in the implementation
with graphene and the two spin polarizations of an electron. From Eq. (\ref{Hmodel}), assuming that the $r$ term is real then $\phi_y=0$. The system is also stable towards a term $U\phi_z\mathbb{I}\otimes s_z$. This term is irrelevant in the semi-metallic phase since an effective hopping term between
bands $2$ and $3$ at the crossing points would require an operator of the form $\sigma^+\otimes(s_y-is_z)$ or equivalently $\sigma^-\otimes(s_y-is_z)$. Here, the operator $s_y$ does not occur to lowest order in the Hamiltonian and in fact is not generated to higher orders in perturbation theory.
Increasing $U$ this is equivalent to progressively rotate the spin eigenstates of the operator $r \mathbb{I}\otimes s_x+U\phi_x \mathbb{I}\otimes s_z$ similarly as the eigenstates (\ref{eigenstates}) using now the basis $|+\rangle_x$ and $|-\rangle_x$.
The stability of the semimetal towards interaction effects can also be understood from the fact that the energy spectrum evolves quadratically close to the two crossing points traducing a topological Fermi liquid and a finite density of states
at zero energy \cite{Semimetal}. This behavior is then distinct from interaction effects in graphene \cite{polarizationgraphene}.

We hope that this work may stimulate further experimental efforts and it is worth mentioning the recent observation of a quantized (thermoelectric) Hall plateau in three-dimension through a nodal line semimetal \cite{Tokyo}. Very recently,
a Fermi-liquid behavior was observed in twisted bilayer graphene semimetals \cite{MITbilayer}.

 \section{Topological Planes or Spheres in a Cube, Geometry and Infinite Series} 
\label{Planks}

From the preceding Sections, we develop applications of the formalism in coupled planes'. We assemble planes and through the geometry, transport and light-matter responses, we present a relation between geometry in a cube, infinite mathematical series from Ramanujan, the $\frac{1}{2}$ in the two-spheres' model and the surface of three-dimensional topological insulators.

Stabilizing three-dimensional quantum Hall phases and topological insulating phases from coupled planes \cite{Orth,Halperin3D,Berneviggraphite} has attracted attention in the community. Very recently, the possibility of a fractional second-order topological insulator from an ensemble of wires in a cube was also identified \cite{JelenaWiresFTI}. Here, we study a network of $j=0,1,...N$ planes ($N+1$ planes) described through a Haldane model \cite{Haldane} with alternating $(-1)^j t_2 e^{i\phi}$ second-nearest neighbors' hopping terms in a cube. Each plane is equivalent to a topological sphere. We suppose the limit of weakly coupled planes with a hopping term or interaction (much) smaller than the topological energy gap in the band structure such that the two-dimensional band structure in each plane (sphere) is stable. One objective of this section will be to link different interpretations of the $\frac{1}{2}$ number, through the Ramanujan infinite series in the planes' model in relation to the spheres and also through surfaces of three-dimensional topological insulators. 

\subsection{Even-Odd effect}

We begin from geometry with a dilute number of planes in a cube. 
The topological number in each plane can be defined through an integration on the Brillouin zone according to previous definitions
\begin{equation}
C_j = \frac{1}{2\pi}\iint d^2 k{\bf F}_j\cdot{\bf e}_z
\end{equation}
with $d^2 k =dk_x d k_y$ and ${\bf e}_z$ the unit vector in the vertical direction. To study geometrical aspects in a three-dimensional space, we re-organize the planes in a cube structure in the $(k_x,k_y,z)$ space with $k_x$ and $k_y$ defined in the reciprocal space such that the integral of the Berry curvature for each square plane agrees with $C_j$ defined above on a torus in Eq. (\ref{pumpingC}). Here, $z$ represents the vertical coordinate in real space measuring the number of planes. Suppose we have a fixed number of planes, $N+1$ in the cube, then we can define an even-odd effect for such a topological effect in agreement with the Green and divergence theorems. 

The Berry curvature in a plane takes the form 
\begin{equation}
\label{Fz}
{\bf F}_j = (-1)^j F_{k_x k_y}\delta_{z j} {\bf e}_z = F_j(k_x,k_y,z){\bf e}_z = F_{z=j}{\bf e}_z
\end{equation}
with the same Berry curvature $F_{k_x k_y}$ for each plane. Equivalently, we can measure $C_j$ and associated observables in each plane, following Sec. \ref{Observables}, from $F_{p_x p_y}(0)\pm F_{\pm p_x p_y}(\pi)$ on the sphere with $p_x$ and $p_y$ measuring deviations for the momentum from the Dirac points. From the divergence theorem, we can define a non-local topological number from top and bottom surfaces
\begin{eqnarray}
\label{number}
&& \hskip -0.5cm C_{top} - C_{bottom} = \frac{1}{2\pi}\int \int d k_x dk_y  \int_{z_{bottom}}^{z_{top}}  \frac{\partial F_{z=j}}{\partial z}dz \\
&=& \frac{1}{2\pi}\int \int d k_x dk_y \left( F_z(k_x, k_y , z_{top}) - F_z(k_x,k_y,z_{bottom})\right).  \nonumber
\end{eqnarray}
Here, top and bottom mean equally up and down.
Also, we can verify Eq. (\ref{number}) in the limit of a dilute number of planes (planks) by calculating $\frac{\partial}{\partial z} [(-1)^j \delta_{zj}] = (-1)^j \delta'_{zj}$. The vertical surfaces of the cube give zero since ${\bf F}=F_{z} {\bf e}_z$ is perpendicular to
the normal vector for these faces. Although the divergence theorem refers to an integration on a closed surface encircling the volume, here $C_{top}-C_{bottom}$ can be defined simply from the horizontal faces.

This is equivalent to
\begin{equation}
C_{top} - C_{bottom} = C((-1)^N - 1)
\end{equation}
with
\begin{equation}
C = \frac{1}{2\pi}\iint dk_x dk_y F_{k_x k_y}=1.
\end{equation}
Here, the bottom surface corresponds to $j=0$ such that $C_{top}-C_{bottom}$ is defined to be real, alternatively $-2$ and $0$ when $N\in \mathbb{N}$. In addition, we can define the total conductivity in response to an electric field for each plane as in Sec. \ref{curvature} such that
\begin{equation}
\label{planessigmaxy}
\sigma_{xy} = \frac{e^2}{h}\sum_{j=0}^N (-1)^j.
\end{equation}
For $N$ odd or a even number $(N+1)$ of planes, the top and bottom surfaces can develop a $\mathbb{Z}_2$ spin Chern number such that with the present definition of the $t_2$ term, $C_{top}-C_{bottom}=-2$ similarly as for the Kane-Mele model, and $\sigma_{xy}=0$ \cite{KaneMele1}. 
The $\mathbb{Z}_2$ symmetry $C_{top}-C_{bottom}=\pm 2$ here characterizes the parity symmetry with respect to the center (middle) of the cube corresponding to modify $z_{bottom}\leftrightarrow z_{top}$. Since $\sigma_{xy}=0$, the system preserves time-reversal symmetry.
The two horizontal facets of the cube develop edge modes moving counter-clockwise and in this way we realize a $\mathbb{Z}_2$ topological insulator from the top and bottom surfaces. For $N$ even or a odd number $(N+1)$ of planes, $C_{top}-C_{bottom} = 0$. Since we start at $j=0$ the counting of the planes, then the total conductivity agrees with that of the quantum Hall effect $\sigma_{xy}=\frac{e^2}{h}$. 

Therefore, for a dilute number of planes we observe a even/odd effect (referring to the number $N$) where the system behaves alternatively as an effective two-dimensional quantum Hall and a quantum spin Hall system.

\subsection{Thermodynamical Limit and Ramanujan Series}
\label{onehalfcharge}

Here, we address the thermodynamic limit of Eq. (\ref{planessigmaxy}) with $N\rightarrow +\infty$ in relation with the Ramanujan alternating series $({\cal R})$, re-written as
\begin{equation}
\label{infiniteseries}
\hbox{lim}_{\epsilon\rightarrow 0} \sum_{j=0}^{+\infty} \left((-1)^ j (1-\epsilon)^j\right) = \frac{1}{1+(1-\epsilon)} = \frac{1}{2}.
\end{equation}
The mathematical regularization is defined in the sense of Abel simply through a term $(1-\epsilon)^j<1$ with $\epsilon$ being any infinitesimal number. The $\frac{1}{2}$ can be understood as follows. The Ramanujan series corresponds to a string of numbers $S=1-1+1-1+...$ such that $S=1-S$ in the thermodynamic limit implying $S=\frac{1}{2}$.

Some questions then arise:  can we measure a halved conductivity through Eq. (\ref{planessigmaxy})? Is there a relation with the sphere model?  There are certainly various ways to address the thermodynamical limit in experiments through a staircase of assembled planes.  In Sec. \ref{planks}, we study in detail one specific realization of this $\frac{1}{2}$ thermodynamical limit related to the sphere model.  It is topologically equivalent to one Dirac point. 

In \ref{GeometryCube}, we also verify that to address the thermodynamical limit from the geometry we can assume an infinite system $z\in [0;+\infty[$ with a Berry curvature $F_z$ being now a continuous function along the $z$-axis, 
\begin{equation}
\label{Fz}
F_z=e^{-i\pi z} F_{k_x k_y}\theta(z),
\end{equation}
 where, as above $F_{k_x k_y}$ does not depend on the $z$ direction.   From Eq. (\ref{number}), then we can verify that in this situation the Heaviside step function gives a $\frac{1}{2}$ precisely at the boundary $z=0$ and integrating $z\in[0;+\infty[$ this is equivalent to say that there is no outer vertical surface in the divergence theorem such that effectively $F_z(top)=0$. In this case, the geometry predicts the occurrence of half quantum numbers from the particular mathematical form of the Berry curvature when approaching $z=0$. A continuous version of Eq. (\ref{number}) with $z\in [0;+\infty[$ can also reveal a $\frac{1}{2}$ topological number on a surface. In the next Sec. \ref{LightSeries}, we aim to relate the $\theta(z)$ function behavior in Eq. (\ref{Fz}) from a plane at $z=0$ (in the light response of the infinite system) from Series and the Riemann-Zeta function $\zeta(s)$. When studying the light response of the $z=0$ plane with topological number $C=1$, the proximity effect with the infinite number of planes will act precisely as if we would renormalize $C=1$ into $1+\zeta(0)=\frac{1}{2}$ at this boundary.
 
This analysis shows that it is indeed possible to predict one-half quantum numbers with specific 
realizations or physical interpretations of the thermodynamical limit. In Sec. \ref{3DQHE}, we relate this physics with surfaces of three-dimensional topological insulators which are also characterized through a similar $\frac{1}{2}$ quantum number.

 \subsection{Transport and Light from Infinite Series}
 \label{LightSeries}
 
 Before addressing a specific physical application, we show that the $\frac{1}{2}$ topological number naturally occurs in observables associated to the resummation of these infinite series. 
 
 Suppose that we apply an electric field along the planes with a small gradient in the vertical direction ${\bf E}={\bf E}_0 (1-\epsilon z)$ then, since the $\epsilon$ term is independent of $k_x$ and $k_y$, we can reproduce the calculation of Sec. \ref{curvature} in a given plane. Using Eqs. (\ref{DeltaP}) and (\ref{chargeE}), we obtain
\begin{equation}
\sigma_{xy}^j = \frac{e^2}{h} (-1)^j (1-\epsilon j) =  \frac{e^2}{h} (-1)^j (1-\epsilon)^j.
\end{equation}
Summing all the currents, this protocol measures $\sigma_{xy} = \sum_{j=0}^{+\infty} \sigma_{xy}^j = \frac{1}{2}\frac{e^2}{h}$ revealing an effective number $C_{eff}=\frac{1}{2}$. 

Here, we describe the light response in the situation of circular polarizations and generalize protocols of Sec. \ref{light}. For $j=2m+1$ (odd), for an identical vector potential in each plane, the right-handed (left-handed) circular light will measure 
\begin{equation}
(*)=\sum_{j=2m+1=1}^{+\infty} C_j^2 = 1+1+1+1 +1+... 
\end{equation}
 from the $K'$ $(K)$ Dirac point related to the inter-band transition probabilities of Eq. (\ref{density}) in time. Similarly, the right-handed (left-handed) circularly polarized light will measure the other planes with $j=2m$ even from the $K$ $(K')$ Dirac point 
 \begin{equation}
 (**)=\sum_{j=2m=0}^{+\infty} C_j^2 = 1+1+1+1+1+... 
 \end{equation}
 For the dilute limit of planes, each light polarization can then reveal the staggered structure of the topological mass in the transverse direction or the number of planes with a positive topological mass $+m$ at the $K$ or $K'$ Dirac point. 

In the thermodynamic limit, if we add the two responses $(*)+(**)$ (supposing that the light intensity can remain perfectly symmetric in all the planes), the infinite sum then diverges in the usual sense. Similarly as the protocol for the conductivity, we can regularize the sum as follows. Suppose that we shine light from the boundary plane at $j=z=0$ with an amplitude $A_0$ for the vector potential and that the amplitude of the vector potential now smoothly evolves with a power law in the bulk of the system such that $A_0\rightarrow A_0\frac{1}{{z}^{s/2}}$ if $z\geq 1$ and $s\geq 0$. In this case, the total light response can be written as
\begin{equation}
\label{S}
C_0^2 + \sum_{j=1}^{+\infty} C_j^2 \frac{1}{j^s} = 1 + \zeta(s),
\end{equation}
with $\zeta(s)$ being the Riemann-Zeta function. In the `ideal' limit where the planes interact quasi-symmetrically with light, then this is similar as if we fix $s\rightarrow 0$ and re-interpret the series for even values of $s$ (such that $s/2$ is an integer in the power-law decay of $A_0(z)$) through
\begin{equation}
\zeta(s=2n) = \frac{(-1)^{n+1} B_{2n}(2\pi)^{2n}}{2(2n!)}.
\end{equation}
In this way, $\zeta(0)$ is defined through the Bernouilli number $B_0=+1$ and $\zeta(0)=-\frac{1}{2}$. The situation is similar as if light reveals
\begin{equation}
\sum_{j=0}^{+\infty} C_j^2 = \sum_{j=0}^{+\infty} |C_j| = \frac{1}{2}.
\end{equation}
The re-interpretation of $C_j^2=|C_j|$ comes from the definition of the photo-induced currents in each plane in Eq. (\ref{photocurrents}). Eq. (\ref{S}) can be viewed as a realization of $S=1-S=\frac{1}{2}$ for the Ramanujan series.
The resummation of the physical responses associated to the planes for $z\geq 1$ give $-\frac{1}{2}$ (in the thermodynamical sense) in the ideal situation where light equally couples to each plane. This is similar to say that from this proximity effect, the plane at $z=0$ reveals an effective $1+\zeta(0)$ effective response at $z=0$ similarly as a $\theta(z)$ function at the boundary $z=0$. 

Eq. (\ref{S}) suggests possible applications for similar ensembles of integer quantum Hall planes.

\subsection{Planks, Geometry and Spheres}
\label{planks}

The infinity can be obtained from the point of view that a cube does not know if there are $N$ or $N+1$ planks. 

Hereafter, we present a specific implementation of the thermodynamical limit related to the sphere model in the sense of producing the same effect as the $\theta(z)$ function in Eq. (\ref{Fz}). Note, that we can equally achieve this point of view with a Semenoff mass $M\epsilon z\sigma_z$ defined as in Sec. \ref{Semenoff} showing a small gradient along $z$ direction with $\epsilon\rightarrow 0$. When we have a small number of planks this term will not affect the alternance between $\mathbb{Z}$ and $\mathbb{Z}_2$ topological numbers. When we increase the number of planks, then we will encounter the situation where the term $M\epsilon z$ will become equal to the band gap (in each plank). The situation will become precisely equivalent to a $\frac{1}{2}$ topological number \ref{Semenoff}. From the point of view of the (total) Berry phases at the two Dirac points,  the system is in a superposition of $N$ and $N+1$ planks.
For this situation, if we increase the number of planks, then a topologically-trivial region, yet a layer in the middle will be topologically equivalent to a $\frac{1}{2}$ topological number.

\begin{center}
\begin{figure}[ht]
\hskip 1.7cm
\includegraphics[width=0.7\textwidth]{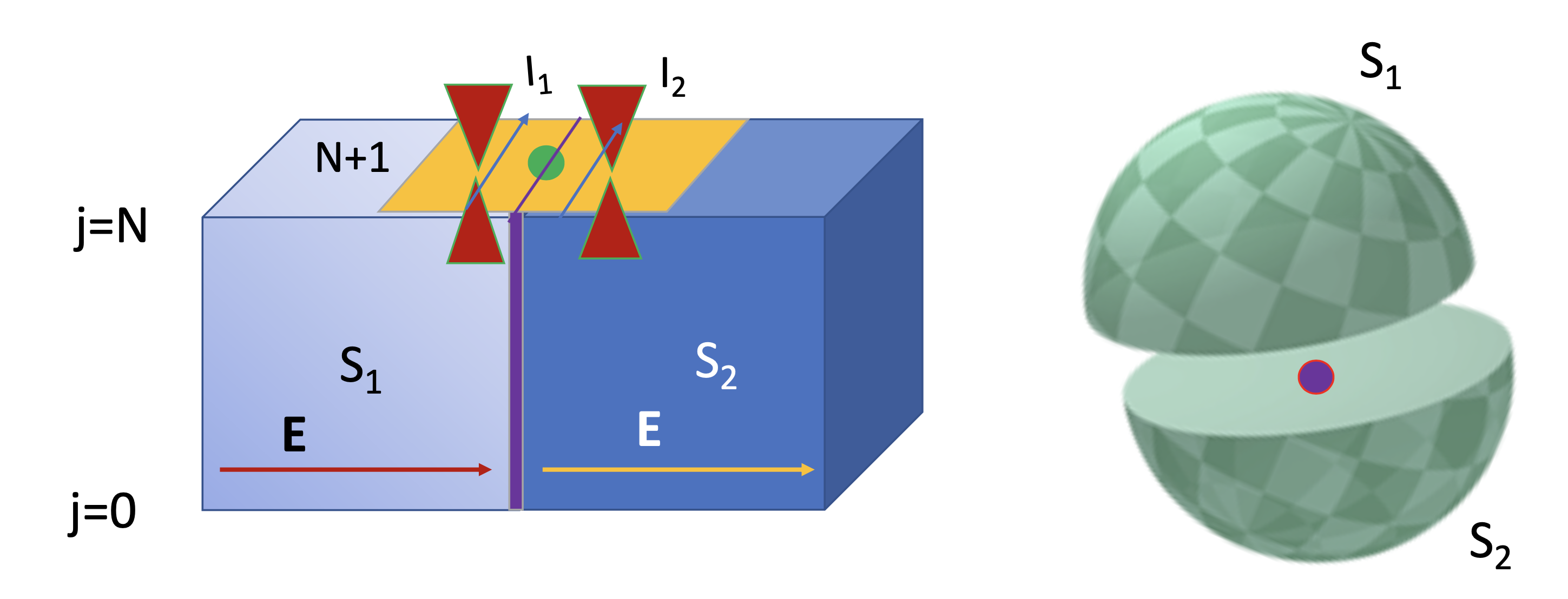}
\caption{Two cubes with an additional plank: The system $S_1$ or $S_2$ is from the top surface in a superposition of a even and odd number of planes. 
For the top yellow layer a measure of the pumped current $I_1$ or $I_2$ is assumed for half of the surface involving precisely one additional Dirac point for $S_1$ or $S_2$.}
\label{Plankhalf}
\end{figure}
\end{center}

 We prepare a cube (system $S_1$) with $j=0,...,N$ corresponding then to $(N+1)$ planes or planks, as described above, and then prepare another identical cube (system $S_2$) with the same number of planes or planks. We assume that there is a thin insulator between the two cubes in purple. Now, we place another `yellow' plane in a slightly different way (see Fig. \ref{Plankhalf}) such that from the point of view of $S_1$ and $S_2$ the last plane is in a superposition of $j=N$ and $j=N+1$ equally. The yellow plane with $j=N+1$ satisfying Eq. (\ref{Fz}) is then equally shared between the two systems $S_1$ and $S_2$. It is similar as if $S_1$ has 1 additional (massive) Dirac point and $S_2$ has also 1 additional Dirac point from the reciprocal space. 

For the $j=(N+1)$ sphere related to the yellow $j=(N+1)$ plane the topological charge in Fig.  \ref{Plankhalf} is equally encircled by the north hemisphere associated to $S_1$ and south hemisphere associated to $S_2$. We can then understand the $\frac{1}{2}$ of Eq. (\ref{infiniteseries}) through the sphere in Fig. \ref{Plankhalf} associated to the yellow plane and quantum transport (see Sec. \ref{ParsevalPlancherel}). The north pole corresponds to the Dirac point in $S_1$ and the south pole corresponds to the Dirac point in $S_2$. Here, $S_1$ and $S_2$ correspond to the two hemispheres with the interface in the equatorial plane where $\theta_c=\frac{\pi}{2}$ precisely. We can then describe the transport similarly as in Sec. \ref{ParsevalPlancherel} through a charge $e$ moving from the north to the equatorial plane along a path in $S_1$. This produces the perpendicular pumped current shown in Fig. \ref{Plankhalf} 
 \begin{equation}
 I_1 = \frac{e}{t} A'_{\varphi}\left(\theta=\frac{\pi}{2}^{-}\right),
 \end{equation}
  where $A'_{\varphi}(\theta=\frac{\pi}{2}^-)=\frac{|C_{N+1}|}{2}$ from the general results on one sphere in Sec. \ref{spin1/2}. The time $t$ corresponds to the time to travel from each pole to the equatorial plane for a charge. 
  
  Similarly, we can evaluate the perpendicular pumped current in $S_2$ associated to a charge $-e$ moving in opposite direction to the electric field towards the equatorial plane resulting in the pumped perpendicular current of Fig. \ref{Plankhalf}
 \begin{equation} 
 I_2=-\frac{e}{t}A'_{\varphi}\left(\theta=\frac{\pi}{2}^+\right)=-\frac{e}{t}\frac{|C_{N+1}|}{2}.
 \end{equation}
 The presence of one additional Dirac point then leads to an additional $\frac{1}{2}$ transverse pumped current in $S_1$ and $S_2$, respectively. This is similar as if we measure the response to circularly polarized light at the $M$ point for the last yellow plane only, 
 see Secs. \ref{lightdipole} and \ref{graphenelight}.
 
 \subsection{Topological Insulators in three dimensions, Quantum Hall effect on a surface and $\theta$ in Electrodynamics}
\label{3DQHE}

Below, we present a relation between the formation of $C=\frac{1}{2}$ in the two spheres' model from three-dimensional topological insulators and the top/bottom surfaces. 

The bulk of the system is described through a three-dimensional band structure which develops a gap as a result of spin-orbit coupling, for instance. The physics of these systems is also related to the axion physics or the $\theta$ term in electrodynamics \cite{SekineNomura}. A three-dimensional topological insulator respects time-reversal symmetry \cite{RMPColloquium}. From electrodynamics, the magnetic field (or spin properties, as described in \ref{timereversal}) is odd under time-reversal symmetry $t\rightarrow -t$ and the electric field is even which allows a specific term in the Lagrangian of the system of the form 
\begin{equation}
\frac{\theta e^2}{4\pi^2 \hbar c}{\bf E}\cdot{\bf B}
\end{equation}
 with $\theta=\pi$ such that this term preserves time-reversal symmetry since $\pi=-\pi$. This term has been in fact derived from microscopic grounds for specific materials such as Bi$_2$Se$_3$ around the $\Gamma$ point from the Fujikawa's method \cite{SekineNomura}. This is also related to topological quantum field theories \cite{QiZhang,Qi}. In this case, on the top and bottom surfaces of one cube the system will develop a metallic state described through one Dirac cone. Assuming we can turn the semi-metallic state into a topologically non-trivial insulator for instance with magnetic dopants or through the quantum Hall effect developing on the top and bottom surfaces, as realized experimentally \cite{Xu,Yoshimi}, then the occurrence of the magneto-electric effect can also be understood within the same geometrical foundations as in Sec. \ref{planks}.  

On the top surface, from the reciprocal space, the system is described similarly as a spin-$\frac{1}{2}$ with a magnetic field ${\bf d}({\bf k})=(v_F k_y, -v_F k_x, m)$ with the mass $m$ induced from a magnetic proximity effect at the $\Gamma$ point. Since we have only one Dirac point, the topological state of the top surface is similar to the one of the north hemisphere associated to the $S_1$ sub-system in the description above. The topological properties of this surface are similar to those for the two-spheres model with $C_j=\frac{1}{2}$ \cite{HH} in Sec. \ref{fractionaltopology} in the sense that the topological properties are described by an hemisphere or one pole only and the mirror disk in the equatorial plane. The north pole corresponding now to a Dirac point centered at $\Gamma$. We have the identification 
\begin{equation}
(v_F k_y, - v_F k_x, m)=(-d\sin\theta\sin\tilde{\varphi},d\sin\theta\cos\tilde{\varphi},m)
\end{equation}
 with $\varphi = \tilde{\varphi}+\frac{\pi}{2}$ compared to Eq. (\ref{correspondence}). The system is topologically equivalent to one hemisphere or a half unit sphere, similarly as for a `meron' or half-Skyrmion \cite{QiZhang} because for wave-vectors sufficiently distant from the $\Gamma$ point then this is as if the polar angle is $\frac{\pi}{2}$.
Applying an electric field similarly as in Fig. \ref{Plankhalf} the pumped transverse current from the north hemisphere then corresponds to $I_1$ associated to a halved quantum Hall conductivity. The halved quantum Hall conductivity can also be understood from the local Eq. (\ref{F}) in the case of one Dirac point reproducing the halved quantized response $\sigma_{xy}=\frac{e^2}{h}\frac{1}{2} \hbox{sgn}(m)$. Related to the possibility to create merons on the surface states of three-dimensional topological insulators from superlattice effects and spin-orbit coupling, see \cite{Cano}.

As found in Ref. \cite{Qi}, the $\frac{1}{2}$ charge can be precisely thought of as a domain wall between a non-trivial three-dimensional topological insulator and the vacuum described through a $\theta=0$ coefficient in the ${\bf E}\cdot{\bf B}$ term.

\section{Applications to Superconductivity and Majorana Fermions}
\label{further}

Here, we show the applicability of the formalism for other systems such as topological superconductors in the Nambu basis related to Majorana fermions which are their own antiparticles \cite{WilczekMajorana,Wilczekclass}. Majorana refers to Ettore Majorana \cite{Majorana}.
The quest of Majorana fermions has engendered a cross-fertilization between areas such as nuclear physics, dark matter, neutrinos and condensed-matter physics \cite{ElliottFranz,Alicea,Beenakker,SatoAndo} with potential applications in quantum information and computing \cite{HalperinMajorana,Kouwenhoven2dots}. In condensed matter physics, it is perhaps important to remind early applications of Majorana fermions \cite{EmeryKivelson,Clarke,SenguptaGeorges,Coleman,KarynMajoranadwave} related to quantum impurity models and solutions of the two-channel Kondo model \cite{NozieresBlandin,AffleckLudwig,AndreiBetheAnsatz}. A magnetic impurity at the edge of a one-dimensional d-wave (or s-wave) superconductor can also induce a Majorana edge mode \cite{KarynMajoranadwave} similar to the Kitaev p-wave superconducting wire \cite{Kitaev}. 

This gives rise to a $\frac{1}{2}\ln 2$ thermodynamical boundary entropy related to the physics of heavy-fermions quantum materials \cite{CoxJarrell}. This entropy has also signatures related to quantum information probes \cite{Alkurtass}. 

The goal in Sec. \ref{pwavewire} is to develop the application of the quantum topometry approach related to physical protocols. In Sec. \ref{SCtopo}, we present additional derivations showing a relation
between the ${\cal I}(\theta)$ function and the $\mathbb{Z}_2$ topological invariant in a Kitaev p-wave superconductor \cite{Kitaev}. In Sec. \ref{SpheresMajorana}, we present a relation between the $\frac{1}{2}$ topological number and Majorana fermions for the two-spheres' model. 
We also address the relation between two interacting spheres and two interacting wires \cite{delPozo}. 

For two-dimensional topological lattice models, the correspondence with the sphere is such that $(k_x,k_y)\rightarrow (\varphi,\theta)$ and for one-dimensional topological superconductors, we 
define $(k,\tilde{\varphi})\rightarrow (\theta,\varphi)$ where the phase of the superfluid reservoir now corresponds to the azimuthal angle. Since all the meridian lines are equivalent on the Bloch sphere, this preserves
the gauge invariance of the topological response with respect to the action of fixing of the macroscopic phase of the superfluid system. 

\subsection{Topological Superconducting Wire and Bloch Sphere}
\label{pwavewire}

We begin with a one-dimensional $p$-wave superconductor described through the Kitaev Hamiltonian \cite{Kitaev}:
\begin{equation}
\label{SCwire}
H = \sum_i \left(-t c^{\dagger}_i c_{i+1}+ \Delta c^{\dagger}_i c^{\dagger}_{i+1}+h.c.\right) - \mu c^{\dagger}_i c_i.
\end{equation}
This model has been suggested for the physics of nanowires with Rashba spin-orbit coupling related to important experimental progress to reveal Majorana particles \cite{Oreg,Lutchyn,Aasen,Delft,Pikulin,MicrosoftQuantum}. This model is also promising for applications in superconducting circuits \cite{GoogleMajorana}. Other recent applications of Majorana fermions in one dimension include carbon nanotubes \cite{Desjardins} and ferromagnetic atomic chains \cite{Yazdani}. On the lattice, the fermionic operators written in the second quantization anticommute such that $\{c_i, c^{\dagger}_i\}=1$. The $p_x$-wave symmetry is reflected through the fact that under the parity symmetry with respect to a bond $[i;i+1]$, the BCS pairing superconducting term is modified as $\Delta\rightarrow -\Delta$. 

The presence of a $\mathbb{Z}_2$ symmetry in the system can be understood through changing locally $c_i\rightarrow -c_i$ and similarly $c_i^{\dagger}\rightarrow -c_i^{\dagger}$. To show the topological aspects in this system, it is judicious to introduce Majorana fermions at each site $\eta_j = \frac{1}{\sqrt{2}}(c_j+c^{\dagger}_j)$ and $\alpha_j =\frac{1}{\sqrt{2}i}(c_j^{\dagger}-c_j)$ such that $\{\eta_j,\eta_j\}=1=\{\alpha_j , \alpha_j\}$ and $2i\eta_j\alpha_j = 1-2c^{\dagger}_j c_j$. In this way, this preserves the anticommutation relations for the fermions $\{c_i,c_i^{\dagger}\}=1$ and $\{c_i,c_i\}=0$.
If we suppose the special value $t=\Delta$ and the half-filled situation with $\mu=0$, the Hamiltonian can be written simply as
\begin{equation}
H = \sum_i 2i t \alpha_{i+1} \eta_i .
\end{equation}
The superconducting phase corresponds to the pairing of Majorana fermions on nearest neighboring sites such that at the two boundaries the system will reveal two zero-energy modes characterized through the operators $\alpha_0$ and $\eta_N$
where $N$ represents the last site of the one-dimensional wire and $0$ the first site \cite{Kitaev}. 

These Majorana modes are usually robust to a small local potential distortion in the sense that the Majorana fermions $\eta_0$ and $\alpha_N$ are gapped such that the density
operator $c^{\dagger}_i c_i$ cannot be a relevant operator in the long wavelength limit.  This model for $t=\Delta$ is equivalent to the transverse field Ising model \cite{Pfeuty,IsingdeGennes} from the Jordan-Wigner transformation such that mapping fermions onto spins is also equivalent to 
\begin{equation}
H = \sum_i \left(J_{\perp} S_{ix} S_{i+1 x} + J_z S_{iz}\right),
\end{equation}
with $J_z=\frac{\mu}{2}$ and $J_{\perp}=t$ implying two quantum phase transitions at $\mu=\mp 2t$. A quantum phase transition refers to the occurrence of a charge density wave on the lattice or to a strong-paired phase \cite{Alicea}. Here, the chemical
potential drives the physics which then favors the pairing of two Majorana fermions at the same site such that there is no free Majorana fermions at the boundary. The topological phase is referred to as a weak-paired phase in the literature \cite{Alicea}. 
In Sec. \ref{SpheresMajorana}, we show that a similar approach allows us to verify that for two spheres the $C_j=\frac{1}{2}$ phase can be understood as one zero-energy Majorana fermion at one pole for each sphere. 

To acquire a further understanding of the topological properties of the system and of the role of the chemical potential $\mu$, we will navigate on the Bloch sphere from
the reciprocal space. Using the Fourier representation
\begin{equation}
c_i = \frac{1}{\sqrt{L}}\sum_k e^{-i k x_i} c_k,
\end{equation}
with $L=Na$ being the length of the wire ($a$ is the lattice spacing), the Hamiltonian $H=\sum_{k\in [-\frac{\pi}{a};+\frac{\pi}{a}]} H(k)$ reads
\begin{equation}
H = \sum_k \epsilon(k) c^{\dagger}_k c_k + \Delta  e^{i\tilde{\varphi}} e^{+i ka} c^{\dagger}_{-k} c^{\dagger}_{+k} +h.c. 
\end{equation}
Here, $\tilde{\varphi}$ represents the macroscopic superfluid phase attached to $\Delta$ or to the superconducting reservoir such that we have shifted $\Delta\rightarrow \Delta e^{i\tilde{\varphi}}$ with $\Delta$ taken to be real.
Then, $\epsilon(k)=-2t\cos(k a) -\mu=\epsilon(-k)$. We can also act with the parity transformation $k\rightarrow -k$ such that $H=\sum_k H(k)$ takes the form
\begin{eqnarray}
H &=& \sum_{-k} \epsilon(k) c^{\dagger}_{-k} c_{-k} - \Delta e^{i\tilde{\varphi}}e^{-ika} c^{\dagger}_{-k} c^{\dagger}_{k}  +h.c. 
\end{eqnarray}
For one specific value of $k$:
\begin{eqnarray}
H(k) &=& \frac{1}{2}(\epsilon(k)c^{\dagger}_k c_k-\epsilon(k)c_{-k}c^{\dagger}_{-k}) \\ \nonumber
&+& i\Delta e^{i\tilde{\varphi}} \sin(ka)c^{\dagger}_{-k}c_k^{\dagger}+h.c. .
\end{eqnarray}
In the Nambu basis $(c_k , c^{\dagger}_{-k})^T$, this is equivalent to
\begin{eqnarray}
\label{matrix}
\hskip -0.2cm
H(k) = \frac{1}{2}\left(
 \begin{array}{cc}
\epsilon(k) & -2i\Delta e^{i\tilde{\varphi}}\sin(ka) \\
2i\Delta e^{-i\tilde{\varphi}} \sin(ka) & -\epsilon(k) \\
\end{array}
\right).
\end{eqnarray}

From the identification with the spin-$\frac{1}{2}$ such that $H(k)=-{\bf d}\cdot \mathbfit{\sigma}$ or
\begin{eqnarray}
H(k) = \left(
 \begin{array}{cc}
- d\cos\theta -\frac{m}{2} & -d\sin\theta e^{-i\varphi} \\
-d\sin\theta e^{i\varphi} & +d\cos\theta + \frac{m}{2} \\
\end{array}
\right)
\label{spinidentification}
\end{eqnarray}
for $t=\Delta$, then this gives rise to 
\begin{eqnarray}
\theta &=& \theta_k = ka \\ \nonumber
-\varphi &=& \varphi_k = \tilde{\varphi} +\frac{\pi}{2}
\end{eqnarray}
with $m$ playing the role of the chemical potential and $t=\Delta=d$. We have effectively a two-dimensional map seeing the superfluid phase of the reservoir as an independent variable which can be tuned through interferometry and SQUID geometry.

For $m=0$, the spin-$\frac{1}{2}$ eigenstates read
\begin{eqnarray}
|\psi_+\rangle &=& \cos\frac{\theta}{2} \left( \begin{array}{c}
1 \\
0 \\
\end{array}\right) - i \sin\frac{\theta}{2}e^{-i\tilde{\varphi}} \left( \begin{array}{c}
0 \\
1 \\
\end{array}\right) \\ \nonumber
|\psi_-\rangle &=& \sin\frac{\theta}{2} \left( \begin{array}{c}
1 \\
0 \\
\end{array}\right) +i\cos\frac{\theta}{2}e^{-i\tilde{\varphi}} \left( \begin{array}{c}
0 \\
1 \\
\end{array}\right).
\end{eqnarray}
For the superconducting wire, eigenstates can be defined through the quasiparticle operator $\tilde{\eta}_k=u_k c_k+v_k c^{\dagger}_{-k}$. The Hamiltonian can be diagonalised accordingly as 
\begin{equation}
\label{etak}
H(k) = E(k) \tilde{\eta}_k^{\dagger} \tilde{\eta}_k
\end{equation}
with $E(k)=\sqrt{\epsilon(k)^2 +4\Delta^2 \sin^2(ka)}$ and the BCS ground state corresponds to a vacuum of quasiparticles $\tilde{\eta}_k |BCS\rangle=0$. We obtain 
\begin{eqnarray}
\tilde{\eta}_k &=& \cos\frac{\tilde{\theta}}{2} c_k + i e^{i\tilde{\varphi}}\sin\frac{\tilde{\theta}}{2} c_{-k}^{\dagger},
\label{eta}
\end{eqnarray}
with $\tilde{\theta}_k=ka+\pi$. 
We can verify $\{ \tilde{\eta}_k , \tilde{\eta}_k^{\dagger} \}=1$. Creating a quasiparticle above the superfluid ground state is related to the $|\psi_-\rangle$ state for the spin-$\frac{1}{2}$.
The $|BCS\rangle$ wavefunction where $|BCS\rangle=\prod_k |BCS\rangle_k$ takes the form
\begin{equation}
\label{BCS}
|BCS\rangle = R\prod_{0<k<\frac{\pi}{a}}\left(\cos\frac{\tilde{\theta}}{2} - i \sin\frac{\tilde{\theta}}{2} e^{i\tilde{\varphi}}c^{\dagger}_k c^{\dagger}_{-k}\right)|0\rangle,
\end{equation}
with $R=(\delta_{\mu<-2t} +(1-\delta_{\mu<-2t})c_0^{\dagger})(\delta_{\mu<2t} +(1-\delta_{\mu<2t})c^{\dagger}_{\pi})$ such that for $\theta=k=0$ we have $|BCS\rangle_0 = c^{\dagger}_0|0\rangle$ and for $\theta=ka=\pi$ we have $|BCS\rangle_{\pi} = |0\rangle$ within the topological phase $-2t<\mu<2t$.

Since the pairing function proportional to $\sin(ka)$ goes to zero at $k=0$ and $ka=\pi$, we can vary $\Delta\neq t$ accordingly such that a quantum
phase transition should only involve the ratio $\frac{\mu}{2t}$.

The occurrence of Majorana fermions can be understood from general relations related to the particle-hole symmetry 
$\omega_x H \omega_x = - H^{\dagger}$ with the charge conjugation matrix
\begin{eqnarray}
\omega_x = \sigma_x =
\left(
 \begin{array}{cc}
0 & 1 \\
1 & 0 \\
\end{array}
\right).
\end{eqnarray}
Usually, for a BCS Hamiltonian, this implies specific relations between the kinetic $T=T^{\dagger}$ and pairing $\Delta = - \Delta^{\dagger}$ operators. This requires that the pairing term is odd when changing $k\rightarrow -k$.  For one-dimensional spin-polarized superconductors, this prerequisite is satisfied through the fact that the most dominant pairing term occurs between nearest neighbors. In addition to particle-hole symmetry or charge conjugation, the system also possesses time-reversal symmetry in a simple way through the fact that the $c$ fermions operators are invariant under time-reversal symmetry defined simply through the transformation $i\rightarrow -i$ (see \ref{timereversal}). The two Majorana fermions $\alpha$ and $\eta$ are precisely transformed accordingly as $\eta\rightarrow \eta$ and $\alpha\rightarrow -\alpha$. 

One may wonder about the role of the superfluid phase producing possibly a complex pairing term in Eq. (\ref{SCwire}). This phase refers to a gauge freedom in the model. On the other hand, we may verify using the Luttinger formalism that we may absorb simply this phase into a re-definition of the superfluid phase $\theta$ associated to the bosonic particles or Cooper pairs forming the superfluid reservoir (see \ref{interactions} and Eq. (\ref{pairingfunction})). Therefore, we can define $\tilde{\varphi}=0$ for simplicity in  (\ref{matrix}). The topological superconducting wire is usually referred to as a BDI phase in the topological classification tables, defined through the square of the time-reversal operator, charge conjugation or particle-hole symmetry and also through chiral symmetry \cite{BernevigNeupert,FidkowskiKitaev}.

Defining the pseudo-spin operator, as introduced by P. W. Anderson \cite{Anderson}
\begin{equation}
{\bf S} = \frac{1}{2}\psi^{\dagger}_k {\mathbfit{\tau}}\psi_k,
\end{equation}
such that $S_z = (c^{\dagger}_k c_k - c_{-k}c^{\dagger}_{-k})$, we introduce an analogy with the sphere where the two variables are then $\theta=ka$ and $\varphi=\tilde{\varphi}+\frac{\pi}{2}$. This mapping is elegant as it allows for a relation with geometry. In particular, for the sphere model we can then use important relations such as Eq. (\ref{polesC}). This implies that to measure the $\mathbb{Z}$ topological invariant one can drive from north to south pole along a particular line represented through fixed $\varphi=\tilde{\varphi}+\frac{\pi}{2}\in[0;2\pi]$. In this representation of $S^2$, $\theta\in[0;\pi]$. In the wire, since we describe half of the Brillouin zone due to the fact that a particle in $[0;\pi]$ is related to a hole in $[-\pi;0]$, we can then also define a $\mathbb{Z}$ topological invariant in a similar way as on the Bloch sphere $C=\frac{1}{2}(\langle S_z(0)\rangle -\langle S_z(\pi)\rangle)$ where here we suppose the (BCS) ground state. 

The north and south poles correspond to $k=0$ and $ka=\pi$ respectively in the Brillouin zone. In our analysis, the spin magnetizations at the pole depend only on the sign of $\epsilon(k)=-2t\cos(k a) -\mu$ at $k=0$ and $k a=\pi$. Therefore, we verify that the topological transition takes place when $\mu=\pm 2t$ with $C=1$ in the topological phase and $C=0$ in the polarized phases. 

Here, we introduce further possible relations between topological properties and observables which will be useful for further identification of these wires.
From the smooth fields in Eq. (\ref{CA'}), for the half-filled situation $\mu=0$ and $\Delta=t$ we also identify a correspondence between the averaged charge and topological number:
\begin{eqnarray}
\langle \hat{Q}\rangle &=& \frac{L}{2\pi}\int_{-\pi}^{+\pi} \langle BCS| c^{\dagger}_k c_k |BCS\rangle dk = \frac{C}{2}L.
\end{eqnarray}
This $\mathbb{Z}$ topological number is similarly described as a Zak phase \cite{Zak} or as the rolling of the phase $\theta_k$ on the whole Brillouin zone \cite{TewariSau,TrifTserkovnyak}:
\begin{equation}
C = \frac{1}{2\pi}\oint d\theta_k.
\end{equation}
 Through $\oint$ we suppose periodic boundary conditions for the Brillouin zone.  On the other hand, we can equivalently write $C = \frac{1}{\pi}\int_0^{\pi} d\theta_k$ similarly as in Eq. (\ref{Zak}).  The Zak phase corresponds to a linear evolution of the phase associated to the polar angle on the sphere. This formulation of the topological invariant is similar for the Su-Schrieffer-Heeger model of polyacetylene \cite{Su}, which has been recently engineered in different quantum platforms \cite{Gadway,Tal,Molenkampcircuit,Rosenthal,Optique}, and for the Rice-Mele model \cite{MunichRiceMele,Spielman}.

Quantum information can also be useful to characterize (topological) properties and charge fluctuations of these wires. Bipartite fluctuations of the charge on a sub-region $A$ of a superconducting wire of length $L$ are described through $F(A)=i_Q L +b\log(L) +{\cal O}(1)$ with \cite{Herviou2017}
\begin{equation}
i_Q = \lim_{L\rightarrow +\infty} \frac{1}{L}\langle \hat{Q}^2\rangle = q \int_{BZ} \frac{dk}{4\pi} \sin^2\theta_k.
\end{equation}
The function $i_Q$ is the quantum Fisher.
At the transition $\mu=-2t$, the matrix (\ref{matrix}) reveals a linear gapless mode around $k=0$ associated to one Majorana fermion which implies $b<0$. A one-dimensional quantum liquid with $U(1)$ charge conservation in contrast would produce $b>0$ \cite{Song}. The quantum phase transition is described through one free (gapless) Majorana fermion in the bulk \cite{Herviou,delPozo}. A similar gapless Majorana chain with central charge $c=\frac{1}{2}$ can be realized in the presence of magnetic impurities \cite{KarynMajorana}. Coupling a cavity or $LC$ circuit to a p-wave superconductor allows for the measure the dynamical susceptibility of the p-wave superconducting wire \cite{Olesia,Matthieu}. It is also relevant to mention theoretical efforts related to the entanglement spectrum \cite{LiHaldane} applied to topological p-wave superconducting wires \cite{Maria1}. A generalization of the Zak phase and finite temperature effects have been recently studied in Ref. \cite{Maria2}.

Various theoretical works have studied the stability of the topological phase for one wire in the presence of weak interactions \cite{PascalDaniel,Stoudenmire,Schuricht,Jelena,Herviou} and also in the presence of a (moderate) inter-wire hopping term \cite{FanKaryn}. The stability of the topological number towards interaction effects can be understood from the Luttinger formalism in \ref{interactions} and it can be viewed as an application of symmetry-topologically-protected phenomenon. The fact that the structure of Majorana fermions remains identical at the edges can first be understood from the fact that there is no gap closing in the system. From renormalization group arguments then the system behaves as if $t\sim \Delta$ in the low-energy fixed point (see \ref{interactions}). To strengthen this conclusion, we can also apply the stochastic approach developed in Secs. \ref{Mott} and \ref{MottKM}. Including the Cooper channel the BCS Hamiltonian becomes modified as
\begin{eqnarray}
H &=& \sum_i (-t +V(\phi_x-i\phi_y))c^{\dagger}_i c_{i+1} + (\Delta + V\phi_{\Delta})c^{\dagger}_i c^{\dagger}_{i+1}  \nonumber \\
&+& h.c. - (\mu+\phi_0) c^{\dagger}_i c_i,
\end{eqnarray}
with $\phi_x+i\phi_y=-\frac{1}{2}\langle c^{\dagger}_i c_{i+1}\rangle$ and the additional pairing channel $\phi_{\Delta}=\phi_{\Delta_x}+ i\phi_{\Delta_y}=\langle c_{i+1}c_i\rangle$. In a similar way as the particle-hole channel introduced in Sec. \ref{Mott}, $\phi_{\Delta_x}+ i\phi_{\Delta_y}=\langle c_{i+1}c_i\rangle$ and $\phi_{\Delta_x}-i\phi_{\Delta_y}=\langle c^{\dagger}_i c^{\dagger}_{i+1}\rangle$. The $\phi_0$ term can be set to zero if we redefine the interaction in a symmetric way from half-filling $V(n_i - \frac{1}{2})(n_{i+1}-\frac{1}{2})$. In the $2\times 2$ matrix, then this renormalizes $2t\rightarrow 2t-2V(\phi_x-i\phi_y)$ and $-2\Delta\rightarrow -2\Delta - 2V\phi_{\Delta}$. From the BCS ground state, we identify $\sum_i \langle BCS| c^{\dagger}_i c_{i+1}+h.c. |BCS\rangle=\frac{N}{2}$ and $\sum_i \langle BCS| c_{i+1} c_i +h.c. |BCS\rangle =-\frac{N}{2}$ with $N$ being the number of sites such that effectively the system stays on a line as if $t=\Delta$. Increasing interactions then a Mott transition can occur when $V\sim t$  for specific fillings \cite{Schuricht}. The study of such a transition line would require to include the $\phi_z=\langle c^{\dagger}_i c_i - c^{\dagger}_{i+1} c_{i+1} \rangle$ channel at larger interaction strengths. Within the topological phase $\phi_z=0$. This approach is related to new insights on disorder effects in topological superconducting wires \cite{DisorderArticleNew}.

\subsection{$\mathbb{Z}_2$ topological invariant and ${\cal I}(\theta)$ function}
\label{SCtopo}

Related to $C^2$ \cite{C2} introduced in Sec. \ref{light} we can equally define a $\mathbb{Z}_2$ formulation of the topological invariant for the $p$-wave superconducting wire related to the Pfaffian definition of Fu-Kane-Mele for topological insulators \cite{FuKaneMajorana}.
From Eq. (\ref{polesC}), we equivalently have
\begin{equation}
\langle S_z(0)\rangle \langle S_z(\pi)\rangle = 1-2C^2 = \mp 1.
\label{wiretopo}
\end{equation}
As a measure of the $\mathbb{Z}_2$ topology, then we can define the quantity from the sphere
\begin{equation}
\label{invariant}
 \langle S_z(0)\rangle \langle S_z(\pi) \rangle  = \Pi_{i=0,\pi} \xi_{i},
\end{equation}
which agrees with the $\mathbb{Z}_2$ index formulated for the wire \cite{Kitaev,SatoAndo}. The variable $\xi_{i}=\langle S_z(i)\rangle$ is defined to have values $\pm 1$ for $i=0,\pi$. Similarly as in the Kane-Mele model, $\xi$ involves the kinetic term at special (symmetry) points in the Brillouin zone. If we are in the weak-paired topological phase this quantity is $-1$ and in the strong-paired phase this is $+1$. We verify that the topological transition takes place when $\mu=\pm 2t$ implying that the function $\langle S_z \rangle$ vanishes at one pole producing a step function in Eq. (\ref{invariant}). 

Related to this definition of the $\mathbb{Z}_2$ topological number, we propose a measure allowing an analogy with the light-matter detection in two dimensions and the function ${\cal I}(\theta)$ in Eq. (\ref{Itheta}) or the function $\alpha(\theta)$ in Eq. (\ref{onehalf}). For $\theta=\frac{\pi}{2}$, we insist on the fact that on the Bloch sphere the light can detect locally this $\mathbb{Z}_2$ number from Eq. (\ref{onehalf}).

Including an AC potential, e.g. $V(t)=V_{ac}\sin(\omega t)$, acting on the Cooper pairs in the BCS reservoir, this gives rise to a time-dependent phase shift for the bosons
\begin{equation}
\langle b\rangle(t) = \langle b(t)\rangle_{V_{ac}=0} \times e^{-\frac{i}{\hbar}\int_0^{t} V(t')dt'}.
\end{equation}
If we develop the phase shift to first order in $V_{ac}$, then 
\begin{equation}
\langle b\rangle(t) \approx \langle b(t)\rangle_{V_{ac}=0}\left(1-\frac{i}{\hbar\omega}V_{ac}\sin(\omega t)\right).
\end{equation}
In the wire model, then we have an additional off-diagonal term in the Nambu basis of the form
\begin{equation}
\label{A0}
\delta {H}(k,t) = A_0\sin(ka)\sin(\omega t)c^{\dagger}_k c^{\dagger}_{-k}+h.c.,
\end{equation}
with $A_0 = \frac{V_{ac}|\langle b\rangle|}{\hbar\omega}$ and the matrix identification $\sigma^+=c^{\dagger}_k c^{\dagger}_{-k}$. In this equation, we assume that $A_0$ is real in Eq. (\ref{A0}).
In the sense of $2\times 2$ matrices in Eq. (\ref{spinidentification}), the term is similar to the one in Sec. \ref{lightdipole} on light-induced dipole transitions where we set $\sin(ka)\sim 1$ for $ka\sim \frac{\pi}{2}$.  

In Eq. (\ref{A0}), $\delta H(k,t)$ must be applied onto the $|BCS\rangle$ state. When evaluating inter-band transitions, we have the identification $c^{\dagger}_k c^{\dagger}_{-k}\rightarrow \cos^2\frac{\tilde{\theta}}{2}\tilde{\eta}^{\dagger}_k \tilde{\eta}^{\dagger}_{-k} + \sin^2\frac{\tilde{\theta}}{2}\tilde{\eta}_{-k}\tilde{\eta}_k$. Close to $ka=\tilde{\theta}=\frac{\pi}{2}$, the factor $\sin(ka)$ in the Hamiltonian remains equal to unity and we identify
\begin{equation}
A_0 \sin(\omega t) c^{\dagger}_k c^{\dagger}_{-k} \rightarrow A_0 \sin(\omega t) \sqrt{C^2-\frac{1}{2}} (\tilde{\eta}^{\dagger}_k \tilde{\eta}^{\dagger}_{-k} +h.c.).
\end{equation}
In this case, the $\mathbb{Z}_2$ topological number $(2C^2-1)$ enters as a prefactor similar to the light response on the Bloch sphere at the equator; see Eq. (\ref{onehalf}). Tuning the chemical potential will result in a quantum phase transition which can be revealed through such time-dependent  protocols \cite{Lightpwavewire}. It is also interesting to mention recent proposals to implement time-dependent protocols such as Kibble-Zurek mechanisms in topological p-wave superconducting wires \cite{KZ}.

\subsection{$C=\frac{1}{2}$ and Majorana Fermions}
\label{SpheresMajorana}

Here, we present another understanding for the occurrence of the fractional topological numbers of Sec. \ref{fractionaltopology} on the sphere through the Majorana fermions of Sec. \ref{pwavewire} which also suggests a relation with two wires.
 
For two spheres, the role of the transverse fields acting on each sphere is to produce an effective interaction $\sigma_{1x}\sigma_{2x}$ (see Eq. (\ref{HamiltonianIsing})) such that from south pole the model is equivalent to 
\begin{equation}
H_{eff} = r\sigma_{1z}\sigma_{2z} - \frac{d^2 \sin^2\theta}{r} \sigma_{1x}\sigma_{2x}.
\end{equation}
At the north pole, from the Jordan-Wigner transformation with two spins we can define $\sigma_{iz}=2c^{\dagger}_i c_i - 1$ with $i=1,2$. Since the superconducting gap is going to zero in the vicinity of the north pole then the ground state satisfies $\sigma_{iz}|GS(0)\rangle =+1=2i\alpha_i\eta_i|GS(0)\rangle$ so that $c^{\dagger}_i c_i|GS(0)\rangle=+1$. In the case of fractional entangled topology, this requires to have $d-M<r<d+M$ implying for the superconducting sphere $M=\frac{m}{2}$ and $t=d=\Delta$ such that $H_{eff}$ is satisfied. In addition we should define the Jordan-Wigner transformation such that $\langle GS(\pi)| \sigma_{iz} |GS(\pi)\rangle =0$ at south pole as long as $t-\frac{\mu}{2}<r<t+\frac{\mu}{2}$, i.e. equally for the two situations $\mu\rightarrow 2t^-$ (such that $\langle c^{\dagger}_{\pi}c_{\pi}\rangle=0$) and $\mu=2t^+$ (such that $\langle c^{\dagger}_{\pi}c_{\pi}\rangle =1$)
for $r>0$. Then, we have $\sigma_{1z} = \frac{1}{i}(c^{\dagger}_1- c_1)$, $\sigma_{1x}= c^{\dagger}_1+c_1$, $\sigma_{2z} = \frac{1}{i}(c^{\dagger}_2- c_2)e^{i\pi c^{\dagger}_1 c_1}$, $\sigma_{2x}= (c^{\dagger}_2+c_2)e^{i\pi c^{\dagger}_1 c_1}$, such that \cite{OneHalfKLH}
\begin{equation}
H_{eff} = -r(c_1+c_1^{\dagger})(c^{\dagger}_2-c_2) -\frac{d^2 \sin^2\theta}{r}(-c_1+c_1^{\dagger})(c_2+c_2^{\dagger}).
\end{equation}
This representation satisfies $\langle GS(\pi)|\sigma_{iz}| GS(\pi)\rangle=0$. This can be re-written in terms of the Majorana fermions as \cite{OneHalfKLH}
\begin{equation}
H_{eff} = -2r i \eta_1\alpha_2 - \frac{2id^2}{r}\sin^2\theta \alpha_1\eta_2.
\end{equation}
The ground state reveals $2i\eta_1 \alpha_2|GS(\pi)\rangle = +|GS(\pi)\rangle$ which then leads to another possible way to write Eq. (\ref{correlation}) for $C_j=\frac{1}{2}$ as
\begin{equation}
\label{corr}
\langle \sigma_{1z}(\pi)\sigma_{2z}(\pi)\rangle = \langle 2i\alpha_2\eta_1\rangle = -1 = -(2C_j)^2.
\end{equation}

This equation is in Table \ref{tableII} and relates the non-local entangled structure of the paired Majorana fermions $\alpha_2$ and $\eta_1$ with the half topological number of a sphere. 

Within this formulation, the Majorana fermions $\alpha_1$ and $\eta_2$ measure the $\ln 2$ entropy associated to the degeneracy between the two states $|\Phi_+\rangle_1 |\Phi_-\rangle_2$ and $|\Phi_-\rangle_1 |\Phi_+\rangle_2$ precisely at the south pole. Yet, if we perform an average on an ensemble of measures, then $\langle GS(\pi)| \sigma_{iz} |GS(\pi)\rangle=0$ such that the result equally reveals the presence of the two possible states $|\Phi_+\rangle_1 |\Phi_-\rangle_2$ and $|\Phi_-\rangle_1 |\Phi_+\rangle_2$. The south pole may be seen as an edge for a one-dimensional chain (path) linking the north and south poles on each sphere. The present device with two spheres presents then one zero-energy Majorana fermion at the edge of each chain. Deviating slightly from $\theta=\pi$ then this will produce the entangled wavefunction preserving $\langle GS(\theta)| \sigma_{iz} |GS(\theta)\rangle=0$. For $\theta\neq \pi$, the two Majorana fermions $\alpha_1$ and $\eta_2$ become entangled corresponding then to the `bulk' of the chain. This system could then have applications in quantum information and in light-matter networks \cite{lightmatter}. Implementing a time-dependent $r_{xy}$ coupling at south pole would correspond to activate in time the state of the non-local qubit formed with the two gapless Majorana fermions at south pole.  In the presence of a fixed ferromagnetic $r_{xy}$ coupling in time, as discussed in Sec. \ref{fractionaltopology}, this reinforces the entangled wavefunction at south pole and stabilizing the $C_j=\frac{1}{2}$ number through $\langle GS(\theta)| \sigma_{iz} |GS(\theta)\rangle=0$. In that case, the Majorana fermions $\alpha_1$ and $\eta_2$ are entangled in a state with fixed parity in time. 

From the identification between charge and spin as formulated in Sec. \ref{Coulomb}, we deduce that the two spheres' model with an Ising interaction is analogous to a two-wires' model with a Coulomb interaction. 
The two wires' superconducting model gives rise to a Double Critical Ising phase (DCI) \cite{Herviou} which presents in fact similar properties as the fractional topological phase (on the sphere at $ka=\pi$) with two gapless Majorana fermions delocalized in the bulk of the system \cite{delPozo}. The Semenoff mass $M$ on the two spheres corresponds simply to the chemical potential for the superconducting wires. The Coulomb interaction between wires precisely favors a charge ordering around $ka=\pi$ along a wire and the Mott phase then corresponds to an antiferromagnetic ordering of the charges on the two wires. Tuning $M$ within the DCI phase, the system yet presents two critical gapless Majorana modes (one per wire) similar to the topological quantum phase transition of one wire \cite{delPozo}. In the DCI phase, the two gapless Majorana fermions correspond then to the two Majorana fermions at south pole on the sphere.

The interacting spheres' model then gives further insight on correlated (topological) superconductors from the interplay with Mott physics. The DMRG approach reveals $\frac{1}{2}$ topological number(s) for the DCI phase in a accurate manner
 \cite{delPozo}; the topological numbers on the lattice being related to correlation functions in real space and represent a topological marker in the presence of disorder \cite{DisorderArticleNew}.
 
 \subsection{p+ip superconductor}

It is important to mention recent progress to realize and engineer topological superconductors in higher dimensions. This includes the possibility to have a $p+ip$ superconductor \cite{ReadGreen} on surface states of Bi$_2$Se$_2$ \cite{FuKane}. Topological superconductivity was for instance reported on Cu$_x$Bi$_2$Se$_3$ \cite{AndoSC}. The $p+ip$ superconductor can also be realized in multi-wires' architectures \cite{Kanewires}, in particular similarly as for the Haldane model through local magnetic fluxes with zero-net flux in a unit cell \cite{FanKaryn}. The physics of the $p+ip$ superconductor is also related to the $\nu=\frac{5}{2}$ fractional quantum Hall effect \cite{MooreRead} and to the Kitaev spin model on the honeycomb lattice in the $B$ phase in the presence of a magnetic field \cite{KitaevHoneycomb,Burnell}. The physics of p-wave superconductors also occurs in He$_3$ \cite{Volovik}, quantum materials such as Sr$_{2}$RuO$_4$ \cite{KallinBerlinsky} and also graphene coupled to a high-$T_c$ superconductor \cite{Angelo}. On the honeycomb lattice, the presence of zero-energy bound states and topological states naturally occur within the BCS theory \cite{DoronKaryn,WilczekGhaemi,Scherer}. A chiral topological superconductor can be engineered from proximity effect in a material described by a quantum anomalous Hall state \cite{QiHughesRaghuZhang}. The honeycomb lattice can also give rise to other interesting phases from interaction effects such as a FFLO $p$-wave superconducting phase \cite{TianhanFFLO} and to $d+id$ topological superconducting phases in the presence of interactions \cite{Annica,AnnicaWeiKaryn,AnnicaKaryn,Schererd,Wolf}.

\begin{figure}[]
\begin{center}
\includegraphics[width=0.55\textwidth]{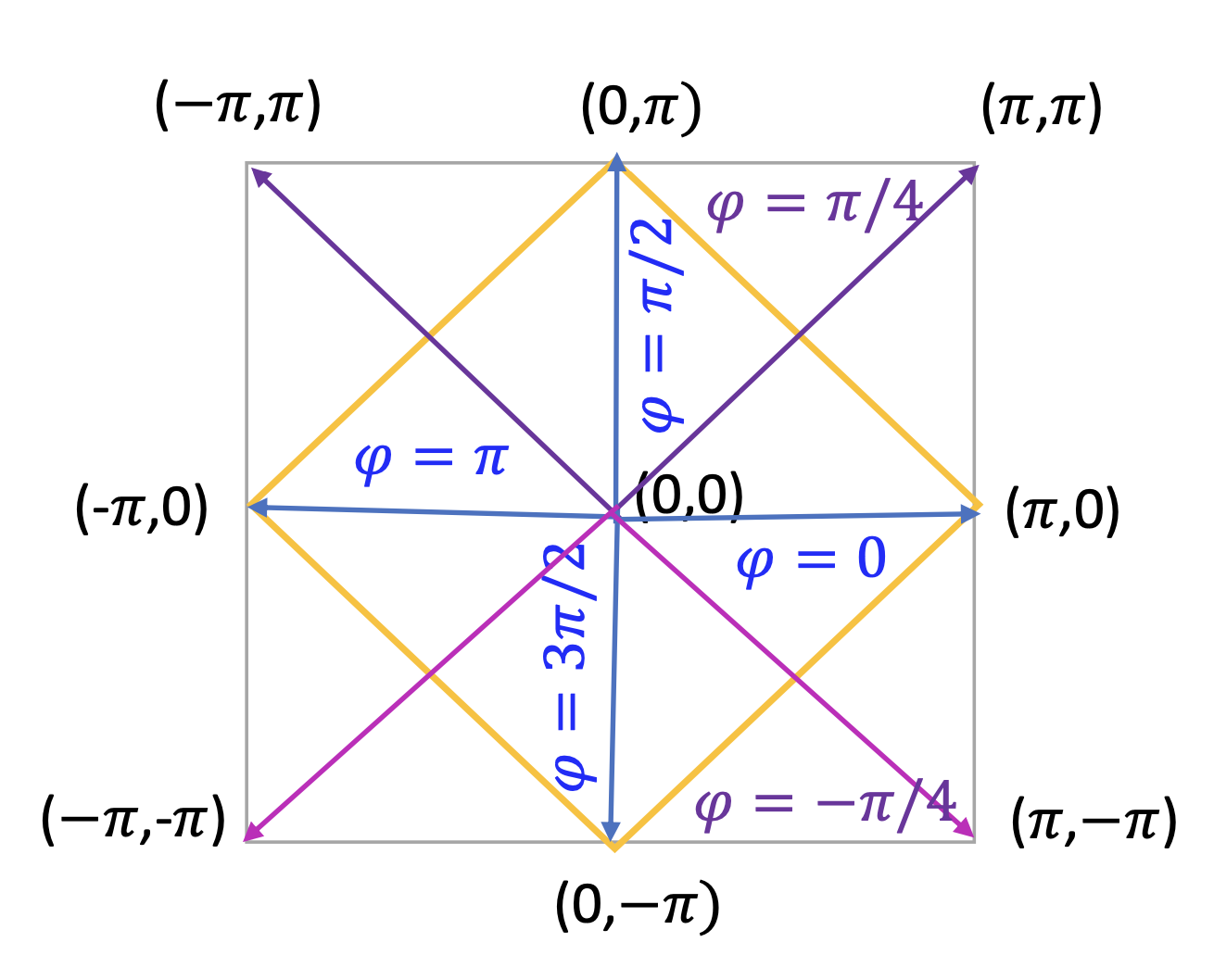}
\caption{Different trajectories in the Brillouin zone represented through the polar angle $\tilde{\varphi}$ or azimuthal angle $\varphi$. The axes are defined as $k_xa$ for the horizontal direction and $k_ya$ for the vertical direction.}
\label{2dtrajectories}
\end{center}
\end{figure}

Here, we show that the local description approach in the reciprocal space can also be applied to a $p+ip$ superconductor. Suppose a $p_x+ip_y$ superconductor on the square lattice in the Nambu basis $(c_{\bf k}, c_{-{\bf k}}^{\dagger})$. Around the $\Gamma=(0,0)$ point in the reciprocal space and the edges of the Brillouin zone at ${\bf k} a=(\pi,\pi)$, we can develop the $2\times 2$ matrix assuming a small deviation ${\bf p}$ from these two points such that ${\bf k}a={\bf p}a+(0,0)a$ and ${\bf k}a =(\pi,\pi)a-{\bf p}a$. In this way, the matrix takes the form
\begin{eqnarray}
\left(
 \begin{array}{cc}
-2t\zeta\cos(pa) -\frac{\mu}{2} & \Delta p a e^{-i\tilde{\varphi}} \\
 \Delta p a e^{i\tilde{\varphi}} & 2t\zeta\cos(pa) +\frac{\mu}{2}  \\
 \end{array}
\right),
\end{eqnarray}
where $p_x+ip_y = p e^{i\tilde{\varphi}}$ with $\zeta=+1$ at the $\Gamma$ point and $\zeta=-1$ at $(\pi,\pi)$. The particular situation $\tilde{\varphi}=\frac{\pi}{4}$ refers to the $p+ip$ superconductor and we also identify the limits of the one-dimensional topological wire corresponding to 
$\tilde{\varphi}=0$ and $\tilde{\varphi}=\frac{\pi}{2}$; see Fig. \ref{2dtrajectories}. This suggests that the $\Gamma$ point can be placed at the north pole on $S^2$ and similarly the $(\pi,\pi)$ point corresponds to the south pole. Assuming that we follow a diagonal path in the reciprocal space joining for instance the $\Gamma$ point to $(\pi,\pi)$ then this corresponds to a line joining the north to the south pole on the Bloch sphere where the polar angle becomes $\theta=p a$. The model is equivalent to the matrix
\begin{eqnarray}
\hskip -0.5cm
\left(
 \begin{array}{cc}
-2t\zeta\cos(pa) -\frac{\mu}{2} & \Delta (\sin(pa)-i\sin(pa)) \\
 \Delta (\sin (pa)+ i\sin(pa)) & 2t\zeta\cos(pa) +\frac{\mu}{2}  \\
 \end{array}
\right),
\end{eqnarray}
with the dressed coordinates $k_{\parallel}=\frac{1}{2}(k_x+k_y)$, $k_{\perp}=\frac{1}{2}(k_x-k_y)$ and along the diagonal $k_x=k_y$ such that $k_{\perp}=0$ and $k_{\parallel}=k_x=k_y=p$.
From the identification with Eq. (\ref{correspondence}), we have effectively $d=2t=-\sqrt{2}\Delta$ with a Semenoff mass $M=\frac{m}{2}=\frac{\mu}{2}$. This is equivalent to an azimuthal angle $\tilde{\varphi}=\varphi=\frac{\pi}{4}$. 
Similarly as for the p-wave superconducting wire, we can define the topological invariant on half of the one-dimensional Brillouin zone due to the particle-hole symmetry.

The topological invariant is defined from the poles of the sphere only. Through the pseudospin-$\frac{1}{2}$ analogy then we have
\begin{equation}
C = \frac{1}{2}(\langle S_z(0,0)\rangle - \langle S_z(\pi,\pi)\rangle),
\end{equation}
which is equivalent to 
\begin{equation}
\label{pipSC}
C = \frac{1}{2}\left(\hbox{sgn}\left(2t+\frac{\mu}{2}\right) - \hbox{sgn}\left(-2t+\frac{\mu}{2}\right)\right).
\end{equation}
This implies that the system is in the topological phase when $-4t<\mu<4t$ and reaches the strong-paired phases when $\mu<-4t$ and $\mu>4t$. This also implies that the $p+ip$ superconductor can be defined through the $\mathbb{Z}_2$ topological number
\begin{equation}
C^2-\frac{1}{2} = -\langle S_z(0,0)\rangle \langle S_z(\pi,\pi)\rangle.
\end{equation}
This topological invariant or $C^2-\frac{1}{2}$ then may be detected from the protocol introduced in Sec. \ref{SCtopo} locally from the diagonal point ${\bf k} a=(\frac{\pi}{2},\frac{\pi}{2})$ on the path.

The occurrence of zero-energy modes at the edges can be verified similarly as in the article of Read and Green \cite{ReadGreen}. Close to the $\Gamma$ point $(0,0)$, looking for solutions of the form $u c_{\bf p} + v c^{\dagger}_{-{\bf p}}\sim u c_{(0,0)}+ v c^{\dagger}_{(0,0)}$
when fixing the chemical potential $\mu\sim -4t$ then this gives rise to the two coupled equations
\begin{eqnarray}
i \frac{\partial u}{\partial t} &=& \left(-2t -\frac{\mu}{2}\right) u + \Delta a (p_x-ip_y) v \\ \nonumber
i \frac{\partial v}{\partial t} &=& \Delta a (p_x+ip_y) u + \left(2t +\frac{\mu}{2}\right)v,
\end{eqnarray}
which admit solutions such that $u=v^*$ corresponding then to a (gapless) chiral Majorana fermion. Fixing $|{\bf p}|\rightarrow 0$ is equivalent to $\theta\rightarrow 0$ on the sphere and to $z=-\frac{H}{2}$ on the cylinder such that the chiral edge mode also occurs
at the edge at the bottom of the cylinder in Fig. \ref{Edges.pdf}. 

Here, we introduce the Green's function of an electron following Wang and Zhang \cite{Wang} and define the vector ${\bf h}=(\sin k_x,\sin k_y, m'+2-\cos k_x-\cos k_y)$ with $a=\Delta=2t=1$ and $m'+2=-M$ acting on the pseudospin-$\frac{1}{2}$ in the reciprocal space. 
At zero frequency, the ${\bf h}$ vector also defines the inverse of the electron Green's function ${\cal G}^{-1}(0,{\bf k})$. The Green's function at zero frequency diverges at specific points in the Brillouin zone. At the quantum phase transition driven
by $m'=0$ corresponding to $\mu=-4t$ then ${\cal G}^{-1}$ shows a zero which can then engender another definition of the transition. The relation with the topology can be precisely verified calculating the topological number in the Brillouin
zone through Eq. (\ref{dvectorsigma}) such that
\begin{equation}
C = \frac{1}{2\pi}\iint F_{xy} d^2 k,
\end{equation}
with $F_{xy} = \frac{1}{2}\epsilon^{abc} n^a \partial_{k_x} n^b \partial_{k_y} n^c$ with ${\bf n}=\frac{{\bf h}}{|{\bf h}|}$. The zeros of the vector ${\bf h}$ also plays a key role in the topological description. Performing a development around ${\bf k}=(0,0)$ such that ${\bf h}\sim (k_x,k_y,m')$ then $F_{xy}=\frac{1}{2}$ if $\mu<-4t$ and $F_{xy}=-\frac{1}{2}$ if $\mu>-4t$. This implies that the change of $C$ at the topological transition can be defined locally as 
\begin{equation}
\Delta C = F_{xy}(0,0,m'\rightarrow 0^-) - F_{xy}(0,0,m'\rightarrow 0^+) = 1
\end{equation}
 which is then equivalent to Eq. (\ref{pipSC}). From the correspondence with the Bloch sphere, we also infer that $C=A_{\varphi}(\pi,\pi) - A_{\varphi}(0,0)$ with $A_{\varphi}(\pi,\pi)=\frac{1}{2}$ and $A_{\varphi}(0,0)=-\frac{1}{2}$ from the eigenstates in Eq. (\ref{eigenstates}). This formalism on the Green's functions also applies in the presence of interactions and uses the fact that the eigenvalues of the inverse of the Green's function are real at zero frequency.

 This formalism may be then developed further for two-dimensional and three-dimensional topological systems in the presence of interactions. For recent developments on the Green's function approach, see \cite{Xu}.
 
\section{Generalized Resonating Valence Bond Theory}
\label{GRVBT}

Here, we elaborate on the smooth fields' formalism related to the fractional entangled geometry for generalized resonating valence bond states \cite{HH}.

Such resonating valence bond states have been shown to play a key role in the understanding of high-$T_c$ cuprates with the presence
of hot spots with preformed pairs forming at the corners of the two-dimensional half-filled Fermi surface of Fig. \ref{2dtrajectories} and a Fermi liquid area surrounding the diagonals \cite{plainvanilla,KarynMaurice}. A possible relation with quantum Hall systems can be revealed  through
the Kalmeyer-Laughlin approach \cite{KalmeyerLaughlin}. 

Assembling coupled spheres, two by two starting from the $C_j=\frac{1}{2}$ state, leads to quantum networks or circuits with superposition of polarized spins at one pole and an Anderson resonating valence bond state at the other pole \cite{AndersonRVB} similar to the Kitaev spin chain \cite{Majorana1}. 
To give a perspective on the fractional topological numbers on the sphere, we show that the formalism of Sec. \ref{fractionaltopology} allows us to justify the fractions observed in spin arrays in a ring geometry for a odd number of sites \cite{HH}. Along the lines of the Affleck-Kennedy-Lieb-Tasaki (AKLT) \cite{AKLT} approach which plays a key role in the Matrix Product States foundations \cite{MPS}, it is possible to find analytically some solutions with generalized resonating valence bond states.  To lighten the definitions, up state equally refers to $|\Phi_+\rangle=\uparrow$ and down state refers to $|\Phi_-\rangle=\downarrow$. For $N$ finite, we fix the parameters at south pole such that the system shows precisely $N$ degenerate ground states. 

For $N=5$: $\downarrow \uparrow \downarrow \uparrow \downarrow$, $\uparrow \downarrow \uparrow \downarrow \downarrow$, $\downarrow \uparrow \downarrow \downarrow \uparrow$, $\uparrow \downarrow \downarrow \uparrow \downarrow$ and $\downarrow \downarrow \uparrow \downarrow \uparrow$. At south pole, the ground state energy is precisely $-(N-2)r-(d-M)$ and lowered with the presence of one pair $\downarrow\downarrow$. The key point here is the presence of the transverse field which allows these $N=5$ states to resonate one onto the other \cite{HH}
\begin{equation}
|GS\rangle = \frac{1}{\sqrt{5}}\left(\downarrow \uparrow \downarrow \uparrow \downarrow + \uparrow \downarrow \uparrow \downarrow \downarrow + \downarrow \uparrow \downarrow \downarrow \uparrow + \uparrow \downarrow \downarrow \uparrow \downarrow + \downarrow \downarrow \uparrow \downarrow \uparrow\right).
\end{equation}
The proof is generalizable to any $N$ for $N$ odd such that the classical antiferromagnetic ground state is frustrated. Here, the resonating valence bond state is formed with the hopping of one bound state $\downarrow \downarrow$ (described through a spin-$\frac{1}{2}$ or `spinon' in the antiferromagnetic N\' eel state) from the transverse field giving rise to terms $- \frac{d^2\sin^2\theta}{r}\sigma_{ix}\sigma_{jx}$ with dominant contributions from $(i,j)$ nearest neighbors. 

At north pole the ferromagnetic ground state is maximally polarized. At south pole, then we have $\langle GS| \sigma_{iz} |GS\rangle = - \frac{1}{N}$ for $N$ odd. This equation simply tells that for a sphere at south pole, the probability to be in the down state is precisely $\frac{N+1}{2N}$ and the probability to be in the up state is $\frac{N-1}{2N}$. The presence of a domain wall and of the transverse field produces a specific class of
resonating valence bonds. 

We can then apply the methodology with Berry fields $A_{j\varphi}$ defined smoothly on the whole surface of the sphere generalizing Eqs. (\ref{eq2}) and (\ref{eq1}). This gives rise to
\begin{equation}
A_{j\varphi}(\pi) = \frac{N-1}{2N} A_{j\varphi}(0) + \frac{N+1}{2N}A_{j\varphi}^{r=0}(\pi).
\end{equation}
Now, we can also use the fact that adiabatically setting $r=0$, we have all the spheres englobing one monopole
\begin{equation}
A_{j\varphi}^{r=0}(\pi) - A_{\varphi}(0) = q =1.
\end{equation}
This equation shows that the presence of a Dirac monopole in the core of each site or each sphere may reveal other forms of states for low-dimensional quantum spin chains \cite{AffleckHaldane}.
Combining these two equations simply lead to
\begin{equation}
C_j = A_{j\varphi}(\pi) - A_{j\varphi}(0) = \frac{N+1}{2N}q,
\end{equation}
which reproduces well the numerical results for $N=3$ $(C_j=\frac{2}{3})$ and $N=5$ $(C_j=\frac{3}{5})$ \cite{HH}. This formula is also in agreement with 
\begin{equation}
C_j = \frac{1}{2}(\langle \sigma_{jz}(0)\rangle - \langle \sigma_{jz}(\pi)\rangle) = \frac{1}{2}\left(1+\frac{1}{N}\right).
\end{equation}
In the thermodynamical limit, for each sphere $C_j\rightarrow \frac{1}{2}$. The physics at south pole is equivalent to a delocalized bound state with an energy spectrum $-2J_{\perp}\cos(ka)$ where $J_{\perp}=\frac{d^2\sin^2\theta}{r}$ associated to a wavefunction $\psi(x)=\frac{1}{\sqrt{N}}e^{i kx}$. The increase of local energy
$2r$ from the formation of the bound state and two aligned spins is absorbed in the definition of the ground state energy $-(N-2)r-(d-M)$. The bound state then is delocalized along the chain from the minimum of energy at $k=0$. From the Majorana fermions representation of Sec. \ref{SpheresMajorana}, we obtain
$i\langle GS| \alpha_1 \eta_2 |GS\rangle_{k\rightarrow 0}\rightarrow 1$ such that the bound state has a probability ${\cal O}(\frac{1}{N})$ to be on each site. In the thermodynamical limit, this is similar to have one gapless Majorana fermion per site
with $\langle GS|\sigma_{iz} |GS\rangle\rightarrow 0$ similarly as for a quantum spin liquid. 

The situation with $N$ even may also reveal various situations of entangled states at south pole with a relation towards $C_j=\frac{1}{2}$ \cite{HH}. For four spheres forming a quantum box model we identify a correspondence between models of spheres with $\frac{1}{2}$-fractional topological numbers and the Kitaev spin model \cite{KitaevHoneycomb} in ladder geometries \cite{HH,Majorana1,Majorana2}. The model also reveals the possibility of Greenberger-Horne-Zeilinger entangled states \cite{GHZ}.

\section{Summary}
\label{Summary}

To summarize, we have elaborated on the formalism related to a (quantum) geometrical approach for topological matter through Berry gauge fields ${\bf A}$ smoothly defined on a whole surface representing for instance the surface of a Bloch sphere in quantum mechanics. Equivalently, the fields ${\bf A}'$ show a discontinuity at the interface (boundary) within the applicability of Stokes' theorem revealing the presence of a topological charge in the core. The definition of ${\bf A}'$ also reveals the presence of a Dirac string transporting the induced information from the core to the poles. The global topological number is defined locally from the poles such that the topological information can be reformulated in an effectively flat metric. From the formalism, we verify the robustness of the induced response towards gentle deformations of the surface, for instance the sphere becoming a cylinder with uniform Berry curvatures. We have shown applicability of this quantum topometry for transport properties in time from Newtonian physics and for the light-matter coupling introducing the functions $\alpha(\theta)$ and ${\cal I}(\theta)$ which are naturally related to the quantum distance and metric in curved space and on the lattice. The theoretical approach was then developed related to topological cystals and energy bands with specific applications of the light-matter interaction and circularly polarized light from the reciprocal or momentum space. The formalism allows us to include interaction effects from the momentum space within a variational stochastic approach. 

Through interaction effects between two Bloch spheres, as a result of a $\mathbb{Z}_2$ inversion symmetry between spins, we have elucidated on the possibility of fractional entangled topology, through a pure state at one pole and an entangled state at the other one pole. One-half of the surface then radiates the Berry curvature and the topological response of the two spheres becomes similar to that of a pair of merons or half-Skyrmions, which may be then engineered in mesoscopic or atomic systems. The meron physics then links with possible solutions of the Yang-Mills equation. We have identified a relation between quantum entanglement and $\frac{1}{2}$ topological numbers. Fractional numbers yet arise in a ring geometry with a odd number of spheres giving rise to generalized entangled states at south pole.

We have described applications of this fractional geometry in relation with topological semimetals in bilayer and monolayer graphene models showing the emergence of a protected topological Fermi liquid in two dimensions. In this way, this formulates an application of quantum entanglement into band theory. The topological semimetals are described through a quantized total quantum Hall conductivity and a half-quantized conductivity per plane or per spin polarization. In three dimensions, we have formulated several understandings of $\frac{1}{2}$ topological numbers in a cube through geometry, Ramanujan alternating infinite series and also transport, responses to circularly polarized light. For three-dimensional topological insulators, it is similarly known that surface states on a cube can be equally described through one Dirac point or a meron. We have then developed the formalism to topological p-wave superconducting systems and Kitaev wires and built a correspondence between Majorana fermions for the two spheres' model and fractional topology.

The two-spheres' model may find further applications related to quantum materials, circuits and quantum information, the production of entangled states locally on the Bloch sphere and it can also be applied in networks related to Matrix Product States developments. Spheres in a quantum bath can also find applications for energy applications through the quantum dynamo effect. Interestingly, the correspondence with classical physics for the smooth vector fields may suggest further applicability of the formalism for the physics of planets related to the quest of Dirac monopoles, black holes and gravitational aspects, axion physics. 
\\

K.L.H. is grateful to discussions and presentations via zoom during the difficult pandemic and Covid isolation period when this work was initiated, in particular at Cambridge, Lisbon, Oxford, Montreal and in person at ENS Paris, Aspen, Dresden, Les Diablerets and UCLondon. K.L.H. also acknowledges students, postdoctoral associates and colleagues for discussions, collaborations and support related to these ideas. This work has benefitted from lectures given at Ecole Polytechnique, four lectures on Geometry and Topology in the Quantum given at University Paris-Saclay, https://www.universite-paris-saclay.fr/lectures, June 2023, and one at UCLondon on quantum theory and applications July 2023. K.L.H. acknowledges  support  from Ecole Polytechnique, CNRS, the Deutsche Forschungsgemeinschaft  (DFG)  under  project number 277974659 and from ANR BOCA. Numerical evaluations on some figures have benefitted from the Pythtb platform.  
\\
This review is dedicated to my family.

\appendix

\section{Geometrical Definitions, Metric, $\alpha(\theta)$ and ${\cal I}(\theta)$ functions}
\label{Berrycurvature}

Here, we develop the formalism on the Berry curvature from the sphere to the lattice, introducing the quantum metric and quantum distance, together with the new geometrical $\alpha(\theta)$ and ${\cal I}(\theta)$ functions.

First, we re-derive useful geometrical relations for a surface parametrized by components ${\bf R}=(R_x,R_y)$. We introduce the Berry connection \cite{Berry}
\begin{equation}
A_{\nu}({\bf R}) = -i\langle \psi |\partial_{\nu} |\psi\rangle.
\end{equation}
Here, $|\psi\rangle$ refers to the ground state or lower-energy state for a spin-$\frac{1}{2}$ model or for a two-band model. 
From $\bm{\nabla}\times{\bf A}$, then we evaluate the Berry curvature \cite{Berry}
\begin{equation}
F_{\mu\nu} = \frac{\partial}{\partial R_{\mu}}A_{\nu} - \frac{\partial}{\partial R_{\nu}}A_{\mu} = \partial_{\mu}A_{\nu} - \partial_{\nu}A_{\mu} = -F_{\nu\mu}.
\end{equation}
We can go step by step. First,
\begin{equation}
\partial_{\mu}A_{\nu}  = -i\langle \partial_{\mu}\psi| \partial_{\nu}\psi\rangle -i \langle \psi| \partial_{\mu}\partial_{\nu}\psi\rangle.
\end{equation}
Therefore, 
\begin{equation}
F_{\mu\nu}({\bf R}) = -i\left(\langle \partial_{\mu}\psi| \partial_{\nu}\psi\rangle-\langle \partial_{\nu}\psi| \partial_{\mu}\psi\rangle\right).
\end{equation}
Now, we can insert the relation $\sum_n |n\rangle\langle n|=1$ including $|n\rangle=|\psi\rangle$ and all the other eigenstates in the energy spectrum such that 
\begin{equation}
\label{Fmunu}
F_{\mu\nu}({\bf R}) = -i\sum_n \left(\langle \partial_{\mu}\psi |n\rangle\langle n| \partial_{\nu}\psi\rangle-\langle \partial_{\nu}\psi| n\rangle\langle n|  \partial_{\mu}\psi\rangle\right).
\end{equation}
If $|\psi\rangle=|n\rangle$, then the result is zero. Therefore, the sum implies $|n\rangle\neq |\psi\rangle$. Playing with the equation
\begin{equation}
H({\bf R}) |n\rangle = E_{n} |n\rangle
\end{equation}
and applying the differential operator $\frac{\partial}{\partial R_{\alpha}}$ on both sides, with $\alpha=\mu$ or $\nu$, then we obtain the identity
\begin{equation}
\langle n| \partial_{\alpha}\psi\rangle = -\frac{\left\langle n \left| \frac{\partial H}{\partial R_{\alpha}}\right| \psi \right\rangle }{(E_n-E_{\psi})}.
\end{equation}
This identity can be verified from the identity $\langle n|\partial_{\alpha} (H|\psi\rangle) = \langle n|\partial_{\alpha} H |\psi\rangle + \langle n| H \partial_{\alpha} |\psi\rangle = E_{\psi} \langle n|\partial_{\alpha}|\psi\rangle$. The second term is equivalent
to $\langle n| H \partial_{\alpha} |\psi\rangle = E_n \langle n| \partial_{\alpha} |\psi\rangle$. If we invert the role of $n$ and $\psi$ then
\begin{equation}
\langle \partial_{\alpha}\psi |n\rangle = \frac{\left\langle \psi \left| \frac{\partial H}{\partial R_{\alpha}}\right| n \right\rangle }{(E_n-E_{\psi})}.
\end{equation}
This is equivalent to
\begin{equation}
\label{munu}
F_{\mu\nu} =  i\sum_{n\neq \psi}
\frac{\left(\left\langle n \left| \frac{\partial H}{\partial R_{\mu}}\right| \psi \right\rangle \left\langle \psi \left| \frac{\partial H}{\partial R_{\nu}}\right| n \right\rangle -  \mu\leftrightarrow \nu\right)}{(E_n-E_{\psi})^2}.
\end{equation}
In the case of a $2\times 2$ matrix Hamiltonian, we have only one $|n\rangle$ excited state.

Here, $F_{\mu\nu}$ can be defined equally on the sphere with $\mu$, $\nu$ representing the angles $\varphi$ and $\theta$ through $F_{\theta\varphi}=\partial_{\theta}A'_{\varphi}-\partial_{\varphi} A'_{\theta}=\frac{\sin\theta}{2}$ with $|\psi\rangle=|\psi_+\rangle$ in Eq. (\ref{eigenstates}) and on the lattice. Following Ref. \cite{Tan}, it is useful to evaluate the quantum distance related to the azimuthal angle through $|\langle \psi_+(\varphi')|\psi_+(\varphi)\rangle|^2$. Defining $\varphi'=\varphi+d\varphi$, then the quantum distance is
\begin{equation}
ds^2 = 1- |\langle \psi_+(\varphi')|\psi_+(\varphi)\rangle|^2 = g_{\varphi\varphi}d\varphi^2
\end{equation}
such that for one sphere with eigenstates in Eq. (\ref{eigenstates}), we can verify that the quantum metric component $g_{\varphi\varphi}$ is related to local topological properties
\begin{equation}
\label{gvarphivarphi}
g_{\varphi\varphi} = \frac{\sin^2\theta}{4} = F_{\theta\varphi}^2.
\end{equation}
This function goes to zero at the poles indicating that the surface is reduced to a point. The eigenstate $|\psi_-\rangle$ reveals a similar form of $g_{\varphi\varphi}$ function.

In Ref. \cite{Tan}, these quantum metric components were recently measured in a superconducting quantum circuit with one artificial atom or spin-$\frac{1}{2}$ either through a periodic drive such that $\theta(t) = \theta_0+2(E/\omega)\cos(\omega t)$ (or after a sudden quench). These authors have also experimentally checked the relation $F_{\theta\varphi}=2\sqrt{det\ g}$ with $g$ being a $2\times 2$ matrix tensor, from another definition of the topological invariant related to the metric and Ricci scalar curvature $R$ \cite{Ma,Kolodrubetz} which is in agreement with the definition
\begin{equation}
C = \frac{1}{4\pi}\int\int_{{\cal M}} \frac{R}{2}\sqrt{det g} d\mu d\nu
\end{equation}
where ${\cal M}=S^2$, $(\mu,\nu)=(\theta,\varphi)$ and the Ricci scalar curvature $R=8$.

From the definition of angles then we also have the general relation \cite{OneHalfKLH}
\begin{equation}
|\langle \psi_+(\theta,\varphi')| \psi_+(\theta,\varphi)\rangle|^2 = \alpha(\theta) + \cos(\varphi'-\varphi)\frac{\sin^2\theta}{2}
\end{equation}
where $\alpha(\theta)$ is precisely the geometrical function introduced regarding the response to circularly polarized light; see Eq. (\ref{alpha}). Equivalently, the function $\alpha(\theta)$ introduced in Eq. (\ref{alpha}) can be re-written as
\begin{equation}
\label{metrictheta}
\alpha(\theta) = C^2 - 2g_{\varphi\varphi}(\theta).
\end{equation}
We have a relation between metric component $g_{\varphi\varphi}(\theta)$ and the response to a circularly polarized field acting on a dipole (spin-$\frac{1}{2}$ or electric dipole) in curved space. 
The response to a circularly polarized field for a dipole then measures the metric component $g_{\varphi\varphi}$. The information hidden in this component is essential on the sphere fiber where we also have
$g_{\theta\theta}=\frac{1}{4}$ and $g_{\theta\varphi}=0$.  In the trivial case $C=0$, the response at any point on the sphere becomes identical to the one at $\theta=0$, revealing the same information as $\alpha(0)$ \cite{OneHalfKLH}. 
For a comparison, for the Bell pair or EPR pair of the two-spheres' model if we measure the quantum distance on one sphere implementing a variation from $\varphi$ to $\varphi'$ for the azimuthal angle on one sphere only then we find instead $g_{\varphi\varphi}=\frac{1}{4}$. This is as if we would fix $\theta=\frac{\pi}{2}$ in the one sphere model from the metric calculation. Selecting $\varphi'-\varphi=\frac{\pi}{2}$ through a circularly polarized field, similarly as a clock, allows for a local measure of the one-half topological number \cite{OneHalfKLH}.  

It is also useful to derive relations on the honeycomb lattice introducing $\mu,\nu$ as $p_x,p_y$. Close to the Dirac point $K$, we can verify results already to linear order ${\cal O}(\theta)$, from Eqs. (\ref{eigenstates}) and (\ref{tan}) the lowest-band wavefunction in the Haldane model $|\psi\rangle = |\psi_+\rangle$ reads:
\begin{equation}
\label{psi+}
|\psi_+\rangle = 
\left(
\begin{matrix}
1 \\
- \frac{1}{2}\frac{\hbar v_F|{\bf p}|}{m} e^{i\tilde{\varphi}}
\end{matrix}
\right),
\end{equation}
with 
\begin{equation}
\tan\theta = \frac{\hbar v_F|{\bf p}|}{m}\approx \sin \theta. 
\end{equation}
Here, we have inserted the form of the lowest-band eigenstate from Eq. (\ref{eigenstates}) and implemented the relation between polar angle around a Dirac point and the azimuthal angle on the sphere $\tilde{\varphi}=\varphi\mp \pi$ from Eq. (\ref{correspondence}). We can then derive useful identities
\begin{equation}
\partial_{p_x}|\psi_+\rangle = 
\left(
\begin{matrix}
0 \\
-\frac{1}{2}\frac{\hbar v_F}{m}
\end{matrix}
\right)
\end{equation}
and
\begin{equation}
\partial_{p_y}|\psi_+\rangle = 
\left(
\begin{matrix}
0 \\
-\frac{i}{2}\frac{\hbar v_F}{m}
\end{matrix}
\right),
\end{equation}
such that 
\begin{equation}
i\partial_{p_y}\langle \psi_+| \partial_{p_x} |\psi_+\rangle = \frac{\hbar^2 v_F^2}{4m^2}.
\end{equation}
Also, we have
\begin{equation}
i\partial_{p_x}\langle \psi_+| \partial_{p_y} |\psi_+\rangle = -\frac{\hbar^2 v_F^2}{4m^2}.
\end{equation}
Defining $F_{p_y p_x} = - F_{p_x p_y} = +i\partial_{p_y}\langle \psi_+| \partial_{p_x} |\psi_+\rangle  - i\partial_{p_x}\langle \psi_+| \partial_{p_y} |\psi_+\rangle$, then to linear order in $\theta$, we have
\begin{equation}
\label{Fpypx}
F_{p_y p_x} = \frac{\hbar^2 v_F^2}{2m^2}.
\end{equation}
This equation is gauge-independent. This relation, formulated as in Eq. (\ref{Fpypx}), was also established in Ref. \cite{Ryu}. The geometrical method presented in Eq. (\ref{swap}) allows us to verify in a simple way the presence of 
$\cos\theta$ as a global prefactor when going to higher orders in $\theta$ in this formula starting from the general sphere eigenstates making a link with $C=1$ at the two Dirac points.  This formula tends to agree with the analysis of Ref. \cite{Meron}. We can also verify from (\ref{psi+}) that the Berry connections $A_{p_x}$ and $A_{p_y}$ are zero at the Dirac points.  

From the definitions, for the diagonal terms, we have $F_{p_x p_x}=0=F_{p_y p_y}$.  We find then judicious to introduce
\begin{equation}
f_{\mu\mu} = \langle \partial_{\mu} \psi| \partial_{\mu} \psi\rangle
\end{equation}
with here $\mu=p_x$ or $p_y$ for the diagonal response. 

We show below that it is related both to the Fubini-Study metric, to the quantum distance on the lattice and also to the response to circularly polarized light through the ${\cal I}(\theta)$ function. It is first useful to generalize Eq. (\ref{munu}) for the $f$ function defining
\begin{equation}
f_{\mu\mu} + f_{\nu\nu} = \langle \partial_{\mu} \psi | \partial_{\mu} \psi\rangle + \langle \partial_{\nu} \psi | \partial_{\nu} \psi\rangle.
\end{equation}
Inserting an eigenstate $|n\rangle$ similarly as for $F_{\mu\nu}$ this gives rise to the identity
\begin{equation}
f_{\mu\mu} + f_{\nu\nu} = \sum_{n\neq \psi} \frac{{\cal I}_{\mu\mu}+{\cal I}_{\nu\nu}}{(E_n-E_{\psi})^2},
\end{equation}
where
\begin{equation}
{\cal I}_{\mu\mu} = \left\langle \psi \left| \frac{\partial H}{\partial R_{\mu}} \right| n\right\rangle \left\langle n \left| \frac{\partial H}{\partial R_{\mu}} \right|\psi \right\rangle,
\end{equation}
and similarly for ${\cal I}_{\nu\nu}$. Introducing $\mu=p_x$ and $\nu=p_y$ then we observe that ${\cal I}_{p_x p_x}+{\cal I}_{p_y p_y}$ function precisely corresponds to ${\cal I}(\theta)$ that we introduced in Eq. (\ref{Itheta}).
This implies that the diagonal function ${\cal I}_{p_x p_x}+{\cal I}_{p_y p_y}$ is also a good measure of topological properties through $C^2$ locally from the Dirac points or from the poles of the sphere, which precisely enters in
the response in time to circularly polarized light in Sec. \ref{light}. The function $|\langle \psi_-| \sigma_x | \psi_+\rangle|=|\langle n|\sigma_x|\psi\rangle|$ can also be measured in principle through the corrections in energy due to the light-matter coupling with a dipole interaction as in Eq. (\ref{energyshift}). At the Dirac points, $(E_n-E_{\psi})^2$ becomes equal to $(2m)^2$. Therefore, the function $f_{p_x p_x}+f_{p_y p_y}$ is a good marker of topological properties locally on the sphere and in the reciprocal space of the topological lattice model. 

We can also evaluate the quantum distance on the lattice \cite{Ryu,BlochMetric} defined in a symmetric way through
\begin{equation}
\langle \psi_+({\bf k}-d{\bf k})| \psi_+({\bf k}+d{\bf k})\rangle.
\end{equation}
At the Dirac points, due to the fact that $A_{p_x}$ and $A_{p_y}$ are zero, then we have
\begin{equation}
\langle \psi_+({\bf K}-d{\bf k})| \psi_+({\bf K}+d{\bf k})\rangle = -\langle \partial_{k_{\mu}} \psi_+ | \partial_{k_{\mu}} \psi_+ \rangle d k_{\mu}^2.
\end{equation}
Therefore, we can access the metric $g_{ij}$ defined as
\begin{equation}
g_{ij} dk_i dk_j = 1-|\langle \psi_+({\bf k}-d{\bf k}) | \psi_+({\bf k}+d{\bf k})\rangle|^2.
\end{equation}
At the Dirac points, then we have the precise identity
\begin{eqnarray}
g_{\mu\mu} &=& 2\hbox{Re}(\langle \partial_{k_{\mu}}\psi_+ | \partial_{k_{\mu}}\psi_+\rangle) \\ \nonumber
&=& 2\hbox{Re}(f_{\mu\mu}) = \frac{1}{2}\frac{\hbar^2 v_F^2}{m^2}C^2.
\end{eqnarray}
This formula is in agreement with the formula of Matsuura and Ryu but the authors did not identify the presence of $C^2$.
This equation also implies that the metric is effectively flat in the vicinity of the poles of the sphere. 

Recently, efforts have been made to relate this quantum metric from the reciprocal or momentum space to a gravitational approach \cite{BlochMetric}. In the sense of the Einstein Field Equation, this metric close to the poles of the sphere or the Dirac point implies a vacuum for the gravitational field assuming a pure quantum state $|\psi\rangle=|\psi_+\rangle$. The response to circularly polarized light also measures the quantum distance on the lattice through ${\cal I}(\theta)$ \cite{C2}.  

\section{Laughlin States and Geometry}
\label{LaughlinQHResponse}

A charged particle in a magnetic field is described through the Hamiltonian 
\begin{equation}
H = \frac{1}{2m}\left(p_x + \frac{eB}{2}y\right)^2 + \frac{1}{2m}\left(p_y - \frac{eB}{2}x\right)^2 = \frac{\hbar\omega_c}{2}\left(\frac{\partial^2}{\partial x^2} + \frac{\partial^2}{\partial y^2} +\frac{1}{4}r^2 -L_z\right)
\end{equation}
with the angular momentum 
\begin{equation}
L_z = -i\left(x\frac{\partial}{\partial y} - y\frac{\partial}{\partial x}\right) = - i\frac{\partial}{\partial{\varphi}}.
\end{equation}
We can introduce the polar coordinates in real space $\tan \varphi = \frac{y}{x}$ and $r^2=x^2+y^2$ defined in units of the cyclotron length. The eigenfunction linked to the lowest Landau level satisfies $H\psi_m = \frac{\hbar\omega_c}{2}\psi_m$.
The associated wavefunction reads
\begin{equation}
\psi_m = {\cal N} r^m e^{i m\varphi} e^{-r^2/4} 
\end{equation}
with ${\cal N}$ a normalization factor and the quantum number $m$ associated to the angular momentum. The system has $N$ electrons occupying a flux quantum and therefore there are $N$ independent $m$ states corresponding to $N_{\phi}$ flux quanta.
We can introduce $z=x+iy$ and since the wavefunction is antisymmetric, it is equally defined in terms of a Vandermonde determinant, such that 
\begin{equation}
\psi = \frac{1}{\sqrt{N!}} \Pi_{j>i} (z_j-z_i) e^{-\frac{1}{4}\sum_{j=1}^N | z_j |^2}.
\end{equation}
Suppose we are on the surface of a cylinder characterized through the radius $r=1$ and perimeter $L=2\pi$, then we can introduce the variable $z=\sum_{j=1}^N z_j=e^{ikL}$ with $k=\frac{N\pi}{L}+\frac{\varphi}{L}$.
The wavefunction is periodic and it can be written in terms of an odd elliptic theta function  \cite{Fradkin2}. The Niu-Thouless-Wu formula \cite{NTW} applies for the Berry phase related to the center of mass or for the quantum Hall conductivity \cite{Fradkin}
\begin{equation}
\label{sigmaxy}
\sigma_{xy} = \frac{e^2}{i\hbar} \oint d\varphi\langle \psi | \frac{\partial}{\partial \varphi} |\psi\rangle =
\frac{e^2}{h}.
\end{equation}
In the wavefunction, this is similar as if we identify on particle at position $z_i$ with the center of mass coordinate $z=e^{i\varphi}$.
The Laughlin wavefunction for the fractional quantum Hall effect at filling factor $\frac{1}{p}=\nu=\frac{1}{3}=\frac{N}{N_{\phi}}$ 
\begin{equation}
\psi = \frac{1}{\sqrt{N!}} \Pi_{j>i} (z_j-z_i)^{p} e^{-\frac{1}{4}\sum_{j=1}^N | z_j |^2}
\end{equation}
One electron is now composed of $p$ fractional charges, which is equivalent to introduce $z^p=\sum_{j=1}^N z_j^p = e^{i p k L}$ modifying $p k = \frac{Np \pi}{L}+\frac{\tilde{\varphi}}{L}$ with $\tilde{\varphi}=p\varphi$. This shift of phase in (\ref{sigmaxy}) gives
\begin{equation}
\sigma_{xy} = \frac{e^2}{i\hbar} \int_0^{\frac{2\pi}{p}} \frac{d\tilde{\varphi}}{p} p \langle \psi | \frac{\partial}{\partial \tilde{\varphi}} |\psi\rangle = \frac{e^2}{i\hbar p} \int_0^{2\pi} d\tilde{\varphi} \langle \psi | \frac{\partial}{\partial \tilde{\varphi}} |\psi\rangle 
 = \frac{e^2}{h} \frac{1}{p}.
\end{equation}
Associated to the phase shift in the many-body wavefunction, this is also equivalent to introduce a fractional Berry function $-\frac{i}{p}\langle \psi | \frac{\partial}{\partial \tilde{\varphi}} |\psi\rangle$.

In the two-spheres' model, the surface is defined in terms of the reciprocal space linked to $(k_{\parallel},k_{\perp})$. At south pole, the measure on one sphere for the transverse pumped current in Sec. \ref{fractionaltopology} is equivalent to project the EPR wavefunction or Bell wavefunction onto $\frac{1}{\sqrt{2}}|\Phi_+\rangle_1 |\Phi_-\rangle_2$ or $\frac{1}{\sqrt{2}}|\Phi_-\rangle_1 |\Phi_+\rangle_2$, such that the Berry phase acquires a $\frac{1}{2}$ prefactor, here at the level of the one-particle wavefunction.

\section{Topometry in Band Theory}
\label{TopometryBandTheory}

Here, we introduce the formalism within the band theory and start with two planes generalizing a proof in Supplemental Material of Ref. \cite{bilayerQSH}. 
Suppose a theory of two planes related to the sphere' model. In the equatorial plane, the theory is equivalent to two Dirac models coupled with a term $r$.
In the equatorial plane, the matrix Hamiltonian is defined through the `hopping'  terms $(d_x,d_y,r)$ (and hereafter we assume a Semenoff mass going to zero)
\begin{equation}
H = 
\begin{pmatrix}
          0   & d_x-id_y & r & 0  \\
          d_x+id_y   & 0 & 0 & r \\
          r & 0 & 0 & d_x-id_y \\
          0 & r & d_x+id_y & 0 
 			  \end{pmatrix}.
\end{equation}			  
The energy eigenvalues are classified as $E_1<E_2<E_3<E_4$ with $E_1 = (-r-|{\bf d}|)$, $E_2 = (r-|{\bf d}|)$, $E_3 = (-r+|{\bf d}|)$ and $E_4 = (r+|{\bf d}|)$. The norm of ${\bf d}$ vector refers ${\bf d}=(d_x,d_y,0)$. 
In the equatorial plane, $d_z=|{\bf d}|\cos\theta=0$. The situation $\theta=\frac{\pi}{2}$ on the sphere corresponds to the $M$ point within the Brillouin zone for this band structure. 

We formulate the proof for the two lowest (occupied) bands at half-filling. A similar proof can be formulated for the two upper bands.
For $r=0$, each of these bands then refers to the lowest band of the Haldane model.

The eigenstates are: 
\begin{eqnarray}
\label{spinorM}
|\Phi_1({\bf M^-})\rangle = |\Phi_1({\bf M^-},\tilde{\varphi})\rangle =
\frac{e^{i\phi_1}}{2}\left(\begin{array}{c}
          -e^{-i\tilde{\varphi}} \\
         1 \\
          e^{-i\tilde{\varphi}}  \\
          -1 
 			  \end{array} \right)	  
\end{eqnarray}		
\begin{eqnarray}	  
			  |\Phi_2({\bf M^-})\rangle = \frac{e^{i\phi_1}}{2} \left(\begin{array}{c}
          - e^{-i\tilde{\varphi}} \\
         1 \\
         - e^{-i\tilde{\varphi}} \\
          1 
 			  \end{array} \right)
\end{eqnarray}
with $\tan\tilde{\varphi}=\frac{d_y}{d_x}$ such that $d_x+i d_y = |{\bf d}|e^{i\tilde{\varphi}}$ and $d_x-i d_y = |{\bf d}|e^{-i\tilde{\varphi}}$. The phase $\phi_1$ is a gauge choice and a function of $\tilde{\varphi}$. We define these eigenstates as $|\Phi_i({\bf M}^-)\rangle$ which means for an angle $\theta=\theta_c^-$ on the sphere and at the point $M-\epsilon$ with $\epsilon\rightarrow 0$ in the Brillouin zone. The two eigenstates are defined to be orthogonal (we will formulate a proof for each band and for simplicity in this Appendix we choose the same Hilbert space representation in terms of global phase for the two bands similarly as for the two spheres). 
 
 Around the two poles on the sphere, we can also introduce the angles $\tilde{\varphi}$ associated to two Dirac cones turning in opposite directions from Eq. (\ref{correspondence}).  Stokes' theorem on the sphere can be interpreted as two circles such that the azimuthal angle would then turn in different directions at $M^-=M-\epsilon$ and $M^+=M+\epsilon$. We propose below a protocol to measure the topological number for a band  from the Dirac points (from the spheres' correspondence) and simultaneously through a swap $\tilde{\varphi}\rightarrow -\tilde{\varphi}$ at the $M$ point.
For this purpose, we introduce 
\begin{eqnarray}
\label{spinorM'}
|\Phi_1({\bf M^+})\rangle = \frac{e^{i\phi_2}}{2}\left(\begin{array}{c}
          -e^{i\tilde{\varphi}} \\
         1 \\
          e^{i\tilde{\varphi}}  \\
          -1 
 			  \end{array} \right)	
\end{eqnarray}
\begin{eqnarray}			    
			  |\Phi_2({\bf M^+})\rangle =  |\Phi_2({\bf M^+},-\tilde{\varphi})\rangle	=  
			  \frac{e^{i\phi_2}}{2} \left(\begin{array}{c}
          - e^{i\tilde{\varphi}} \\
         1 \\
         - e^{i\tilde{\varphi}} \\
          1 
 			  \end{array} \right).
\end{eqnarray}
At this stage, the phase $\phi_2$ is a gauge choice related to the equatorial plane in south hemisphere.

We address here the situation with two symmetric planes. Topological properties are encoded through the functions $d_z^1$ and $d_z^2$ at $K$ and $K'$ points referring to the poles of the sphere. At these points, the matrix also takes a simple form
\begin{eqnarray}
\label{matrixdz}
H =
\begin{pmatrix}
          d_z^1  &  0 & r & 0  \\
          0   & -d_z^1 & 0 & r \\
          r & 0 & d_z^2 & 0 \\
          0 & r & 0 & -d_z^2
 			  \end{pmatrix}.
\end{eqnarray}			  
The energy eigenvalues can be solved accordingly. 
\begin{equation}
E_i' = \frac{1}{2}\left(\mp (d_z^1+d_z^2)\mp \sqrt{(d_z^1-d_z^2)^2+4r^2}\right).
\end{equation}
For the situation with $d_z^1=d_z^2=d_z>0$ and $d_z>r$, the eigenenergies are
\begin{equation}
E_1' = -d_z-r,\hskip 0.2cm E_2' = -d_z+r .
\end{equation}
This corresponds to the energetics at one pole or the $K$ Dirac point. 		
At $K$ point or for a vector ${\bf K}$, the energy eigenstates read
\begin{eqnarray}
\hskip -0.5cm
\label{spinorK}
|\Phi_1({\bf K})\rangle = \frac{e^{i\phi_1'}}{\sqrt{2}}\left(\begin{array}{c}
         0  \\
         1 \\
         0  \\
          -1 
 			  \end{array} \right)
			  \hskip 0.2cm			  
			  |\Phi_2({\bf K})\rangle = \frac{e^{i\phi_1'}}{\sqrt{2}} \left(\begin{array}{c}
         0  \\
         1 \\
         0 \\
          1 
 			  \end{array} \right).
\end{eqnarray}
At vector ${\bf K}'$, since $d_z\rightarrow -d_z$ the two lowest bands (with the same energies $E_1'$ and $E_2'$) are now described through 
\begin{eqnarray}
\hskip -0.6cm
\label{spinorK'}
		  |\Phi_1({\bf K}')\rangle = \frac{e^{i\phi_2'}}{\sqrt{2}}\left(\begin{array}{c}
         1  \\
          0  \\
          -1 \\
           0
 \end{array} \right)
 \hskip 0.2cm
   |\Phi_2({\bf K}')\rangle = \frac{e^{i\phi_2'}}{\sqrt{2}}\left(\begin{array}{c}
         1  \\
         0 \\
          1 \\
           0
 \end{array} \right).
\end{eqnarray}
The forms of these eigenstates are identical to those of bands $3$ and $4$ at the $K$ point. The eigenstates at $K$ and $K'$ are characterized through a flip of polarization encoding the information $d_z\rightarrow -d_z$. We introduce two phases $\phi_1'$ and $\phi_2'$ at $K$ and $K'$ to encode the topological information. This topological information for a band will then fix a relation between these phases.

We define the topological number of each band $C=C_1=C_2$ in two equivalent ways (similarly as on the sphere). From the mapping onto the sphere(s) (and the knowledge of the limit $r\rightarrow 0$) we can define
\begin{eqnarray}
\label{phi'}
C = A_{\tilde{\varphi}}({\bf K}') - A_{\tilde{\varphi}}({\bf K}) = \frac{\partial(\phi_2' - \phi_1')}{\partial\tilde{\varphi}}.
\end{eqnarray}
From the $M$ point, we can equivalently introduce 
\begin{equation}
-i\langle \Phi_1({\bf M}^-)|\frac{\partial}{\partial{\tilde{\varphi}}} |\Phi_1({\bf M}^-)\rangle = \frac{\partial \phi_1}{\partial\tilde{\varphi}} - \frac{1}{2}.
\end{equation}
Below, we show from the geometry on the sphere that the topological number can be equivalently measured through
\begin{eqnarray}
\label{bandsmoothdef}
C &=& -i\langle \Phi_1({\bf M}^-)|\frac{\partial}{\partial{\tilde{\varphi}}} |\Phi_1({\bf M}^-)\rangle \\ \nonumber
&+& i\langle \Phi_1({\bf M}^+)|\frac{\partial}{\partial{\tilde{\varphi}}} |\Phi_1({\bf M}^+)\rangle \\ \nonumber
&=& \frac{\partial \phi_1}{\partial\tilde{\varphi}} - \frac{1}{2} - \frac{\partial \phi_2}{\partial\tilde{\varphi}} - \frac{1}{2}.
\end{eqnarray}

The wavefunction at $M-\epsilon$ can be reformulated as
\begin{eqnarray}
\label{raccordement}
|\Phi_1({\bf M^{-})}\rangle = \frac{e^{i\phi_1}}{2}
\left(\begin{array}{c}
         0  \\
         1 \\
         0  \\
          -1 
 			  \end{array} \right)
+
\frac{e^{i\phi_1}}{2}
\left(\begin{array}{c}
         e^{-i\tilde{\varphi}}  \\
         0 \\
         e^{-i\tilde{\varphi}}  \\
          0 
 			  \end{array} \right).
\end{eqnarray}	
The identification of the first part in $|\Phi_1\rangle$ with the wavefunction at $K$ leads to $\phi_1=\phi_1'$ and the identification of the second part with the wavefunction at $K'$ leads to $\phi_2'=\phi_1-\tilde{\varphi}$. 
This is in agreement with $C=-1$ with the definition of the ${\bf d}$-vector. Through the swap of $\tilde{\varphi}\rightarrow -\tilde{\varphi}$ at $M+\epsilon$ then this is equivalent to measure $C=-1$ from $M$. Indeed, 
through the definition of eigenstates in Eqs. (\ref{spinorM}) and (\ref{spinorM'}) with the two additional phases $\phi_1$ and $\phi_2$, this is also equivalent to introduce the relations with the sphere' definitions
\begin{equation}
\label{eq1new}
-i\langle \Phi_1({\bf M}^-)|\frac{\partial}{\partial{\tilde{\varphi}}} |\Phi_1({\bf M}^-)\rangle = A_{\tilde{\varphi}}({\bf K}) - A_{\tilde{\varphi}}({\bf K}) =0,
\end{equation}
and
\begin{equation}
\label{eq2new}
-i\langle \Phi_1({\bf M}^+)|\frac{\partial}{\partial{\tilde{\varphi}}} |\Phi_1({\bf M}^+)\rangle = A_{\tilde{\varphi}}({\bf K}) - A_{\tilde{\varphi}}({\bf K}') = - C.
\end{equation}
Here, we find it judicious to do a correspondence with the lowest-eigenstate $|\psi_-\rangle$ of the spin-$\frac{1}{2}$ model  in Eq. (\ref{eigenstates})(for the definition of the matrix in Eq. (\ref{matrixdz}), we have flipped ${\bf d}\rightarrow -{\bf d}$) then this is equivalent to 
$\phi_1=\phi_1'=\frac{\tilde{\varphi}}{2}=A_{\tilde{\varphi}}({\bf K})$ and  $\phi_1-\tilde{\varphi}=\phi_2'=-\frac{\tilde{\varphi}}{2}=A_{\tilde{\varphi}}({\bf K}')$ leading to $C=-1$. Eq. (\ref{bandsmoothdef}) is gauge invariant.

\section{Photo-Induced Currents and Conductivity}
\label{lightconductivity}

The photocurrents are responses to the electric field related to light. The electric field takes the form ${\bf E}=e^{i\frac{\pi}{2}}A_0\omega e^{-i\omega t}({\bf e}_x\mp i{\bf e}_y)$ such that $\hbox{Re}{\bf E}=-(|A_0|\omega)(\sin\omega t,\mp \cos\omega t,0)$. If we suppose $A_0<0$, then at short times, the physics is analogous to the effect of an electric field ${\bf E}=\mp \omega |A_0| {\bf e}_{\varphi}$ with the unit vector tangent to the azimuthal angle in the equatorial plane ${\bf e}_{\varphi}\sim{\bf e}_y$ and $\pm$ refers to the right-handed 
$(+)$ and left-handed $(-)$ polarizations respectively as defined in Sec. \ref{electricfield}. The two light polarizations produce photocurrents turning in different directions. 

Now, we calculate the photo-currents. We begin with the continuity equation 
\begin{equation}
\bm{\nabla}\cdot{\bf J} +\frac{\partial \hat{n}}{\partial t}=0.
\end{equation}
Due to the structure of the $2\times 2$ matrix Hamiltonian, then $\hat{n}_a(t) = \frac{1}{2}\left(\mathbb{I} + \sigma_z\right)$ and $\hat{n}_b(t) = \frac{1}{2}\left(\mathbb{I} - \sigma_z\right)$. Here, $\mathbb{I}$ refers to the identity matrix. Then, starting from the reciprocal space, we can write
\begin{equation}
\frac{d\hat{n}_a}{dt} = -\frac{d\hat{n}_b}{dt},
\end{equation}
such that for transport properties, the current density in this model can be defined from 
\begin{equation}
\hat{J}(t) = \frac{d}{dt}(\hat{n}_a({\bf k},t)-\hat{n}_b({\bf k},t)).
\end{equation}
On the lattice, we can approximate $\frac{\partial \hat{J}_i}{\partial x_i}\sim \frac{\hat{J}_i}{a}$ with the lattice spacing set to unity $a=1$ and from Fourier transform we can equivalently evaluate the current density from the reciprocal space.
In this sense, the current refers to the current from an electrical dipole measuring the charge polarization between a $a$ and $b$ state for a given ${\bf k}$.  Transferring one $a$ particle (from a lower energy band) to a $b$ particle (to a upper energy band) at the $K$ point through light will induce a current.

Now, we use the definitions related to the Ehrenfest theorem. We
have
\begin{equation}
\langle \sigma_z \rangle (t) = \langle \psi(t) | \sigma_z | \psi(t)\rangle = \langle \psi(0) | e^{\frac{i H t}{\hbar}} \sigma_z e^{\frac{-i H t}{\hbar}} |\psi(0)\rangle.
\end{equation}
Therefore, we can assume equivalently that the operator $\sigma_z(t)$ now evolves in time such that 
\begin{equation}
\sigma_z(t) = e^{\frac{i H t}{\hbar}} \sigma_z e^{\frac{-i H t}{\hbar}},
\end{equation}
and that the wave-functions are at fixed time $t=0$. From the Ehrenfest theorem, 
\begin{equation}
\frac{d}{dt}\langle \sigma_z\rangle(t) = \frac{i}{\hbar}\langle \psi(t) | [H,\sigma_z] |\psi(t)\rangle,
\end{equation}
this is equivalent to 
\begin{equation}
\frac{d}{dt}\sigma_z(t) = \frac{i}{\hbar} [H,\sigma_z(t)].
\end{equation}
For a charge $e=1$, we have
\begin{equation}
\hat{J}(t) = \frac{1}{2}\frac{d}{dt}\sigma_z(t).
\end{equation}
From the form of the Hamiltonian, including the light-matter coupling we obtain
\begin{equation}
\hat{J}(t)=v_F\left((p_x+A_x(t))\sigma_y - (\zeta p_y +A_y(t))\sigma_x\right).
\end{equation}

When we evaluate the current response to second-order in $A_0$ from Fermi golden's rule, it is the same as keeping just the $A_x(t)$ and $A_y(t)$ in this equation \cite{Klein}. Therefore, we see that it is the same
to select a Dirac point from a light polarization setting $p_x=p_y=0$ or to perform an average on all the wave-vectors \cite{Goldman}. Therefore, we can equally calculate the response at a Dirac point leading to 
\begin{equation}
\label{Kstructure}
\hat{J}_{\pm,\zeta}(t) = \frac{1}{2\hbar}A_0e^{-i\omega t}\left(\frac{\partial H}{\partial(\zeta p_y)} \pm i \frac{\partial H}{\partial p_x}\right)+h.c.
\end{equation}
Now, we can apply the Fermi golden rule on the current density. This results in
\begin{eqnarray}
\tilde{\Gamma}_{\pm} &=& \frac{2\pi}{\hbar} \frac{A_0^2}{2\hbar^2} \left| \left\langle \psi_- \left | \left( \pm i\frac{\partial H}{\partial p_x}  + \frac{\partial H}{\partial p_y}\right) \right |\psi_+\right\rangle \right|^2 \nonumber \\
&\times& \delta(E_-(0)-E_+(0)-\hbar\omega).
\end{eqnarray}
Here, we have taken into account the structure of the eigenstates $|\psi_+(0)\rangle = - |\psi_-(\pi)\rangle$ and $|\psi_-(0)\rangle=|\psi_+(\pi)\rangle$.
For the currents we must rather evaluate $\tilde{\Gamma}_{+}(K)-\tilde{\Gamma}_{-}(K')$. 
This leads to
\begin{eqnarray}
&&\frac{\tilde{\Gamma}_{+}(K,\omega)-\tilde{\Gamma}_{-}(K',\omega)}{2} = -\frac{2\pi}{\hbar}\frac{A_0^2}{(\hbar v_F)^2} \\ \nonumber
&\times& m^2\left(F_{p_y p_x}(0)-F_{-p_y p_x}(\pi)\right)\delta(E_-(0)-E_+(0)-\hbar\omega).
\end{eqnarray}
We recall that $E_+(0)$ refers to the lowest eigenenergy. 
This equation shows that the photo-induced currents at the Dirac points are related to $C$ and therefore to the quantum Hall conductivity through Eq. (\ref{F}). If we integrate on frequencies, this leads to
\begin{equation}
\label{photocurrents}
\left| \frac{\tilde{\Gamma}_+(K) - \tilde{\Gamma}_-(K')}{2} \right| =  \frac{2\pi}{\hbar} A_0^2 |C|.
\end{equation}
When measuring the variation of population in time, since $dN/dt^2<0$ then $|C|$ should occur in the response.

\section{Quantum Hall Response in Graphene}
\label{Halldrift}

For completeness, here we provide a simple understanding of the quantum Hall response in graphene \cite{Zhang,Novoselov} related to the drift velocity \cite{Nascimbene}, starting from the ${\cal O}$ operators in Eqs. (\ref{O}). We have the identity
\begin{equation}
{\cal O}^{\dagger}{\cal O} = \frac{1}{2}(\hat{r}^2 -\partial_r^2 +[\hat{r},\partial_r])=\frac{1}{2}(\hat{r}^2-\partial_r^2-1).
\end{equation}
Therefore, Eqs. (\ref{O}) lead to
\begin{eqnarray}
\frac{\hbar\omega_c^*}{2}\left(\hat{r}^2 -\partial_r^2 +[\hat{r},\partial_r]\right)\Phi_A(y)=\frac{E^2}{\hbar\omega_c^*}\Phi_A(y).
\end{eqnarray}
This is equivalent to define an effective Hamiltonian 
\begin{equation}
 \hat{H}_{eff}\Phi_A(r) = \left(\frac{E^2}{\hbar\omega_c^*}+\frac{\hbar\omega_c^*}{2}\right)\Phi_A(r) = \hbar\omega_c^*\left(N+\frac{1}{2}\right)\Phi_A(r)
 \end{equation}
 building then an analogy with the Schr\" odinger equation of a charged particle in a magnetic field, $\hat{H}_{eff}=\frac{\hbar\omega_c^*}{2}\left(\hat{r}^2 -\partial_r^2\right)$, and with quantum Hall physics in MOSFETS \cite{Hall}. 
 
 Now, we can add an electric field along $y$ direction corresponding to a potential term $-eV(y)= e E y$ in the Hamiltonian. Introducing the mass $m=\frac{\hbar}{\omega_c^* l_B^2}$, we can re-write $\hat{H}_{eff}$ as
\begin{equation}
\hat{H}_{eff} = \frac{p_y^2}{2m} +eEy + \frac{1}{2}m(\omega_c^*)^2\left(y-l_B^2 k\right)^2.
\end{equation}
We can absorb the effect of the electric field completing the square such that it modifies the oscillator coordinate 
\begin{equation}
k \rightarrow k - \frac{eE}{m\omega_c^{*2} l_B^2}.
\end{equation}
This is equivalent to a drift velocity in $x$ direction
\begin{equation}
\langle v_x\rangle = \frac{\hbar k}{m} = -\frac{E}{B}.
\end{equation}
In the absence of an electric field, the plane wave associated to the $x$ direction has the same chance to go in one direction or the other implying that $k$ may be positive or negative.
This velocity can be justified from physical arguments starting with the classical situation. If we include both a Coulomb and Lorentz force for a charge $q$, $q({\bf E}+{\bf v}\times{\bf B})$ then this is equivalent to modify the electric field along $y$ direction such that $E_y \rightarrow E_y - v_x B$. 

In the quantum theory, we can adjust $v_x$ into $\langle v_x\rangle$.
We can now precisely verify the prefactor in the quantum Hall conductivity from the current density $j_x=n e |\langle v_x\rangle |$ transverse to the applied electric field. We have a general relation between the number of particles $N_e$ and the density $n$ through $N_e = n S$ with $S=L_xL_y$ being the area of the two-dimensional plane. We assume $(2N+1)$ filled energy Landau levels at a Dirac point. The number of accessible states in each Landau level is
\begin{equation}
{\cal N} = L_x \int_0^{|k|_{max}} \frac{d|k|}{2\pi}. 
\end{equation}
The wave-vector $k$ has the dimension of $y/l_B^2$ with $y_{max}=L_y$ such that $|k|_{max}=L_y/l_B^2$. We discuss here the effect of the drift velocity where $k<0$. We can re-write ${\cal N}$ 
\begin{equation}
{\cal N} = \frac{S B e}{2\pi \hbar} = \frac{\Phi}{\Phi_0},
\end{equation}
where $\Phi=SB$ corresponds to the magnetic flux in the area and $\Phi_0=\frac{h}{e}$ is the flux quantum. The total number of states is equal to the number of flux quanta in the sample. Within the definitions, then we have 1
state per Landau level if the flux is equal to the flux quantum. We have
\begin{equation}
N_e = 2(2N+1){\cal N}
\end{equation}
particles in the system. For each Landau level, we can have ${\cal N}$ electrons. We have $2(2N+1)$ Landau levels if we take into account the fact that the $K$ and $K'$ points double the number of Landau levels. 
Therefore, 
\begin{equation}
j_{\perp} = j_x = \frac{2(2N+1){\cal N}}{S} \frac{E}{B}e = \frac{2(2N+1) e^2}{h} E,
\end{equation}
which leads to a quantized quantum Hall conductivity
\begin{equation}
\sigma_{xy} = \pm 2(2N+1)\frac{e^2}{h}.
\end{equation}
The $\pm $ signs refer to particle and hole plateaus. 

\section{Time-Reversal Symmetry}
\label{timereversal}

From the time-dependent quantum equation $i\hbar\frac{\partial}{\partial t}\psi = H\psi$.
if we modify $t\rightarrow -t$, this also requires to modify $i\rightarrow -i$ to preserve the validity of this equation. This has deep consequences, such as the momentum is also reversed ${\bf p}=-i\hbar\bm{\nabla}\rightarrow -{\bf p}$. Therefore, we can represent
the effect of time reversal through an operator, for instance, in the reciprocal space
\begin{equation}
\Theta |\psi({\bf k})\rangle = |\psi(-{\bf k})\rangle^*.
\end{equation}
The symbol $^*$ means that we change $i\rightarrow -i$ in the phase factors (and all the factors) associated to the wave-function. We can equivalently absorb the effect of $\Theta$ in a re-definition of the Hamiltonian $\Theta^{-1} H \Theta$
such that when calculating averaged values of observables on $\psi$ the effect is identical. For the Haldane model, changing $i\rightarrow -i$ in the phase factor of the $t_2$ term then the Hamiltonian is not time-reversal invariant.

In the Kane-Mele situation, the wave-function is a spinor, and the effect of time-reversal must be slightly modified. As shown in Fig. \ref{KaneMeleSpectrum}, if we change time $t\rightarrow -t$,
then we also have $K\rightarrow K'$ such that for a fixed energy the $\uparrow$ particles flip to $\downarrow$ and vice-versa. This can be accounted for similarly as a rotation acting in the Hilbert space of spin degrees of freedom.
The mass term at the Dirac points takes the form
\begin{equation}
H_{t_2}^{KM}=-h_z({\bf k}) \sigma_z\otimes(\left |\uparrow\rangle \langle \uparrow | - |\downarrow\rangle \langle \downarrow | \right).
\end{equation}
To define the time-reversal operator, this requires that 
\begin{equation}
U^{-1} H_{t_2}^{KM} U = + h_z(-{\bf k}) \sigma_z\otimes(\left |\uparrow\rangle \langle \uparrow | - |\downarrow\rangle \langle \downarrow | \right),
\end{equation}
with $h_z(-{\bf k})=-h_z({\bf k})$ such that the (total) Hamiltonian is invariant under time-reversal symmetry. We can then define $U$ as a rotation perpendicular to the $z$ axis 
\begin{equation}
U=i(\mathbb{I}\otimes s_y)\Theta
\end{equation}
 with the identity matrix acting in the sublattice space. 
In this way, we have $-is_y|\uparrow\rangle=|\downarrow\rangle$, $-is_y|\downarrow\rangle=-|\uparrow\rangle$ and similarly $\langle \uparrow | (i s_y)= \langle \downarrow |$ and $\langle \downarrow | (i s_y)= \langle \uparrow |(-1)$. Under time-reversal, $s_z\rightarrow -s_z$. Also, we have the interesting property that $U^2=-1$ for a spin-$\frac{1}{2}$ particle and also for topological insulators of the $\hbox{AII}$ category. In comparison, for topological spinless models such as for one-dimensional topological superconducting wires, time-reversal symmetry may be only defined through $\Theta$ such that $\Theta^2=\mathbb{I}$.

Another way to see the time-reversal symmetry is to use the form from Eq. (\ref{classification})
\begin{equation}
H({\bf k})=d_1({\bf k})\Gamma_1 + d_{12}({\bf k})\Gamma_{12} + d_{15}({\bf k})\Gamma_{15},
\end{equation}
such that
\begin{equation}
H(-{\bf k}) = d_1({\bf k})\Gamma_1 + d_{12}(-{\bf k})\Gamma_{12}-d_{15}(-{\bf k})\Gamma_{15}.
\end{equation}
This is equivalent to modify $K\rightarrow K'$. The time-reversal symmetry of the Hamiltonian can be formulated as follows. Here, formally we should write down $d_{12}(K')=-d_{12}(K)$, but as mentioned previously this simply means here that we should modify $\tilde{\varphi}\rightarrow -\tilde{\varphi}$ between the two Dirac points. We remind here that close to the two Dirac points $d_1({\bf k})=v_F|{\bf k}|\cos\tilde{\varphi}$, $d_{12}=v_F|{\bf k}|\sin\tilde{\varphi}$ where ${\bf k}$ measures a small deviation from the Dirac points. Now, we also have $d_{15}({\bf k})=-m\zeta$ such that $-d_{15}(-{\bf k})=d_{15}({\bf k})$. The $-$ sign in $-d_{15}(-{\bf k})\Gamma_{15}$, at the origin of the time-reversal symmetry of the Hamiltonian, is equivalent to modify $s_z\rightarrow -s_z$ and therefore corresponds indeed to modify $|\uparrow\rangle\rightarrow|\downarrow\rangle$ when time $t\rightarrow -t$. The time-reversal symmetry has other important applications such as the two-fold degeneracy of the energy spectrum related to the Kramers degeneracy.

\section{Geometry in the Cube}
\label{GeometryCube}

Here, we elaborate further on the geometry in a cube related to a Berry curvature of the form
\begin{equation}
F_z = (-1)^z F_{k_x k_y} \theta(z),
\end{equation}
and to the divergence theorem in the continuum limit assuming a dense limit of planes or an infinite number of planes. The goal is to show that the divergence theorem in this case may reveal a halved topological quantum number when $z\in [0:+\infty[$.
The presence of the $\theta(z)$ Heaviside function means that we have a face boundary at $z=0$ where $F_z$ jumps to zero in a step form in the vicinity to the vacuum. Setting the limit $z\rightarrow +\infty$ is similar to have $F_{k_x k_y}=0$ at the top surface in Eq. (\ref{number}) in the sense that there is no `flux' coming out through this region.
We verify below that the system behaves then as a topological system with a fractional topological number $\frac{1}{2}$ arising from the step $\theta(z)$ function at $z=0$. 
Here, simply we can write down
\begin{equation}
\label{curvaturez}
\frac{\partial F_z}{\partial z} = -i \pi e^{-i \pi z} F_{k_x k_y}\theta(z) + e^{-i \pi z} F_{k_x k_y} \delta(z).
\end{equation}
Integrating $\frac{\partial F_z}{\partial z}$ with respect to $z$ as in Eq. (\ref{number}) with
$z\in[0;+\infty[$, the second term at $z=0$ gives $\frac{1}{2}F_{k_x k_y}$. This precisely corresponds to the surface term at $z=0$. We can then verify that the real part of the first term gives $0$. Here, we can use the
identities $\int_0^L \sin(\pi z)dz = \frac{1}{\pi}(1-\cos(\pi L)) = \frac{2}{\pi} \sin^2\frac{\pi L}{2}$ and $\sin\frac{\pi L}{2}= \frac{\pi}{2}\int_0^{L} \cos\left(\frac{\pi x}{2}\right)dx$ which converges to $0$ when $L\rightarrow +\infty$ from the definition of the Dirac $\delta$ function
in the sense of Fourier transforms and distributions.

The second term in Eq. (\ref{curvaturez}) then leads to
\begin{equation}
\label{surface}
\frac{1}{2\pi}\iint dk_x dk_y F_{k_x k_y} \int_0^{+\infty} \cos(\pi z) \delta(z) dz = \frac{1}{2}.
\end{equation}
The continuum limit in $z$ direction then produces the same $\frac{1}{2}$ number as Eq. (\ref{infiniteseries}). 

This can also be verified from the evaluation of $F_z(k_x, k_y , z_{top}) - F_z(k_x,k_y,z_{bottom})$. Here, $F_z(k_x,k_y,z_{bottom})-F_z(k_x, k_y , z_{top})=\left(\cos(\pi L)-\frac{1}{2}\right)F_{k_x k_y}$ with $\hbox{lim}_{L\rightarrow +\infty}\cos(\pi L) = \hbox{lim}_{L\rightarrow +\infty}(\cos(\pi L) -1 + 1) = \hbox{lim}_{L\rightarrow +\infty} -2\sin^2\frac{\pi L}{2} +1 =+1$ again in the sense of Fourier transforms and distributions. This leads to the same conclusion as above that $F_z(k_x, k_y , z_{top}) - F_z(k_x,k_y,z_{bottom})=\frac{1}{2}$. It is important to mention that from Eq. (\ref{number}), the primitive of $\frac{\partial F_z}{\partial z}$ with respect to $z$ can be re-defined as $F_z(k_x, k_y , z) + a$ with $a$ being a number. On the other hand, from Eq. (\ref{surface}) we identify that the $\frac{1}{2}$ comes precisely from the boundary square at $z=0$. 
The surface becomes topologically equivalent to one Dirac point through the $\theta(z)$ function.

\section{Interactions in a wire}
\label{interactions}

Here, in a Luttinger formalism, we verify the stability of the topological phase in the Kitaev model. For this purpose, we begin simply with the Jordan-Wigner transformation between fermions and spins $c_i = S^-_i e^{i\pi\sum_{p<i} \pi n_i}$, $c_i^{\dagger} = S^+_i e^{-i\pi\sum_{p<i} \pi n_i}$ and $2 c^{\dagger}_i c_i -1=S_{iz}$ where $n_i=c^{\dagger}_i c_i$ represents the number of particles at site $i$. In the continuum limit $c_i\rightarrow c(x)$ and we have the dimensional correspondence $\sum_i c^{\dagger}_i c_i = \int \frac{dx}{a} c^{\dagger}(x) c(x)$. In the long-wavelength limit, spin operators located at sites $i$ and $j\neq i$ commute such that $[S_i^+,S_j^-]=0$. This algebra is similar to bosons leading to 
\begin{equation}
c(x) = b(x) e^{\pm i\pi\int^x n(x')dx'},
\end{equation}
with the bosonic superfluid operator $b(x)=\sqrt{\rho(x)}e^{i\theta(x)}$ and the $\pm$ signs come from the fact that $e^{i\pi}=e^{-i\pi}$. Taking the continuum limit, subtleties occur. In particular, since the system is infinite and we assume spinless fermions such that $(c^{\dagger}(x))^2=0$ then the number of particles in
the wire is also infinite. Since the density of bosons satisfy $b^{\dagger}(x) b(x) = c^{\dagger}(x)c(x) \propto \frac{1}{a}$ from the dimensional analysis, then this implies that 
\begin{equation}
\hbox{lim}_{L\rightarrow +\infty}\int_0^{L} \langle c^{\dagger}(x) c(x)\rangle dx \approx \frac{L}{a}\rightarrow +\infty. 
\end{equation}
This requires a proper regularization for the low-energy theory. Close to the two Fermi points located at $+k_F$ and $-k_F$ in the one-dimensional band structure, then a common approach is to substract the infinite number of particles such that the ground state corresponds to the `vacuum' 
such that all infinite observables are regularized through $\hat{{\cal O}} = \hat{\cal O} -  \langle GS| \hat{\cal O} |GS\rangle$. Since the physical observables will correspond to smooth deformations or fluctuations of the density of the particles around the two Fermi points, then this implies that the number of particles located quite away from the Fermi points will not modify substantially the low-energy theory. This has the important consequence that we can fix the mean density of particles modulo a global phase such that $b^{\dagger}(x) b(x) \sim \frac{1}{a}$.

Following the usual normalization in the literature for the density of particles, then we reach 
\begin{equation}
c(x) = \frac{1}{\sqrt{2\pi a}} e^{i\theta(x)} e^{\pm i \pi\int^x n(x')dx'}.
\end{equation}
Now, we can introduce right- and left-moving particles around these two Dirac points (with respectively a positive and negative momentum)
\begin{equation}
c(x) = c_R(x) e^{+i k_F x} + c_L(x) e^{-i k_F x},
\end{equation}
allowing to fix appropriately the sign $\pm$ for each mover and decompose the density as 
\begin{equation}
n(x) = c^{\dagger}(x) c(x) = b^{\dagger}(x) b(x) = \rho_0 +\frac{\partial_x \phi}{\pi}.
\end{equation}
In this way, we verify the standard form for the Haldane Luttinger theory \cite{HaldaneLuttinger}
\begin{equation}
c_p = \frac{1}{\sqrt{2\pi a}}e^{i(\theta(x) + p\phi(x))}
\end{equation}
with $p=\pm 1$ for right and left-movers. The motion of particles is similar to that of a vibrating quantum string and similar to the harmonic oscillator we have the commutation relations $[\frac{\phi(x)}{\pi},\theta(y) ] = iH(x-y)$ with the Heaviside step function being denoted here $H$.
This is equivalent to $[\partial_x\phi(x),\theta(y)] = i\pi\delta(x-y)$. To ensure that $\{ c_L, c_R\} = 0$ then this requires to introduce Klein factors $U_p$ with $U_L U_R=-i$. We remind here that to evaluate $c_L c_R$, this formally means $c_L(x-a) c_R(x)$ and requires
the introduction of the Baker-Campbell-Hausdorff formula.

In this way, a Dirac Hamiltonian in one dimension (coming from the fact that we linearize the energy spectrum around the two Fermi points) takes the form
\begin{eqnarray}
H_0 &=& -iv_F \int dx (c^{\dagger}_R(x)\partial_x c_R(x) - c^{\dagger}_L(x)\partial_x c_L(x)) \\ \nonumber
&=& \frac{v_F}{2\pi}\int dx ( \partial_x\phi(x))^2 +(\partial_x\theta(x))^2
\end{eqnarray}
with the Fermi velocity $v_F= 2 ta\sin(k_F a)\sim 2ta$. 

Interactions can be easily introduced within this formalism such that 
\begin{equation}
H_{Int} = \int dx V n(x) n(x+a) \sim \int dx \frac{V}{(2\pi)^2}(\partial_x\phi(x))^2,
\end{equation}
leading to the Luttinger theory
\begin{equation}
H=H_0 +H_{Int} = \frac{v}{2\pi}\int dx \frac{1}{K}(\partial_x\phi(x))^2 + K(\partial_x\theta(x))^2.
\end{equation}
We have the identifications
\begin{eqnarray}
vK &=& v_F  \\ \nonumber
\frac{v}{K} &=& v_F +\frac{V}{2\pi}.
\end{eqnarray}
The first equality reveals usually the Galil\' ee invariance. For repulsive interactions, $V>0$ implying $K<1$ and for attractive interactions, $V<0$ implying $K>1$. For free electrons, we have $K=1$. Including a superconducting pairing term and assuming $\sin(k_F a)\sim 1$ close to half-filling in
the Fourier transform leads to
\begin{equation}
\label{pairingfunction}
\Delta c^{\dagger}_L(x) c^{\dagger}_R(x) +h.c. = \frac{\Delta}{2\pi a}\cos(2\theta(x)),
\end{equation}
and here we have applied precisely $U_L U_R =-i$. To visualize the effect of this term on ground-state properties we can write down
\begin{eqnarray}
\langle \cos(2\theta(x))\rangle &=& \hbox{Tr}(e^{-\beta H}\cos(2\theta(x)) \\ \nonumber
&\approx& \Delta \int_{\frac{a}{v}}^{\beta} d\tau \langle \cos(2\theta(x,\tau))\cos(2\theta(x,0))\rangle_{H_0}
\end{eqnarray}
in imaginary time formalism with $\beta=1/(k_B T)$. The key property here is that the Green's function of the superfluid phase $\theta$ can be calculated from the Gaussian model $H_0$ and for free electrons $K=1$ this results in
\begin{equation}
\langle \cos(2\theta(x,\tau))\cos(2\theta(x,0))\rangle_{H_0} = e^{-[\theta(\tau)-\theta(0)]^2} = \frac{a^2}{v^2}\frac{1}{\tau^2}.
\end{equation}
In this way, we verify the correspondence
\begin{equation}
\langle \cos(2\theta(x))\rangle \approx \Delta \frac{a}{v},
\end{equation}
traducing the formation of the superconducting gap in the band structure and such that $\Delta \frac{a}{v}$ represents the fraction of the particles participating in the superfluid or Bardeen-Cooper-Schrieffer (BCS) ground state. 

The stability of the BCS topological phase regarding interactions can be understood related to the phenomenon of charge fractionalization in one dimension \cite{Safi,Pham,Steinberg,fractionalcharges}. In the presence of interactions in one dimension, an electron gives rise to fractional charges such that
$N_R + N_L =N$ and $v(N_R - N_L)=v_F J$ where $N$ corresponds to the number of injected electrons and $J$ measures the difference $(N_R-N_L)$ between the number of electrons going to the right and to the left. In this way, when we inject an electron at $+k_F$ this corresponds to $N=+1$ and $J=+1$ resulting in $N_R=\frac{1+K}{2}$ and $N_L = \frac{1-K}{2}$. If we inject an electron at $-k_F$, this corresponds to $N=+1$ and $J=-1$ such that $N_R=\frac{1-K}{2}$ and $N_L=\frac{1+K}{2}$. Therefore, when we inject a Cooper pair from the superconducting reservoir, we inject both an electron at $+k_F$ and $-k_F$ such that in average we have a total charge $N_R^{tot}=+1$ moving to the right and a total charge $N_L^{tot}=-1$ moving to the left, the two charges remaining entangled through the BCS mechanism. In this way, moderate (repulsive) interactions in the wire will not alter the formation of Cooper pairs and therefore the superfluid state. From renormalization group arguments, we justify below that the pairing term flows to values of the order
of the kinetic term or $t$ justifying why the Majorana fermions structure remains similar.

The effect of the interactions can be understood from the change of variables $\theta=\frac{1}{\sqrt{K}}\tilde{\theta}$ and $\phi={\sqrt{K}}\tilde{\phi}$ such that the Hamiltonian for the variables $\tilde{\phi}$ and $\tilde{\theta}$ is identical to $H_0$ and such that
the pairing term becomes $\cos(2\theta)=\cos(\frac{2\tilde{\theta}}{\sqrt{K}})$. The dressing of the superfluid phase with interactions implies that the superflow acquires a different velocity. Using standard renormalization group techniques, developing the partition function to second-order in $\Delta$ we obtain the equation
\begin{equation}
\label{flow}
\frac{d\Delta}{dl} = \left(2-\frac{1}{K}\right)\Delta,
\end{equation}
with $l = \ln\left(\frac{L}{a}\right)$. For $K=1$, solving this equation leads to a typical length scale defined as
\begin{equation}
\label{flow2}
\frac{a\Delta}{v} = \frac{a}{L},
\end{equation}
with $\Delta(a)=\Delta$ at which the pairing term flows to strong couplings and becomes renormalized to a value close to the hopping term $\Delta(L)\sim \frac{v}{a}$.
 In the presence of interactions, $\Delta(l)$ yet flows to strong couplings as long as $K>\frac{1}{2}$ to the same typical value $\frac{v}{a}$ at a typical length scale $L\sim a \left(\frac{v}{a\Delta}\right)^{\frac{1}{2-\frac{1}{K}}}$. 
 This equation is obtained from the same equality as for $K=1$, $\frac{\Delta}{\Delta(L)} =\frac{a}{L}$, with $\Delta(L)=\frac{v}{a}=\Delta(\frac{L}{a})^{2-\frac{1}{K}}$.
 
  For long wavelengths corresponding to lengths larger than $L$, the `effective' theory is analogous to the topological superconducting wire with $t=\Delta$, which is another way to interpret the stability of the superconducting phase.
 This can be viewed as an application of symmetry-protected topological phase as the interaction preserves the $\mathbb{Z}_2$ symmetry, $c\rightarrow -c$.


\bibliographystyle{elsarticle-num}
\biboptions{sort&compress} 
\journal{Physics Reports}


\end{document}